%% file: dml.tex
\documentclass[12pt,a4paper]{article}
\usepackage[utf8]{inputenc}
\usepackage[margin=1in]{geometry}

\usepackage[T1]{fontenc}
\usepackage[utf8]{inputenc}

\usepackage{graphicx}
\usepackage{amsmath}
\usepackage{amsthm}
\usepackage{multirow}
\usepackage{amssymb}
\usepackage{fancyvrb}
\usepackage{bbm}
\usepackage{moreverb}
\usepackage{pdfpages}
\usepackage[normalem]{ulem}
\usepackage{array}
\usepackage{float}
\usepackage{lipsum}
\usepackage{setspace}
\usepackage{indentfirst}
\usepackage{tikz}
\usepackage{mathrsfs}
\usepackage{hyperref}
\usepackage{apacite}
\usepackage{rotating}
\usepackage{pdflscape}
\usepackage{cleveref}
\usepackage[font=small]{caption}
\usepackage{fancyhdr}
\usepackage{enumerate}
\usepackage{natbib}

\AtBeginDocument{

}

\usetikzlibrary{arrows,positioning,matrix,arrows.meta,decorations.pathreplacing}

\newcommand\redsout{\bgroup\markoverwith{\textcolor{red}{\rule[0.5ex]{4pt}{1.5pt}}}\ULon}

\newtheorem{assumption}{Assumption}
\newtheorem{theorem}{Theorem}
\newtheorem{remark}{Remark}
\newtheorem{step}{Step}
\newtheorem{condition}{Condition}
\newtheorem{lemma}{Lemma}

\makeatletter
\let\ps@plain\ps@fancy
\makeatother
\title{\Large Efficient estimation of weighted cumulative treatment effects by double/debiased machine learning}

\author{Shenbo Xu$^1$ \and Bang Zheng$^2$ \and Bowen Su$^3$ \and Stan N. Finkelstein$^4$ \\
\and Roy E. Welsch$^5$ \and Kenney Ng$^6$ \and Ioanna Tzoulaki$^7$ \and Zach Shahn$^8$}

\date{\small %
    $^1$ Institute for Data, Systems, and Society, Massachusetts Institute of Technology. \\ Email: \href{mailto:xushenbo@mit.edu}{xushenbo@mit.edu} \\%
    $^2$ Ageing Epidemiology Research Unit, Imperial College London. \\
    Email: \href{mailto:b.zheng17@imperial.ac.uk}{b.zheng17@imperial.ac.uk} \\%
    $^3$ School of Public Health, Imperial College London. \\ Email: \href{mailto:b.su@imperial.ac.uk}{b.su@imperial.ac.uk} \\%
    $^4$ Institute for Data, Systems, and Society, Massachusetts Institute of Technology. \\ Email: \href{mailto:snf@mit.edu}{snf@mit.edu} \\%
    $^5$ Sloan School of Management, Massachusetts Institute of Technology. \\ Email: \href{mailto:rwelsch@mit.edu}{rwelsch@mit.edu} \\%
    $^6$ Center for Computational Health, IBM Research. \\ Email: \href{mailto:Kenney.Ng@us.ibm.com}{Kenney.Ng@us.ibm.com} \\%
    $^7$ School of Public Health, Imperial College London. \\ Email: \href{mailto:i.tzoulaki@imperial.ac.uk}{i.tzoulaki@imperial.ac.uk} \\%
    $^8$ School of Public Health, The City University of New York. \\ Email: \href{mailto:zachary.shahn@sph.cuny.edu}{zachary.shahn@sph.cuny.edu} \\ [2ex] %
    \today
}

\begin{document}

% \begin{center}
% {\large Efficient estimation of weighted cumulative treatment effects by double/debiased machine learning} % Self-Controlled Cohort Study for Large Scale Hypothesis-Generating Nonparametric Data-Driven High-Throughput Drug Screening
% \end{center}

\maketitle

% estimator for counterfactual restricted mean time lost (RMTL) curve and
% of cumulative treatment effects
% allows for more meaningful target estimands with the most overlap

\textbf{Abstract}: In empirical studies with time-to-event outcomes, investigators often leverage observational data to conduct causal inference on the effect of exposure when randomized controlled trial data is unavailable. Model misspecification and lack of overlap are common issues in observational studies, and they often lead to inconsistent and inefficient estimators of the average treatment effect. Estimators targeting overlap weighted effects have been proposed to address the challenge of poor overlap, and methods enabling flexible machine learning for nuisance models address model misspecification. However, the approaches that allow machine learning for nuisance models have not been extended to the setting of weighted average treatment effects for time-to-event outcomes when there is poor overlap. In this work, we propose a class of one-step cross-fitted double/debiased machine learning estimators for the weighted cumulative causal effect as a function of restriction time. We prove that the proposed estimators are consistent, asymptotically linear, and reach semiparametric efficiency bounds under regularity conditions. Our simulations show that the proposed estimators using nonparametric machine learning nuisance models perform as well as established methods that require correctly-specified parametric nuisance models, illustrating that our estimators mitigate the need for oracle parametric nuisance models. We apply the proposed methods to real-world observational data from a UK primary care database to compare the effects of anti-diabetic drugs on cancer clinical outcomes. \\


\section{Introduction}

Two well documented challenges in estimating average treatment effects (ATEs) from observational data--even when the standard causal assumptions of consistency, positivity, and exchangeability hold--are poor overlap between treatment groups and so-called nuisance model misspecification. Poor overlap occurs when one (binary) treatment assignment is much likelier than the other at certain values of covariates that need to be adjusted for in the analysis. If a treatment value cannot occur at all at certain covariate values, then the positivity assumption is violated and effect estimates may be biased. But even if positivity is not violated and overlap is merely poor, ATE estimators may have undesirable statistical properties. Methods have been proposed to address the issue of poor overlap by ``moving the goalposts'' \citep{crump2006moving} towards the population with equipoise and targeting a weighted average treatment effect (WATE) in which subjects with propensity scores near zero or one are down-weighted. This idea is similar to importance sampling in the reinforcement learning literature \citep{hanna2019importance} which was introduced to the statistics community by \citet{kloek1978bayesian}. Some authors have even argued that these alternative estimands are more policy relevant (in addition to having more stable estimators) than the ATE. We will consider WATE estimators in this work.

For time to event outcomes with censoring, multiple measures have been defined to quantify the effect of exposure, such as number-needed-to-treat, survival functions, median survival times, and restricted mean survival time (RMST) \citep{kloecker2020uses}. In this work, we focus on the RMST which is the expected event-free survival time through the follow-up time of interest $\tau$, the `restriction time’. This quantity has the appealing properties that it both summarizes survival outcomes in a single number and retains a causally and clinically meaningful interpretation. Past work has proposed estimators for WATEs on survival outcomes, including the RMST. \citet{cheng2022addressing} proposed inverse probability censoring weighted (IPCW) estimators for weighted survival functions and \citet{zeng2021propensity} studied jack-knifed survival functions, RMSTs, and mean survival times for WATE. However, these methods all depend on correctly specified parametric nuisance models.

In other settings, approaches have been developed to solve the issue of potential misspecification of parametric nuisance regression models for the outcome, treatment, and censoring process required to compute causal effect estimators. Nonparametric machine learning models are inherently correctly specified. However, simply plugging machine learning estimated nuisance parameters into standard effect estimators can lead to plug-in bias and incorrect confidence intervals due to slow convergence rates. Based on semiparametric theory, estimators based on efficient influence functions, such as debiased machine learning (DML) and targeted learning estimators, have been proposed to ensure $n^{1/2}$ consistency and asymptotic normality if the outcome model and the treatment model are both consistent (e.g. if they are both estimated with nonparametric machine learning), and the \textit{product} of their convergence rates is at least $n^{1/2}$ \citep{chernozhukov2018double,van2006targeted}. This property is known as rate double-robustness and the nonparametric efficiency bound can be attained when all nuisances are consistently estimated. However, double/debiased machine learning for WATEs in a survival setting has not yet been developed. That is the primary gap we fill here.

We also extend our estimators to the competing risks setting. When there are multiple causes of failure but only a subset is of interest, the other causes of failure are referred to as competing events because they can prevent the event of interest from occurring. In such scenarios, researchers often consider marginal, cause-specific, subdistribution, and composite cumulative incidence functions (CIFs) \citep{young2020causal}. Here, we consider cause-specific restricted mean time lost (RMTLs) as a cumulative measure that summarizes CIFs which can roughly be interpreted as expected time lost from the cause or causes of interest until restriction time $\tau$ \citep{andersen2013decomposition}. We are not aware of any estimator for the weighted cause-specific RMTL or the weighted cumulative treatment effect that is defined as the contrast between weighted treatment-specific RMTLs.

The organization of the paper is as follows. In Section 2, we introduce notation and formally define our weighted RMST and RMTL estimands of interest. In section 3, we derive a class of cross-fitted DML estimators for these estimands from their influence functions. In Section 4, we derive the asymptotic linearity of the proposed estimators and show the proposed estimators achieve semi-parametric efficiency bounds. We also provide methods for constructing pointwise confidence intervals and simultaneous confidence bands across a range of restriction times. In Section 5, we present a simulation study illustrating that our DML estimators with nuisance functions estimated via machine learning perform as well as estimators using correctly specified parametric nuisance models. In Section 6, we apply the proposed estimators to compare the effects of two first-line anti-hyperglycemia treatments on cancer clinical outcomes among patients with type II diabetes using real-world observational primary care data from the UK. In Section 7, we conclude.

%% \Bka\  papers do not contain a `contents of the paper' paragraph.

\section{Data and Estimands}

\subsection{Notations}

We observe pretreatment confounders $X \subseteq \mathcal{R}^d$, binary exposure $A \in\{0,1\}$. Let $T^{a}_{j} > 0$ denote the potential failure time of cause $j$ had, possibly contrary to fact, treatment been set to $a$ and $T^{a}_{j} = \infty$ if failure from cause $j$ would not occur under treatment assignment $a$. Define $T^{a}=\operatorname{min}_{j}\{T^{a}_{j}\}$ as the uncensored potential survival time from 0 to the first event $j$ and $J^{a}=\operatorname{argmin}_{j}\{T^{a}_{j}\}$ as the corresponding potential cause of failure. We define $T=T^A$ and $J=J^A$ as the failure time and the cause of failure had $A=a$, respectively. We consider censoring as an external coarsening process at time $C$, the censored failure time $\widetilde{T}=\operatorname{min}\{T, C\}$, an event indicator as $\Delta=I(T \leq C)$ of whether failure is censored, and $\widetilde{J} = J\Delta$ as the observed cause of failure where $\widetilde{J} \in \{0,1,\ldots,j^*\}$. The structure of observed data unit is $O=(X, A, \widetilde{T}, \widetilde{J})\sim P_0$, where $P_0$ is the unknown true data-generating distribution. When there is only one type of failure, i.e. $j^*=1$, the observed data become $O=(X, A, \widetilde{T}, \Delta)$. We assume that we observe $n$ independent and identically distributed (i.i.d.) observations $O_1, \ldots, O_n$ sampled from $P_0$ with probability $1/n$ and denote $P_n$ as the empirical probability distribution. We denote the observed cause-specific counting process as $N_j(t)=I(\widetilde{T} \leq t, \widetilde{J}=j)$, all-cause counting process as $N(t)=I(\widetilde{T} \leq t)$, censoring process as $N^C(t)=I(\widetilde{T} \leq t, \Delta=0)=(1-\Delta) N(t)$, and at-risk process as $R(t)=I(\widetilde{T} \geq t)$. % and denote censoring time as $C$. We define $T=T^{A}$ and let $J=J^{A}$ be the failure time and cause of failure had $A=a$, respectively. Then, we denote $\widetilde{T}=\min \{T, C\}$ as the observed survival time, $\Delta=I(T \leq C)$ as the event indicator, and $\widetilde{J}=J \Delta$ as the observed cause of failure. where $j=1, \ldots, j^*$ and $j^*$ is the total number of causes

\subsection{Estimands and identification}

In the absence of competing risks, we take our causal estimand to be the weighted average difference in restricted mean survival time (RMST) under alternative treatments, i.e.
$$
\begin{aligned}
\psi_{0}^{\mathrm{RMST}, h}(\tau)=&\psi_{0}^{\mathrm{RMST}, h}(\tau, a=1)-\psi_{0}^{\mathrm{RMST}, h}(\tau, a=0) \\
=&\frac{E_{0}[h_{0}(X)\{\min(T^{a=1},\tau)-\min(T^{a=0},\tau)\}]}{E_{0}\{h_{0}(X)\}}
\end{aligned}
$$

\noindent where $h_{0}(X)$ is a prespecified `tilting function'. $h_0(X)$ is typically chosen to downweight subjects with propensity scores near $0$ or $1$~\citep{li2018balancing}. If there are multiple competing causes of failure ($j* > 1$), our estimand is the weighted average cause-specific restricted mean time lost (RMTL)
$$
\begin{aligned}
\psi^{\mathrm{RMTL}, h}_{j, 0}(\tau)=&\psi_{j, 0}^{\mathrm{RMTL}, h}(\tau, a=1)-\psi_{j, 0}^{\mathrm{RMTL}, h}(\tau, a=0) \\
=&\frac{E_{0}[h_{0}(X)\{(\tau-\min (T^{a=1}, \tau)) I(J^{a=1}=j)-(\tau-\min(T^{a=0}, \tau)) I(J^{a=0}=j)\}]}{E_{0}\{h_{0}(X)\}}
\end{aligned}
$$

The following assumptions are required to identify causal effects, for $a \in\{0,1\}$:

\begin{assumption}\label{assumption:positivity}
(Positivity) $P_{0}(A=a \mid X) \geq1/\epsilon$ and $P_{0}(C \geq t \mid A=a, X)\geq1/\epsilon$ almost surely for $t \in[0, \tau]$ and $\epsilon \in(1, \infty)$;
\end{assumption}

\begin{assumption}\label{assumption:SUTVA_survival}
(SUTVA) $T=T^{A}$, $T_{i}^{(A_{1}, \ldots, A_{n})}=T_{i}^{(A_{1}^{\prime}, \ldots, A_{n}^{\prime})}$, if $A_{i}=A_{i}^{\prime}$, $\forall i$;
\end{assumption}

\begin{assumption}\label{assumption:ignorability_survival}
(Conditional ignorability) $T^{a} \perp\!\!\!\perp A \mid X$;
\end{assumption}

\begin{assumption}\label{assumption:depcens_survival}
(Conditionally independent censoring) $T^{a} \perp\!\!\!\perp C \mid A, X$.
\end{assumption}

\begin{assumption}\label{assumption:SUTVA_competing}
(Competing risks SUTVA) $(T^{a}, J^{a})=A (T^{a=1}, J^{a=1})+(1-A) (T^{a=0}, J^{a=0})$, $(T^{(A_{1}, \ldots, A_{n})}, J^{(A_{1}, \ldots, A_{n})})=(T^{(A_{1}^{\prime}, \ldots, A_{n}^{\prime})}, J^{(A_{1}^{\prime}, \ldots, A_{n}^{\prime})})$, if $A_{i}=A_{i}^{\prime}$, $\forall i$;
\end{assumption}

\begin{assumption}\label{assumption:ignorability_competing}
(Competing risks conditional ignorability) $(T^{a}, J^{a}) \perp\!\!\!\perp A \mid X$;
\end{assumption}

\begin{assumption}\label{assumption:depcens_competing}
(Competing risks conditionally independent censoring) $(T^{a}, J^{a}) \perp\!\!\!\perp C \mid A, X$.
\end{assumption}

% \begin{assumption}\label{assumption:depcens}
% $C \perp\!\!\!\perp A \mid X$;
% \end{assumption}

Assumptions~\ref{assumption:positivity}--\ref{assumption:depcens_survival} are required for identification of (weighted) RMST estimands in the absence of competing risks, and assumptions~\ref{assumption:positivity} and~\ref{assumption:SUTVA_competing}--\ref{assumption:depcens_competing} are required for identification of (weighted) RMTL estimands in the presence of competing risks. Assumption \ref{assumption:positivity} requires that every unit has positive probability of receiving each treatment and remaining uncensored through follow up. Assumption~\ref{assumption:SUTVA_survival} and \ref{assumption:SUTVA_competing} are called stable unit treatment value (SUTVA) assumptions. They state that observed outcomes are equal to the potential outcomes corresponding to observed treatments (consistency) and that one unit's treatment does not impact another's outcomes (no interference). The conditional ignorability and conditionally independent censoring Assumptions~\ref{assumption:ignorability_survival}--\ref{assumption:depcens_survival} and \ref{assumption:ignorability_competing}--\ref{assumption:depcens_competing} state that observed baseline covariates are sufficient to adjust for confounding.

Under Assumptions~\ref{assumption:positivity}-\ref{assumption:depcens_survival}, we have
$$
\psi_{0}^{\mathrm{RMST}, h}(\tau, a)=\frac{E_{0}[h_{0}(X)\{\mathrm{RMST}_{0}(\tau \mid a, X)\}]}{E_{0}\{h_{0}(X)\}}
$$

\noindent and under Assumptions~\ref{assumption:positivity} and~\ref{assumption:SUTVA_competing}-\ref{assumption:depcens_competing}
$$
\psi_{j, 0}^{\mathrm{RMTL}, h}(\tau, a)=\frac{E_{0}[h_{0}(X)\{\mathrm{RMTL}_{j, 0}(\tau \mid a, X)\}]}{E_{0}\{h_{0}(X)\}}
$$

\noindent where $\mathrm{RMST}_{0}(\tau \mid a, X)=E_{0} \{\min(T,\tau) \mid A=a, X\}$ and $\mathrm{RMTL}_{j, 0}(\tau \mid a, X)=E_{0} [\{\tau-\min(T, \tau)\}I(J=j) \mid A=a, X]$.

\section{Proposed methods}

\subsection{Efficient influence function}

In this section, we first derive the EIFs for our estimands~\eqref{eq:rmst_ueif_tilt} and~\eqref{eq:rmtlj_ueif_tilt}. Then, we construct cross-fitted double debiased estimators based on these EIFs, followed by a simple precedure to incorporate shape constraints. We provide sandwich estimators for variances and ensure that variance monotonically increases with respect to restriction time $\tau$. % we first derive the untilted centered EIFs from semiparametric theory on missing data, detailed in the Supplementary Material.; for the tilted uncentered EIFs. Then, we average the tilted uncentered EIFs to obtain weighted treatment-specific RMSTs and cause-specific RMTLs, 

We introduce several quantities before we proceed. Define the propensity score as $\pi(a \mid X)=P(A=a \mid X)$, the balancing weight as $w^{h}(a, X)=h(X)I(A=a)/\pi(a \mid X)$, $S(t \mid a, X)=P(T>t \mid A=a, X)$, $F_{j}(t \mid a, X)=P(T \leq t, J=j \mid A=a, X)$, $G(t \mid a, X)=P(C>t \mid A=a, X)$. The all-cause martingale, cause-specific martingale, and censoring martingale can be defined as $\mathrm{d} M(t \mid a, X)=\mathrm{d} N(t)-R(t) \mathrm{d}\Lambda(t \mid a, X)$, $\mathrm{d} M_{j}(t \mid a, X)=\mathrm{d} N_j(t)-R(t) \mathrm{d}\Lambda_{j}(t \mid a, X)$, and $\mathrm{d} M^C(t \mid a, X)=\mathrm{d} N^C(t)-R(t) \mathrm{d}\Lambda^C(t \mid a, X)$, respectively. Also denote $\eta_{0}^{\Lambda}=\{\pi_{0}(a \mid X), \Lambda_{0}^C(t \mid a, X), \Lambda_{0}(t \mid a, X)\}$, $\eta_{0}^{\Lambda, h}=\{h_{0}(X), \eta_{0}^{\Lambda}\}$, $\eta_{j, 0}^{\Lambda}=\{\pi_{0}(a \mid X), \Lambda_{0}^C(t \mid a, X), \Lambda_{0}(t \mid a, X), \Lambda_{j, 0}(t \mid a, X)\}$, and $\eta_{j, 0}^{\Lambda, h}=\{h_{0}(X), \eta_{j, 0}^{\Lambda}\}$.

In Theorem 1, we provide EIFs for $\psi_{0}^{\mathrm{RMST}, h}(\tau, a)$ and $\psi_{j, 0}^{\mathrm{RMTL}, h}(\tau, a)$.

\begin{theorem} \label{theorem:eif}
Assume $S_{0}(t \mid a, X)>1/\epsilon$, $0 \leq h_{\mathrm{min}}\leq h_{0}(X)\leq h_{\mathrm{max}} \leq \infty$, where $\epsilon \in (1, \infty)$ and let $\Delta(\tau)=I(C > \tau \wedge T-)$, then $\psi_{0}^{\mathrm{RMST}, h}(\tau, a)$ is a pathwise differentiable quantity with tilted uncentered EIF as $\phi_{0}^{\mathrm{RMST}, h}(\tau, a; \eta_{0}^{\Lambda, h})=[E_{0}\{h_{0}(X)\}]^{-1} \{h_{0}(X)\phi^{\mathrm{RMST}}_{0}(\tau, a; \eta_{0}^{\Lambda})\}$ and tilted centered EIF
\begin{equation} \label{eq:rmst_ueif_tilt}
\varphi_{0}^{\mathrm{RMST}, h}(\tau, a; \eta_{0}^{\Lambda, h})=[E_{0}\{h_{0}(X)\}]^{-1} [h_{0}(X)\{\phi_{0}^{\mathrm{RMST}}(\tau, a; \eta_{0}^{\Lambda})-\psi_{0}^{\mathrm{RMST}, h}(\tau, a)\}]
\end{equation}

\noindent where the untilted nonparametric uncentered EIF is
\begin{equation} \label{eq:rmst_ueif}
\begin{aligned}
\phi_{0}^{\mathrm{RMST}}(\tau, a; \eta_{0}^{\Lambda})&=\frac{I(A=a)\Delta(\tau)\min(\widetilde{T}, \tau)}{\pi_{0}(a \mid X)G_{0}(\tau \wedge \widetilde{T}- \mid a, X)}+\left\{1-\frac{I(A=a)}{\pi_{0}(a \mid X)}\right\} \mathrm{RMST}_{0}(\tau \mid a, X) \\
&+\frac{I(A=a)}{\pi_{0}(a \mid X)} \int_{0}^{\tau \wedge \widetilde{T}} \frac{t \mathrm{d} M_{0}^{C}(t \mid a, X)}{G_{0}(t- \mid a, X)} \\
&+\frac{I(A=a)}{\pi_{0}(a \mid X)}\int_{0}^{\tau \wedge \widetilde{T}} \frac{\{\mathrm{RMST}_{0}(\tau \mid a, X)-\mathrm{RMST}_{0}(t \mid a, X)\} \mathrm{d} M_{0}^{C}(t \mid a, X)}{S_{0}(t \mid a, X)G_{0}(t- \mid a, X)}
\end{aligned}
\end{equation}

% and the untilted nonparametric uncentered EIF is
% $$
% \begin{aligned}
% \phi^{\mathrm{RMST}, np}(\tau, a)=&\mathrm{RMST}(\tau \mid a, X)-\frac{I(A=a)\mathrm{RMST}(\tau \mid a, X)}{\pi(a \mid X)}\int_0^{\tau \wedge \widetilde{T}} \frac{\mathrm{d} M(t \mid a, X)}{S(t \mid a, X)G(t- \mid a, X)} \\
% &+\frac{I(A=a)}{\pi(a \mid X)}\int_0^{\tau \wedge \widetilde{T}} \frac{\mathrm{RMST}(t \mid a, X)\mathrm{d} M(t \mid a, X)}{S(t \mid a, X)G(t- \mid a, X)}
% \end{aligned}
% $$

$\psi_{j, 0}^{\mathrm{RMTL}, h}(\tau, a)$ is a pathwise differentiable quantity with tilted uncentered EIF $\phi_{j, 0}^{\mathrm{RMTL}, h}(\tau, a; \eta_{j, 0}^{\Lambda, h})=[E_{0}\{h_{0}(X)\}]^{-1} [h_{0}(X)\phi_{j, 0}^{\mathrm{RMTL}}(\tau, a; \eta_{j, 0}^{\Lambda})]$ and tilted centered EIF
\begin{equation} \label{eq:rmtlj_ueif_tilt}
\varphi_{j, 0}^{\mathrm{RMTL}, h}(\tau, a; \eta_{j, 0}^{\Lambda, h})=[E_{0}\{h_{0}(X)\}]^{-1} [h_{0}(X)\{\phi_{j, 0}^{\mathrm{RMTL}}(\tau, a; \eta_{j, 0}^{\Lambda})-\psi_{j, 0}^{\mathrm{RMTL}, h}(\tau, a)\}]
\end{equation}

\noindent where the untilted nonparametric uncentered EIF is
\begin{equation} \label{eq:rmtlj_ueif}
\begin{aligned}
\phi_{j, 0}^{\mathrm{RMTL}}(\tau, a; \eta_{j, 0}^{\Lambda})=&\frac{I(A=a) (\tau-\min(\widetilde{T}, \tau))I(\widetilde{J}=j)}{\pi_{0}(a \mid X) G_{0}(\widetilde{T}- \mid a, X)}+\left\{1-\frac{I(A=a)}{\pi_{0}(a \mid X)}\right\}\mathrm{RMTL}_{j, 0}(\tau \mid a, X) \\
&-\frac{I(A=a)}{\pi_{0}(a \mid X)} \int_{0}^{\tau \wedge \widetilde{T}} \frac{(\tau-t) F_{j, 0}(t \mid a, X) \mathrm{d} M_{0}^{C}(t \mid a, X)}{S_{0}(t \mid a, X)G_{0}(t- \mid a, X)} \\
&+\frac{I(A=a)}{\pi_{0}(a \mid X)} \int_{0}^{\tau \wedge \widetilde{T}} \frac{\{\mathrm{RMTL}_{j, 0}(\tau \mid a, X)-\mathrm{RMTL}_{j, 0}(t \mid a, X)\} \mathrm{d} M_{0}^{C}(t \mid a, X)}{S_{0}(t \mid a, X)G_{0}(t- \mid a, X)}
\end{aligned}
\end{equation}

% and the untilted nonparametric uncentered EIF is
% $$
% \begin{aligned}
% \phi_j^{\mathrm{RMTL}, np}(\tau, a)=&\mathrm{RMTL}_j(\tau \mid a, X)+\frac{I(A=a)}{\pi(a \mid X)} \int_0^{\tau \wedge \widetilde{T}} \frac{(\tau-t)\mathrm{d}M_j(t \mid a, X)}{G(t- \mid a, X)} \\
% &-\frac{I(A=a)}{\pi(a \mid X)} \int_0^{\tau \wedge \widetilde{T}}\frac{\{\mathrm{RMTL}_j(\tau \mid a, X)-\mathrm{RMTL}_j(t \mid a, X)\} \mathrm{d}M(t \mid a, X)}{S(t \mid a, X)G(t- \mid a, X)} \\
% &+\frac{I(A=a)}{\pi(a \mid X)} \int_0^{\tau \wedge \widetilde{T}}\frac{(\tau-t)F_j(t \mid a, X)\mathrm{d}M(t \mid a, X)}{S(t \mid a, X)G(t- \mid a, X)}
% \end{aligned}
% $$
\end{theorem}

The first terms in~\eqref{eq:rmst_ueif} and~\eqref{eq:rmtlj_ueif} are the inverse probability weighted complete case terms, representing individuals who take the treatment of interest $A=a$ with observed event of interest before restriction time, $\Delta(\tau)=1$ or $\widetilde{J}=j$. The empirical mean of the first terms are often called IPCW estimators \citep{robins1992recovery}. As the IPCW estimators only consider complete cases, there is a loss in efficiency of these estimators, which becomes severe when censoring is heavy and/or treatment assignment is highly unbalanced. To improve efficiency, the second terms in~\eqref{eq:rmst_ueif} and~\eqref{eq:rmtlj_ueif}, similar to the augmentation term in doubly-robust estimators for continuous outcomes~\citep{robins1994estimation}, capture information for all participants through conditional expectations of outcomes, regardless of exposure group or censoring status. The remaining Lebesgue-Stieltjes censoring martingale integral terms further restore information from censored observations $\Delta=0$ in the exposure group of interest $A=a$.
  
% \begin{proposition}
% \label{proposition1}
% $\phi^{\mathrm{RMST}, sp}(\tau, a)=\phi^{\mathrm{RMST}, np}(\tau, a)$ and $\phi_j^{\mathrm{RMTL}, sp}(\tau, a)=\phi_j^{\mathrm{RMTL}, np}(\tau, a)$
% \end{proposition}

% Because of Proposition~\ref{proposition1}, it is sufficient to consider either two within each equality and we choose $\phi^{\mathrm{RMST}}(\tau, a)=\phi^{\mathrm{RMST}, np}(\tau, a)$ and $\phi_j^{\mathrm{RMTL}}(\tau, a)=\phi_j^{\mathrm{RMTL}, np}(\tau, a)$ as their representations are more compact than their semiparametric counterpart.

\subsection{Debiased machine learning estimators}

We describe a short version of the estimation procedure with four steps: (i) partition the data with size $n$ into $K=2$ disjoint sets $\mathcal{V}_1$ and $\mathcal{V}_2$; (ii) construct $\widehat{\eta}_{1}^{\Lambda, h}$ or $\widehat{\eta}_{j, 1}^{\Lambda, h}$ trained on $\mathcal{V}_2$ and estimated on $\mathcal{V}_1$. Reverse the training and estimation set to obtain $\widehat{\eta}_{2}^{\Lambda, h}$ or $\widehat{\eta}_{j, 2}^{h}$; (iii) For $k=1,2$, compute $\widehat{\psi}_{k}^{\mathrm{RMST}, h}(\tau, a)=P_{n, k}\{\widehat{\phi}_{k}^{\mathrm{RMST}, h}(\tau, a; \widehat{\eta}_{k}^{\Lambda, h})\}$ or $\widehat{\psi}_{j, k}^{\mathrm{RMTL}, h}(\tau, a)=P_{n, k}\{\widehat{\phi}_{j, k}^{\mathrm{RMTL}, h}(\tau, a; \widehat{\eta}_{j, k}^{\Lambda, h})\}$ by plugging in $\widehat{\eta}_{k}^{\Lambda, h}$ or $\widehat{\eta}_{j, k}^{\Lambda, h}$, where $P_{n, k}$ is the empirical probability distribution of subset $\mathcal{V}_{k}$ and the full expression for $\widehat{\phi}_{k}^{\mathrm{RMST}, h}(\tau, a; \widehat{\eta}_{k}^{\Lambda, h})$ and $\widehat{\phi}_{j, k}^{\mathrm{RMTL}, h}(\tau, a; \widehat{\eta}_{j, k}^{\Lambda, h})$ are given as~\eqref{eq:survival_estimator} and~\eqref{eq:competing_estimator} in the Supplementary Material; (iv) compute $\widehat{\psi}^{\mathrm{RMST}, h}(\tau, a)=1/2 \{\widehat{\psi}_{k=1}^{\mathrm{RMST}, h}(\tau, a)+\widehat{\psi}_{k=2}^{\mathrm{RMST}, h}(\tau, a)\}$ or $\widehat{\psi}_{j}^{\mathrm{RMTL}, h}(\tau, a)=1/2 \{\widehat{\psi}^{\mathrm{RMTL}, h}_{j, k=1}(\tau, a)+\widehat{\psi}^{\mathrm{RMTL}, h}_{j, k=2}(\tau, a)\}$  and correct for shape constraints thereafter. The details of the implementation procedure of the proposed cross-fitted DML estimators are described in the Supplementary Material.

% and $\widehat{\psi}_{j}^{h}(\tau, a)=P_n \{\widehat{\phi}_{i, j}^{h}(\tau, a)\}$.

To reduce sampling variation in \eqref{eq:survival_estimator} and \eqref{eq:competing_estimator}, balancing weights used in their numerators are averaged to be their denominators to control for numerical instability \citep{hajek1971comment}. However, we do not further regulate numerical instability for the IPCW term using $P_{n, k}[\{\Delta(\tau)/\widehat{G}_k(\tau \wedge \widetilde{T}- \mid a, X)\}]^{-1}$ in~\eqref{eq:survival_estimator} or $P_{n, k}[\{\Delta/\widehat{G}_k(\widetilde{T}- \mid a, X)\}]^{-1}$ in~\eqref{eq:competing_estimator}. Note that the one-step estimators and the estimating equation estimators are equivalent as the $\varphi_{0}^{\mathrm{RMST}, h}(\tau, a; \eta_{0}^{\Lambda, h})$ and $\varphi_{j, 0}^{\mathrm{RMTL}, h}(\tau, a; \eta_{j, 0}^{\Lambda, h})$ are linear in $\psi_{0}^{\mathrm{RMST}, h}(\tau, a)$ and $\psi_{j, 0}^{\mathrm{RMST}, h}(\tau, a)$ \citep{kennedy2022semiparametric}.

\begin{remark} \label{remark:cpl}
Time-to-event nuisance functions can only be attained from conditional hazards $\mathrm{d}\widehat{\Lambda}_{k}^C(t \mid a, X)$, $\mathrm{d}\widehat{\Lambda}_{k}(t \mid a, X)$, $\mathrm{d}\widehat{\Lambda}_{j, k}(t \mid a, X)$, where $\widehat{S}_{k}(t \mid a, X)=\exp\{-\widehat{\Lambda}_{k}(t \mid a, X)\}$, $\mathrm{R}\widehat{\mathrm{MS}}\mathrm{T}_{k}(\tau \mid a, X)=\int_{0}^{\tau} \widehat{S}_{k}(t \mid a, X) \mathrm{d}t$, $\widehat{F}_{j, k}(t \mid a, X)=\int_{0}^{t} \widehat{S}_{k}(u \mid a, X) \mathrm{d}\widehat{\Lambda}_{j, k}(u \mid a, X)$, $\mathrm{R}\widehat{\mathrm{MT}}\mathrm{L}_{j, k}(\tau \mid a, X)=\int_{0}^{\tau} \widehat{F}_{j, k}(t \mid a, X) \mathrm{d}t$, and $\widehat{G}_{k}(t \mid a, X)=\exp\{-\widehat{\Lambda}_{k}^{C}(t \mid a, X)\}$. Direct modelling of counting processes generally leads to inconsistent estimates of conditional time-to-event distributions. Details are outlined in the Supplementary Material.
\end{remark}

\begin{remark} \label{remark:shape_constraints}
$\psi_{0}^{\mathrm{RMST}, h}(\tau, a)$ and $\psi_{j, 0}^{\mathrm{RMTL}, h}(\tau, a)$ are naturally concave and convex functions with respect to $\tau$ as their derivatives are monotonically non-increasing survival functions and non-decreasing cause-specific cumulative incidence functions. Aside from shape constraints on point estimates, asymptotic standard errors $\sigma_{0}^{\mathrm{RMST}, h}(\tau, a)$, $\sigma_{0}^{\mathrm{RMST}, h}(\tau)$, $\sigma_{j, 0}^{\mathrm{RMTL}, h}(\tau, a)$ and $\sigma_{j, 0}^{\mathrm{RMTL}, h}(\tau)$ are strictly non-decreasing as censoring is always monotonically coarsened with respect to follow-up time.
\end{remark}

However, the proposed estimators are not guaranteed to be shape-restricted in any finite sample since estimates are separately made at each time point without any global constraints. In this paper, we ensure convexity/concavity of point estimates using Grenander-type estimators \citep{westling2020unified} and monotonicity of asymptotic variance using projection-type estimators \citep{daouia2013projection}. Since shape-corrected estimators are guaranteed to be no further from the truth than the unrestricted estimators in all finite samples, and both estimators are asymptotically equivalent, we focus on large-sample properties of the uncorrected estimators.

\begin{remark} \label{remark:multiple_treatment}
Note that the proposed estimators can easily be extended to multiple treatments $A \in \{0, 1, 2, ...\}$. One can simply cross-fit multi-class classifiers for $\widehat{\pi}_{k}(a \mid X)$ and hazard models to obtain $\widehat{\Lambda}_{k}(t \mid a, X)$, $\widehat{\Lambda}_{k}^C(t \mid a, X)$, and perhaps $\widehat{\Lambda}_{j, k}(t \mid a, X)$ for every treatment $a$.
\end{remark}

\begin{remark} \label{remark:survival_times}
The proposed method allows for continuous, discrete, and mixed survival times. One can simply replace continuous time hazards and martingale integrals by their discrete or mixed survival time equivalents.
\end{remark}

% Here, we describe an procedure for the proposed cross-fitted DML estimators.

\section{Asymptotic properties}

\subsection{Consistency and asymptotic linearity}

We consider large-sample properties of the proposed estimators here. We begin with conditions for pointwise and uniform consistency.

% The following theorem shows that the proposed estimators are consistent.

\begin{theorem} \label{theorem:consistency}
(Consistency): Under regularity conditions given in the Supplementary Material, the proposed estimators $\widehat{\psi}^{\mathrm{RMST}, h}(\tau, a)$ and $\widehat{\psi}_{j}^{\mathrm{RMTL}, h}(\tau, a)$ converge in probability respectively to $\psi^{\mathrm{RMST}, h}(\tau, a)$ and $\psi_{j}^{\mathrm{RMTL}, h}(\tau, a)$. Under slightly stronger regularity conditions, also in the Supplementary Material, the convergence is uniform.
\end{theorem}

The following theorem shows that the proposed estimators are asymptotically linear both pointwise and uniformly, which enables construction of both pointwise and curvewise confidence intervals.

\begin{theorem} \label{theorem:asymptotic}
(Asymptotic linearity) Under regularity conditions given in the Supplementary Material, the proposed estimators $\widehat{\psi}^{\mathrm{RMST}, h}(\tau, a)$ and $\widehat{\psi}_{j}^{\mathrm{RMTL}, h}(\tau, a)$ are asymptotically linear for given $\tau$. Under slightly stronger regularity conditions, also in the Supplementary Material, the asymptotic linearity is uniform over $\tau$.

(Efficiency bounds) Furthermore, under $\operatorname{var}\{\phi_{0}^{\mathrm{RMST}}(\tau, 1; \eta_{0}^{\Lambda, h}) \mid X\}=\operatorname{var}\{\phi_{0}^{\mathrm{RMST}}(\tau, 0; \eta_{0}^{\Lambda, h}) \mid X\}$ and $\operatorname{var}\{\phi_{j, 0}^{\mathrm{RMTL}}(\tau, 1; \eta_{j, 0}^{\Lambda, h}) \mid X\}=\operatorname{var}\{\phi_{j, 0}^{\mathrm{RMTL}}(\tau, 0; \eta_{j, 0}^{\Lambda, h}) \mid X\}$, the variance is minimized across the class of balancing IPTW estimators when $h_{0}(X)\propto\operatorname{var}_{0}(a \mid X)$
% gives the smallest asymptotic variance among the class of balancing IPTW estimators for
\end{theorem}

Theorem~\ref{theorem:asymptotic} calls for the nuisance functions in~\eqref{eq:survival_estimator} and~\eqref{eq:competing_estimator} to converge at $n^{-1/2}$ rate. When not all nuisance functions are estimated consistently, the resulting variance estimators will be biased in unknown directions with unknown magnitude despite the point estimates being consistent owing to the double robustness property. In such scenarios, the bootstrap procedure may be able to reclaim proper coverage at the cost of significant computing time \citep{bai2013doubly}. %  and space requirement from training learning algorithms for nuisance functions in each bootstrap

\subsection{Asymptotic inference}

We can perform asymptotic inference based on Theorem 3. The pointwise confidence intervals (CIs) with significance level $\alpha$ are defined as
$$
P_0\left(\widehat{l}^{\mathrm{RMST}, h}(\tau, a) \leq \widehat{\psi}^{\mathrm{RMST}, h, lcm}(\tau, a) \leq \widehat{u}^{\mathrm{RMST}, h}(\tau, a) \right) = 1-\alpha
$$

Here, $\widehat{l}^{\mathrm{RMST}, h}(\tau, a)$ and $\widehat{u}^{\mathrm{RMST}, h}(\tau, a)$ are two stochastic processes, and $z_{q}$ is the $q$-quantile of the standard normal distribution. The Wald type symmetric CIs can be constructed as $\widehat{\psi}^{\mathrm{RMST}, h, lcm}(\tau, a) \pm z_{1-\alpha/2} \widehat{\sigma}^{\mathrm{RMST}, h, +}(\tau, a)/\sqrt{n}$. Unlike the confidence intervals which only describe the estimated curves at a single time point, simultaneous confidence bands encapsulate the entire estimated curves over a user-specified follow-up time, which can be used to evaluate equivalence or noninferiority claims in comparative effectiveness studies \citep{zhao2016restricted}.

Under Condition~\ref{condition4} and~\ref{condition11}, simultaneous confidence bands can be constructed at $(1-\alpha)$ coverage level on the interval $\left[\tau_l, \tau_u\right]$ such that
$$
P_0\left(\widehat{l}^{\mathrm{RMST}, h}(\tau, a) \leq \widehat{\psi}^{\mathrm{RMST}, h, lcm}(\tau, a) \leq \widehat{u}^{\mathrm{RMST}, h}(\tau, a) \quad \forall \tau \in\left[\tau_l, \tau_u\right]\right) \geq 1-\alpha
$$

We refer to \citet{chen2021optband} for the development of various types of simultaneous confidence bands for survival functions. Here, we consider an increasing-width confidence band $\widehat{\psi}^{\mathrm{RMST}, h, lcm}(\tau, a) \pm \widehat{c}_{\alpha} \widehat{\sigma}^{\mathrm{RMST}, h, +}(\tau, a)/\sqrt{n}$, where $\widehat{c}_{\alpha}$ is the $(1-\alpha)$-quantile of the supremum of the absolute value of sample paths of a zero mean Gaussian process over $[\tau_l, \tau_u]$ with cross-fitted covariance function
$$
\widehat{\sigma}^{\mathrm{RMST}, h}(u, t, a)=\frac{P_n\{\widehat{\varphi}^{\mathrm{RMST}, h}(u, a; \widehat{\eta}_{k}^{\Lambda, h})\} \widehat{\varphi}^{\mathrm{RMST}, h}(t, a; \widehat{\eta}_{k}^{\Lambda, h})\}}{\widehat{\sigma}^{\mathrm{RMST}, h, +}(u, a) \widehat{\sigma}^{\mathrm{RMST}, h, +}(t, a)}
$$

In practice, we choose $\tau_l \geq \operatorname{argmin}_{\tau} \{\widehat{\sigma}^{\mathrm{RMST}, h, +}(\tau, a) > 0 \}$ to avoid zero denominators in the covariance estimator and $\tau_u \leq \tau_{\mathrm{max}}$. We discuss how to address non-positive semidefinite covariance estimators in the Supplementary Material.

\section{Simulation study}

\subsection{Data generating processes}

\begin{figure}[H]
% The arguments in the next line are {height}{optional width}{used only by OUP for typesetting} for figure empty box eg 
%\figurebox{20pc}{25pc}{}
%if actual size of graphics need plese see below command
%\figurebox{}{}{}[fig1]
%need to reducing the figure size use below command
%\figuresize{.8}%
\includegraphics[scale=0.5]{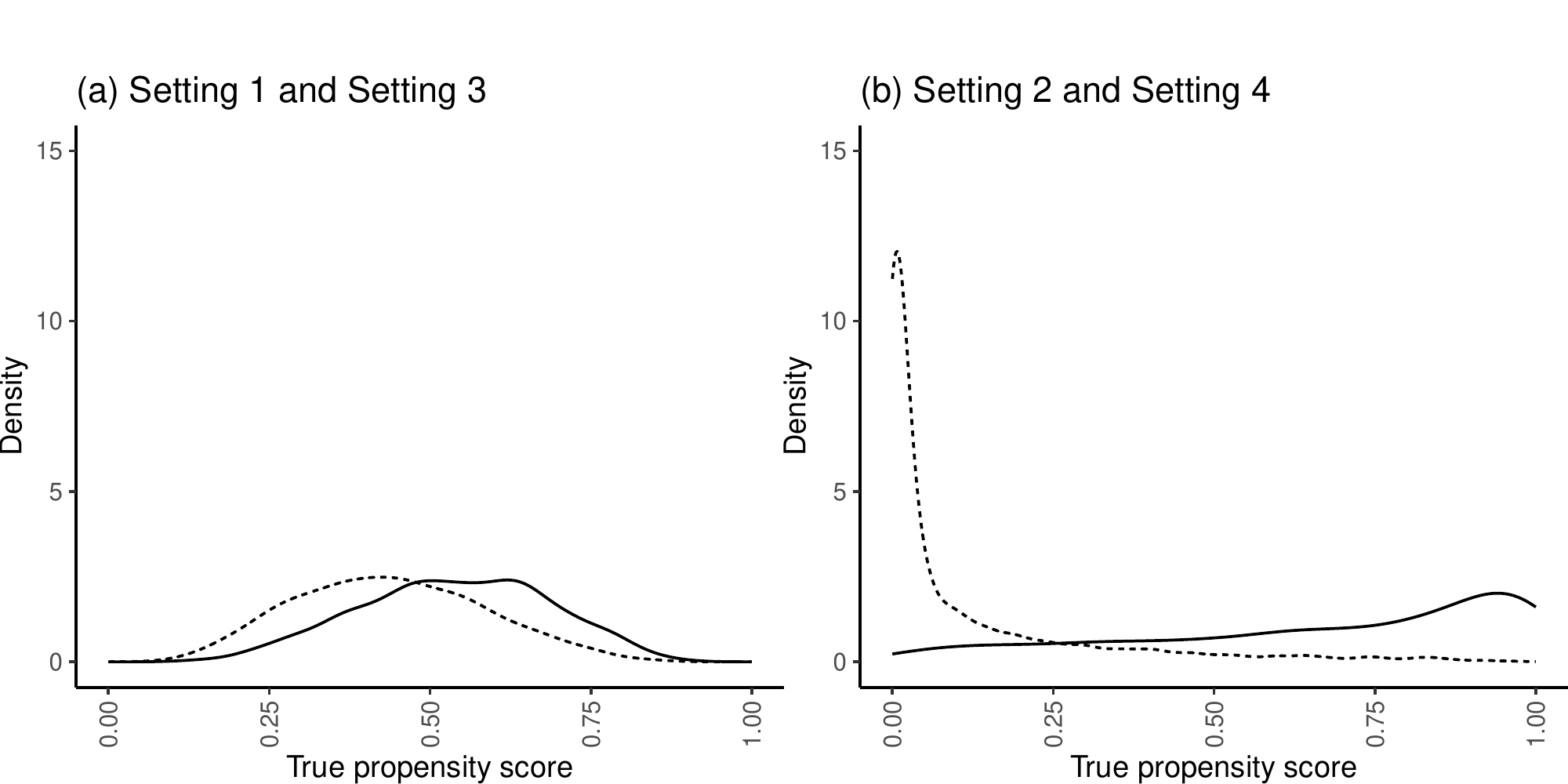}
\centering
% note that files may not be rotated
\caption{Density of true propensity score, note that the treatment models for the good overlap settings (\textit{Setting} 1 and \textit{Setting} 3) are identical, and the treatment models for the poor overlap settings (\textit{Setting} 2 and \textit{Setting} 4) are identical; dotted: treatment; solid: control.}
\label{fig:sim_ps}
\end{figure}

We studied four data generating processes to evaluate the finite-sample performance of comparator and proposed estimators on the ATE and WATE. The simulations resemble observational studies under four settings: setting 1 - good overlap and no competing risks; setting 2 - poor overlap and no competing risks; setting 3 - good overlap with competing risks; setting 4 - poor overlap with competing risks. We generate $1000$ replications for each setting. The details of the data generating processes and summary statistics of the simulated datasets can be found in the Supplementary Material. The distributions of the true propensity scores by exposure arm are shown in Fig.~\ref{fig:sim_ps}. We can observe that \textit{Setting} 1 and \textit{Setting} 3 share sufficient overlap while \textit{Setting} 2 and \textit{Setting} 4 have poor overlap.

% data-generating process exhibits long tails with deficient overlap.

\subsection{Estimands and estimators}

We consider estimands corresponding to four types of balancing weights summarized in Table~\ref{tab:balance_weight}. Specification of $h(x)$ defines the target parameter and weights. We consider outcome regression, IPCW, and doubly robust comparator estimators for the survival functions and cause-specific cumulative incidence functions. All nuisance models in the comparator estimators are correctly specified parametrically - a logistic regression model for the exposure and Cox proportional hazards models for all conditional time-to-event distributions. All standard errors of comparator estimators are obtained from nonparametric bootstraps. The details of these estimators are provided in the Supplementary Material. For the proposed estimators, we proceed as if the true data-generating mechanism is not known and cross-fit all of the nuisance models via adaptive machine learning. Specifically, we use the SuperLearner \citep{polley2017super} and survSuperLearner \citep{westling2020survsuper} R \citep{R:2010} packages to learn stacked ensembles of machine learning nuisance models. The specific learners in our ensembles are described in the Supplementary Material.

\begin{table}[H]
\centering
\caption{Estimands, tilting functions, and their origins. ATO: Average treatment effect among the overlap population~\citep{li2018balancing}; ATM: Average treatment effect among the matched population~\citep{li2013weighting}; ATEN: Average treatment effect of entropy weighted population~\citep{zhou2020propensity}}
\begin{tabular}{l l l} %\\
 \\ \hline
Estimands & Tilting function $h(x)$ & Origin \\ \hline
ATE & 1 & Survey \\
ATO & $\pi(0 \mid x)\pi(1 \mid x)$ & Gini index \\
ATM & $\min_{a=0, 1} \left\{\pi(a \mid x) \right\}$ & Misclassification error \\
ATEN & $-\sum_{a=0, 1}\pi(a \mid x)\log \pi(a \mid x)$ & Cross-entropy or deviance \\ \hline
\end{tabular}
\label{tab:balance_weight}
\end{table}

\subsection{Simulation results}

\begin{figure}[H]
\centering
% The arguments in the next line are {height}{optional width}{used only by OUP for typesetting} for figure empty box eg 
%\figurebox{20pc}{25pc}{}
%if actual size of graphics need plese see below command
%\figurebox{}{}{}[fig1]
%need to reducing the figure size use below command
%\figuresize{.8}%
\includegraphics[scale=0.6]{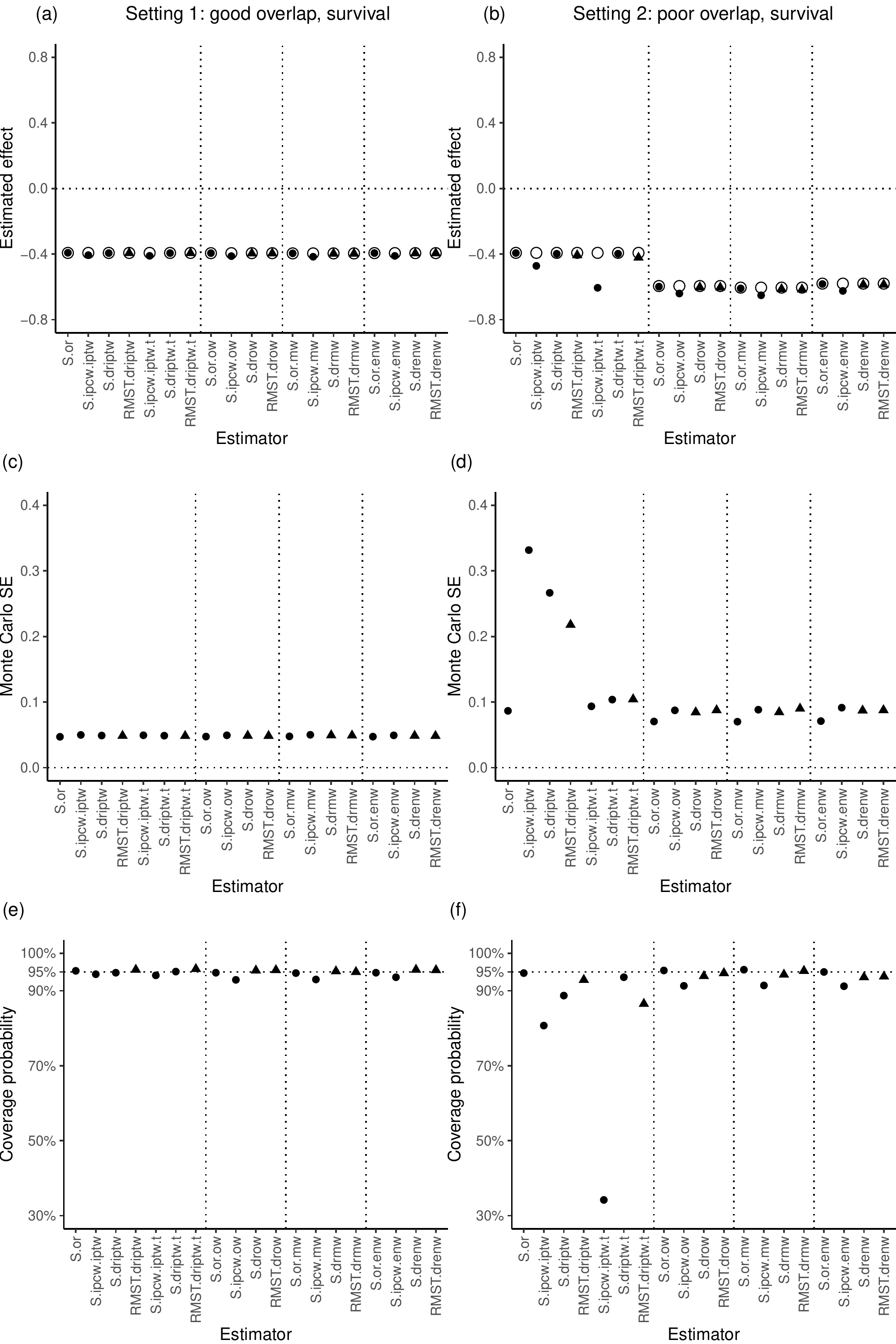}
% note that files may not be rotated
\caption{Estimated cumulative treatment effect, Monte Carlo standard errors, and coverage probabilities from different estimators grouped by estimand at $\tau=4$. Estimands in each plot are separated by vertical dotted lines. The order of the estimands from left to right in each plot is: ATE, ATO, ATM, ATEN. Hollow circles in (a) and (b) represent true effects; solid circles represent the comparator estimators; and solid triangles represent the proposed estimators. S.or: outcome regression estimator; S.ipcw.iptw: IPCW estiamtor; S.dr: doubly-robust estimator; ow: overlap weight; mw: matching weight; enw: entropy weights; RMST. prefix: the proposed estimators; .t suffix: 99th percentile Winsorized IPTW.}
\label{fig:sim_survival_main}
\end{figure}

Simulation results from the survival settings (\textit{Setting} 1 and \textit{Setting} 2) are shown here in Fig.~\ref{fig:sim_survival_main} and results from the competing risks setting (\textit{Setting} 3 and \textit{Setting} 4) can be found in Fig.~\ref{fig:sim_competing_main} in the Supplementary Material. The survival settings and the competing risks settings have similar interpretations and conclusions. % The average standard errors from \textit{Setting} 1-4 are also available in Fig.~\ref{fig:sim_survival_ase} and Fig.~\ref{fig:sim_competing_ase} from the Supplementary Material.

For each of the comparator estimators and the proposed estimators, we report their finite sample performance from data generating processes with sufficient and poor overlap at restriction time $\tau=4$. Figure~\ref{fig:sim_survival_main} (a) and (b) show the estimated weighted cumulative treatment effects and true effects grouped by estimand. The true ATE and WATE are similar when the overlap is adequate (Fig.~\ref{fig:sim_survival_main} (a)), while the difference between ATE and WATE is large when the overlap is poor (Fig.~\ref{fig:sim_survival_main} (b)). The bias (hollow vs solid) is generally quite low under adequate overlap (Fig.~\ref{fig:sim_survival_main} (a)) but is occasionally high under poor overlap (Fig.~\ref{fig:sim_survival_main} (b)). For each estimand, even with correctly specified nuisance models, the IPCW estimators have larger finite sample bias compared with other estimators, and this is exacerbated as overlap deteriorates. When overlap is limited, Winsorizing the IPTW at its 99\% percentile significantly increases the bias. In addition, it can be observed that estimators with (correctly specified) outcome models have less bias than those without. In all settings, the proposed estimators appear to be approximately unbiased. % , indicating that correctly-specified outcome models may serve as estimate stabilizers

Figure~\ref{fig:sim_survival_main} (c) and (d) show Monte Carlo standard errors grouped by estimand at restriction time $\tau=4$. The Monte Carlo standard errors agree nicely with estimated standard errors. When overlap is sufficient, the standard errors are similar across estimands and estimators. However, estimators for WATE are more efficient than estimators for ATE when overlap is limited, while Winsorization improves efficiency for ATE (at the cost of bias, as we saw in Fig.~\ref{fig:sim_survival_main} (b)). Within each estimand in the limited overlap setting, it is obvious that the outcome regression estimators are the most efficient, followed by the doubly robust estimators and the proposed estimators, while the IPCW estimators have larger finite sample variance. This demonstrates that a proper augmentation term can restore large amount of information which leads to efficiency gains.

% \begin{figure}[H]
% % The arguments in the next line are {height}{optional width}{used only by OUP for typesetting} for figure empty box eg 
% %\figurebox{20pc}{25pc}{}
% %if actual size of graphics need plese see below command
% %\figurebox{}{}{}[fig1]
% %need to reducing the figure size use below command
% %\figuresize{.8}%
% \figuresize{0.6}
% \figurebox{20pc}{25pc}{}[sim_survival_se.pdf]
% % note that files may not be rotated
% \caption{Standard errors of different estimators grouped by estimand at $\tau=4$. Solid circles represent standard errors of the comparator estimators; and solid triangles represent standard errors of the proposed estimators. Other symbols and notations are similar to Fig.~\ref{fig:sim_survival_effect}.}
% \label{fig:sim_survival_se}
% \end{figure}

% \begin{figure}[H]
% % The arguments in the next line are {height}{optional width}{used only by OUP for typesetting} for figure empty box eg 
% %\figurebox{20pc}{25pc}{}
% %if actual size of graphics need plese see below command
% %\figurebox{}{}{}[fig1]
% %need to reducing the figure size use below command
% %\figuresize{.8}%
% \figuresize{0.6}
% \figurebox{20pc}{25pc}{}[sim_survival_cp.pdf]
% % note that files may not be rotated
% \caption{Coverage probabilities of different estimators grouped by estimand at $\tau=4$. Solid circles represent coverage probabilities of the comparator estimators; and solid triangles represent coverage probabilities of the proposed estimators. Other symbols and notations are similar to Fig.~\ref{fig:sim_survival_effect}.}
% \label{fig:sim_survival_cp}
% \end{figure}

\noindent This is expected because the doubly robust estimators and the proposed estimators are the most efficient among the class of IPTW estimators, but they are slightly less efficient than the outcome regression estimators \citep{robins1994estimation,bai2013doubly}. Of course, this increased efficiency would be at the cost of bias if the outcome model is misspecified, a concern which is greatly mitigated by the data adaptive nuisance model estimation in our estimators.

% . The proposed estimator for the ATE is slightly more efficient than the doubly robust comparator. A potential reason is that parametric regression may fail to model the exposure consistently for all observations under poor overlap and unbalanced treatment. On the other hand, data-adaptive estimation may relieve inconsistent estimation of propensity scores in such scenarios.

Figure~\ref{fig:sim_survival_main} (e) and (f) show the empirical coverage of 95\% pointwise confidence intervals grouped by estimand at restriction time $\tau=4$. When overlap is sufficient, Fig.~\ref{fig:sim_survival_main} (e) reveals that the comparator and the proposed estimators have empirical coverage probabilities quite close to the nominal value of 0.95. However, when overlap is poor, the IPCW estimators are undercovered more than other estimators within each estimand, shown in Fig.~\ref{fig:sim_survival_main} (f). Winsorizing the IPTW lowers the coverage since it has larger bias and smaller estimated variance. When overlap is limited, the outcome regression estimators, the doubly robust estimators, and the proposed estimators all have decent coverage close to 0.95 for WATEs.

% The IPCW-IPTW estimator has lower pointwise coverage than other balancing weighted estimators as both the conditional censoring probabilities and the propensity scores are inversely weighted in the IPCW-IPTW estimator; The coverage probability also drops if we Winsorize IPTW in the doubly robust estimators and the proposed estimators, indicating Winsorizating extreme IPTW is not an appropriate procedure in such settings; As in Fig.~\ref{fig:sim_survival_main} (d), the doubly robust estimator for the ATE is slightly undercovered because its point estimates have larger variation than its bootstrapped standard errors, compared with the Monte Carlo standard errors and the estimated standard errors of the proposed estimator as in Fig.~\ref{fig:sim_survival_main} (d) and Fig.~\ref{fig:sim_survival_ase} (b).

To summarize, we have demonstrated through simulation that our proposed estimators perform competitively with comparator estimators, even when the comparator estimators are given access to correct parametric nuisance model specifications and our methods must estimate the nuisance models via machine learning. This suggests that our proposed estimator is a much safer choice than alternatives, as the decreased risk of bias from model misspecification under our approach does not come at the cost of decreased efficiency.

% Based on simulation studies and asymptotic properties, we advocate for the proposed double/debiased machine learning approach to estimate the average treatment effect among the overlap population (ATO) to obtain clinical and statistical optimality in observational studies, especially under limited overlap and unbalanced treatment.

% Details on the candidate learners are presented in Supplementary Material.

\section{Application}

In this section, we apply the proposed methods to data from the Clinical Research Practice Datalink (CPRD) to assess the effects of first-line anti-hyperglycemia monotherapy on: (a) cancer incidence among type II diabetic patients with no history of cancer; and (b) on all-cause deaths among type II diabetic patients with cancer at time of initiation. The CPRD is a continuing general practice primary care database containing more than 60 million patients among 674 practices in the UK \citep{herrett2015data}. The study population included all type II diabetic patients aged over 50 before April 30, 2018. They were followed up until the earliest of death, leaving practice/CPRD database, or the end of data inclusion on April 30, 2018. The Independent Scientific Advisory Committee (ISAC) authorizes access to CPRD data with protocol 20\_000207.

For simplicity and coherence, we follow \citet{tsilidis2014metformin} in our exposure and outcome definitions and covariate selection. We compare metformin initiators ($A=1$) versus sulphonylurea initiators ($A=0$). When the clinical outcome is all-cause mortality among cancer patients, then there is only one type of failure with no competing risks. However, when the outcome of interest is cancer incidence, mortality becomes a competing event. In both studies, we took baseline to be the time of initial first-line anti-hypoglycemia drug prescription. The failure times are subject to right censoring due to loss to follow-up or administrative end of follow up on April 30, 2018. The pretreatment confounders $X$ include gender, prescription age, prescription year, body mass index (BMI), Hemoglobin A1C (HbA1c), heart failure, coronary heart disease, atrial fibrillation, stroke, hypertension, peripheral vascular disease, chronic kidney disease, chronic obstructive pulmonary disease, smoking, index for multiple deprivation (IMD), and region. Compared with sulphonylurea, metformin starters are generally younger, heavier, and initiated later in calendar time. %  in this study. When the outcome of interest is cancer incidence, mortality becomes a natural competing event; Confounders, selected by epidemiologists and physicians based on subject matter knowledge, are expected to be common causes of treatment decision, event process, and censoring distribution; More details of cohort demographics are available in \citet{tsilidis2014metformin}.

We apply the proposed methods to estimate IPTW and overlap weighted RMST $\psi_{0}^{\mathrm{RMST}, h}(\tau, a)$ and cause-specific RMTL $\psi_{j, 0}^{\mathrm{RMTL}, h}(\tau, a)$. When Assumption~\ref{assumption:positivity}-\ref{assumption:depcens_survival} hold, the counterfactual RMST curves can be interpreted as the average survival time until restriction time $\tau$ had all participants taken treatment $a$. If Assumption~\ref{assumption:positivity} and~\ref{assumption:SUTVA_competing}-\ref{assumption:depcens_competing} hold, the counterfactual RMTL curves correspond to the average time lost from cause $j$ at restriction time $\tau$ had all patients had exposure $a$.

\begin{figure}[H]
\centering
% The arguments in the next line are {height}{optional width}{used only by OUP for typesetting} for figure empty box eg 
%\figurebox{20pc}{25pc}{}
%if actual size of graphics need plese see below command
%\figurebox{}{}{}[fig1]
%need to reducing the figure size use below command
%\figuresize{.8}%
\includegraphics[scale=0.5]{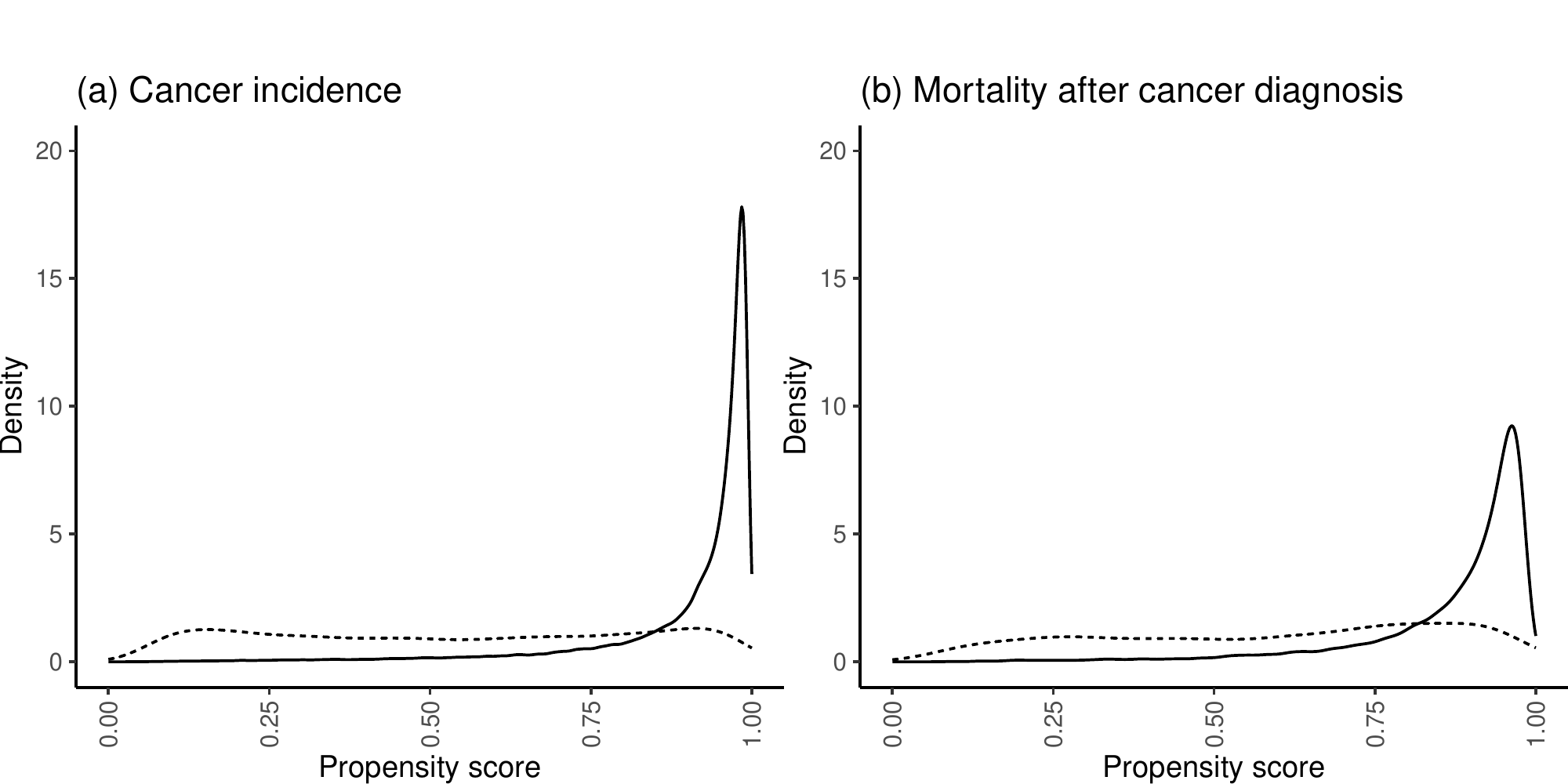}
% note that files may not be rotated
\caption{Density of estimated propensity score; dotted: sulphonylurea; solid: metformin; (a) cancer incidence: $n$ (sulphonylurea)$=15,595$ and $n$ (metformin)$=88,091$; (b) mortality after cancer diagnosis: $n$ (sulphonylurea)$=2,110$ and $n$ (metformin)$=9,763$.}
\label{fig:application_ps}
\end{figure}

As in our simulations, nuisance functions are again obtained by super learners \citep{van2007super}. We specify the ensemble nuisance learners in the Supplementary Material. The nonparametric densities of cross-fitted propensity scores by treatment arm for both cancer incidence and all-cause mortality after cancer diagnosis are shown in Fig.~\ref{fig:application_ps}. The covariate overlap between treatment groups is limited for both settings and the overlap for cancer incidence is poorer.

% The results from this real data application are not directly comparable with previous work mainly due to lack of utilization of similar approaches in the  community. In contrast, we take competing events, dependent censoring, and poor common support into consideration. However, we still compare our conclusions with observational studies using similar exposure definition on the same database CPRD.

% Though DIPTW estimators behaved analogously to other balancing weights for marginal/cause-specific/subdistribution CIFs/RMTLs of cancer when they are not significant, DIPTW estimators revealed more tilted estimates than other balancing weighted estimators on composite effect of cancer and death, death after cancer, and all cause mortality.

\begin{figure}[H]
\centering
% The arguments in the next line are {height}{optional width}{used only by OUP for typesetting} for figure empty box eg 
%\figurebox{20pc}{25pc}{}
%if actual size of graphics need plese see below command
%\figurebox{}{}{}[fig1]
%need to reducing the figure size use below command
%\figuresize{.8}%
\includegraphics[scale=0.6]{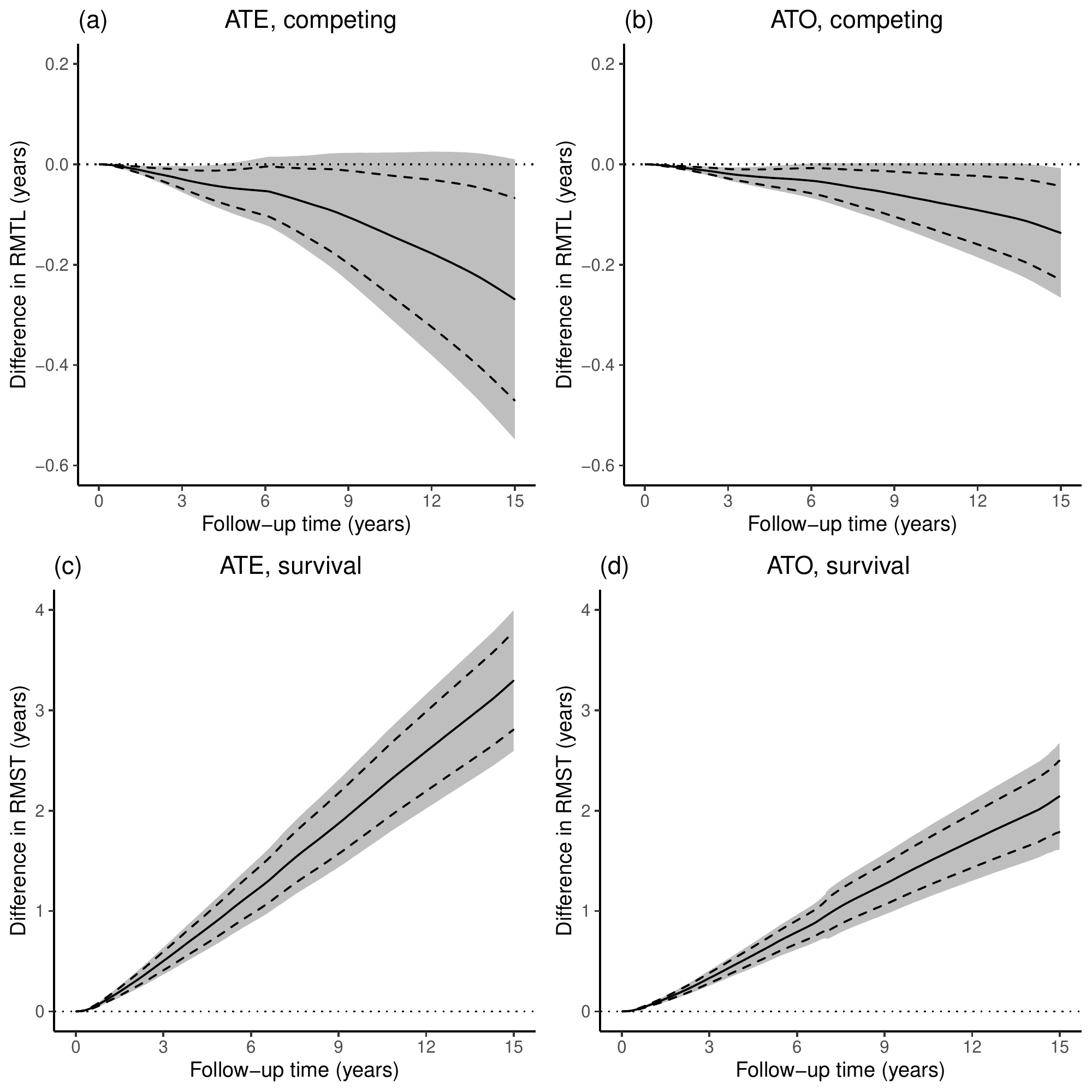}
% note that files may not be rotated
\caption{Curves of estimated cumulative treatment effect on the CPRD data; (a) cause-specific ATE of cancer incidence; (b) cause-specific ATO of cancer incidence; (c) ATE of mortality after cancer diagnosis; and (d) ATO of mortality after cancer diagnosis. In all figures, dotted line: $y=0$; solid lines: point estimates; dashed lines: 95\% pointwise confidence intervals; shaded area: 95\% uniform simultaneous confidence bands.}
\label{fig:app_effect}
\end{figure}

Figure~\ref{fig:app_effect} shows the cumulative treatment effects until restriction time $\tau=15$ years. The first row shows the estimated cause-specific DR-IPTW and DR-OW RMTLs of cancer incidence. Estimates of  DR-IPTW and DR-OW RMSTs of all-cause mortality after cancer diagnosis are shown in the second row. These effects are the difference between metformin initiators and sulphonylurea initiators, respectively. The simultaneous confidence bands are constructed from $10,000$ realizations of sample paths of Gaussian processes with $[\tau_l, \tau_u]=[1/365.25, 7779/365.25]$ years for cancer incidence and $[\tau_l, \tau_u]=[2/365.25, 4533/365.25]$ years, which correspond to the maximum follow-up interval that the estimated standard errors are strictly positive. Our estimates indicate that starting metformin is protective over sulphonylurea on all cause deaths among cancer patients while its advantage on cancer incidence is marginal, which aligns with the conclusions from \citet{morgan2014association,tsilidis2014metformin}. In each analysis, the point estimate of the WATE was closer to the null and had lower standard error than the point estimate of the ATE. We provide additional results in the Supplementary Material.

\section{Discussion}

Aside from estimators derived from EIFs of causal estimands, there exists a plug-in estimator for RMST derived from doubly robust estimator for the cumulative hazards \citep{zhang2012double}. However, this Nelson-Aalen type estimator cannot be modified to target a WATE as the product integral of balancing weighted cumulative hazards do not correspond to the survival functions/CIFs of interest.

Unlike the one-step estimator which targets the entire survival distribution \citep{cai2020one,rytgaard2021one}, the proposed estimators may still violate the constraints that the sum of RMST and RMTL from all causes is equal to $\tau$. One possible direction of research is to ensure that this constraint and shape constraints are satisfied without parametric modeling assumptions and even when hazard models are misspecified.

Other estimands in the competing risks setting include the survivor average causal effect (SACE) \citep{tchetgen2014identification}, separable effects \citep{stensrud2021generalized,stensrud2022separable,stensrud2022conditional}. Future work can develop DML estimators for overlap weighted versions of these estimands. We conjecture that the regularity conditions for separable effects will be equivalent to ours as both estimators involve analogous nuisance functions \citep{martinussen2021estimation}.

% Apart from survival functions, is another causal estimand based on a set of more restrictive assumptions \citep{Young2020ACF}. In the competing risks setting, new causal estimands, such as separable direct effects and separable indirect effects, may be considered to assess the cumulative treatment effects not mediated by the competing events and the effects only through competing events in future studies \citep{stensrud2022separable,stensrud2021generalized}. % and it should not be a major indicator for the effectiveness since the subset of the principal stratum of survivors can't be determined

Indeed, a drawback of balancing weights is that its consistency relies on the consistent estimation of the tilting function, which is often a bounded transformation of the propensity score. Though unsatisfying, this assumption is omnipresent for WATEs such that over-extrapolation of causal effects away from the covariate distribution can be mitigated.

The proposed estimators and their regularity conditions also apply to targeted learning, and owing to cross-fitting, there is no restriction on the complexity of nuisance parameters. However, both DML/TMLE estimators can fail to be regular and asymptotically linear (RAL) when one nuisance function is not consistently estimated. To draw doubly robust inference under inconsistent estimation of at most one nuisance parameter \citep{benkeser2017doubly}, drift-corrected estimators may be constructed by Gaussianizing the drift terms following \citet{diaz2019statistical}. Another potential direction is to achieve valid inference under inconsistent estimation of nuisance functions \citep{smucler2019unifying}.

In spite of alleviation of model misspecification by adaptive estimation, the proposed method is still subject to unmeasured confounding. Sensitivity analysis \citep{ciocuanea2022sensitivity}, instrumental variables \citep{ding2017instrumental}, and recently-developed proximal inference \citep{tchetgen2020introduction} may be extended to time-to-event outcomes with limited overlap in future studies. Besides binary exposure, continuous treatment \citep{kennedy2017non} and multiply robust time-varying treatment \citep{luedtke2017sequential} in our settings are attractive and meaningful topics of future research.

\section*{Acknowledgement}

This work was supported by IBM Research and National Institutes of Health. The authors thank Prof. Lee-Jen Wei at Harvard University and Prof. Ted Westling at the University of Massachusetts Amherst for their explanations and guidance. The authors are also grateful to the editors and the reviewers for their insightful suggestions.

% (awarded to R.E.W. and S.N.F.), (awarded to Mark W. Albers), The computing was performed on Big Data and Analytical Unit servers at Imperial College London, Satori Supercomputer at Massachusetts Institute of Technology, and IBM Cloud Pak for Data.

\section*{Supplementary material}\label{SM}

Supplementary material includes technical details of extended proofs, additional simulation information, and numerical results from application.

\bibliographystyle{apacite}
\bibliography{paper-ref}

\newpage
\
\newpage

\include{appendix}

\end{document}

%% file: appendix.tex
\appendix

\footnotesize

\begin{center}
{\Large \textbf{Supplementary material to ``Efficient estimation of weighted cumulative treatment effects by double/debiased machine learning''}}
\end{center}

\section{Proposed methods}

\subsection{Implementation procedure of the proposed DML estimators}

\begin{step} \label{step1}
Sample-splitting. We randomly split the data $O_1, \ldots, O_n$ into $K$ disjoint validation sets $\mathcal{V}_{1}, \ldots, \mathcal{V}_{K}$ with sizes $n_1, \ldots, n_K$, where $K \in\{2,3, \ldots,\lfloor n / 2\rfloor\}$. For each $k = 1, \ldots, K$, we define training set $\overline{\mathcal{V}}_{k}=\{O_i: i \notin \mathcal{V}_{k}\}$ and treatment-specific training set $\overline{\mathcal{V}}_{a, k}=\{O_i: i \in \overline{\mathcal{V}}_{k}, A_i=a\}$.
\end{step}

\begin{step} \label{step2}
Adaptive estimation. For every $k = 1, \ldots, K$, (1) train the propensity model on $\overline{\mathcal{V}}_{k}$ and predict $\widehat{\pi}_{k}(a \mid X)$ on $\mathcal{V}_{k}$; (2) train the time-to-event models on $\overline{\mathcal{V}}_{a=0, k}$ and $\overline{\mathcal{V}}_{a=1, k}$ separately, and evaluate $\widehat{\Lambda}_{k}(t \mid a, X)$, $\widehat{\Lambda}_{k}^C(t \mid a, X)$, and perhaps $\widehat{\Lambda}_{j, k}(t \mid a, X)$ for $a=0,1$ using $\mathcal{V}_{k}$ at unique observed survival times $\widetilde{T}$ in $\mathcal{V}_{k}$ to capture jump information from observed counting processes $\mathrm{d} N(t)$, $\mathrm{d} N^C(t)$, and perhaps $\mathrm{d} N_j(t)$. Counting process martingales can be obtained by directly plugging in corresponding cumulative hazards.
\end{step}

\begin{step} \label{step3}
Asymptotically efficient estimators. Construct $K$ estimators $\widehat{\phi}^{\mathrm{RMST}, h}_{k}(\tau, a; \widehat{\eta}_{k}^{\Lambda, h})$ or $\widehat{\phi}^{\mathrm{RMTL}, h}_{j, k}(\tau, a; \widehat{\eta}_{j, k}^{\Lambda, h})$ by plugging estimated nuisance functions from Step~\ref{step2} into
\begin{equation} \label{eq:survival_estimator}
\begin{aligned}
\widehat{\phi}_{k}^{\mathrm{RMST}, h}(\tau, a; \widehat{\eta}_{k}^{\Lambda, h})&=[P_{n, k} \{\widehat{h}_k(X)\}]^{-1} \widehat{h}_k(X) \mathrm{R}\widehat{\mathrm{MS}}\mathrm{T}_k(\tau \mid a, X)+[P_{n, k} \{\widehat{w}_{k}^{h}(a, X)\}]^{-1} \widehat{w}_{k}^{h}(a, X) \\
&\times \left\{\frac{\Delta(\tau)\min(\widetilde{T}, \tau)}{\widehat{G}_k(\tau \wedge \widetilde{T}- \mid a, X)}-\mathrm{R}\widehat{\mathrm{MS}}\mathrm{T}_{k}(\tau \mid a, X)+\int_{0}^{\tau \wedge \widetilde{T}} \frac{t \mathrm{d} \widehat{M}_{k}^{C}(t \mid a, X)}{\widehat{G}_{k}(t- \mid a, X)} \right. \\
&\left.\int_{0}^{\tau \wedge \widetilde{T}} \frac{\{\mathrm{R}\widehat{\mathrm{MS}}\mathrm{T}_{k}(\tau \mid a, X)-\mathrm{R}\widehat{\mathrm{MS}}\mathrm{T}_{k}(t \mid a, X)\} \mathrm{d} M_{0}^{C}(t \mid a, X)}{S_{k}(t \mid a, X)G_{k}(t- \mid a, X)} \right\}
\end{aligned}
\end{equation}

\noindent or % whereas $F_j(t \mid a, x)$ is required for competing risks, such that
\begin{equation} \label{eq:competing_estimator}
\begin{aligned}
\widehat{\phi}_{j, k}^{\mathrm{RMTL}, h}(\tau, a; \widehat{\eta}_{j, k}^{\Lambda, h})&=[P_{n, k} \{\widehat{h}_k(X)\}]^{-1} \widehat{h}_k(X) \mathrm{R}\widehat{\mathrm{MT}}\mathrm{L}_{j, k}(\tau \mid a, X)+ [P_{n, k} \{\widehat{w}_{k}^{h}(a, X)\}]^{-1} \widehat{w}_{k}^{h}(a, X) \\
&\times \left\{\frac{(\tau-\min(\widetilde{T}, \tau))I(\widetilde{J}=j)}{\widehat{G}_{k}(\widetilde{T}- \mid a, X)}-\mathrm{R}\widehat{\mathrm{MT}}\mathrm{L}_{j, k}(\tau \mid a, X) \right. \\
&-\int_{0}^{\tau \wedge \widetilde{T}} \frac{(\tau-t) \widehat{F}_{j, k}(t \mid a, X) \mathrm{d} \widehat{M}_{k}^{C}(t \mid a, X)}{\widehat{S}_{k}(t \mid a, X)\widehat{G}_{k}(t- \mid a, X)} \\
&+\left. \int_{0}^{\tau \wedge \widetilde{T}} \frac{\{\mathrm{R}\widehat{\mathrm{MT}}\mathrm{L}_{j, k}(\tau \mid a, X)-\mathrm{R}\widehat{\mathrm{MT}}\mathrm{L}_{j, k}(t \mid a, X)\} \mathrm{d} \widehat{M}_{k}^{C}(t \mid a, X)}{\widehat{S}_{k}(t \mid a, X)\widehat{G}_{k}(t- \mid a, X)} \right\}
\end{aligned}
\end{equation}

Combine $k$-folds together as $\widehat{\phi}^{\mathrm{RMST}, h}(\tau, a; \widehat{\eta}^{\Lambda, h})$ or $\widehat{\phi}^{\mathrm{RMTL}, h}_{j}(\tau, a; \widehat{\eta}_{j}^{\Lambda, h})$ where $\widehat{\eta}^{\Lambda, h}=\{\widehat{\eta}_{k=1}^{\Lambda, h}, \ldots, \widehat{\eta}_{k=K}^{\Lambda, h}\}$ and $\widehat{\eta}_{j}^{\Lambda, h}=\{\widehat{\eta}_{j, k=1}^{\Lambda, h}, \ldots, \widehat{\eta}_{j, k=K}^{\Lambda, h}\}$.
\end{step}

\begin{step} \label{step4}
Shape-corrected point estimators. Compute $\widehat{\psi}^{\mathrm{RMST}, h}(\tau, a)=P_n \{\widehat{\phi}^{\mathrm{RMST}, h}(\tau, a; \widehat{\eta}^{\Lambda, h})\}$ and $\widehat{\psi}_{j}^{\mathrm{RMTL}, h}(\tau, a)=P_n \{\widehat{\phi}^{\mathrm{RMTL}, h}_{j}(\tau, a; \widehat{\eta}_{j}^{\Lambda, h})\}$. For each $a \in \{0,1\}$, we project $\widehat{\psi}^{\mathrm{RMST}, h}(\tau, a)$ onto a space of concave functions using least concave majorant to obtain $\widehat{\psi}^{\mathrm{RMST}, h, lcm}(\tau, a)$ and the bounded $\widehat{\psi}^{\mathrm{RMTL}, h}_{j}(\tau, a)$ onto a space of convex functions using greatest convex minorant to acquire $\widehat{\psi}^{\mathrm{RMTL}, h, gcm}_{j}(\tau, a)$. Then
$$
\begin{aligned}
\widehat{\psi}^{\mathrm{RMST}, h, lcm}(\tau)=&\widehat{\psi}^{\mathrm{RMST}, h, lcm}(\tau, 1)-\widehat{\psi}^{\mathrm{RMST}, h, lcm}(\tau, 0)\\
\widehat{\psi}_{j}^{\mathrm{RMTL}, h, gcm}(\tau)=&\widehat{\psi}_{j}^{\mathrm{RMTL}, h, gcm}(\tau, 1)-\widehat{\psi}_{j}^{\mathrm{RMTL}, h, gcm}(\tau, 0)
\end{aligned}
$$
\end{step}

\begin{step} \label{step5}
Shape-corrected estimators for asymptotic standard errors. % the bounded, we simply bound $\widehat{\psi}^{h}(\tau, a)$ and $\widehat{\psi}^{h}_{j}(\tau, a)$ by $[0, \tau]$. Then
$$
\begin{aligned}
&\widehat{\varphi}^{\mathrm{RMST}, h}(\tau, a; \widehat{\eta}^{\Lambda, h})=\widehat{\phi}^{\mathrm{RMST}, h}(\tau, a; \widehat{\eta}^{\Lambda, h})-\widehat{\psi}^{\mathrm{RMST}, h, lcm}(\tau, a) \\
&\widehat{\varphi}_{j}^{\mathrm{RMTL}, h}(\tau, a; \widehat{\eta}_{j}^{\Lambda, h})=\widehat{\phi}_{j}^{\mathrm{RMTL}, h}(\tau, a; \widehat{\eta}_{j}^{\Lambda, h})-\widehat{\psi}_{j}^{\mathrm{RMTL}, h, gcm}(\tau, a) \\
% &\widehat{\phi}^{\mathrm{RMST}, h}(\tau; O)=\widehat{\phi}^{\mathrm{RMST}, h}(\tau, 1; O)-\widehat{\phi}^{\mathrm{RMST}, h}(\tau, 0; O) \\
% &\widehat{\phi}_{j}^{\mathrm{RMTL}, h}(\tau; O)=\widehat{\phi}_{j}^{\mathrm{RMTL}, h}(\tau, 1; O)-\widehat{\phi}_{j}^{\mathrm{RMTL}, h}(\tau, 0; O) \\
&\widehat{\varphi}^{\mathrm{RMST}, h}(\tau; \widehat{\eta}^{\Lambda, h})=\widehat{\varphi}^{\mathrm{RMST}, h}(\tau, 1; \widehat{\eta}^{\Lambda, h})-\widehat{\varphi}^{\mathrm{RMST}, h}(\tau, 0; \widehat{\eta}^{\Lambda, h}) \\
&\widehat{\varphi}_{j}^{\mathrm{RMTL}, h}(\tau; \widehat{\eta}_{j}^{\Lambda, h})=\widehat{\varphi}_{j}^{\mathrm{RMTL}, h}(\tau, 1; \widehat{\eta}_{j}^{\Lambda, h})-\widehat{\varphi}_{j}^{\mathrm{RMTL}, h}(\tau, 0; \widehat{\eta}_{j}^{\Lambda, h}) \\
% &\widehat{\varphi}^{\mathrm{RMST}, h}(\tau)=\widehat{\phi}^{\mathrm{RMST}, h}(\tau)-\widehat{\psi}^{\mathrm{RMST}, h}(\tau) \text{ and } \widehat{\varphi}_{j}^{\mathrm{RMTL}, h}(\tau)=\widehat{\phi}_{j}^{\mathrm{RMTL}, h}(\tau)-\widehat{\psi}_{j}^{\mathrm{RMTL}, h}(\tau) \\
% &\widehat{\psi}^{\mathrm{RMST}, h}(\tau)=\widehat{\psi}^{\mathrm{RMST}, h}(\tau, 1)-\widehat{\psi}^{\mathrm{RMST}, h}(\tau, 0) \text{ and } \widehat{\psi}_{j}^{\mathrm{RMTL}, h}(\tau)=\widehat{\psi}_{j}^{\mathrm{RMTL}, h}(\tau, 1)-\widehat{\psi}_{j}^{\mathrm{RMTL}, h}(\tau, 0) \\
&\widehat{\sigma}^{\mathrm{RMST}, h}(\tau, a)=[P_n \{\widehat{\varphi}^{\mathrm{RMST}, h}(\tau, a; \widehat{\eta}^{\Lambda, h})\}^{2}]^{1/2}, \quad \widehat{\sigma}_{j}^{\mathrm{RMTL}, h}(\tau, a)=[P_n \{\widehat{\varphi}_{j}^{\mathrm{RMTL}, h}(\tau, a; \widehat{\eta}_{j}^{\Lambda, h})\}^{2}]^{1/2} \\
&\widehat{\sigma}^{\mathrm{RMST}, h}(\tau)=[P_n \{\widehat{\varphi}^{\mathrm{RMST}, h}(\tau; \widehat{\eta}^{\Lambda, h})\}^{2}]^{1/2}, \quad \widehat{\sigma}_{j}^{\mathrm{RMTL}, h}(\tau)=[P_n \{\widehat{\varphi}_{j}^{\mathrm{RMTL}, h}(\tau; \widehat{\eta}_{j}^{\Lambda, h})\}^{2}]^{1/2}
\end{aligned}
$$

When parametric models are employed to estimate nuisance parameters, the asymptotic expansion of the resulting estimator becomes the original plug-in influence function plus a function related to the influence functions of the nuisance models \citep{ozenne2020estimation}. The sandwich variance of the plug-in estimator may be underestimated when this additional term that reflects nuisance model uncertainty is omitted.

We project $\widehat{\sigma}^{\mathrm{RMST}, h}(\tau, a)$, $\widehat{\sigma}^{\mathrm{RMST}, h}(\tau)$, $\widehat{\sigma}_{j}^{\mathrm{RMTL}, h}(\tau, a)$ and $\widehat{\sigma}_{j}^{\mathrm{RMTL}, h}(\tau)$ onto a space of non-decreasing functions by taking their cumulative maximum and denote as
$$
\begin{aligned}
&\widehat{\sigma}^{\mathrm{RMST}, h, +}(\tau, a)=\max_{t \leq \tau} \widehat{\sigma}^{\mathrm{RMST}, h}(t, a), \quad \widehat{\sigma}_{j}^{\mathrm{RMTL}, h, +}(\tau, a)=\max_{t \leq \tau} \widehat{\sigma}_{j}^{\mathrm{RMTL}, h}(t, a) \\
&\widehat{\sigma}^{\mathrm{RMST}, h, +}(\tau)=\max_{t \leq \tau} \widehat{\sigma}^{\mathrm{RMST}, h}(t), \quad \widehat{\sigma}_{j}^{\mathrm{RMTL}, h, +}(\tau)=\max_{t \leq \tau} \widehat{\sigma}_{j}^{\mathrm{RMTL}, h}(t)
\end{aligned}
$$

Eventually, the finite sample standard errors of $\psi^{\mathrm{RMST}, h}(\tau, a)$, $\psi^{\mathrm{RMST}, h}(\tau)$, $\psi_{j}^{\mathrm{RMTL}, h}(\tau, a)$, and $\psi_{j}^{\mathrm{RMTL}, h}(\tau)$ can be calculated as $\widehat{\sigma}^{\mathrm{RMST}, h, +}(\tau, a)/\sqrt{n}$, $\widehat{\sigma}^{\mathrm{RMST}, h, +}(\tau)/\sqrt{n}$, $\widehat{\sigma}_{j}^{\mathrm{RMTL}, h, +}(\tau, a)/\sqrt{n}$, and $\widehat{\sigma}_{j}^{\mathrm{RMTL}, h, +}(\tau)/\sqrt{n}$.
\end{step}

\subsection{Remark~\ref{remark:cpl}}

Below are some identities between conditional uncensored time-to-event distributions and conditional censored time-to-event distributions weighted by IPCW.
$$
\begin{aligned}
S(t \mid a, x)=&\frac{P(C > t \mid A=a, X=x) P(T > t \mid A=a, X=x)}{P(C > t \mid A=a, X=x)} \\
=&\frac{P(\widetilde{T} > t \mid A=a, X=x)}{G(t \mid a, x)}=E\left\{\frac{I(\widetilde{T} > t)}{G(t \mid a, x)} \mid A=a, X=x\right\} \\
S(t \mid a, x)=&\frac{E(\Delta \mid A=a, X=x)E\{I(T > t) \mid A=a, X=x\}}{G(\widetilde{T}- \mid a, x)} \\
=&\frac{E\{\Delta I(\widetilde{T} > t) \mid A=a, X=x\}}{G(\widetilde{T}- \mid a, x)}=E\left\{\frac{\Delta I(\widetilde{T} > t)}{G(\widetilde{T}- \mid a, x)}\mid A=a, X=x \right\} \\
\mathrm{RMST}(\tau \mid a, x)=&\frac{E\{\Delta(\tau) \mid A=a, X=x\} E\{\min(T, \tau) \mid A=a, X=x\}}{G(\tau \wedge\widetilde{T}- \mid a, x)} \\
=&\frac{E\{\Delta(\tau) \min(\widetilde{T}, \tau) \mid A=a, X=x\}}{G(\tau \wedge\widetilde{T}- \mid a, x)}=E\left\{\frac{\Delta(\tau) \min(\widetilde{T}, \tau)}{G(\tau \wedge\widetilde{T}- \mid a, x)} \mid A=a, X=x \right\} \\
F_j(t \mid a, x)=&\frac{E(\Delta \mid A=a, X=x) E\{I(T \leq t, J=j) \mid A=a, X=x\}}{G(\widetilde{T}- \mid a, x)} \\
=&\frac{E\{I(\widetilde{T} \leq t, \widetilde{J}=j) \mid A=a, X=x\}}{G(\widetilde{T}- \mid a, x)}=E\left\{\frac{I(\widetilde{T} \leq t, \widetilde{J}=j)}{G(\widetilde{T}- \mid a, x)} \mid A=a, X=x\right\}
\end{aligned}
$$

$$
\begin{aligned}
\mathrm{RMTL}_j(\tau \mid a, x)=&\frac{E(\Delta \mid A=a, X=x)E[\{\tau-\min(T, \tau)\}I(J=j) \mid A=a, X=x]}{G(\widetilde{T}- \mid a, x)} \\
=&\frac{E[\{\tau-\min(\widetilde{T}, \tau)\}I(\widetilde{J}=j) \mid A=a, X=x]}{G(\widetilde{T}- \mid a, x)} \\
=&E \left[\frac{\{\tau-\min(\widetilde{T}, \tau)\}I(\widetilde{J}=j)}{G(\widetilde{T}- \mid a, x)}\mid A=a, X=x\right]
\end{aligned}
$$

Build pooled classification or regression models by taking observable $I(\widetilde{T} > t)$, $I(\widetilde{T} > t)/G(t \mid A, X)$, $\Delta I(\widetilde{T} > t)$, $\Delta I(\widetilde{T} > t)/G(\widetilde{T}- \mid A, X)$, $\Delta(\tau) \min(\widetilde{T}, \tau)$, $\Delta(\tau) \min(\widetilde{T}, \tau)/G(\tau \wedge\widetilde{T}- \mid A, X)$, $I(\widetilde{T} \leq t, \widetilde{J}=j)$, $I(\widetilde{T} \leq t, \widetilde{J}=j)/G(\widetilde{T}- \mid A, X)$, $\{\tau-\min(\widetilde{T}, \tau)\}I(\widetilde{J}=j)$, and $\{\tau-\min(\widetilde{T}, \tau)\}I(\widetilde{J}=j)/G(\widetilde{T}- \mid A, X)$ as response and making $A$, $X$ and $t$ or $\tau$ as features has three major drawbacks.

First of all, converting data from a wide format to a long format with $t$ or $\tau$ as pivot only allows discrete-time survival times \citep{stitelman2010collaborative}. Coarsening continous survival times into discrete often leads to loss of information and accuracy. Second, training the long format requires much more space and can be very time-consuming. Last, but not the least, most standard statistical learning algorithms require data are i.i.d. However, this type of ``counting process learning'' keeps all rows even after a subject leaves the risk set, which leaves strong dependency between rows in the long format. On the other hand, discrete-time conditional hazard type classification satisfies the i.i.d assumption. By definition,
$$
\begin{aligned}
&\lambda(t \mid a, x)=P(T=t \mid T \geq t, A=a, X=x)=\frac{P(T=t, T \geq t, A=a, X=x)}{P(T \geq t, A=a, X=x)} \\
=&\frac{P(T=t, A=a, X=x)P(C \geq t, A=a, X=x)}{P(T \geq t, A=a, X=x)P(C \geq t, A=a, X=x)}=\frac{P(\widetilde{T}=t, T \leq C, A=a, X=x)}{P(\widetilde{T} \geq t, A=a, X=x)} \\
=&\frac{P(\widetilde{T}=t, \Delta, \widetilde{T} \geq t, A=a, X=x)}{P(\widetilde{T} \geq t, A=a, X=x)}=P(\widetilde{T}=t, \Delta \mid \widetilde{T} \geq t, A=a, X=x)
\end{aligned}
$$

The condition in the last probability above assures units who left the risk set are not trained any more such that for all units at risk, whether each unit has the event at each time point becomes i.i.d. Censoring hazards and cause-specific hazards follow a similar derivation and reasoning.

\section{Simulation study}

\subsection{Data generating processes}

Within each replicate, consider six covariates $X_1$-$X_6$ following practices from \citet{cheng2022addressing}. The covariates $X_4$, $X_5$, and $X_6$ are drawn independently from a Bernoulli distribution with 50\% probability. The covariates $X_1$, $X_2$, and $X_3$ are generated from a multivariate normal distribution with mean zero, unit variance, and $0.5$ pairwise correlation.

\textit{Setting} 1 (good overlap, no competing risks): We consider a sample size of $n=4000$. The exposure is generated from a Bernoulli distribution with propensity score
$$\operatorname{expit}(-(0.3+0.2 X_1+0.3 X_2+0.3 X_3-0.2 X_4-0.3 X_5-0.2 X_6))$$

We generate potential survival times $T^{a=0}$ and $T^{a=1}$ as well as potential censoring times $C^{a=0}$ and $C^{a=1}$ from exponential distribution with hazards
$$
\begin{aligned}
\lambda^{a=0}(t \mid X)&=0.12 \exp \{0.1+0.1 X_1-0.2 X_2+0.2 X_3+0.1 X_4+0.8 X_5-0.2 X_6\} \\
\lambda^{a=1}(t \mid X)&=0.15 \exp \{0.17+0.2 X_1-0.1 X_2+0.4 X_3+0.2 X_4+0.3 X_5+0.4 X_6\} \\
\lambda^{C}(t \mid A=0, X)&=0.06 \exp \{0.1+0.4 X_1-0.7 X_2-0.4 X_3-0.5 X_4+0.8 X_5-0.6 X_6\} \\
\lambda^{C}(t \mid A=1, X)&=0.08 \exp \{0.5 X_1-0.6 X_2+0.2 X_3+0.6 X_4+0.9 X_5-0.5 X_6\}
\end{aligned}
$$

\textit{Setting} 2 (poor overlap, no competing risks): We consider a sample size of $n=4000$. The exposure is generated from a Bernoulli distribution with propensity score
$$\operatorname{expit}(-(-1+X_1+1.5 X_2+1.5 X_3-X_4-1.5X_5-X_6))$$

Potential survival and censoring times are generated the same as \textit{Setting} 1.

\textit{Setting} 3 (good overlap, with competing risks): We consider a sample size of $n=4000$. The exposure is simulated the same as \textit{Setting} 1. Potential failure times of two causes and censoring times are generated with hazards
$$
\begin{aligned}
\lambda_{j=1}^{a=0}(t \mid X)&= 0.12\exp \{0.1+ 0.1X_1-0.2 X_2+0.2 X_3+0.1 X_4+0.8 X_5-0.2 X_6\} \\
\lambda_{j=1}^{a=1}(t \mid X)&= 0.15\exp \{0.17+ 0.2X_1-0.1 X_2+0.4 X_3+0.2 X_4+0.3 X_5+0.4 X_6\} \\
\lambda_{j=2}^{a=0}(t \mid X)&= 0.1\exp \{0.12-0.1 X_1+0.3 X_2+0.1 X_3+0.2 X_4-0.4 X_5+0.5 X_6\} \\
\lambda_{j=2}^{a=1}(t \mid X)&= 0.08\exp \{0.1-0.2 X_1-0.1 X_2+0.2 X_3+0.3 X_4+0.3 X_5-0.3 X_6\} \\
\lambda^{C}(t \mid A=0, X)&= 0.12\exp \{0.1+0.4 X_1-0.7 X_2-0.4 X_3-0.5 X_4+0.8 X_5-0.6 X_6\} \\
\lambda^{C}(t \mid A=1, X)&= 0.14\exp \{0+0.5 X_1-0.6 X_2+0.2 X_3+0.6 X_4+0.9 X_5-0.5 X_6\}
\end{aligned}
$$

Cause 1 is of our interest.

\textit{Setting} 4 (poor overlap, with competing risks): We consider a sample size of $n=4000$. The exposure is simulated the same as \textit{Setting} 2. Potential failure times of two causes and censoring times are generated the same as \textit{Setting} 3. Cause 1 is of our interest.

\subsection{Summary statistics}

Summary statistics, including treatment prevalence, censoring probability, average rate of violation of proportional hazards assumption from $1000$ simulations, and event rate before $\tau=4$, of \textit{Setting} 1-4 is tabulated in Table~\ref{tab:sim_summary}.

\begin{table}[H]
\centering
\caption{Summary of data generating mechanisms \textit{Setting} 1-\textit{Setting} 4. \%PH violation: percentage of violation of proportional hazards (PH) assumption, 4 hazards models ($\lambda^{a=0}, \lambda^{a=1}, \lambda^{a=0, C}, \lambda^{a=1, C}$) for \textit{Setting} 1 and \textit{Setting} 2 and 6 hazards models ($\lambda^{a=0}_{j=1}, \lambda^{a=1}_{j=1}, \lambda^{a=0}, \lambda^{a=1}, \lambda^{a=0, C}, \lambda^{a=1, C}$) for \textit{Setting} 3 and \textit{Setting} 4. \\
Event rate before $\tau=4$: number of event-of-interest that happened before $\tau=4$ divided by $n$.}
\begin{tabular}{l r r r r} %\\
 \\ \hline
& \textit{Setting} 1 & \textit{Setting} 2 & \textit{Setting} 3 & \textit{Setting} 4 \\ \hline
Percentage of treatment & 49\% & 25\% & 49\% & 25\% \\
Censoring probability & 35.7\% & 35.4\% & 35.9\% & 33.7\% \\
\%PH violation & 33.3\% & 29.8\% & 8.2\% & 13.7\% \\
Event rate before $\tau=4$ & 48.7\% & 50.2\% & 36.2\% & 37.6\% \\ \hline
\end{tabular}
\label{tab:sim_summary}
\end{table}

% Event rate at $\tau=1,2,3$: 20.8\%, 32.5\%, 41.7\%, 
% Event rate at $\tau=1,2,3$: 21.4\%, 35.0\%, 44.2\%, 
% Event rate at $\tau=1,2,3$: 18.5\%, 28.3\%, 33.6\%, 
% Event rate at $\tau=1,2,3$: 19.6\%, 29.3\%, 34.4\%,

\subsection{Comparator estimators}

Some of these comparator estimators haven't been explicitly proposed, and we spell out these estimators here for completeness. In the survival setting, the g-formula estimator using outcome regression is
$$
\widehat{\psi}^{S, \text{OR}, h}(t, a)=[P_{n}\{\widehat{h}(X)\}]^{-1} P_{n}[\widehat{h}(X)\{\widehat{S}(t \mid a, X)\}]
$$

\noindent and the IPCW estimator \citep{cheng2022addressing} is
$$
\widehat{\psi}^{S, \text{IPCW}, h}(t, a)=[P_{n} \{\widehat{w}^{h}(a, X)\}]^{-1} P_{n} \left\{\frac{\widehat{w}^{h}(a, X)I(\widetilde{T} > t)}{\widehat{G}(t \mid a, X)} \right\}
$$

Another IPCW estimator can be constructed as
$$
\widehat{\psi}^{S, \text{IPCW}, h}(t, a)=[P_{n} \{\widehat{w}^{h}(a, X)\}]^{-1} P_{n} \left\{\frac{\widehat{w}^{h}(a, X)\Delta I(\widetilde{T} > t)}{\widehat{G}(\widetilde{T}- \mid a, X)} \right\}
$$

The doubly robust estimator takes the form
$$
\begin{aligned}
\widehat{\psi}^{S, \text{DR}, h}(t, a)&=[P_{n} \{\widehat{h}(X)\}]^{-1} P_{n}[\widehat{h}(X) \widehat{S}(t \mid a, X)]+[P_{n} \{\widehat{w}^{h}(a, X)\}]^{-1} P_{n} \left[ \widehat{w}^{h}(a, X) \right. \\
&\times \left. \left\{\frac{I(\widetilde{T}>t)}{\widehat{G}(t \mid a, X)}-\widehat{S}(t \mid a, X)+\widehat{S}(t \mid a, X)\int_{0}^{t \wedge \widetilde{T}} \frac{\mathrm{d} \widehat{M}^{C}(u \mid a, X)}{\widehat{S}(u \mid a, X)\widehat{G}(u- \mid a, X)} \right\} \right]
\end{aligned}
$$

This is a right continuous function with respect to $t$ while \citet{bai2013doubly} provided a left continuous version. As the weighted group-specific survival functions are monotonically non-increasing, we take their cumulative minimum and denote as $\widehat{\psi}^{S, h, -}(t, a)=\min_{t \leq \tau} \widehat{\psi}^{S, h}(t, a)$. The causal contrasts are $\widehat{\psi}^{S, h}(t)=\widehat{\psi}^{S, h, -}(t, 1)-\widehat{\psi}^{S, h, -}(t, 0)$. Then $\widehat{\psi}^{\mathrm{RMST}, h}(\tau)$ can be obtained by integrating each estimator above, $\int_{0}^{\tau} \widehat{\psi}^{S, h}(t) \mathrm{d}t$.

In the competing risks setting, the g-formula estimator using outcome regressions for the cause-specific CIF is,
$$
\widehat{\psi}_{j}^{F, \text{OR}, h}(t, a)=[P_{n}\{\widehat{h}(X)\}]^{-1} P_{n}[\widehat{h}(X)\{\widehat{F}_{j}(t \mid a, X)\}]
$$

and the IPCW estimator is
$$
\widehat{\psi}_{j}^{F, \text{IPCW}, h}(t, a)=[P_{n} \{\widehat{w}^{h}(a, X)\}]^{-1} P_{n} \left\{\frac{\widehat{w}^{h}(a, X) I(\widetilde{T} \leq t, \widetilde{J}=j)}{\widehat{G}(\widetilde{T}- \mid a, X)} \right\}
$$

The doubly robust estimator takes the form
$$
\begin{aligned}
\widehat{\psi}_{j}^{F, \text{DR}, h}(t, a)&=[P_{n} \{\widehat{h}(X)\}]^{-1} P_{n}[\widehat{h}(X) \widehat{F}_{j}(t \mid a, X)]+ [P_{n} \{\widehat{w}^{h}(a, X)\}]^{-1} P_{n}\left[\widehat{w}^{h}(a, X) \right. \\
&\times \left\{\frac{I(\widetilde{T} \leq t, \widetilde{J}=j)}{\widehat{G}(\widetilde{T}- \mid a, X)}-\widehat{F}_{j}(t \mid a, X)+\widehat{F}_{j}(t \mid a, X)\int_{0}^{t \wedge \widetilde{T}} \frac{\mathrm{d} \widehat{M}^{C}(u \mid a, X)}{\widehat{S}(u \mid a, X) \widehat{G}(u- \mid a, X)} \right. \\
&\left.\left. -\int_{0}^{t \wedge \widetilde{T}} \frac{\widehat{F}_{j}(u \mid a, X) \mathrm{d} \widehat{M}^{C}(u \mid a, X)}{\widehat{S}(u \mid a, X) \widehat{G}(u- \mid a, X)} \right\}\right]
\end{aligned}
$$

\noindent which is similar to \citet{ozenne2020estimation}. As the weighted group-specific cause-specific CIFs are monotonically non-decreasing, we take their cumulative maximum and denote as $\widehat{\psi}_{j}^{F, h, +}(t, a)=\max_{t \leq \tau} \widehat{\psi}_{j}^{F, h}(t, a)$. The causal contrasts are $\widehat{\psi}_{j}^{F, h}(t)=\widehat{\psi}_{j}^{F, h, +}(t, 1)-\widehat{\psi}_{j}^{F, h, +}(t, 0)$. Then $\widehat{\psi}_{j}^{\mathrm{RMTL}, h}(\tau)$ can be obtained by integrating each estimator above, $\int_{0}^{\tau} \widehat{\psi}_{j}^{F, h}(t) \mathrm{d}t$.

\subsection{Additional simulation results}

\begin{figure}[H]
\centering
% The arguments in the next line are {height}{optional width}{used only by OUP for typesetting} for figure empty box eg 
%\figurebox{20pc}{25pc}{}
%if actual size of graphics need plese see below command
%\figurebox{}{}{}[fig1]
%need to reducing the figure size use below command
%\figuresize{.8}%
\includegraphics[scale=0.55]{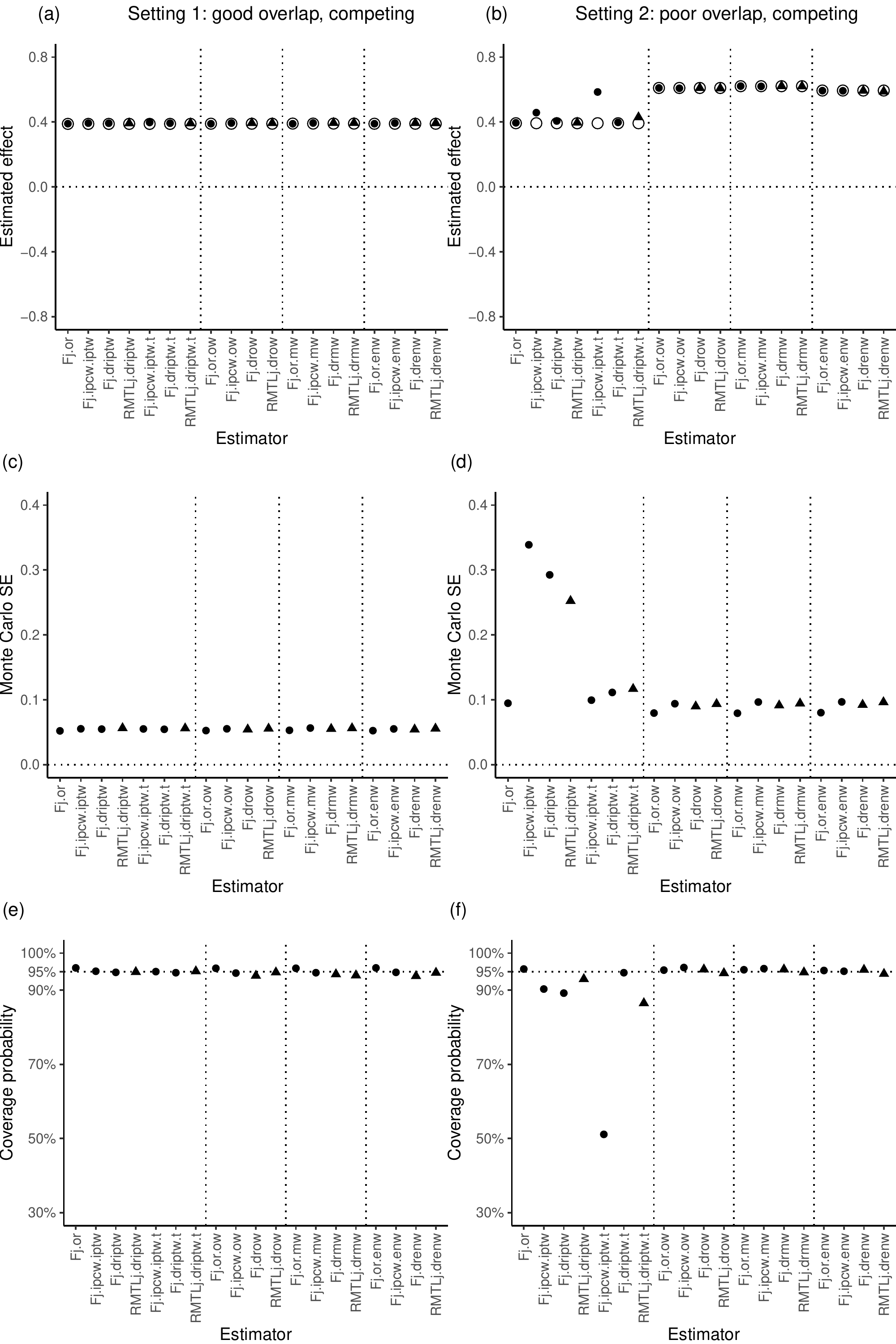}
% note that files may not be rotated
\caption{Estimated cumulative treatment effect, Monte Carlo standard errors, and coverage probabilities from different estimators grouped by estimand at $\tau=4$. Estimands in each plot are separated by vertical dotted lines. The order of the estimands from left to right in each plot is: ATE, ATO, ATM, ATEN. Hollow circles in (a) and (b) represent true effects; solid circles represent the comparator estimators; and solid triangles represent the proposed estimators. S.or stands for the outcome regression estimator; S.ipcw.iptw stands for the IPCW estiamtor; S.dr stands for the doubly-robust estimator; .t implies the IPTW is Winsorized at its 99th percentile; RMST. represents the proposed estimators; ow, mw, and enw stand for overlap weight, matching weight, and entropy weights, respectively.}
\label{fig:sim_competing_main}
\end{figure}

\begin{figure}[H]
\centering
% The arguments in the next line are {height}{optional width}{used only by OUP for typesetting} for figure empty box eg 
%\figurebox{20pc}{25pc}{}
%if actual size of graphics need plese see below command
%\figurebox{}{}{}[fig1]
%need to reducing the figure size use below command
%\figuresize{.8}%
\includegraphics[scale=0.6]{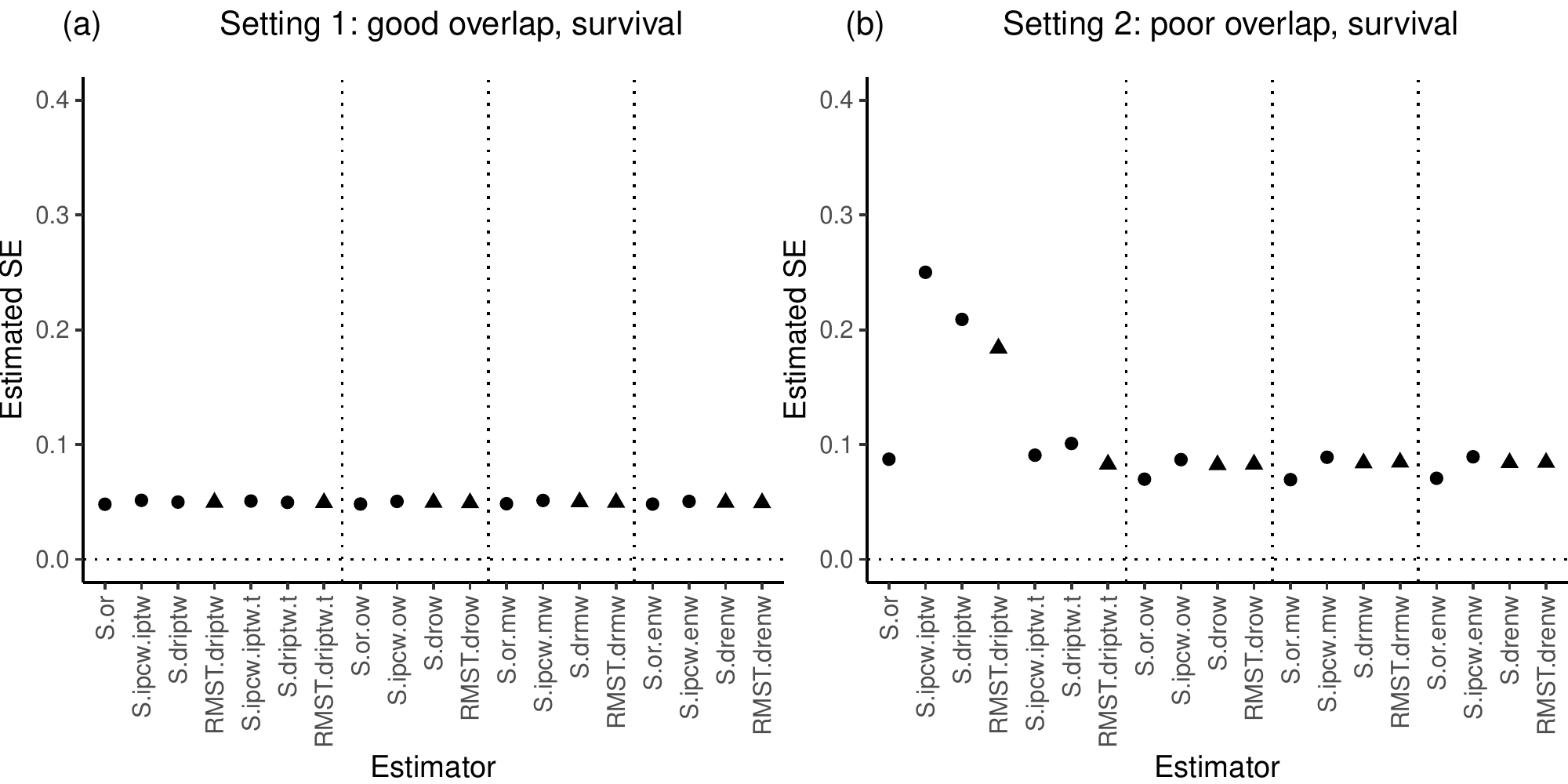}
% note that files may not be rotated
\caption{Average standard errors of different estimators grouped by estimand at $\tau=4$. Solid circles represent average bootstrap standard errors of the comparator estimators; and solid triangles represent average estimated standard errors of the proposed estimators. Other symbols and notations are similar to Fig.~\ref{fig:sim_survival_main}.}
\label{fig:sim_survival_ase}
\end{figure}

\begin{figure}[H]
\centering
% The arguments in the next line are {height}{optional width}{used only by OUP for typesetting} for figure empty box eg 
%\figurebox{20pc}{25pc}{}
%if actual size of graphics need plese see below command
%\figurebox{}{}{}[fig1]
%need to reducing the figure size use below command
%\figuresize{.8}%
\includegraphics[scale=0.55]{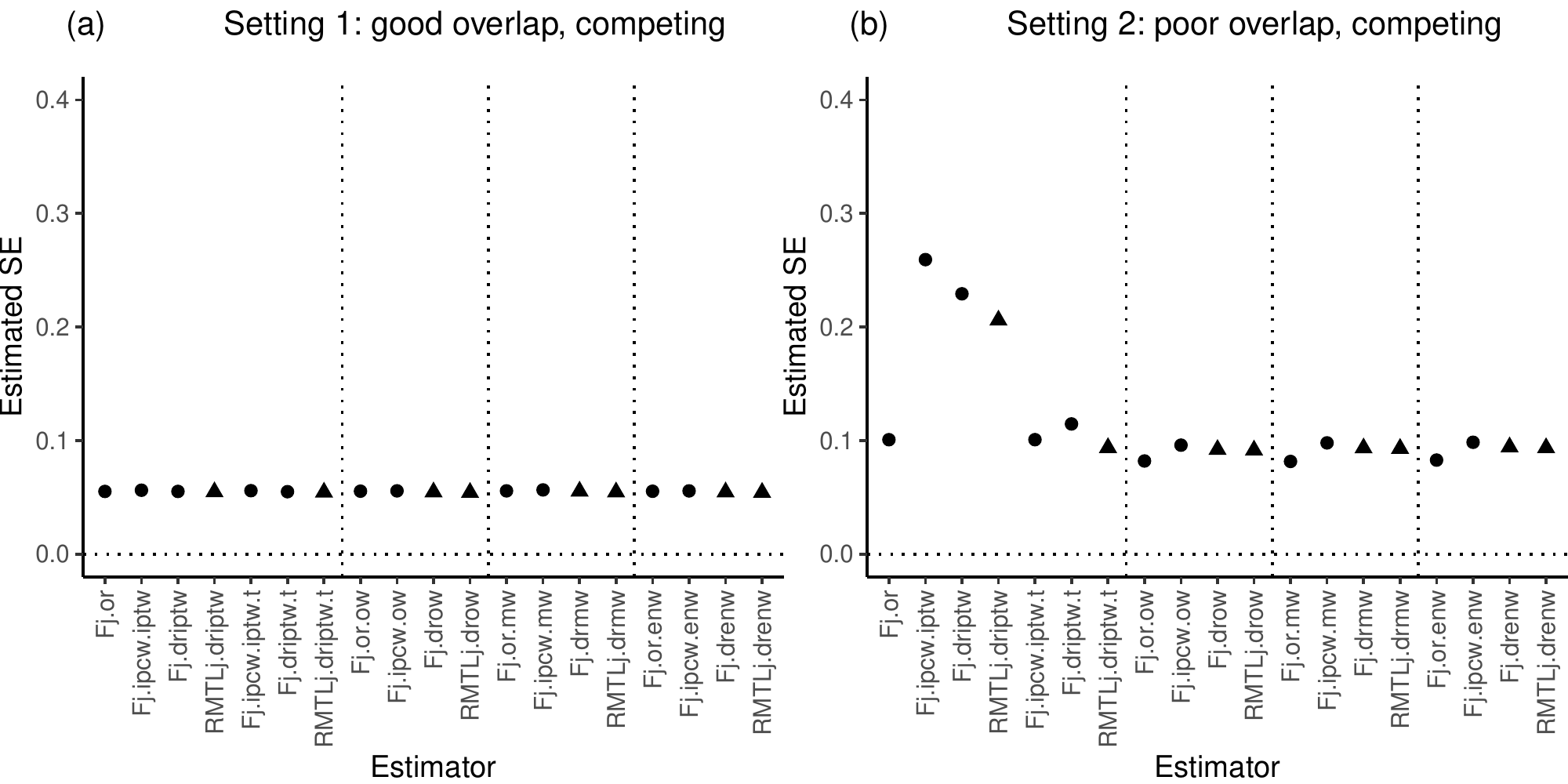}
% note that files may not be rotated
\caption{Average standard errors of different estimators grouped by estimand at $\tau=4$. Solid circles represent average bootstrap standard errors of the comparator estimators; and solid triangles represent average estimated standard errors of the proposed estimators. Other symbols and notations are similar to Fig.~\ref{fig:sim_competing_main}.}
\label{fig:sim_competing_ase}
\end{figure}

% \begin{figure}[H]
% % The arguments in the next line are {height}{optional width}{used only by OUP for typesetting} for figure empty box eg 
% %\figurebox{20pc}{25pc}{}
% %if actual size of graphics need plese see below command
% %\figurebox{}{}{}[fig1]
% %need to reducing the figure size use below command
% %\figuresize{.8}%
% \figuresize{0.55}
% \figurebox{20pc}{25pc}{}[sim_competing_cp.pdf]
% % note that files may not be rotated
% \caption{Coverage probabilities of different estimators grouped by estimand at $\tau=4$. Solid circles represent coverage probabilities of the comparator estimators; and solid triangles represent coverage probabilities of the proposed estimators. Other symbols and notations are similar to Fig.~\ref{fig:sim_competing_effect}.}
% \label{fig:sim_competing_cp}
% \end{figure}

\subsection{Candidate learners}

To speed up training, we use two-fold cross-validation throughout this paper and select relatively fast learners among all possible candidate learners. For the propensity score, we employed marginal mean (SL.mean), feed-forward neural networks (SL.nnet), kernel k-nearest neighbors (SL.kernelKnn), classification trees (SL.rpartPrune), extreme gradient boosting (SL.xgboost), random forests (SL.ranger), logistic regression (SL.glm), forward/backwards stepwise logistic regression (SL.step), generalized additive model (SL.gam), regularized logistic regression (SL.glmnet), and multivariate adaptive regression splines (SL.earth). In terms of conditional failure-time distributions, we considered the Kaplan-Meier estimator (survSL.km), piecewise constant hazard regression (survSL.pchreg), survival random forest (survSL.rfsrc), Cox proportional hazards regression (survSL.coxph), and regression assuming event and censoring times follow conditional exponential (survSL.expreg), Weibull (survSL.weibreg), and log-logistic (survSL.loglogreg) distributions.

\section{Application}

\subsection{Dealing with non-positive semidefinite covariance estimator}

When the sample size is large, the observed survival times can be close at earlier survival times. Under such circumstances, the covariance estimator may fail to be positive semidefinite (PSD) such that sample paths can't be generated directly. One may alternatively consider generating sample paths by perturbing centered EIFs with a standard normal random variable $\xi$ as
$$
\xi \frac{\widehat{\varphi}^{\mathrm{RMST}, h}(t, a; \widehat{\eta}^{\Lambda, h})}{\widehat{\sigma}^{\mathrm{RMST}, h, +}(t, a)} \sim \mathcal{N} \left(0, \frac{\widehat{\varphi}^{\mathrm{RMST}, h}(t, a; \widehat{\eta}^{\Lambda, h})\{\widehat{\varphi}^{\mathrm{RMST}, h}(t, a; \widehat{\eta}^{\Lambda, h})\}^{T}}{\{\widehat{\sigma}^{\mathrm{RMST}, h, +}(t, a)\}^2}\right)
$$

\noindent such that the PSD requirement for the cross-fitted covariance function can be avoided. Aside from perturbing the centered EIFs, \citet{lin1993checking} proposed to approximate the limiting distribution of certain martingale integrals by perturbing corresponding counting processes, i.e. $\mathrm{d} M(t \mid A, X)=\xi I(\widetilde{T} = t, \Delta)$, $\mathrm{d} M^C(t \mid A, X)=\xi I(\widetilde{T} = t, \Delta=0)$, and $\mathrm{d} M_j(t \mid A, X)=\xi I(\widetilde{T} = t, \widetilde{J} = j)$. Besides perturbation methods, the non-PSD estimated covariance may be converted to its nearest PSD matrix \citep{higham1988computing}, but the asymptotic equivalence of this transformation has not been proved yet. The inference of causal contrasts between treatment-specific RMSTs and cause-specific RMTLs can be established similarly by replacing corresponding shape-corrected estimates.

% Besides differences, we can conduct inference on other differentiable contrasts using the delta method, such as ratio.

\subsection{Results}

As piecewise constant hazard regression (survSL.pchreg) is too time-consuming for large datasets, we remove it from the simulation candidate library for all conditional time-to-event distributions. We estimate the propensity scores using the simulation candidate library without kernel k-nearest neighbors (SL.kernelKnn) and generalized additive model (SL.gam).

\begin{table}[H]
\centering
\caption{Results (in years) of estimated counterfactual RMTL/RMST and cumulative treatment effect between metformin initiators and sulphonylurea initiators on the CPRD data at $\tau=10$ years. SE: standard error; CI: confidence interval; CB: confidence band.}
\begin{tabular}{l c c c} %\\
 \\ \hline
 & Point estimates (SE) & 95\% pointwise CI & 95\% uniform CB \\ \hline
\multicolumn{2}{l}{Cancer incidence ATE} \\
Sulphonylurea (RMTL) & 0.859 (0.056) & (0.750, 0.968) & (0.716, 1.002) \\
Metformin (RMTL) & 0.730 (0.008) & (0.713, 0.746) & (0.706, 0.754) \\
Metformin-Sulphonylurea & -0.129 (0.056) & (-0.239, -0.019) & (-0.282, 0.023) \\ \\
\multicolumn{2}{l}{Cancer incidence ATO} \\
Sulphonylurea (RMTL) & 0.855 (0.023) & (0.809, 0.901) & (0.787, 0.922) \\
Metformin (RMTL) & 0.785 (0.013) & (0.758, 0.811) & (0.748, 0.822) \\
Metformin-Sulphonylurea & -0.070 (0.027) & (-0.122, -0.018) & (-0.142, 0.002) \\
(difference in RMTL) \\ \\
\multicolumn{2}{l}{Mortality after cancer ATE} \\
Sulphonylurea (RMST) & 5.998 (0.166) & (5.672, 6.324) & (5.539, 6.457) \\
Metformin (RMST) & 8.114 (0.039) & (8.037, 8.191) & (8.003, 8.226) \\
Metformin-Sulphonylurea & 2.116 (0.17) & (1.782, 2.451) & (1.636, 2.597) \\
difference in (RMST) \\ \\
\multicolumn{2}{l}{Mortality after cancer diagnosis ATO} \\
Sulphonylurea (RMST) & 6.260 (0.112) & (6.041, 6.479) & (5.945, 6.574) \\
Metformin (RMST) & 7.679 (0.068) & (7.545, 7.813) & (7.495, 7.863) \\
Metformin-Sulphonylurea & 1.419 (0.115) & (1.194, 1.644) & (1.084, 1.754) \\
(difference in RMST) \\ \hline
\end{tabular}
\label{tab:application}
\end{table}

\pagebreak

\begin{figure}[H]
\centering
% The arguments in the next line are {height}{optional width}{used only by OUP for typesetting} for figure empty box eg 
%\figurebox{20pc}{25pc}{}
%if actual size of graphics need plese see below command
%\figurebox{}{}{}[fig1]
%need to reducing the figure size use below command
%\figuresize{.8}%
\includegraphics[scale=0.5]{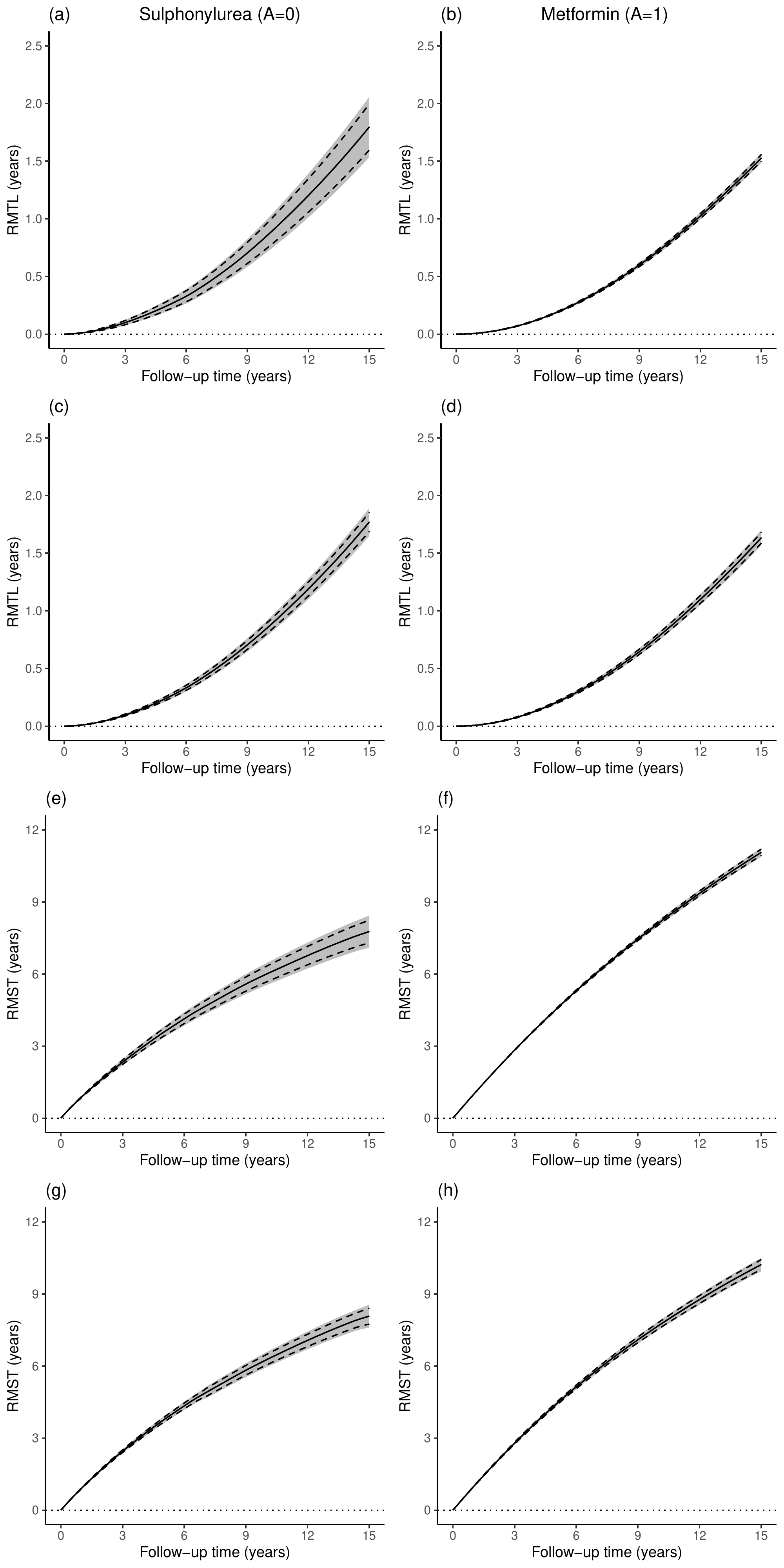}
% note that files may not be rotated
\caption{Curves of estimated counterfactual RMTL/RMST on the CPRD data; first column (a), (c), (e), and (g): sulphonylurea initiators; second column (b), (d), (f), and (h): metformin initiators; first row (a) and (b): IPTW treatment-specific cause-specific RMTL of cancer incidence; second row (c) and (d): OW treatment-specific cause-specific RMTL of cancer incidence; (e) and (f): IPTW treatment-specific RMST of mortality after cancer diagnosis; (g) and (h) OW treatment-specific RMST of mortality after cancer diagnosis. In all figures, dotted line: $y=0$; solid lines: point estimates; dashed lines: 95\% pointwise confidence intervals; shaded area: 95\% uniform simultaneous confidence bands.}
\label{fig:app_treatment_specific}
\end{figure}

\section{Proof of Theorems}

\subsection{Theorem~\ref{theorem:eif}}
% \subsection{Derivation of $\varphi^{\mathrm{RMST}, h, sp}(\tau, a)$}

Proof of Theorem~\ref{theorem:eif}: following \citet{robins1992recovery,tsiatis2006semiparametric}, without additional assumptions on the time-to-event distributions, $\psi_{0}^{\mathrm{RMST}, h}(\tau, a)$ can be formed as augmented inverse probability weighted complete case (AIPWCC) estimators with estimating function
$$
\begin{aligned}
& m^{\text{AIPWCC}}\{\psi_{0}^{\mathrm{RMST}, h}(\tau, a); \eta_{0}^{\Lambda, h}\} \\
% =&\frac{w_{0}^{h}(a, x) \Delta}{G_{0}(T^{a}- \mid a, x)}\{\min(T^{a}, \tau)-\psi_{0}^{\mathrm{RMST}, h}(\tau, a)\}+\{h_{0}(x)-w_{0}^{h}(a, x)\} \\
% &\times E\{\min(T^{a}, \tau)-\psi_{0}^{\mathrm{RMST}, h}(\tau, a) \mid A=a, X\}+w_{0}^{h}(a, x) \int_{0}^{\infty} \frac{\mathrm{d} M_{0}^{C}(t \mid a, x)}{G_{0}(t- \mid a, x)} \\
% &\times E\{\min(T^{a}, \tau)-\psi_{0}^{\mathrm{RMST}, h}(\tau, a) \mid A=a, X, T^{a} > t\} \\
=&\frac{w_{0}^{h}(a, X) \Delta}{G_{0}(\widetilde{T}- \mid a, X)}\{\min(\widetilde{T}, \tau)-\psi_{0}^{\mathrm{RMST}, h}(\tau, a)\}+\{h_{0}(X)-w_{0}^{h}(a, X)\} \\
&\times E_{0}\{\min(T, \tau)-\psi_{0}^{\mathrm{RMST}, h}(\tau, a) \mid A=a, X\}+w_{0}^{h}(a, X) \int_{0}^{\infty} \frac{\mathrm{d} M_{0}^{C}(t \mid a, X)}{G_{0}(t- \mid a, X)} \\
&\times E_{0}\{\min(T, \tau)-\psi_{0}^{\mathrm{RMST}, h}(\tau, a) \mid A=a, X, T>t\}
\end{aligned}
$$

Note that
$$
\begin{aligned}
&\int_{0}^{\infty} \frac{\mathrm{d} M_{0}^{C}(t \mid a, X)}{G_{0}(t- \mid a, X)}E_{0}\{\min(T, \tau)-\psi_{0}^{\mathrm{RMST}, h}(\tau, a) \mid A=a, X, T > t \} \\
=&\int_{0}^{\infty} \frac{\mathrm{d} M_{0}^{C}(t \mid a, X)}{G_{0}(t- \mid a, X)} [(t \leq \tau)[E_{0}\{\min(T, \tau) \mid A=a, X, T>t\}-\psi_{0}^{\mathrm{RMST}, h}(\tau, a) \\
&+(t > \tau)\{\tau-\psi_{0}^{\mathrm{RMST}, h}(\tau, a)\}] \\
=&\int_{0}^{\tau \wedge \widetilde{T}} \frac{\mathrm{d} M_{0}^{C}(t \mid a, X)}{G_{0}(t- \mid a, X)} E_{0}\{\min(T, \tau) \mid A=a, X, T>t\}+\tau\int_{\tau+}^{\infty} \frac{\mathrm{d} M_{0}^{C}(t \mid a, X)}{G_{0}(t- \mid a, X)} \\
&-\psi_{0}^{\mathrm{RMST}, h}(\tau, a)\int_{0}^{\infty} \frac{\mathrm{d} M_{0}^{C}(t \mid a, X)}{G_{0}(t- \mid a, X)}
\end{aligned}
$$

\noindent where
$$
\begin{aligned}
&E_{0}\{\min(T, \tau) \mid A=a, X, T > t \}=\int_t^\infty \min(t, \tau) \mathrm{d} P_{0}(T \leq u \mid T > t, A=a, X) \\
=& \int_t^\infty \min(t, \tau) \mathrm{d} \left\{\frac{P_{0}(T > t, A=a, X, T \leq u)}{P_{0}(T > t, A=a, X)}\right\}= \int_t^\infty \min(t, \tau) \mathrm{d} \left\{\frac{P_{0}(t < T \leq u \mid A=a, X)}{P_{0}(T > t \mid A=a, X)}\right\} \\
=& \int_t^\infty \min(t, \tau) \mathrm{d} \left\{\frac{S_{0}(t \mid a, X)-S_{0}(u \mid a, X)}{S_{0}(t \mid a, X)}\right\}= \frac{\int_t^\infty \min(t, \tau) \{-\mathrm{d} S_{0}(u \mid a, X)\}}{S_{0}(t \mid a, X)} \\
=& \frac{-\int_t^\tau u \mathrm{d} S_{0}(u \mid a, X)+\tau S_{0}(\tau \mid a, X)}{S_{0}(t \mid a, X)}
=t+\frac{\int_t^\tau S_{0}(u \mid a, X) \mathrm{d}t}{S_{0}(t \mid a, X)} \\
=&t+\frac{\mathrm{RMST}_{0}(\tau \mid a, X)-\mathrm{RMST}_{0}(t \mid a, X)}{S_{0}(t \mid a, X)}
\end{aligned}
$$

Note that
$$
\begin{aligned}
&\frac{\Delta}{G_{0}(\widetilde{T}- \mid a, X)}+\frac{I(\widetilde{T} > \tau)}{G_{0}(\tau \mid a, X)}-\frac{\Delta I(\widetilde{T} > \tau) }{G_{0}(\widetilde{T}- \mid a, X)}=\frac{\Delta I(\widetilde{T} \leq \tau) }{G_{0}(\tau \wedge \widetilde{T}- \mid a, X)}+\frac{I(\widetilde{T} > \tau)}{G_{0}(\tau \wedge \widetilde{T}- \mid a, X)} \\
=&\frac{I(C > \tau \wedge T-)}{G_{0}(\tau \wedge \widetilde{T}- \mid a, X)}=\frac{\Delta(\tau)}{G_{0}(\tau \wedge \widetilde{T}- \mid a, X)}
\end{aligned}
$$

Therefore,
$$
\begin{aligned}
&m^{\text{AIPWCC}}\{\psi_{0}^{\mathrm{RMST}, h}(\tau, a); \eta_{0}^{\Lambda, h}\} \\
=&\frac{w_{0}^{h}(a, X) \Delta \min(\widetilde{T}, \tau)}{G_{0}(\widetilde{T}- \mid a, X)}-\frac{w_{0}^{h}(a, X) \Delta \psi_{0}^{\mathrm{RMST}, h}(\tau, a)}{G_{0}(\widetilde{T}- \mid a, X)}+\{h_{0}(X)-w_{0}^{h}(a, X)\}\mathrm{RMST}_{0}(\tau \mid a, X) \\
&-\{h_{0}(X)-w_{0}^{h}(a, X)\}\psi_{0}^{\mathrm{RMST}, h}(\tau, a)+w_{0}^{h}(a, X) \int_{0}^{\tau \wedge \widetilde{T}} \frac{\mathrm{d} M_{0}^{C}(t \mid a, X)}{G_{0}(t- \mid a, X)} \\
&\times \left\{t+\frac{\mathrm{RMST}_{0}(\tau \mid a, X)-\mathrm{RMST}_{0}(t \mid a, X)}{S_{0}(t \mid a, X)}\right\}+\frac{w_{0}^{h}(a, X)\tau I(\widetilde{T} > \tau)}{G_{0}(\tau \mid a, X)} \\
&-\frac{w_{0}^{h}(a, X)\tau \Delta I(\widetilde{T} > \tau)}{G_{0}(\widetilde{T}- \mid a, X)}-w_{0}^{h}(a, X)\psi_{0}^{\mathrm{RMST}, h}(\tau, a)+\frac{w_{0}^{h}(a, X)\Delta \psi_{0}^{\mathrm{RMST}, h}(\tau, a)}{G_{0}(\widetilde{T}- \mid a, X)} \\
=&\frac{w_{0}^{h}(a, X)\Delta \min(\widetilde{T}, \tau)}{G_{0}(\widetilde{T}- \mid a, X)}+\frac{w_{0}^{h}(a, X)I(\widetilde{T} > \tau)\min(\widetilde{T}, \tau)}{G_{0}(\tau \mid a, X)}-\frac{w_{0}^{h}(a, X)\Delta I(\widetilde{T} > \tau) \min(\widetilde{T}, \tau)}{G_{0}(\widetilde{T}- \mid a, X)} \\
&+\{h_{0}(X)-w_{0}^{h}(a, X)\}\mathrm{RMST}_{0}(\tau \mid a, X)+w_{0}^{h}(a, X) \\
&\times\int_{0}^{\tau \wedge \widetilde{T}} \frac{\mathrm{d} M_{0}^{C}(t \mid a, X)}{G_{0}(t- \mid a, X)} \left\{t+\frac{\mathrm{RMST}_{0}(\tau \mid a, X)-\mathrm{RMST}_{0}(t \mid a, X)}{S_{0}(t \mid a, X)}\right\}-h_{0}(X)\psi_{0}^{\mathrm{RMST}, h}(\tau, a) \\
=&\frac{w_{0}^{h}(a, X)\Delta(\tau) \min(\widetilde{T}, \tau)}{G_{0}(\tau \wedge \widetilde{T}- \mid a, X)}+\{h_{0}(X)-w_{0}^{h}(a, X)\}\mathrm{RMST}_{0}(\tau \mid a, X)+w_{0}^{h}(a, X) \\
&\times\int_{0}^{\tau \wedge \widetilde{T}} \frac{\mathrm{d} M_{0}^{C}(t \mid a, X)}{G_{0}(t- \mid a, X)}\left\{t+\frac{\mathrm{RMST}_{0}(\tau \mid a, X)-\mathrm{RMST}_{0}(t \mid a, X)}{S_{0}(t \mid a, X)}\right\}-h_{0}(X)\psi_{0}^{\mathrm{RMST}, h}(\tau, a) \\
\end{aligned}
$$

The influence function of the fully augmented M-estimating equation is the efficient influence function (EIF),
$$
\begin{aligned}
\varphi_{0}^{\mathrm{RMST}, h}(\tau, a; \eta_{0}^{\Lambda, h})&=-\left\{E_{0}\left[\frac{\partial m^{\text{AIPWCC}}\{\psi_{0}^{\mathrm{RMST}, h}(\tau, a); \eta_{0}^{\Lambda, h}\}}{\partial \psi_{0}^{\mathrm{RMST}, h}(\tau, a)\}}\right]\right\}^{-1} m^{\text{AIPWCC}}\{\psi_{0}^{\mathrm{RMST}, h}(\tau, a); \eta_{0}^{\Lambda, h}\} \\
=&[E_{0}\{h_{0}(X)\}]^{-1} \left[w_{0}^{h}(a, X)\left\{\frac{\Delta(\tau)\min(\widetilde{T}, \tau)}{G_{0}(\tau \wedge \widetilde{T}- \mid a, X)}-\mathrm{RMST}_{0}(\tau \mid a, X) \right. \right. \\
&\left.\left.+\int_{0}^{\tau \wedge \widetilde{T}} \frac{t \mathrm{d} M_{0}^{C}(t \mid a, X)}{G_{0}(t- \mid a, X)} +\int_{0}^{\tau \wedge \widetilde{T}} \frac{\{\mathrm{RMST}_{0}(\tau \mid a, X)-\mathrm{RMST}_{0}(t \mid a, X)\} \mathrm{d} M_{0}^{C}(t \mid a, X)}{S_{0}(t \mid a, X)G_{0}(t- \mid a, X)}\right\} \right. \\
&\left.+h_{0}(X)\mathrm{RMST}_{0}(\tau \mid a, X)-h_{0}(X)\psi_{0}^{\mathrm{RMST}, h}(\tau, a) \right]
\end{aligned}
$$

In the competing risks setting, the doubly robust estimating function for $\psi_{j, 0}^{\mathrm{RMTL}, h}(\tau, a)$ is
%\begin{equation} \label{eq:RMTLj_sp}
$$
\begin{aligned}
& m^{\text{AIPWCC}}\{\psi_{j, 0}^{\mathrm{RMTL}, h}(\tau, a); \eta_{j, 0}^{\Lambda, h}\} \\
% =&\frac{w_{0}^{h}(a, X) \Delta}{G_{0}(T^{a}- \mid a, X)}[\{\tau-\min(T^{a}, \tau)\}I(J^{a}=j)-\psi_{j, 0}^{\mathrm{RMTL}, h}(\tau, a)]+\{h_{0}(x)-w_{0}^{h}(a, X)\} \\
% &\times E[\{\tau-\min(T^{a}, \tau)\} I(J^{a}=j)-\psi_{j, 0}^{\mathrm{RMTL}, h}(\tau, a) \mid A=a, X]+w_{0}^{h}(a, X)\int_{0}^{\infty} \frac{\mathrm{d} M_{0}^{C}(t \mid a, X)}{G_{0}^{a}(t- \mid a, X)} \\
% &\times E[\{\tau-\min(T^{a}, \tau)\}I(J^{a}=j)-\psi_{j, 0}^{\mathrm{RMTL}, h}(\tau, a) \mid A=a, X, T^{a} > t]
=&\frac{w_{0}^{h}(a, X)}{G_{0}(\widetilde{T}- \mid a, X)}[\{\tau-\min(\widetilde{T}, \tau)\} I(\widetilde{J}=j)-\Delta \psi_{j, 0}^{\mathrm{RMTL}, h}(\tau, a)]+\{h_{0}(X)-w_{0}^{h}(a, X)\} \\
&\times E_{0}[\{\tau-\min(T, \tau)\} I(J=j)-\psi_{j, 0}^{\mathrm{RMTL}, h}(\tau, a) \mid A=a, X]+w_{0}^{h}(a, X) \int_{0}^{\infty} \frac{\mathrm{d} M_{0}^{C}(t \mid a, X)}{G_{0}(t- \mid a, X)} \\
&\times E_{0}[\{\tau-\min(T, \tau)\} I(J=j)-\psi_{j, 0}^{\mathrm{RMTL}, h}(\tau, a) \mid A=a, X, T > t]
\end{aligned}
$$

\noindent where
$$
\begin{aligned}
&E_{0}[\{\tau-\min(T, \tau)\} (J=j) \mid A=a, X, T > t] \\
=& \int_{t}^\infty \{\tau-\min(u, \tau)\} \mathrm{d} P_{0}(T \leq u, J=j \mid A=a, X, T > t) \\
=& \int_t^\infty \{\tau-\min(u, \tau)\} \mathrm{d} \left\{\frac{P_{0}(t < T \leq u, J=j, A=a, X)}{P_{0}(A=a, X, T > t)}\right\} \\
=& \int_t^\infty \{\tau-\min(u, \tau)\} \mathrm{d} \left\{\frac{P_{0}(t < T \leq u, J=j \mid A=a, X)}{P_{0}(T >t \mid A=a, X)}\right\} \\
=& \int_t^\infty \{\tau-\min(u, \tau)\} \mathrm{d} \left\{\frac{F_{j, 0}(u \mid a, X)-F_{j, 0}(t \mid a, X)}{S_{0}(t \mid a, X)}\right\}=\frac{\int_t^\infty \{\tau-\min(u, \tau)\} \mathrm{d} F_{j, 0}(u \mid a, X)}{S_{0}(t \mid a, X)} \\
=& \frac{\int_t^\tau (\tau-u) \mathrm{d} F_{j, 0}(u \mid a, X)}{S_{0}(t \mid a, X)}=\frac{(t-\tau) F_{j, 0}(t \mid a, X)+\mathrm{RMTL}_{j, 0}(\tau \mid a, X)-\mathrm{RMTL}_{j, 0}(t \mid a, X)}{S_{0}(t \mid a, X)} \\
\end{aligned}
$$

Note that
$$
\begin{aligned}
&\int_{0}^{\infty} \frac{\mathrm{d} M_{0}^{C}(t \mid a, X)}{G_{0}(t- \mid a, X)}E_{0}[\{\tau-\min(T, \tau)\} I(J=j)-\psi_{j, 0}^{\mathrm{RMTL}, h}(\tau, a) \mid A=a, X, T > t] \\
=&\int_{0}^{\infty} \frac{\mathrm{d} M_{0}^{C}(t \mid a, X)}{G_{0}(t- \mid a, X)} [(t \leq \tau)(E_{0}[\{\tau-\min(T, \tau)\}I(J=j) \mid A=a, X, T>t]-\psi_{j, 0}^{\mathrm{RMTL}, h}(\tau, a)) \\
& +(t > \tau)\{-\psi_{j, 0}^{\mathrm{RMTL}, h}(\tau, a)\} ] \\
=&\int_{0}^{\tau \wedge \widetilde{T}} \frac{\{(t-\tau) F_{j, 0}(t \mid a, X)+\mathrm{RMTL}_{j, 0}(\tau \mid a, X)-\mathrm{RMTL}_{j, 0}(t \mid a, X)\} \mathrm{d} M_{0}^{C}(t \mid a, X)}{S_{0}(t \mid a, X) G_{0}(t- \mid a, X)} \\
&-\psi_{j, 0}^{\mathrm{RMTL}, h}(\tau, a)+\frac{\Delta \psi_{j, 0}^{\mathrm{RMTL}, h}(\tau, a)}{G_{0}(\widetilde{T}- \mid a, X)}
\end{aligned}
$$

Consequently, the estimating function becomes % \eqref{eq:RMTLj_sp}
$$
\begin{aligned}
&m^{\text{AIPWCC}}\{\psi_{j, 0}^{\mathrm{RMTL}, h}(\tau, a); \eta_{j, 0}^{\Lambda, h}\} \\
=&\frac{w_{0}^{h}(a, X) \{\tau-\min(\widetilde{T}, \tau)\}I(\widetilde{J}=j)}{G_{0}(\widetilde{T}- \mid a, X)}-\frac{w_{0}^{h}(a, X) \Delta \psi_{j, 0}^{\mathrm{RMTL}, h}(\tau, a)}{G_{0}(\widetilde{T}- \mid a, X)}+\{h_{0}(X)-w_{0}^{h}(a, X)\} \\
&\times \mathrm{RMTL}_{j, 0}(\tau \mid a, X)-\{h_{0}(X)-w_{0}^{h}(a, X)\}\psi_{j, 0}^{\mathrm{RMTL}, h}(\tau, a)+w_{0}^{h}(a, X) \\
&\times \int_{0}^{\tau \wedge \widetilde{T}} \frac{\{(t-\tau) F_{j, 0}(t \mid a, X)+\mathrm{RMTL}_{j, 0}(\tau \mid a, X)-\mathrm{RMTL}_{j, 0}(t \mid a, X)\} \mathrm{d} M_{0}^{C}(t \mid a, X)}{S_{0}(t \mid a, X) G_{0}(t- \mid a, X)} \\
&-w_{0}^{h}(a, X)\psi_{j, 0}^{\mathrm{RMTL}, h}(\tau, a)+\frac{w_{0}^{h}(a, X) \Delta \psi_{j, 0}^{\mathrm{RMTL}, h}(\tau, a)}{G_{0}(\widetilde{T}- \mid a, X)} \\
=&\frac{w_{0}^{h}(a, X)\{\tau-\min(\widetilde{T}, \tau)\}I(\widetilde{J}=j)}{G_{0}(\widetilde{T}- \mid a, X)}+\{h_{0}(X)-w_{0}^{h}(a, X)\} \mathrm{RMTL}_{j, 0}(\tau \mid a, X)+w_{0}^{h}(a, X) \\
&\times \int_{0}^{\tau \wedge \widetilde{T}} \frac{\{(t-\tau) F_{j, 0}(t \mid a, X)+\mathrm{RMTL}_{j, 0}(\tau \mid a, X)-\mathrm{RMTL}_{j, 0}(t \mid a, X)\} \mathrm{d} M_{0}^{C}(t \mid a, X)}{S_{0}(t \mid a, X) G_{0}(t- \mid a, X)} \\
&-h_{0}(X) \psi_{j, 0}^{\mathrm{RMTL}, h}(\tau, a)
\end{aligned}
$$

The centered EIF is
$$
\begin{aligned}
&\varphi_{j, 0}^{\mathrm{RMTL}, h}(\tau, a; \eta_{j, 0}^{\Lambda, h}) \\
=&-\left\{E_{0}\left[\frac{\partial m^{\text{AIPWCC}}\{\psi_{j, 0}^{\mathrm{RMTL}, h}(\tau, a); \eta_{j, 0}^{\Lambda, h}\}}{\partial \psi_{j, 0}^{\mathrm{RMTL}, h}(\tau, a)}\right]\right\}^{-1} \partial m^{\text{AIPWCC}}\{\psi_{j, 0}^{\mathrm{RMTL}, h}(\tau, a); \eta_{j, 0}^{\Lambda, h}\} \\
=&[E_{0} \{h_{0}(X)\}]^{-1} \left(w_{0}^{h}(a, X)\left[\frac{\{\tau-\min(\widetilde{T}, \tau)\}I(\widetilde{J}=j)}{G_{0}(\widetilde{T}- \mid a, X)}-\mathrm{RMTL}_{j, 0}(\tau \mid a, X)\right.\right. \\
&\left.\left. +\int_{0}^{\tau \wedge \widetilde{T}} \frac{\{(t-\tau) F_{j, 0}(t \mid a, X)+\mathrm{RMTL}_{j, 0}(\tau \mid a, X)-\mathrm{RMTL}_{j, 0}(t \mid a, X)\} \mathrm{d} M_{0}^{C}(t \mid a, X)}{S_{0}(t \mid a, X) G_{0}(t- \mid a, X)} \right] \right. \\
&\left.+h_{0}(X)\mathrm{RMTL}_{j, 0}(\tau \mid a, X)-h_{0}(X)\psi_{j, 0}^{\mathrm{RMTL}, h}(\tau, a)\right)
\end{aligned}
$$

\subsection{Lemma~\ref{lemma1}}
% \subsection{Derivation of $\phi^{\mathrm{RMST}, sp}(\tau, a)=\phi^{\mathrm{RMST}, np}(\tau, a)$}

\begin{lemma} \label{lemma1}
$$
\begin{aligned}
\phi_{0}^{\mathrm{RMST}}(\tau, a; \eta_{0}^{\Lambda})=&\mathrm{RMST}_{0}(\tau \mid a, X)-\frac{I(A=a)\mathrm{RMST}_{0}(\tau \mid a, X)}{\pi_{0}(a \mid X)}\int_0^{\tau \wedge \widetilde{T}} \frac{\mathrm{d} M_{0}(t \mid a, X)}{S_{0}(t \mid a, X)G_{0}(t- \mid a, X)} \\
&+\frac{I(A=a)}{\pi_{0}(a \mid X)}\int_0^{\tau \wedge \widetilde{T}} \frac{\mathrm{RMST}_{0}(t \mid a, X)\mathrm{d} M_{0}(t \mid a, X)}{S_{0}(t \mid a, X)G_{0}(t- \mid a, X)}
\end{aligned}
$$

\noindent and
$$
\begin{aligned}
\phi_{j, 0}^{\mathrm{RMTL}}(\tau, a; \eta_{j, 0}^{\Lambda})=&\mathrm{RMTL}_{j, 0}(\tau \mid a, X)+\frac{I(A=a)}{\pi_{0}(a \mid X)} \int_0^{\tau \wedge \widetilde{T}} \frac{(\tau-t)\mathrm{d}M_{j, 0}(t \mid a, X)}{G_{0}(t- \mid a, X)} \\
&-\frac{I(A=a)}{\pi_{0}(a \mid X)} \int_0^{\tau \wedge \widetilde{T}}\frac{\{\mathrm{RMTL}_{j, 0}(\tau \mid a, X)-\mathrm{RMTL}_{j, 0}(t \mid a, X)\} \mathrm{d}M_{0}(t \mid a, X)}{S_{0}(t \mid a, X)G_{0}(t- \mid a, X)} \\
&+\frac{I(A=a)}{\pi_{0}(a \mid X)} \int_0^{\tau \wedge \widetilde{T}}\frac{(\tau-t)F_{j, 0}(t \mid a, X)\mathrm{d}M_{0}(t \mid a, X)}{S_{0}(t \mid a, X)G_{0}(t- \mid a, X)}
\end{aligned}
$$
\end{lemma}

Proof of Lemma~\ref{lemma1}: we start from the RHS of the first equation above
$$
\begin{aligned}
&\mathrm{RMST}_{0}(\tau \mid a, X)-\frac{I(A=a)\mathrm{RMST}_{0}(\tau \mid a, X)}{\pi_{0}(a \mid X)}  \int_0^{\tau \wedge \widetilde{T}} \frac{\mathrm{d} M_{0}(t \mid a, X)}{S_{0}(t \mid a, X)G_{0}(t- \mid a, X)} \\
&+\frac{I(A=a)}{\pi_{0}(a \mid X)}\int_0^{\tau \wedge \widetilde{T}} \frac{\mathrm{d} M_{0}(t \mid a, X) \mathrm{RMST}_{0}(t \mid a, X)}{S_{0}(t \mid a, X)G_{0}(t- \mid a, X)}-h_{0}(x) \psi_{0}^{\mathrm{RMST}, h}(\tau, a) \\
=&\mathrm{RMST}_{0}(\tau \mid a, X)-\frac{I(A=a)\mathrm{RMST}_{0}(\tau \mid a, X)}{\pi_{0}(a \mid X)} \int_0^\tau \frac{\mathrm{d} N(t)}{S_{0}(t \mid a, X)G_{0}(t- \mid a, X)} \\
&+\frac{I(A=a)\mathrm{RMST}_{0}(\tau \mid a, X)}{\pi_{0}(a \mid X)} \int_0^{\tau \wedge \widetilde{T}} \frac{\mathrm{d} \Lambda_{0}(t \mid a, X)}{S_{0}(t \mid a, X)G_{0}(t- \mid a, X)}+\frac{I(A=a)}{\pi_{0}(a \mid X)} \\
&\times \int_0^\tau \frac{\mathrm{RMST}_{0}(t \mid a, X) \mathrm{d} N(t)}{S_{0}(t \mid a, X)G_{0}(t- \mid a, X)}-\frac{I(A=a)}{\pi_{0}(a \mid X)} \int_0^{\tau \wedge \widetilde{T}} \frac{\mathrm{RMST}_{0}(t \mid a, X) \mathrm{d} \Lambda_{0}(t \mid a, X)}{S_{0}(t \mid a, X)G_{0}(t- \mid a, X)}
\end{aligned}
$$

\noindent where
$$
\int_0^\tau \frac{\mathrm{d} N(t)}{S_{0}(t \mid a, X)G_{0}(t- \mid a, X)}=\frac{\Delta I(\widetilde{T} \leq \tau)}{S_{0}(\tau \wedge \widetilde{T} \mid a, X)G_{0}(\tau \wedge \widetilde{T}- \mid a, X)}
$$

\noindent and
$$
\int_0^\tau \frac{\mathrm{RMST}_{0}(t \mid a, X) \mathrm{d} N(t)}{S_{0}(t \mid a, X)G_{0}(t- \mid a, X)}=\frac{\Delta I(\widetilde{T} \leq \tau) \mathrm{RMST}_{0}(\tau \wedge \widetilde{T} \mid a, X)}{S_{0}(\tau \wedge \widetilde{T} \mid a, X)G_{0}(\tau \wedge \widetilde{T}- \mid a, X)}
$$

Note that $\mathrm{d}\Lambda_{0}^C(u- \mid a, X)=P_{0}(u- \leq T<(u+\mathrm{d} u)- \mid C \geq u-, A=a, X)=P_{0}(u \leq T<u+\mathrm{d} u \mid C \geq u, A=a, X)=\mathrm{d}\Lambda_{0}^C(u \mid a, X)$, $\mathrm{d}\Lambda_{0}(u- \mid a, X)=\mathrm{d}\Lambda_{0}(u \mid a, X)$, and $\mathrm{d}\Lambda_{j, 0}(u- \mid a, X)=\mathrm{d}\Lambda_{j, 0}(u \mid a, X)$. Thus
$$
\begin{aligned}
&\int_0^{\tau \wedge \widetilde{T}} \frac{\mathrm{d} \Lambda_{0}(t \mid a, X)}{S_{0}(t \mid a, X)G_{0}(t- \mid a, X)}=\int_0^{\tau \wedge \widetilde{T}} \frac{1}{G_{0}(t- \mid a, X)} \mathrm{d} \left\{\frac{1}{S_{0}(t \mid a, X)}\right\} \\
=&\left. \frac{1}{S_{0}(t \mid a, X)G_{0}(t- \mid a, X)} \right|_{0}^{\tau \wedge \widetilde{T}}-\int_0^{\tau \wedge \widetilde{T}} \frac{1}{S_{0}(t \mid a, X)} \mathrm{d} \left\{\frac{1}{G_{0}(t- \mid a, X)}\right\} \\
=&\frac{1}{S_{0}(\tau \wedge \widetilde{T} \mid a, X) G_{0}(\tau \wedge \widetilde{T}- \mid a, X)}-1-\int_0^{\tau \wedge \widetilde{T}} \frac{\mathrm{d} \Lambda_{0}^{C}(t- \mid a, X)}{S_{0}(t \mid a, X)G_{0}(t- \mid a, X)} \\
=&\frac{1}{S_{0}(\tau \wedge \widetilde{T} \mid a, X) G_{0}(\tau \wedge \widetilde{T}- \mid a, X)}-1+\int_0^{\tau \wedge \widetilde{T}} \frac{\mathrm{d} M_{0}^{C}(t \mid a, X)}{S_{0}(t \mid a, X)G_{0}(t- \mid a, X)}-\int_0^{\tau} \frac{\mathrm{d} N^{C}(t \mid a, X)}{S_{0}(t \mid a, X)G_{0}(t- \mid a, X)} \\
=&\frac{1}{S_{0}(\tau \wedge \widetilde{T} \mid a, X) G_{0}(\tau \wedge \widetilde{T}- \mid a, X)}-1+\int_0^{\tau \wedge \widetilde{T}} \frac{\mathrm{d} M_{0}^{C}(t \mid a, X)}{S_{0}(t \mid a, X)G_{0}(t- \mid a, X)} \\
&-\frac{(1-\Delta) I(\widetilde{T} \leq \tau)}{S_{0}(\tau \wedge \widetilde{T}- \mid a, X) G_{0}(\tau \wedge \widetilde{T}- \mid a, X)} \\
=&\frac{\Delta(\tau)}{S_{0}(\tau \wedge \widetilde{T} \mid a, X) G_{0}(\tau \wedge \widetilde{T}- \mid a, X)}-1+\int_0^{\tau \wedge \widetilde{T}} \frac{\mathrm{d} M_{0}^{C}(t \mid a, X)}{S_{0}(t \mid a, X)G_{0}(t- \mid a, X)}
\end{aligned}
$$

Since
$$
\begin{aligned}
\mathrm{d} \left\{ \frac{\mathrm{RMST}_{0}(t \mid a, X)}{G_{0}(t- \mid a, X)} \right\} &=\frac{S_{0}(t \mid a, X)G_{0}(t- \mid a, X) \mathrm{d} t + G_{0}(t- \mid a, X) \mathrm{d} \Lambda_{0}^{C}(t- \mid a, X) \mathrm{RMST}_{0}(t \mid a, X)}{G_{0}^2(t- \mid a, X)} \\
&=\frac{S_{0}(t \mid a, X) \mathrm{d} t}{G_{0}(t- \mid a, X)}+\frac{\mathrm{RMST}_{0}(t \mid a, X)\mathrm{d} \Lambda_{0}^{C}(t \mid a, X)}{G_{0}(t- \mid a, X)}
\end{aligned}
$$

\noindent and
$$
\begin{aligned}
&\int_0^{\tau \wedge \widetilde{T}} \frac{\mathrm{RMST}_{0}(t \mid a, X) \mathrm{d} \Lambda_{0}(t \mid a, X)}{S_{0}(t \mid a, X)G_{0}(t- \mid a, X)}=\int_0^{\tau \wedge \widetilde{T}} \frac{\mathrm{RMST}_{0}(t \mid a, X)}{G_{0}(t- \mid a, X)} \mathrm{d} \left\{\frac{1}{S_{0}(t \mid a, X)}\right\} \\
=&\left. \frac{\mathrm{RMST}_{0}(t \mid a, X)}{S_{0}(t \mid a, X)G_{0}(t- \mid a, X)} \right|_{0}^{\tau \wedge \widetilde{T}}-\int_0^{\tau \wedge \widetilde{T}} \frac{1}{S_{0}(t \mid a, X)} \mathrm{d} \left\{\frac{\mathrm{RMST}_{0}(t \mid a, X)}{G_{0}(t- \mid a, X)}  \right\} \\
=& \frac{\mathrm{RMST}_{0}(\tau \wedge \widetilde{T} \mid a, X)}{G_{0}(\tau \wedge \widetilde{T}- \mid a, X)S_{0}(\tau \wedge \widetilde{T} \mid a, X)}-\int_0^{\tau \wedge \widetilde{T}} \frac{\mathrm{d} t}{G_{0}(t- \mid a, X)}-\int_0^{\tau \wedge \widetilde{T}} \frac{\mathrm{RMST}_{0}(t \mid a, X) \mathrm{d} \Lambda_{0}^{C}(t \mid a, X)}{S_{0}(t \mid a, X)G_{0}(t- \mid a, X)} \\
=&\frac{\mathrm{RMST}_{0}(\tau \wedge \widetilde{T} \mid a, X)}{G_{0}(\tau \wedge \widetilde{T}- \mid a, X)S_{0}(\tau \wedge \widetilde{T} \mid a, X)}-\frac{\tau \wedge \widetilde{T}}{G_{0}(\tau \wedge \widetilde{T}- \mid a, X)}+\int_0^{\tau \wedge \widetilde{T}} \frac{t \mathrm{d} \Lambda_{0}^{C}(t \mid a, X)}{G_{0}(t- \mid a, X)} \\
&+\int_0^{\tau \wedge \widetilde{T}} \frac{\mathrm{RMST}_{0}(t \mid a, X)\mathrm{d} M_{0}^{C}(t \mid a, X)}{S_{0}(t \mid a, X)G_{0}(t- \mid a, X)}-\int_0^{\tau} \frac{\mathrm{RMST}_{0}(t \mid a, X) \mathrm{d} N^{C}(t)}{S_{0}(t \mid a, X)G_{0}(t- \mid a, X)}
\end{aligned}
$$

$$
\begin{aligned}
=&\frac{\mathrm{RMST}_{0}(\tau \wedge \widetilde{T} \mid a, X)}{G_{0}(\tau \wedge \widetilde{T}- \mid a, X)S_{0}(\tau \wedge \widetilde{T} \mid a, X)}-\frac{\tau \wedge \widetilde{T}}{G_{0}(\tau \wedge \widetilde{T}- \mid a, X)}-\int_0^{\tau \wedge \widetilde{T}} \frac{t \mathrm{d} M_{0}^{C}(t \mid a, X)}{G_{0}(t- \mid a, X)} \\
&+\int_0^{\tau} \frac{t \mathrm{d} N^{C}(t \mid a, X)}{G_{0}(t- \mid a, X)}+\int_0^{\tau \wedge \widetilde{T}} \frac{\mathrm{RMST}_{0}(t \mid a, X)\mathrm{d} M_{0}^{C}(t \mid a, X)}{S_{0}(t \mid a, X)G_{0}(t- \mid a, X)}-\frac{\mathrm{RMST}_{0}(\tau \wedge \widetilde{T} \mid a, X) I(\widetilde{T} \leq \tau) (1-\Delta)}{G_{0}(\tau \wedge \widetilde{T}- \mid a, X)S_{0}(\tau \wedge \widetilde{T} \mid a, X)} \\
=&\frac{\mathrm{RMST}_{0}(\tau \wedge \widetilde{T} \mid a, X)}{G_{0}(\tau \wedge \widetilde{T}- \mid a, X)S_{0}(\tau \wedge \widetilde{T} \mid a, X)}-\frac{\tau \wedge \widetilde{T}}{G_{0}(\tau \wedge \widetilde{T}- \mid a, X)}-\int_0^{\tau \wedge \widetilde{T}} \frac{t \mathrm{d} M_{0}^{C}(t \mid a, X)}{G_{0}(t- \mid a, X)} \\
&+\frac{(1-\Delta) I(\widetilde{T} \leq \tau) (\tau \wedge \widetilde{T})}{G_{0}(\tau \wedge \widetilde{T}- \mid a, X)}+\int_0^{\tau \wedge \widetilde{T}} \frac{\mathrm{RMST}_{0}(t \mid a, X)\mathrm{d} M_{0}^{C}(t \mid a, X)}{S_{0}(t \mid a, X)G_{0}(t- \mid a, X)}-\frac{(1-\Delta) I(\widetilde{T} \leq \tau) \mathrm{RMST}_{0}(\tau \wedge \widetilde{T} \mid a, X)}{G_{0}(\tau \wedge \widetilde{T}- \mid a, X)S_{0}(\tau \wedge \widetilde{T} \mid a, X)} \\
=&\frac{\Delta(\tau)\mathrm{RMST}_{0}(\tau \wedge \widetilde{T} \mid a, X)}{G_{0}(\tau \wedge \widetilde{T}- \mid a, X)S_{0}(\tau \wedge \widetilde{T} \mid a, X)}-\frac{\Delta(\tau) \min(\widetilde{T}, \tau)}{G_{0}(\tau \wedge \widetilde{T}- \mid a, X)}-\int_0^{\tau \wedge \widetilde{T}} \frac{t \mathrm{d} M_{0}^{C}(t \mid a, X)}{G_{0}(t- \mid a, X)} \\
&+\int_0^{\tau \wedge \widetilde{T}} \frac{\mathrm{RMST}_{0}(t \mid a, X)\mathrm{d} M_{0}^{C}(t \mid a, X)}{S_{0}(t \mid a, X)G_{0}(t- \mid a, X)}
\end{aligned}
$$

Substitute previous equations into the RHS of the first equation in Lemma~\ref{lemma1} gives,
$$
\begin{aligned}
&\mathrm{RMST}_{0}(\tau \mid a, X)-\frac{I(A=a) \Delta I(\widetilde{T} \leq \tau) \mathrm{RMST}_{0}(\tau\mid a, X)}{\pi_{0}(a \mid X)S_{0}(\tau \wedge \widetilde{T} \mid a, X)G_{0}(\tau \wedge \widetilde{T}- \mid a, X)}+\frac{I(A=a)\mathrm{RMST}_{0}(\tau \mid a, X)}{\pi_{0}(a \mid X)} \\
&\times \left\{\frac{\Delta (\tau)}{S_{0}(\tau \wedge \widetilde{T} \mid a, X)G_{0}(\tau \wedge \widetilde{T}- \mid a, X)}-1+\int_0^{\tau \wedge \widetilde{T}} \frac{\mathrm{d} M_{0}^{C}(t \mid a, X)}{S_{0}(t \mid a, X)G_{0}(t- \mid a, X)} \right\} \\
&+ \frac{I(A=a) \Delta I(\widetilde{T} \leq \tau) \mathrm{RMST}_{0}(\tau \wedge \widetilde{T} \mid a, X)}{\pi_{0}(a \mid X)S_{0}(\tau \wedge \widetilde{T} \mid a, X)G_{0}(\tau \wedge \widetilde{T}- \mid a, X)}-\frac{I(A=a)}{\pi_{0}(a \mid X)} \left\{\frac{\Delta(\tau)\mathrm{RMST}_{0}(\tau \wedge \widetilde{T} \mid a, X)}{S_{0}(\tau \wedge \widetilde{T} \mid a, X)G_{0}(\tau \wedge \widetilde{T}- \mid a, X)} \right. \\
& \left. -\frac{\Delta(\tau)\min(\widetilde{T}, \tau)}{G_{0}(\tau \wedge \widetilde{T}- \mid a, X)}-\int_0^{\tau \wedge \widetilde{T}} \frac{t \mathrm{d} M_{0}^{C}(t \mid a, X)}{G_{0}(t- \mid a, X)}+\int_0^{\tau \wedge \widetilde{T}} \frac{\mathrm{RMST}_{0}(t \mid a, X)\mathrm{d} M_{0}^{C}(t \mid a, X)}{S_{0}(t \mid a, X)G_{0}(t- \mid a, X)} \right\}
\end{aligned}
$$

$$
\begin{aligned}
=&\mathrm{RMST}_{0}(\tau \mid a, X)-\frac{I(A=a) \Delta I(\widetilde{T} \leq \tau)\mathrm{RMST}_{0}(\tau\mid a, X)}{\pi_{0}(a \mid X)S_{0}(\tau \wedge \widetilde{T} \mid a, X)G_{0}(\tau \wedge \widetilde{T}- \mid a, X)}+\frac{I(A=a)\Delta (\tau)\mathrm{RMST}_{0}(\tau \mid a, X)}{\pi_{0}(a \mid X)S_{0}(\tau \wedge \widetilde{T} \mid a, X)G_{0}(\tau \wedge \widetilde{T}- \mid a, X)} \\
&-\frac{I(A=a)\mathrm{RMST}_{0}(\tau \mid a, X)}{\pi_{0}(a \mid X)}+\frac{I(A=a)\mathrm{RMST}_{0}(\tau \mid a, X)}{\pi_{0}(a \mid X)}\int_0^{\tau \wedge \widetilde{T}} \frac{\mathrm{d} M_{0}^{C}(t \mid a, X)}{S_{0}(t \mid a, X)G_{0}(t- \mid a, X)} \\
&+\frac{I(A=a)\Delta I(\widetilde{T} \leq \tau) \mathrm{RMST}_{0}(\tau \wedge \widetilde{T} \mid a, X)}{\pi_{0}(a \mid X)S_{0}(\tau \wedge \widetilde{T} \mid a, X)G_{0}(\tau \wedge \widetilde{T}- \mid a, X)}-\frac{I(A=a)\Delta(\tau)\mathrm{RMST}_{0}(\tau \wedge \widetilde{T} \mid a, X)}{\pi_{0}(a \mid X)S_{0}(\tau \wedge \widetilde{T} \mid a, X)G_{0}(\tau \wedge \widetilde{T}- \mid a, X)} \\
&+\frac{I(A=a)\Delta(\tau)\min(\widetilde{T}, \tau)}{\pi_{0}(a \mid X)G_{0}(\tau \wedge \widetilde{T}- \mid a, X)}+\frac{I(A=a)}{\pi_{0}(a \mid X)}\int_0^{\tau \wedge \widetilde{T}} \frac{t \mathrm{d} M_{0}^{C}(t \mid a, X)}{G_{0}(t- \mid a, X)} \\
& -\frac{I(A=a)}{\pi_{0}(a \mid X)}\int_0^{\tau \wedge \widetilde{T}} \frac{\mathrm{RMST}_{0}(t \mid a, X)\mathrm{d} M_{0}^{C}(t \mid a, X)}{S_{0}(t \mid a, X)G_{0}(t- \mid a, X)} \\
=&\frac{I(A=a) \Delta(\tau) \min(\widetilde{T}, \tau)}{\pi_{0}(a \mid X) G_{0}(\tau \wedge \widetilde{T}- \mid a, X)}+\left\{1-\frac{I(A=a)}{\pi_{0}(a \mid X)}\right\}\mathrm{RMST}_{0}(\tau \mid a, X)+\frac{I(A=a)}{\pi_{0}(a \mid X)} \\
&\times\int_{0}^{\tau \wedge \widetilde{T}} \frac{\mathrm{d} M_{0}^{C}(t \mid a, X)}{G_{0}(t- \mid a, X)}\left\{t+\frac{\mathrm{RMST}_{0}(\tau \mid a, X)-\mathrm{RMST}_{0}(t \mid a, X)}{S_{0}(t \mid a, X)}\right\} \\
=&\phi_{0}^{\mathrm{RMST}, h}(\tau, a; \eta_{0}^{\Lambda})
\end{aligned}
$$

Note that $\mathrm{RMST}_{0}(\tau \wedge \widetilde{T} \mid a, X) \neq \mathrm{RMST}_{0}(\tau \mid a, X)$ and
$$
\begin{aligned}
&-\Delta I(\widetilde{T} \leq \tau)\mathrm{RMST}_{0}(\tau \mid a, X)+\Delta(\tau)\mathrm{RMST}_{0}(\tau \mid a, X)+\mathrm{RMST}(\tau \wedge \widetilde{T} \mid a, X)-\Delta(\tau)\mathrm{RMST}_{0}(\tau \wedge \widetilde{T} \mid a, X) \\
=&-\Delta I(\widetilde{T} \leq \tau)\mathrm{RMST}_{0}(\tau \mid a, X)+\Delta I(\widetilde{T} \leq \tau)\mathrm{RMST}_{0}(\tau \mid a, X)+I(\widetilde{T} > \tau)\mathrm{RMST}_{0}(\tau \mid a, X) \\
&+\Delta I(\widetilde{T} \leq \tau)\mathrm{RMST}_{0}(\tau \wedge \widetilde{T} \mid a, X)-\Delta I(\widetilde{T} \leq \tau)\mathrm{RMST}_{0}(\tau \wedge \widetilde{T} \mid a, X)-I(\widetilde{T} > \tau)\mathrm{RMST}_{0}(\tau \wedge \widetilde{T} \mid a, X)=0
\end{aligned}
$$

% \subsection{Derivation of $\phi_j^{h, sp}(\tau, a)=\phi_j^{h, np}(\tau, a)$}

In the competing risks setting, we transform several conditional event martingale terms
$$
\begin{aligned}
&\int_0^{\tau \wedge \widetilde{T}} \frac{(\tau-t)\mathrm{d}M_{j, 0}(t \mid a, X)}{G_{0}(t- \mid a, X)}=\tau \int_0^{\tau} \frac{\mathrm{d}M_{j, 0}(t \mid a, X)}{G_{0}(t- \mid a, X)}-\int_0^{\tau} \frac{t\mathrm{d}M_{j, 0}(t \mid a, X)}{G_{0}(t- \mid a, X)} \\
=&\frac{\tau I(\widetilde{T} \leq \tau, \widetilde{J}=j)}{G_{0}(\widetilde{T}- \mid a, X)}-\tau \int_0^{\tau \wedge \widetilde{T}} \frac{\mathrm{d}\Lambda_{j, 0}(t \mid a, X)}{G_{0}(t- \mid a, X)}-\frac{\widetilde{T} I(\widetilde{T} \leq \tau, \widetilde{J}=j)}{G_{0}(\widetilde{T}- \mid a, X)}+\int_0^{\tau \wedge \widetilde{T}} \frac{t \mathrm{d}\Lambda_{j, 0}(t \mid a, X)}{G_{0}(t- \mid a, X)} \\
=&\frac{\{\tau-\min(\widetilde{T}, \tau)\}I(\widetilde{J}=j)}{G_{0}(\widetilde{T}- \mid a, X)}-\tau \int_0^{\tau \wedge \widetilde{T}} \frac{\mathrm{d}\Lambda_{j, 0}(t \mid a, X)}{G_{0}(t- \mid a, X)}+\int_0^{\tau \wedge \widetilde{T}} \frac{t \mathrm{d}\Lambda_{j, 0}(t \mid a, X)}{G_{0}(t- \mid a, X)}
\end{aligned}
$$

\noindent and
$$
\begin{aligned}
&\mathrm{d} \left\{\frac{\mathrm{RMTL}_{j, 0}(t \mid a, X)}{G_{0}(t- \mid a, X)}\right\}=\frac{F_{j, 0}(t \mid a, X)G_{0}(t- \mid a, X) \mathrm{d}t+\mathrm{RMTL}_{j, 0}(t \mid a, X)G_{0}(t- \mid a, X)\mathrm{d}\Lambda_{0}^C(t- \mid a, X)}{G_{0}^2(t- \mid a, X)} \\
=&\frac{F_{j, 0}(t \mid a, X)\mathrm{d}t}{G_{0}(t- \mid a, X)}+\frac{\mathrm{RMTL}_{j ,0}(t \mid a, X)\mathrm{d}\Lambda_{0}^C(t \mid a, X)}{G_{0}(t- \mid a, X)}
\end{aligned}
$$

\noindent and
$$
\begin{aligned}
&\int_0^{\tau \wedge \widetilde{T}}\frac{\mathrm{RMTL}_{j, 0}(t \mid a, X) \mathrm{d}M_{0}(t \mid a, X)}{S_{0}(t \mid a, X)G_{0}(t- \mid a, X)} \\
=&\int_0^\tau \frac{\mathrm{RMTL}_{j, 0}(t \mid a, X) \mathrm{d}N(t)}{S_{0}(t \mid a, X)G_{0}(t- \mid a, X)}-\int_0^{\tau \wedge \widetilde{T}}\frac{\mathrm{RMTL}_{j, 0}(t \mid a, X) \mathrm{d}\Lambda_{0}(t \mid a, X)}{S_{0}(t \mid a, X)G_{0}(t- \mid a, X)} \\
=&\frac{\Delta I(\widetilde{T} \leq \tau) \mathrm{RMTL}_{j, 0}(\widetilde{T} \mid a, X)}{S_{0}(\widetilde{T} \mid a, X)G_{0}(\widetilde{T}- \mid a, X)}-\int_0^{\tau \wedge \widetilde{T}}\frac{\mathrm{RMTL}_{j, 0}(t \mid a, X)}{G_{0}(t- \mid a, X)} \mathrm{d}\left\{ \frac{1}{S_{0}(t \mid a, X)} \right\} \\
=&\frac{\Delta I(\widetilde{T} \leq \tau) \mathrm{RMTL}_{j, 0}(\tau \wedge \widetilde{T} \mid a, X)}{S_{0}(\tau \wedge \widetilde{T} \mid a, X)G_{0}(\tau \wedge \widetilde{T}- \mid a, X)}-\left. \frac{\mathrm{RMTL}_{j, 0}(t \mid a, X)}{S_{0}(t \mid a, X)G_{0}(t- \mid a, X)} \right|_0^{\tau \wedge \widetilde{T}}+\int_0^{\tau \wedge \widetilde{T}}\frac{1}{S_{0}(t \mid a, X)} \mathrm{d} \left\{\frac{\mathrm{RMTL}_{j, 0}(t \mid a, X)}{G_{0}(t- \mid a, X)}\right\} \\
=&\frac{\Delta I(\widetilde{T} \leq \tau) \mathrm{RMTL}_{j, 0}(\tau \wedge \widetilde{T} \mid a, X)}{S_{0}(\tau \wedge \widetilde{T} \mid a, X)G_{0}(\tau \wedge \widetilde{T}- \mid a, X)}-\frac{\mathrm{RMTL}_{j, 0}(\tau \wedge \widetilde{T} \mid a, X)}{S_{0}(\tau \wedge \widetilde{T} \mid a, X)G_{0}(\tau \wedge \widetilde{T}- \mid a, X)}+\int_0^{\tau \wedge \widetilde{T}}\frac{F_{j, 0}(t \mid a, X)\mathrm{d}t}{S_{0}(t \mid a, X)G_{0}(t- \mid a, X)} \\
&+\int_0^{\tau \wedge \widetilde{T}}\frac{\mathrm{RMTL}_{j, 0}(t \mid a, X)\mathrm{d}\Lambda_{0}^C(t \mid a, X)}{S_{0}(t \mid a, X)G_{0}(t- \mid a, X)} \\
=&\frac{\Delta I(\widetilde{T} \leq \tau) \mathrm{RMTL}_{j, 0}(\tau \wedge \widetilde{T} \mid a, X)}{S_{0}(\tau \wedge \widetilde{T} \mid a, X)G_{0}(\tau \wedge \widetilde{T}- \mid a, X)}-\frac{\mathrm{RMTL}_{j, 0}(\tau \wedge \widetilde{T} \mid a, X)}{S_{0}(\tau \wedge \widetilde{T} \mid a, X)G_{0}(\tau \wedge \widetilde{T}- \mid a, X)}+\int_0^{\tau \wedge \widetilde{T}}\frac{F_{j, 0}(t \mid a, X)\mathrm{d}t}{S_{0}(t \mid a, X)G_{0}(t- \mid a, X)} \\
&-\int_0^{\tau \wedge \widetilde{T}}\frac{\mathrm{RMTL}_{j, 0}(t \mid a, X)\mathrm{d}M_{0}^C(t \mid a, X)}{S_{0}(t \mid a, X)G_{0}(t- \mid a, X)}+\frac{(1-\Delta)I(\widetilde{T} \leq \tau) \mathrm{RMTL}_{j, 0}(\tau \wedge \widetilde{T} \mid a, X)}{S_{0}(\tau \wedge \widetilde{T} \mid a, X)G_{0}(\tau \wedge \widetilde{T}- \mid a, X)} \\
=&-\frac{I(\widetilde{T} > \tau) \mathrm{RMTL}_{j, 0}(\tau \wedge \widetilde{T} \mid a, X)}{S_{0}(\tau \wedge \widetilde{T} \mid a, X)G_{0}(\tau \wedge \widetilde{T}- \mid a, X)}+\int_0^{\tau \wedge \widetilde{T}}\frac{F_{j, 0}(t \mid a, X)\mathrm{d}t}{S_{0}(t \mid a, X)G_{0}(t- \mid a, X)} \\
&-\int_0^{\tau \wedge \widetilde{T}}\frac{\mathrm{RMTL}_{j, 0}(t \mid a, X)\mathrm{d}M_{0}^C(t \mid a, X)}{S_{0}(t \mid a, X)G_{0}(t- \mid a, X)}
\end{aligned}
$$

\noindent and
$$
\begin{aligned}
&\int_0^{\tau \wedge \widetilde{T}}\frac{t F_{j, 0}(t \mid a, X)\mathrm{d}M_{0}(t \mid a, X)}{S_{0}(t \mid a, X)G_{0}(t- \mid a, X)} \\
=&\int_0^{\tau \wedge \widetilde{T}}\frac{t F_{j, 0}(t \mid a, X)\mathrm{d}N(t)}{S_{0}(t \mid a, X)G_{0}(t- \mid a, X)}-\int_0^{\tau \wedge \widetilde{T}}\frac{t F_{j, 0}(t \mid a, X)}{G_{0}(t- \mid a, X)} \mathrm{d}\left\{ \frac{1}{S_{0}(t \mid a, X)} \right\} \\
=&\frac{\Delta I(\widetilde{T} \leq \tau) (\tau \wedge \widetilde{T}) F_{j, 0}(\tau \wedge \widetilde{T} \mid a, X)}{S_{0}(\tau \wedge \widetilde{T} \mid a, X)G_{0}(\tau \wedge \widetilde{T}- \mid a, X)}-\left.\frac{t F_{j, 0}(t \mid a, X)}{S_{0}(t \mid a, X)G_{0}(t- \mid a, X)}\right|_0^{\tau \wedge \widetilde{T}}+\int_0^{\tau \wedge \widetilde{T}}\frac{1}{S_{0}(t \mid a, X)} \mathrm{d}\left\{\frac{t F_{j, 0}(t \mid a, X)}{G_{j, 0}(t- \mid a, X)}\right\} \\
=&\frac{\Delta I(\widetilde{T} \leq \tau) (\tau \wedge \widetilde{T}) F_{j, 0}(\tau \wedge \widetilde{T} \mid a, X)}{S_{0}(\tau \wedge \widetilde{T} \mid a, X)G_{0}(\tau \wedge \widetilde{T}- \mid a, X)}-\frac{(\tau \wedge \widetilde{T}) F_{j, 0}(\tau \wedge \widetilde{T} \mid a, X)}{S_{0}(\tau \wedge \widetilde{T} \mid a, X)G_{0}(\tau \wedge \widetilde{T}- \mid a, X)} \\
&+\int_0^{\tau \wedge \widetilde{T}}\frac{1}{S_{0}(t \mid a, X)} \mathrm{d}\left\{\frac{t F_{j, 0}(t \mid a, X)}{G_{0}(t- \mid a, X)}\right\} \\
=&\frac{\Delta I(\widetilde{T} \leq \tau) (\tau \wedge \widetilde{T}) F_{j, 0}(\tau \wedge \widetilde{T} \mid a, X)}{S_{0}(\tau \wedge \widetilde{T} \mid a, X)G_{0}(\tau \wedge \widetilde{T}- \mid a, X)}-\frac{(\tau \wedge \widetilde{T}) F_{j, 0}(\tau \wedge \widetilde{T} \mid a, X)}{S_{0}(\tau \wedge \widetilde{T} \mid a, X)G_{0}(\tau \wedge \widetilde{T}- \mid a, X)}+\int_0^{\tau \wedge \widetilde{T}}\frac{F_{j, 0}(t \mid a, X)\mathrm{d}t}{S_{0}(t \mid a, X)G_{0}(t- \mid a, X)} \\
&+\int_0^{\tau \wedge \widetilde{T}}\frac{t \mathrm{d}\Lambda_{j, 0}(t \mid a, X)}{G_{0}(t- \mid a, X)}-\int_0^{\tau \wedge \widetilde{T}}\frac{t F_{j, 0}(t \mid a, X)\mathrm{d}M_{0}^C(t- \mid a, X)}{S_{0}(t \mid a, X)G_{0}(t- \mid a, X)}+\frac{(1-\Delta)I(\widetilde{T} \leq \tau)(\tau \wedge \widetilde{T}) F_{j, 0}(\tau \wedge \widetilde{T} \mid a, X)}{S_{0}(\tau \wedge \widetilde{T} \mid a, X)G_{0}(\tau \wedge \widetilde{T}- \mid a, X)} \\
=&-\frac{I(\widetilde{T} > \tau)(\tau \wedge \widetilde{T}) F_{j, 0}(\tau \wedge \widetilde{T} \mid a, X)}{S_{0}(\tau \wedge \widetilde{T} \mid a, X)G_{0}(\tau \wedge \widetilde{T}- \mid a, X)}+\int_0^{\tau \wedge \widetilde{T}}\frac{F_{j, 0}(t \mid a, X)\mathrm{d}t}{S_{0}(t \mid a, X)G_{0}(t- \mid a, X)}+\int_0^{\tau \wedge \widetilde{T}}\frac{t \mathrm{d}\Lambda_{j, 0}(t \mid a, X)}{G_{0}(t- \mid a, X)} \\
&-\int_0^{\tau \wedge \widetilde{T}}\frac{t F_{j, 0}(t \mid a, X)\mathrm{d}M_{0}^C(t- \mid a, X)}{S_{0}(t \mid a, X)G_{0}(t- \mid a, X)}
\end{aligned}
$$

\noindent and
$$
\begin{aligned}
&\mathrm{d}\left\{\frac{t F_{j, 0}(t \mid a, X)}{G_{0}(t- \mid a, X)}\right\} \\
=&\frac{G_{0}(t- \mid a, X)\{F_{j, 0}(t \mid a, X) + t S_{0}(t \mid a, X) \mathrm{d}\Lambda_{j, 0}(t \mid a, X)\} + t F_{j, 0}(t \mid a, X)G_{0}(t- \mid a, X)\mathrm{d}\Lambda_{0}^C(t- \mid a, X)}{G_{0}^2(t- \mid a, X)} \\
=&\frac{F_{j, 0}(t \mid a, X)}{G_{0}(t- \mid a, X)}+\frac{t S_{0}(t \mid a, X) \mathrm{d}\Lambda_{j, 0}(t \mid a, X)}{G_{0}(t- \mid a, X)}+\frac{t F_{j, 0}(t \mid a, X)\mathrm{d}\Lambda_{0}^C(t- \mid a, X)}{G_{0}(t- \mid a, X)}
\end{aligned}
$$

Substitute all preparation equalities into the RHS of the second equation in Lemma~\ref{lemma1} gives, % estimating function of $\phi_{j}^{\mathrm{RMST}, h, np}(\tau, a)$,
$$
\begin{aligned}
&\frac{I(A=a)}{\pi_{0}(a \mid X)} \int_0^{\tau \wedge \widetilde{T}} \frac{(\tau-t)\mathrm{d}M_{j, 0}(t \mid a, X)}{G_{0}(t- \mid a, X)}-\frac{I(A=a)}{\pi_{0}(a \mid X)} \\
&\times \int_0^{\tau \wedge \widetilde{T}}\frac{\{\mathrm{RMTL}_{j, 0}(\tau \mid a, X)-\mathrm{RMTL}_{j, 0}(t \mid a, X)\}\mathrm{d}M_{0}(t \mid a, X)}{S_{0}(t \mid a, X)G_{0}(t- \mid a, X)}+\frac{I(A=a)}{\pi_{0}(a \mid X)} \\
&\times \int_0^{\tau \wedge \widetilde{T}}\frac{(\tau-t)F_{j, 0}(t \mid a, X)\mathrm{d}M_{0}(t \mid a, X)}{S_{0}(t \mid a, X)G_{0}(t- \mid a, X)}+\mathrm{RMTL}_{j, 0}(\tau \mid a, X)
\end{aligned}
$$

$$
\begin{aligned}
=&\frac{I(A=a)\{\tau-\min(\widetilde{T}, \tau)\}I(\widetilde{J}=j)}{\pi_{0}(a \mid X)G_{0}(\widetilde{T}- \mid a, X)}-\frac{I(A=a)\tau}{\pi_{0}(a \mid X)}\int_0^{\tau \wedge \widetilde{T}} \frac{\mathrm{d}\Lambda_{j, 0}(t \mid a, X)}{G_{0}(t- \mid a, X)} \\
&+\frac{I(A=a)}{\pi_{0}(a \mid X)}\int_0^{\tau \wedge \widetilde{T}} \frac{t \mathrm{d}\Lambda_{j, 0}(t \mid a, X)}{G_{0}(t- \mid a, X)}+\frac{I(A=a)I(\widetilde{T} > \tau)\mathrm{RMTL}_{j, 0}(\tau \mid a, X)}{\pi_{0}(a \mid X)S_{0}(\tau \wedge \widetilde{T}- \mid a, X)G_{0}(\tau \wedge \widetilde{T}- \mid a, X)} \\
&-\frac{I(A=a)\mathrm{RMTL}_{j, 0}(\tau \mid a, X)}{\pi_{0}(a \mid X)}+\frac{I(A=a)\mathrm{RMTL}_{j, 0}(\tau \mid a, X)}{\pi_{0}(a \mid X)} \\
&\times \int_0^{\tau \wedge \widetilde{T}} \frac{\mathrm{d} M_{0}^C(t \mid a, X)}{S_{0}(t \mid a, X)G_{0}(t- \mid a, X)}-\frac{I(A=a)I(\widetilde{T} > \tau) \mathrm{RMTL}_{j, 0}(\tau \wedge \widetilde{T} \mid a, X)}{\pi_{0}(a \mid X)S_{0}(\tau \wedge \widetilde{T} \mid a, X)G_{0}(\tau \wedge \widetilde{T}- \mid a, X)} \\
&+\frac{I(A=a)}{\pi_{0}(a \mid X)}\int_0^{\tau \wedge \widetilde{T}}\frac{F_{j, 0}(t \mid a, X)\mathrm{d}t}{S_{0}(t \mid a, X)G_{0}(t- \mid a, X)}-\frac{I(A=a)}{\pi_{0}(a \mid X)} \\
&\times \int_0^{\tau \wedge \widetilde{T}}\frac{\mathrm{RMTL}_{j, 0}(t \mid a, X)\mathrm{d}M_{0}^C(t \mid a, X)}{S_{0}(t \mid a, X)G_{0}(t- \mid a, X)}-\frac{I(A=a)\tau I(\widetilde{T} > \tau) F_{j, 0}(\tau \mid a, X)}{\pi_{0}(a \mid X)S_{0}(\tau \wedge \widetilde{T} \mid a, X) G_{0}(\tau \wedge \widetilde{T}- \mid a, X)} \\
&+\frac{I(A=a)\tau}{\pi_{0}(a \mid X)}\int_{0}^{\tau \wedge \widetilde{T}} \frac{\mathrm{d} \Lambda_{j, 0}(t \mid a, X)}{G_{0}(t- \mid a, X)}-\frac{I(A=a)\tau}{\pi_{0}(a \mid X)}\int_{0}^{\tau \wedge \widetilde{T}} \frac{F_{j, 0}(t \mid a, X)\mathrm{d} M_{0}^{C}(t \mid a, X)}{S_{0}(t \mid a, X)G_{0}(t- \mid a, X)} \\
&+\frac{I(A=a)I(\widetilde{T} > \tau)(\tau \wedge \widetilde{T}) F_{j, 0}(\tau \wedge \widetilde{T} \mid a, X)}{\pi_{0}(a \mid X)S_{0}(\tau \wedge \widetilde{T} \mid a, X)G_{0}(\tau \wedge \widetilde{T}- \mid a, X)}-\frac{I(A=a)}{\pi_{0}(a \mid X)}\int_0^{\tau \wedge \widetilde{T}}\frac{F_{j, 0}(t \mid a, X)\mathrm{d}t}{S_{0}(t \mid a, X)G_{0}(t- \mid a, X)} \\
&-\frac{I(A=a)}{\pi_{0}(a \mid X)}\int_0^{\tau \wedge \widetilde{T}}\frac{t \mathrm{d}\Lambda_{j, 0}(t \mid a, X)}{G_{0}(t- \mid a, X)}+\frac{I(A=a)}{\pi_{0}(a \mid X)}\int_0^{\tau \wedge \widetilde{T}}\frac{t F_{j, 0}(t \mid a, X)\mathrm{d}M_{0}^C(t- \mid a, X)}{S_{0}(t \mid a, X)G_{0}(t- \mid a, X)}\\
&+\mathrm{RMTL}_{j, 0}(\tau \mid a, X) \\
=&\frac{I(A=a) \{\tau-\min(\widetilde{T}, \tau)\}I(\widetilde{J}=j)}{\pi_{0}(a \mid X) G_{0}(\widetilde{T}- \mid a, X)}+\left\{1-\frac{I(A=a)}{\pi_{0}(a \mid X)}\right\} \mathrm{RMTL}_{j, 0}(\tau \mid a, X)+\frac{I(A=a)}{\pi_{0}(a \mid X)} \\
&\times \int_{0}^{\tau \wedge \widetilde{T}} \frac{\{(t-\tau) F_{j, 0}(t \mid a, X)+\mathrm{RMTL}_{j, 0}(\tau \mid a, X)-\mathrm{RMTL}_{j, 0}(t \mid a, X)\} \mathrm{d} M_{0}^{C}(t \mid a, X)}{S_{0}(t \mid a, X)G_{0}(t- \mid a, X)} \\
=&\phi_{j, 0}^{\mathrm{RMTL}, h}(\tau, a; \eta_{j, 0}^{\Lambda})
\end{aligned}
$$

% \section{Appendix }
\subsection{Lemma~\ref{lemma2}}

\begin{lemma} \label{lemma2}
$$
\begin{aligned}
&P_{0} \{\widehat{\phi}^{\mathrm{RMST}, h}(\tau, a; \widehat{\eta}^{\Lambda, h})\}-\psi_{0}^{\mathrm{RMST}, h}(\tau, a) \\
=&[P_n\{\widehat{w}^{h}(a, X)\}]^{-1} \left(E_{0}\left[\widehat{h}(X) \mathrm{R}\widehat{\mathrm{MS}}\mathrm{T}(\tau \mid a, X) \int_0^{\tau} \frac{S_{0}	(t \mid a, X)}{\widehat{S}(t \mid a, X)} \left\{\frac{\pi_{0}(a \mid X) G_{0}(t- \mid a, X)}{\widehat{\pi}(a \mid X) \widehat{G}(t- \mid a, X)}-1 \right\} \right.\right. \\
&\left.\times \{\mathrm{d}\widehat{\Lambda}(t \mid a, X)-\mathrm{d}\Lambda_{0}(t \mid a, X)\} \right] \\
&\left.-E_{0}\left[\widehat{h}(X) \int_0^{\tau} \frac{\mathrm{R}\widehat{\mathrm{MS}}\mathrm{T}(t \mid a, X) S_{0}(t \mid a, X)}{\widehat{S}(t \mid a, X)} \left\{\frac{\pi_{0}(a \mid X) G_{0}(t- \mid a, X)}{\widehat{\pi}(a \mid X) \widehat{G}(t- \mid a, X)}-1 \right\} \{\mathrm{d}\widehat{\Lambda}(t \mid a, X)-\mathrm{d}\Lambda_{0}(t \mid a, X)\} \right] \right) \\
&+([P_n \{\widehat{h}(X)\}]^{-1}-[P_n\{\widehat{w}^{h}(a, X)\}]^{-1}) E_{0}\{\widehat{h}(X)\mathrm{R}\widehat{\mathrm{MS}}\mathrm{T}(\tau \mid a, X)\} \\
&+([P_n\{\widehat{w}^{h}(a, X)\}]^{-1}-[E_{0} \{h_{0}(X)\}]^{-1}) E_{0}\{\widehat{h}(X)\mathrm{RMST}_{0}(\tau \mid a, X)\} \\
&+[E \{h_{0}(x)\}]^{-1} E_{0}[\{\widehat{h}(X)-h_{0}(X)\}\mathrm{RMST}_{0}(\tau \mid a, X)]
\end{aligned}
$$

$$
\begin{aligned}
&P \{\widehat{\phi}_{j}^{\mathrm{RMTL}, h}(\tau, a; \widehat{\eta}_{j}^{\Lambda, h})\}-\psi_{j, 0}^{\mathrm{RMTL}, h}(\tau, a) \\
=&[P_n\{\widehat{w}^{h}(a, X)\}]^{-1} \left(-E_{0}\left[\widehat{h}(X) \tau\int_0^{\tau}S_{0}(t \mid a, X)\left\{\frac{\pi_{0}(a \mid X)G_{0}(t- \mid a, X)}{\widehat{\pi}(a \mid X)\widehat{G}(t- \mid a, X)}-1 \right\} \right.\right. \\
&\left.\times\{\mathrm{d}\widehat{\Lambda}_{j}(t \mid a, X)-\mathrm{d}\Lambda_{j, 0}(t \mid a, X)\}\right]+E_{0}\left[\widehat{h}(X) \int_0^{\tau} t S_{0}(t \mid a, X)\left\{\frac{\pi_{0}(a \mid X)G_{0}(t- \mid a, X)}{\widehat{\pi}(a \mid X)\widehat{G}(t- \mid a, X)}-1 \right\} \right. \\
&\left.\times\{\mathrm{d}\widehat{\Lambda}_{j}(t \mid a, X)-\mathrm{d}\Lambda_{j, 0}(t \mid a, X)\}\right]+E_{0}\left[\widehat{h}(X) \mathrm{R}\widehat{\mathrm{MT}}\mathrm{L}_{j}(\tau \mid a, X) \int_0^{\tau} \frac{S_{0}(t \mid a, X)}{\widehat{S}(t \mid a, X)} \right. \\
&\left.\times\left\{\frac{\pi(a \mid X) G_{0}(t- \mid a, X)}{\widehat{\pi}(a \mid X) \widehat{G}(t- \mid a, X)}-1 \right\} \{\mathrm{d}\widehat{\Lambda}(t \mid a, X)-\mathrm{d}\Lambda_{0}(t \mid a, X)\}  \right] \\
&-E_{0}\left[\widehat{h}(X) \int_0^{\tau} \frac{\mathrm{R}\widehat{\mathrm{MT}}\mathrm{L}_{j}(t \mid a, X) S_{0}(t \mid a, X)}{\widehat{S}(t \mid a, X)} \left\{\frac{\pi_{0}(a \mid X) G_{0}(t- \mid a, X)}{\widehat{\pi}(a \mid X) \widehat{G}(t- \mid a, X)}-1 \right\} \{\mathrm{d}\widehat{\Lambda}(t \mid a, X)-\mathrm{d}\Lambda_{0}(t \mid a, X)\} \right] \\
&-E_{0}\left[\widehat{h}(X) \tau\int_0^{\tau}\frac{\widehat{F}_{j}(t \mid a, X)S_{0}(t \mid a, X)}{\widehat{S}(t \mid a, X)}\left\{\frac{\pi_{0}(a \mid X)G_{0}(t- \mid a, X)}{\widehat{\pi}(a \mid X)\widehat{G}(t- \mid a, X)}-1 \right\}\{\mathrm{d}\widehat{\Lambda}(t \mid a, X)-\mathrm{d}\Lambda_{0}(t \mid a, X)\}\right] \\
&\left.+E_{0}\left[\widehat{h}(X) \int_0^{\tau}\frac{t\widehat{F}_{j}(t \mid a, X)S_{0}(t \mid a, X)}{\widehat{S}(t \mid a, X)}\left\{\frac{\pi_{0}(a \mid X)G_{0}(t- \mid a, X)}{\widehat{\pi}(a \mid X)\widehat{G}(t- \mid a, X)}-1 \right\}\{\mathrm{d}\widehat{\Lambda}(t \mid a, X)-\mathrm{d}\Lambda_{0}(t \mid a, X)\}\right]\right) \\
&+([P_n\{\widehat{h}(X)\}]^{-1}-[P_n\{\widehat{w}^{h}(a, X)\}]^{-1}) E_{0}\{\widehat{h}(X)\mathrm{R}\widehat{\mathrm{MT}}\mathrm{L}_{j}(\tau \mid a, X)\} \\
&+([P_n\{\widehat{w}^{h}(a, X)\}]^{-1}-[E_{0} \{h_{0}(X)\}]^{-1}) E_{0}\{\widehat{h}(X)\mathrm{RMTL}_{j, 0}(\tau \mid a, X)\} \\
&+[E\{h_{0}(X)\}]^{-1} E_{0}[\{\widehat{h}(X)-h_{0}(X)\}\mathrm{RMTL}_{j, 0}(\tau \mid a, X)]
\end{aligned}
$$
\end{lemma}

Proof of Lemma~\ref{lemma2}:
$$
\begin{aligned}
&P \{\widehat{\phi}^{\mathrm{RMST}, h}(\tau, a; \widehat{\eta}^{\Lambda, h})\}-\psi_{0}^{\mathrm{RMST}, h}(\tau, a) \\
=&[P_n\{\widehat{w}^{h}(a, X)\}]^{-1} E_{0}\left[\frac{\widehat{h}(X)\pi_{0}(a \mid x)}{\widehat{\pi}(a \mid X)} E_{0}\left\{-\mathrm{R}\widehat{\mathrm{MS}}\mathrm{T}(\tau \mid a, X) \int_0^{\tau \wedge \widetilde{T}} \frac{\mathrm{d} \widehat{M}(t \mid a, X)}{\widehat{S}(t \mid a, X)\widehat{G}(t- \mid a, X)}\right.\right. \\
&\left.\left.+\int_0^{\tau \wedge \widetilde{T}} \frac{\mathrm{R}\widehat{\mathrm{MS}}\mathrm{T}(t \mid a, X)\mathrm{d} \widehat{M}(t \mid a, X)}{\widehat{S}(t \mid a, X)\widehat{G}(t- \mid a, X)} \mid A=a, X \right\} \right] \\
&+[P_n \{\widehat{h}(X)\}]^{-1} E_{0}[\widehat{h}(X)\{\mathrm{R}\widehat{\mathrm{MS}}\mathrm{T}(\tau \mid a, X)\}]-[E_{0}\{h_{0}(X)\}]^{-1} E_{0}[h_{0}(X)\mathrm{RMST}_{0}(\tau \mid a, X)] \\
=&[P_n\{\widehat{w}^{h}(a, X)\}]^{-1} E_{0}\left[\frac{\widehat{h}(X)\pi_{0}(a \mid X)}{\widehat{\pi}(a \mid X)} E_{0}\left\{-\mathrm{R}\widehat{\mathrm{MS}}\mathrm{T}(\tau \mid a, X) \int_0^{\tau \wedge \widetilde{T}} \frac{\mathrm{d} \widehat{M}(t \mid a, X)}{\widehat{S}(t \mid a, X)\widehat{G}(t- \mid a, X)}\right.\right. \\
&\left.\left.+\int_0^{\tau \wedge \widetilde{T}} \frac{\mathrm{R}\widehat{\mathrm{MS}}\mathrm{T}(t \mid a, X)\mathrm{d} \widehat{M}(t \mid a, X)}{\widehat{S}(t \mid a, X)\widehat{G}(t- \mid a, X)} \mid A=a, X \right\}+\widehat{h}(X) \{\mathrm{R}\widehat{\mathrm{MS}}\mathrm{T}(\tau \mid a, X)-\mathrm{RMST}_{0}(\tau \mid a, X)\} \right] \\
&+[P_n \{\widehat{h}(X)\}]^{-1} E_{0}[\widehat{h}(X)\{\mathrm{R}\widehat{\mathrm{MS}}\mathrm{T}(\tau \mid a, X)\}]-[E_{0}\{h_{0}(X)\}]^{-1} E_{0}[h_{0}(X)\mathrm{RMST}_{0}(\tau \mid a, X)] \\
&+([P_n \{\widehat{h}(X)\}]^{-1}-[P_n\{\widehat{w}^{h}(a, X)\}]^{-1}) E_{0}\{\widehat{h}(X)\mathrm{R}\widehat{\mathrm{MS}}\mathrm{T}(\tau \mid a, X)\} \\
&+([P_n\{\widehat{w}^{h}(a, X)\}]^{-1}-[E_{0} \{h_{0}(X)\}]^{-1}) E_{0}\{\widehat{h}(X)\mathrm{RMST}_{0}(\tau \mid a, X)\} \\
&+[E_{0}\{h_{0}(X)\}]^{-1} E_{0}[\{\widehat{h}(X)-h_{0}(X)\}\mathrm{RMST}_{0}(\tau \mid a, X)]
\end{aligned}
$$

For the first conditional event martingale integral term,
$$
\begin{aligned}
&E_{0} \left\{ \int_0^{\tau \wedge \widetilde{T}} \frac{\mathrm{d} \widehat{M}(t \mid a, X)}{\widehat{S}(t \mid a, X)\widehat{G}(t- \mid a, X)} \mid A=a, X \right\} \\
=&-\int_0^{\tau} \frac{S_{0}(t- \mid a, X) G_{0}(t- \mid a, X) \{\mathrm{d}\widehat{\Lambda}(t \mid a, X)-\mathrm{d}\Lambda_{0}(t \mid a, X) \}}{\widehat{S}(t \mid a, X) \widehat{G}(t- \mid a, X)} % =-\int_0^{\tau} \frac{G(t- \mid a, X)}{\widehat{G}(t- \mid a, X)} \mathrm{d}\left\{\frac{S(t\mid a, X)}{\widehat{S}(t \mid a, X)}-1\right\}
\end{aligned}
$$

\noindent and the second conditional event martingale integral term
$$
\begin{aligned}
&E_{0} \left\{ \int_0^{\tau \wedge \widetilde{T}} \frac{\mathrm{R}\widehat{\mathrm{MS}}\mathrm{T}(t \mid a, X)\mathrm{d} \widehat{M}(t \mid a, X)}{\widehat{S}(t \mid a, X)\widehat{G}(t- \mid a, X)}\mid A=a, X\right\} \\
=&E_{0}\left\{ \frac{\Delta I(\widetilde{T} \leq \tau) \mathrm{R}\widehat{\mathrm{MS}}\mathrm{T}(\widetilde{T} \mid a, X)}{\widehat{S}(\widetilde{T} \mid a, X) \widehat{G}(\widetilde{T}- \mid a, X)}-\int_0^{\tau \wedge \widetilde{T}} \frac{\mathrm{R}\widehat{\mathrm{MS}}\mathrm{T}(t \mid a, X) \mathrm{d}\widehat{\Lambda}(t \mid a, X)}{\widehat{S}(t \mid a, X) \widehat{G}(t- \mid a, X)} \mid A=a, X\right\} \\
%=&\int_0^{\tau} \frac{S(t- \mid a, X) G(t- \mid a, X) \mathrm{RMST}_{0}(t \mid a, X) \mathrm{d}\Lambda(t \mid a, X)}{S_{0}\left(t \mid a, x\right) G_{0}\left(t- \mid a, X\right)}-\int_0^{\tau} \frac{S(t- \mid a, X) G(t- \mid a, X) \mathrm{RMST}_{0}(t \mid a, X) \mathrm{d}\Lambda_{0}\left(t \mid a, x\right)}{S_{0}\left(t \mid a, X\right) G_{0}\left(t- \mid a, X\right)} \\
=&-\int_0^{\tau} \frac{S_{0}(t- \mid a, X) G_{0}(t- \mid a, X) \mathrm{R}\widehat{\mathrm{MS}}\mathrm{T}(t \mid a, X) \{\mathrm{d}\widehat{\Lambda}(t \mid a, X)-\mathrm{d}\Lambda_{0}(t \mid a, X) \}}{\widehat{S}(t \mid a, X) \widehat{G}(t- \mid a, X)} \\
%=&-\int_0^{\tau} \frac{G(t- \mid a, X) \mathrm{R}\widehat{\mathrm{MS}}\mathrm{T}(t \mid a, X)}{\widehat{G}(t- \mid a, X)} \mathrm{d}\left\{\frac{S(t\mid a, X)}{\widehat{S}(t \mid a, X)}-1\right\}
\end{aligned}
$$

The Duhamel equation \citep{gill1990survey} gives
$$
S_{0}(t\mid a, X)-\widehat{S}(t \mid a, X)=\widehat{S}(t \mid a, X) \int_{0}^{t} \frac{S_{0}(u-\mid a, X)}{\widehat{S}(u \mid a, X)}\{\mathrm{d}\widehat{\Lambda}(u \mid a, X)-\mathrm{d} \Lambda_{0}(u \mid a, X)\}
$$

\noindent such that
$$
\begin{aligned}
&\mathrm{R}\widehat{\mathrm{MS}}\mathrm{T}(t \mid a, X)-\mathrm{RMST}_{0}(t \mid a, X)=\int_{0}^{\tau} \{\widehat{S}(t \mid a, X)-S_{0}(t \mid a, X) \} \mathrm{d}t \\
=&-\int_{0}^{\tau} \left[\widehat{S}(t \mid a, X) \int_0^t \frac{S_{0}(u-\mid a, X)}{\widehat{S}(u \mid a, X)}\{\mathrm{d}\widehat{\Lambda}(u \mid a, X)-\mathrm{d}\Lambda_{0}(u \mid a, X)\} \right] \mathrm{d}t \\
=&-\mathrm{R}\widehat{\mathrm{MS}}\mathrm{T}(\tau \mid a, X) \int_{0}^{\tau} \frac{S_{0}(t-\mid a, X)}{\widehat{S}(t \mid a, X)}\{\mathrm{d}\widehat{\Lambda}(t \mid a, X)-\mathrm{d}\Lambda_{0}(t \mid a, X)\} \\
&+ \int_{0}^{\tau} \frac{\mathrm{R}\widehat{\mathrm{MS}}\mathrm{T}(t \mid a, X) S_{0}(t-\mid a, X)}{\widehat{S}(t \mid a, X)}\{\mathrm{d}\widehat{\Lambda}(t \mid a, X)-\mathrm{d}\Lambda_{0}(t \mid a, X)\}
\end{aligned}
$$

Substitute the conditional event martingale integral terms and $\mathrm{R}\widehat{\mathrm{MS}}\mathrm{T}(t \mid a, X)-\mathrm{RMST}_{0}(t \mid a, X)$ into $P_{0} \{\widehat{\phi}^{\mathrm{RMST}, h}(\tau, a; \widehat{\eta}^{\Lambda, h})\}-\psi_{0}^{\mathrm{RMST}, h}(\tau, a)$ yields the equality as presented in Lemma~\ref{lemma2}.

Then, we consider the competing risks setting
$$
\begin{aligned}
&P_{0} \{\widehat{\phi}_{j}^{\mathrm{RMTL}, h}(\tau, a; \widehat{\eta}_{j}^{\Lambda, h})\}-\psi_{j, 0}^{\mathrm{RMTL}, h}(\tau, a) \\
=&[P_n\{\widehat{w}^{h}(a, X)\}]^{-1} E_{0}\left[\frac{\widehat{h}(X)\pi_{0}(a \mid X)}{\widehat{\pi}(a \mid X)} E_{0}\left\{\tau\int_0^{\tau \wedge \widetilde{T}} \frac{\mathrm{d}\widehat{M}_j(t \mid a, X)}{\widehat{G}(t- \mid a, X)}-\int_0^{\tau \wedge \widetilde{T}} \frac{t\mathrm{d}\widehat{M}_j(t \mid a, X)}{\widehat{G}(t- \mid a, X)} \right.\right. \\
&-\mathrm{R}\widehat{\mathrm{MT}}\mathrm{L}_{j}(\tau \mid a, X) \int_0^{\tau \wedge \widetilde{T}} \frac{\mathrm{d} \widehat{M}(t \mid a, X)}{\widehat{S}(t \mid a, X)\widehat{G}(t- \mid a, X)}+\int_0^{\tau \wedge \widetilde{T}} \frac{\mathrm{R}\widehat{\mathrm{MT}}\mathrm{L}_{j}(t \mid a, X)\mathrm{d} \widehat{M}(t \mid a, X)}{\widehat{S}(t \mid a, X)\widehat{G}(t- \mid a, X)} \\
&\left.\left. +\tau\int_0^{\tau \wedge \widetilde{T}}\frac{\widehat{F}_j(t \mid a, X)\mathrm{d}\widehat{M}(t \mid a, X)}{\widehat{S}(t \mid a, X)\widehat{G}(t- \mid a, X)}-\int_0^{\tau \wedge \widetilde{T}}\frac{t \widehat{F}_j(t \mid a, X)\mathrm{d}\widehat{M}(t \mid a, X)}{\widehat{S}(t \mid a, X)\widehat{G}(t- \mid a, X)} \mid A=a, X \right\} \right] \\
&+[P_n \{\widehat{h}(X)\}]^{-1} E [\widehat{h}(X)\{\mathrm{R}\widehat{\mathrm{MT}}\mathrm{L}_{j}(\tau \mid a, X)\}]-[E_{0} \{h_{0}(X)\}]^{-1} E_{0}[h_{0}(X)\mathrm{RMTL}_{j, 0}(\tau \mid a, X)]
\end{aligned}
$$

$$
\begin{aligned}
=&[P_n\{\widehat{w}^{h}(a, X)\}]^{-1} E_{0}\left[\frac{\widehat{h}(X)\pi_{0}(a \mid X)}{\widehat{\pi}(a \mid X)} E_{0}\left\{\tau\int_0^{\tau \wedge \widetilde{T}} \frac{\mathrm{d}\widehat{M}_j(t \mid a, X)}{\widehat{G}(t- \mid a, X)}-\int_0^{\tau \wedge \widetilde{T}} \frac{t\mathrm{d}\widehat{M}_j(t \mid a, X)}{\widehat{G}(t- \mid a, X)} \right.\right. \\
&-\mathrm{R}\widehat{\mathrm{MT}}\mathrm{L}_{j}(\tau \mid a, X) \int_0^{\tau \wedge \widetilde{T}} \frac{\mathrm{d} \widehat{M}(t \mid a, X)}{\widehat{S}(t \mid a, X)\widehat{G}(t- \mid a, X)}+\int_0^{\tau \wedge \widetilde{T}} \frac{\mathrm{R}\widehat{\mathrm{MT}}\mathrm{L}_{j}(t \mid a, X)\mathrm{d} \widehat{M}(t \mid a, X)}{\widehat{S}(t \mid a, X)\widehat{G}(t- \mid a, X)} \\
&\left.\left. +\tau\int_0^{\tau \wedge \widetilde{T}}\frac{\widehat{F}_j(t \mid a, X)\mathrm{d}\widehat{M}(t \mid a, X)}{\widehat{S}(t \mid a, X)\widehat{G}(t- \mid a, X)}-\int_0^{\tau \wedge \widetilde{T}}\frac{t \widehat{F}_j(t \mid a, X)\mathrm{d}\widehat{M}(t \mid a, X)}{\widehat{S}(t \mid a, X)\widehat{G}(t- \mid a, X)} \mid A=a, X \right\} \right. \\
&\left.+\widehat{h}(X) \{\mathrm{R}\widehat{\mathrm{MS}}\mathrm{T}(\tau \mid a, X)-\mathrm{RMST}(\tau \mid a, X)\} \right] \\
&+([P_n \{\widehat{h}(X)\}]^{-1}-[P_n\{\widehat{w}^{h}(a, X)\}]^{-1}) E_{0}\{\widehat{h}(X)\mathrm{R}\widehat{\mathrm{MT}}\mathrm{L}_{j}(\tau \mid a, X)\} \\
&+([P_n\{\widehat{w}^{h}(a, X)\}]^{-1}-[E_{0} \{h_{0}(X)\}]^{-1}) E_{0}\{\widehat{h}(X)\mathrm{RMTL}_{j, 0}(\tau \mid a, X)\} \\
&+[E_{0} \{h(X)\}]^{-1} E_{0}[\{\widehat{h}(X)-h_{0}(X)\}\mathrm{RMTL}_{j, 0}(\tau \mid a, X)]
\end{aligned}
$$

We derive equalities on conditional cause-specific martingale integrals
$$
\begin{aligned}
&E_{0} \left\{ \int_0^{\tau \wedge \widetilde{T}} \frac{\mathrm{d}\widehat{M}_{j}(t \mid a, X)}{\widehat{G}(t- \mid a, X)} \mid A=a, X\right\}=-\int_0^{\tau} \frac{S_{0}(t \mid a, X) G_{0}(t- \mid a, X) \{\mathrm{d}\widehat{\Lambda}_{j}\left(t \mid a, X\right)-\mathrm{d}\Lambda_{j, 0}\left(t \mid a, X\right) \}}{\widehat{G}(t- \mid a, X)} \\
&E_{0} \left\{ \int_0^{\tau \wedge \widetilde{T}} \frac{t \mathrm{d}\widehat{M}_{j}(t \mid a, X)}{\widehat{G}(t- \mid a, X)} \mid A=a, X\right\}=-\int_0^{\tau} \frac{t S_{0}(t \mid a, X) G_{0}(t- \mid a, X) \{\mathrm{d}\widehat{\Lambda}_{j}\left(t \mid a, X\right)-\mathrm{d}\Lambda_{j, 0}\left(t \mid a, X\right) \}}{\widehat{G}(t- \mid a, X)}
\end{aligned}
$$

Similar for the conditional event martingale integrals,
$$
\begin{aligned}
&E_{0} \left\{ \int_0^{\tau \wedge \widetilde{T}} \frac{\mathrm{R}\widehat{\mathrm{MT}}\mathrm{L}_{j}(t \mid a, X)\mathrm{d} \widehat{M}(t \mid a, X)}{\widehat{S}(t \mid a, X)\widehat{G}(t- \mid a, X)}\mid A=a, X\right\} \\
=&-\int_0^{\tau} \frac{S_{0}(t \mid a, X) G_{0}(t- \mid a, X) \mathrm{R}\widehat{\mathrm{MT}}\mathrm{L}_{j}(t \mid a, X) \{\mathrm{d}\widehat{\Lambda}(t \mid a, X)-\mathrm{d}\Lambda_{0}(t \mid a, X) \}}{\widehat{S}(t \mid a, X) \widehat{G}(t- \mid a, X)} \\
&E_{0} \left\{ \int_0^{\tau \wedge \widetilde{T}}\frac{\widehat{F}_{j}(t \mid a, X)\mathrm{d}\widehat{M}(t \mid a, X)}{\widehat{S}(t \mid a, X)\widehat{G}(t- \mid a, X)} \mid A=a, X\right\} \\
=&-\int_0^{\tau} \frac{S_{0}(t \mid a, X) G_{0}(t- \mid a, X) \widehat{F}_{j}(t \mid a, X) \{\mathrm{d}\widehat{\Lambda}(t \mid a, X)-\mathrm{d}\Lambda_{0}(t \mid a, X) \}}{\widehat{S}(t \mid a, X) \widehat{G}(t- \mid a, X)} \\
&E_{0} \left\{ \int_0^{\tau \wedge \widetilde{T}}\frac{t \widehat{F}_{j}(t \mid a, X)\mathrm{d}\widehat{M}(t \mid a, X)}{\widehat{S}(t \mid a, X)\widehat{G}(t- \mid a, X)} \mid A=a, X\right\} \\
=&-\int_0^{\tau} \frac{t S_{0}(t \mid a, X) G_{0}(t- \mid a, X) \widehat{F}_{j}(t \mid a, X) \{\mathrm{d}\widehat{\Lambda}(t \mid a, X)-\mathrm{d}\Lambda_{0}(t \mid a, X) \}}{\widehat{S}(t \mid a, X) \widehat{G}(t- \mid a, X)}
\end{aligned}
$$

We wish to derive $\mathrm{R}\widehat{\mathrm{MT}}\mathrm{L}_{j, 0}(\tau \mid a, X)-\mathrm{RMTL}_{j}(\tau \mid a, X)$ in terms of conditional hazards differences. To begin with, we consider
$$
\begin{aligned}
&\widehat{F}_{j}(t \mid a, X)-F_{j, 0}(t \mid a, X)=\int_{0}^{t} \widehat{S}(u- \mid a, X) \mathrm{d}\widehat{\Lambda}_{j}(u \mid a, X)-\int_{0}^{t} S_{0}(u- \mid a, X) \mathrm{d}\Lambda_{j, 0}(u \mid a, X) \\
% =&\int_{0}^{t} \widehat{S}(u- \mid a, x) \mathrm{d}\widehat{\Lambda}_{j}(u \mid a, x)-\int_{0}^{t} S(u- \mid a, x) \mathrm{d}\widehat{\Lambda}_{j}(u \mid a, x)+\int_{0}^{t} S(u- \mid a, x) \mathrm{d}\widehat{\Lambda}_{j}(u \mid a, x) \\
% &-\int_{0}^{t} S(u- \mid a, x) \mathrm{d}\Lambda_{j}(u \mid a, x) \\
=&\int_{0}^{t} \{\widehat{S}(u- \mid a, X)-S_{0}(u- \mid a, X)\} \mathrm{d}\widehat{\Lambda}_{j}(u \mid a, X)+\int_{0}^{t} S_{0}(u- \mid a, X) \{\mathrm{d}\widehat{\Lambda}_{j}(u \mid a, X)-\mathrm{d}\Lambda_{j, 0}(u \mid a, X)\} \\
=&\int_{0}^{t} \left[-\widehat{S}(u- \mid a, X) \int_{0}^{u} \frac{S_{0}(s- \mid a, X)}{\widehat{S}(s- \mid a, X)} \{\mathrm{d}\widehat{\Lambda}(s- \mid a, X)-\mathrm{d}\Lambda_{0}(s- \mid a, X)\} \right] \mathrm{d}\widehat{\Lambda}_{j}(u \mid a, X) \\
&+\int_{0}^{t} S_{0}(u- \mid a, X) \{\mathrm{d}\widehat{\Lambda}_{j}(u \mid a, X)-\mathrm{d}\Lambda_{j, 0}(u \mid a, X)\} \\
=&\int_{0}^{t} \left[- \int_{0}^{u} \frac{S_{0}(s- \mid a, X)}{\widehat{S}(s- \mid a, X)} \{\mathrm{d}\widehat{\Lambda}(s \mid a, X)-\mathrm{d}\Lambda_{0}(s \mid a, X)\} \right] \mathrm{d} \widehat{F}_{j}(u \mid a, X) \\
&+\int_{0}^{t} S_{0}(u- \mid a, X) \{\mathrm{d}\widehat{\Lambda}_{j}(u \mid a, X)-\mathrm{d}\Lambda_{j, 0}(u \mid a, X)\} \\
=&-\widehat{F}_{j}(t \mid a, X) \int_{0}^{t} \frac{S_{0}(u- \mid a, X)}{\widehat{S}(u- \mid a, X)} \{\mathrm{d}\widehat{\Lambda}(u \mid a, X)-\mathrm{d}\Lambda_{0}(u \mid a, X)\}+\int_{0}^{t} \frac{\widehat{F}_{j}(u \mid a, X) S_{0}(u- \mid a, X)}{\widehat{S}(u- \mid a, X)} \\
&\times \{\mathrm{d}\widehat{\Lambda}(u \mid a, X)-\mathrm{d}\Lambda_{0}(u \mid a, X)\}+\int_{0}^{t} S_{0}(u- \mid a, X) \{\mathrm{d}\widehat{\Lambda}_{j}(u \mid a, X)-\mathrm{d}\Lambda_{j, 0}(u \mid a, X)\}
\end{aligned}
$$

Since the integral of the first term
$$
\begin{aligned}
&\int_{0}^{\tau} \left[ \widehat{F}_{j}(t \mid a, X) \int_{0}^{t} \frac{S_{0}(u- \mid a, X)}{\widehat{S}(u- \mid a, X)} \{\mathrm{d}\widehat{\Lambda}(u \mid a, X)-\mathrm{d}\Lambda_{0}(u \mid a, X)\} \right] \mathrm{d}t \\
=&\mathrm{R}\widehat{\mathrm{MT}}\mathrm{L}_{j}(\tau \mid a, X) \int_{0}^{\tau} \frac{S_{0}(t- \mid a, X)}{\widehat{S}(t- \mid a, X)} \{\mathrm{d}\widehat{\Lambda}(t \mid a, X)-\mathrm{d}\Lambda_{0}(t \mid a, X)\} \\
&-\int_{0}^{\tau} \frac{\mathrm{R}\widehat{\mathrm{MT}}\mathrm{L}_{j}(t \mid a, X)S_{0}(t- \mid a, X)}{\widehat{S}(t- \mid a, X)} \{\mathrm{d}\widehat{\Lambda}(t \mid a, X)-\mathrm{d}\Lambda_{0}(t \mid a, X)\}
\end{aligned}
$$

\noindent the integral of the second term
$$
\begin{aligned}
&\int_{0}^{\tau} \left[ \int_{0}^{t} \frac{\widehat{F}_{j}(u \mid a, X) S_{0}(u- \mid a, X)}{\widehat{S}(u- \mid a, X)} \{\mathrm{d}\widehat{\Lambda}(u \mid a, X)-\mathrm{d}\Lambda_{0}(u \mid a, X)\} \right] \mathrm{d}t \\
=&\tau \int_{0}^{\tau} \frac{\widehat{F}_{j}(t \mid a, X) S_{0}(t- \mid a, X)}{\widehat{S}(t- \mid a, X)} \{\mathrm{d}\widehat{\Lambda}(t \mid a, X)-\mathrm{d}\Lambda_{0}(t \mid a, X)\} \\
&-\int_{0}^{\tau} \frac{t \widehat{F}_{j}(t \mid a, X) S_{0}(t- \mid a, X)}{\widehat{S}(t- \mid a, X)} \{\mathrm{d}\widehat{\Lambda}(t \mid a, X)-\mathrm{d}\Lambda_{0}(t \mid a, X)\}
\end{aligned}
$$

\noindent the integral of the third term
$$
\begin{aligned}
&\int_{0}^{\tau} \left[ \int_{0}^{t} S_{0}(u- \mid a, X) \{\mathrm{d}\widehat{\Lambda}_{j}(u \mid a, X)-\mathrm{d}\Lambda_{j, 0}(u \mid a, X)\} \right] \mathrm{d}t \\
=&\int_{0}^{\tau} t S_{0}(t- \mid a, X) \{\mathrm{d}\Lambda_{j, 0}(t \mid a, X)-\mathrm{d}\widehat{\Lambda}_{j}(t \mid a, X)\}-\tau \int_{0}^{\tau} S_{0}(t- \mid a, X) \{\mathrm{d}\Lambda_{j, 0}(t \mid a, X)-\mathrm{d}\widehat{\Lambda}_{j}(t \mid a, X)\}
\end{aligned}
$$

To make use of the Duhamel equation, we also assume $\widehat{S}$ is continuous, i.e. $\widehat{S}(t- \mid a, X)=\widehat{S}(t \mid a, X)$, we have
$$
\begin{aligned}
&\mathrm{R}\widehat{\mathrm{MT}}\mathrm{L}_{j}(\tau \mid a, X)-\mathrm{RMTL}_{j, 0}(\tau \mid a, X)=\int_{0}^{\tau} \{\widehat{F}_{j}(t \mid a, X)-F_{j, 0}(t \mid a, X)\} \mathrm{d}t \\
=&-\int_{0}^{\tau} \frac{S_{0}(t- \mid a, X)\{\mathrm{R}\widehat{\mathrm{MT}}\mathrm{L}_{j}(\tau \mid a, X)-\mathrm{R}\widehat{\mathrm{MT}}\mathrm{L}_{j}(t \mid a, X)\}}{\widehat{S}(t \mid a, X)} \{\mathrm{d}\widehat{\Lambda}(t \mid a, X)-\mathrm{d}\Lambda_{0}(t \mid a, X)\} \\
&+\int_{0}^{\tau} \frac{(\tau-t)\widehat{F}_{j}(t \mid a, X) S_{0}(t- \mid a, X)}{\widehat{S}(t \mid a, X)} \{\mathrm{d}\widehat{\Lambda}(t \mid a, X)-\mathrm{d}\Lambda_{0}(t \mid a, X)\} \\
&+\int_{0}^{\tau} (\tau-t)S_{0}(t- \mid a, X) \{\mathrm{d}\widehat{\Lambda}_{j}(t \mid a, X)-\mathrm{d}\Lambda_{j, 0}(t \mid a, X)\}
\end{aligned}
$$

After replacing conditional martingale integral terms and $\mathrm{R}\widehat{\mathrm{MT}}\mathrm{L}_{j}(t \mid a, X)-\mathrm{RMTL}_{j, 0}(t \mid a, X)$ by conditional hazards differences, we have the equality for $P_{0} \{\widehat{\phi}_{j}^{\mathrm{RMTL}, h}(\tau, a; \widehat{\eta}_{j}^{\Lambda, h})\}-\psi_{j, 0}^{\mathrm{RMTL}, h}(\tau, a)$ as presented.

\subsection{Lemma~\ref{lemma3}}

\begin{lemma} \label{lemma3}

If Condition~\ref{condition2} holds, then for every $a$, $\tau$, and $k$, two universal constant vectors $C_{\iota}^{\mathrm{RMST}, h}$ and $C_{j, \iota}^{\mathrm{RMTL}, h}$ ensure that
$$
\begin{aligned}
&[P_{0}\{\widehat{\phi}_{k}^{\mathrm{RMST}, h}(\tau, a; \widehat{\eta}_{k}^{\Lambda, h})-\phi_{\infty}^{\mathrm{RMST}, h}(\tau, a; \eta_{\infty}^{\Lambda, h})\}^{2}]^{1/2} \leq \sum_{\iota=1}^{6} C_{\iota}^{\mathrm{RMST}, h} \mathcal{B}_{\iota, k}^{\mathrm{RMST}, h}(\widehat{\eta}_{k}^{\Lambda, h}, \eta_{\infty}^{\Lambda, h}) \\
&[P_{0}\{\sup_{t \in [0, \tau]}|\widehat{\phi}_{k}^{\mathrm{RMST}, h}(t, a; \widehat{\eta}_{k}^{\Lambda, h})-\phi_{\infty}^{\mathrm{RMST}, h}(t, a; \eta_{\infty}^{\Lambda, h})|\}^{2}]^{1/2} \\
& \leq C_{1}^{\mathrm{RMST}, h} \widetilde{\mathcal{B}}_{1, k}^{\mathrm{RMST}, h}(\widehat{\eta}_{k}^{\Lambda, h}, \eta_{\infty}^{\Lambda, h})+\sum_{\iota=2}^{6} C_{\iota}^{\mathrm{RMST}, h} \mathcal{B}_{\iota, k}^{\mathrm{RMST}, h}(\widehat{\eta}_{k}^{\Lambda, h}, \eta_{\infty}^{\Lambda, h}) \\
&[P_{0}\{\widehat{\phi}_{j, k}^{\mathrm{RMTL}, h}(\tau, a; \widehat{\eta}_{j, k}^{\Lambda, h})-\phi_{j, \infty}^{\mathrm{RMTL}, h}(\tau, a; \eta_{\infty}^{\Lambda, h})\}^{2}]^{1/2} \leq \sum_{\iota=1}^{9} C_{j, \iota}^{\mathrm{RMTL}, h} \mathcal{B}_{j, \iota, k}^{\mathrm{RMTL}, h}(\widehat{\eta}_{j, k}^{\Lambda, h}, \eta_{j, \infty}^{\Lambda, h}) \\
&[P_{0}\{\sup_{t \in [0, \tau]}|\widehat{\phi}_{j, k}^{\mathrm{RMTL}, h}(t, a; \widehat{\eta}_{j, k}^{\Lambda, h})-\phi_{j, \infty}^{\mathrm{RMTL}, h}(t, a; \eta_{\infty}^{\Lambda, h})|\}^{2}]^{1/2} \\
\leq& C_{j, 1}^{\mathrm{RMTL}, h} \widetilde{\mathcal{B}}_{j, 1, k}^{\mathrm{RMTL}, h}(\widehat{\eta}_{j, k}^{\Lambda, h}, \eta_{j, \infty}^{\Lambda, h})+\sum_{\iota=2}^{9} C_{j, \iota}^{\mathrm{RMTL}, h} \mathcal{B}_{j, \iota, k}^{\mathrm{RMTL}, h}(\widehat{\eta}_{j, k}^{\Lambda, h}, \eta_{j, \infty}^{\Lambda, h})
\end{aligned}
$$

\noindent where
$$
\begin{aligned}
&\{\mathcal{B}_{1, k}^{\mathrm{RMST}, h}(\widehat{\eta}_{k}^{\Lambda, h}, \eta_{\infty}^{\Lambda, h})\}^2=E_{0} \left[\sup_{t \in [0, \tau]} \left| \frac{\mathrm{R}\widehat{\mathrm{MS}}\mathrm{T}_{k}(\tau \mid a, X)}{\widehat{S}_{k}(t \mid a, X)}-\frac{\mathrm{RMST}_{\infty}(\tau \mid a, X)}{S_{\infty}(t \mid a, X)} \right| \right]^2 \\
&\{\widetilde{\mathcal{B}}_{1, k}^{\mathrm{RMST}, h}(\widehat{\eta}_{k}^{\Lambda, h}, \eta_{\infty}^{\Lambda, h})\}^2=E_{0} \left[\sup_{t \in [0, \tau]} \sup_{u \in [0, t]} \left| \frac{\mathrm{R}\widehat{\mathrm{MS}}\mathrm{T}_{k}(t \mid a, X)}{\widehat{S}_{k}(u \mid a, X)}-\frac{\mathrm{RMST}_{\infty}(t \mid a, X)}{S_{\infty}(u \mid a, X)} \right| \right]^2 \\
&\{\mathcal{B}_{2, k}^{\mathrm{RMST}, h}(\widehat{\eta}_{k}^{\Lambda, h}, \eta_{\infty}^{\Lambda, h})\}^2=E_{0} |[P_{n, k}\{\widehat{h}_{k}(X)\}]^{-1}\widehat{h}_{k}(X)-[E_{0}\{h_{\infty}(X)\}]^{-1}h_{\infty}(X)|^2 \\
&\{\mathcal{B}_{3, k}^{\mathrm{RMST}, h}(\widehat{\eta}_{k}^{\Lambda, h}, \eta_{\infty}^{\Lambda, h})\}^2=([P_{n, k}\left\{\widehat{w}_{k}^{h}(a, X)\right\}]^{-1}-[E_{0}\{h_{\infty}(X)\}]^{-1})^2 \\
&\{\mathcal{B}_{4, k}^{\mathrm{RMST}, h}(\widehat{\eta}_{k}^{\Lambda, h}, \eta_{\infty}^{\Lambda, h})\}^2=E_{0} \left|\widehat{w}_{k}^{h}(a, X)-w_{\infty}^{h}(a, X) \right|^2 \\
&\{\mathcal{B}_{5, k}^{\mathrm{RMST}, h}(\widehat{\eta}_{k}^{\Lambda, h}, \eta_{\infty}^{\Lambda, h})\}^2=E_{0} \left[\sup_{t \in [0, \tau]} \left| \frac{1}{\widehat{G}_{k}(t- \mid a, X)}-\frac{1}{G_{\infty}(t- \mid a, X)} \right| \right]^2
\end{aligned}
$$

$$
\begin{aligned}
&\{\mathcal{B}_{6, k}^{\mathrm{RMST}, h}(\widehat{\eta}_{k}^{\Lambda, h}, \eta_{\infty}^{\Lambda, h})\}^2=E_{0} \left[\sup_{t \in [0, \tau]} \left| \frac{\mathrm{R}\widehat{\mathrm{MS}}\mathrm{T}_{k}(t \mid a, X)}{\widehat{S}_{k}(t \mid a, X)}-\frac{\mathrm{RMST}_{\infty}(t \mid a, X)}{S_{\infty}(t \mid a, X)} \right| \right]^2
\end{aligned}
$$

and
$$
\begin{aligned}
&\{\mathcal{B}_{j, 1, k}^{\mathrm{RMTL}, h}(\widehat{\eta}_{j, k}^{\Lambda, h}, \eta_{j, \infty}^{\Lambda, h})\}^2=\max_{k} E_{0} \left[\sup_{t \in [0, \tau]} \left| \frac{\mathrm{R}\widehat{\mathrm{MT}}\mathrm{L}_{j, k}(\tau \mid a, X)}{\widehat{S}_{k}(t \mid a, X)}-\frac{\mathrm{RMTL}_{j, \infty}(\tau \mid a, X)}{S_{\infty}(t \mid a, X)} \right| \right]^2 \stackrel{P}{\longrightarrow} 0 \\
&\{\mathcal{B}_{j, 2, k}^{\mathrm{RMTL}, h}(\widehat{\eta}_{j, k}^{\Lambda, h}, \eta_{j, \infty}^{\Lambda, h})\}^2=\max_{k} E_{0} |[P_{n, k}\{\widehat{h}_{k}(X)\}]^{-1}\widehat{h}_{k}(X)-[E_{0}\{h_{\infty}(X)\}]^{-1}h_{\infty}(X))|^2 \stackrel{P}{\longrightarrow} 0 \\
&\{\mathcal{B}_{j, 3, k}^{\mathrm{RMTL}, h}(\widehat{\eta}_{j, k}^{\Lambda, h}, \eta_{j, \infty}^{\Lambda, h})\}^2=\max_{k} ([P_{n, k}\{\widehat{w}_{k}^{h}(a, X)\}]^{-1}-[E_{0}\{h_{\infty}(X)\}]^{-1})^2 \stackrel{P}{\longrightarrow} 0 \\
&\{\mathcal{B}_{j, 4, k}^{\mathrm{RMTL}, h}(\widehat{\eta}_{j, k}^{\Lambda, h}, \eta_{j, \infty}^{\Lambda, h})\}^2=\max_{k} E_{0} \left|\widehat{w}_{k}^{h}(a, X)-w_{\infty}^{h}(a, X)\right|^2 \stackrel{P}{\longrightarrow} 0 \\
&\{\mathcal{B}_{j, 5, k}^{\mathrm{RMTL}, h}(\widehat{\eta}_{j, k}^{\Lambda, h}, \eta_{j, \infty}^{\Lambda, h})\}^2=\max_{k} E_{0} \left[\sup_{t \in [0, \tau]} \left| \frac{1}{\widehat{G}_{k}(t- \mid a, X)}-\frac{1}{G_{\infty}(t- \mid a, X)} \right| \right]^2 \stackrel{P}{\longrightarrow} 0 \\
&\{\mathcal{B}_{j, 6, k}^{\mathrm{RMTL}, h}(\widehat{\eta}_{j, k}^{\Lambda, h}, \eta_{j, \infty}^{\Lambda, h})\}^2=\max_{k} E_{0} \left[\sup_{t \in [0, \tau]} \left| \widehat{\Lambda}_{j, k}(t \mid a, X)-\Lambda_{j, \infty}(t \mid a, X) \right| \right]^2 \stackrel{P}{\longrightarrow} 0 \\
&\{\mathcal{B}_{j, 7, k}^{\mathrm{RMTL}, h}(\widehat{\eta}_{j, k}^{\Lambda, h}, \eta_{j, \infty}^{\Lambda, h})\}^2=\max_{k} E_{0} \left[\sup_{t \in [0, \tau]} \left| \frac{\mathrm{R}\widehat{\mathrm{MT}}\mathrm{L}_{j, k}(t \mid a, X)}{\widehat{S}_{k}(t \mid a, X)}-\frac{\mathrm{RMTL}_{j, \infty}(t \mid a, X)}{S_{\infty}(t \mid a, X)} \right| \right]^2 \stackrel{P}{\longrightarrow} 0 \\
&\{\mathcal{B}_{j, 8, k}^{\mathrm{RMTL}, h}(\widehat{\eta}_{j, k}^{\Lambda, h}, \eta_{j, \infty}^{\Lambda, h})\}^2=\max_{k} E_{0} \left[\sup_{t \in [0, \tau]} \left| \frac{\widehat{F}_{j, k}(t \mid a, X)}{\widehat{S}_{k}(t \mid a, X)}-\frac{F_{j, \infty}(t \mid a, X)}{S_{\infty}(t \mid a, X)} \right| \right]^2 \stackrel{P}{\longrightarrow} 0 \\
&\{\mathcal{B}_{j, 9, k}^{\mathrm{RMTL}, h}(\widehat{\eta}_{j, k}^{\Lambda, h}, \eta_{j, \infty}^{\Lambda, h})\}^2=\max_{k} E_{0} \left[\sup_{t \in [0, \tau]} \left| \widehat{\Lambda}_{k}(t \mid a, X)-\Lambda_{\infty}(t \mid a, X) \right| \right]^2 \stackrel{P}{\longrightarrow} 0
\end{aligned}
$$

Proof of Lemma~\ref{lemma3}: We expand $\widehat{\phi}_{k}^{\mathrm{RMST}, h}(\tau, a; \widehat{\eta}_{k}^{\Lambda, h})-\phi_{\infty}^{\mathrm{RMST}, h}(\tau, a; \eta_{\infty}^{\Lambda, h})=\sum_{\iota=1}^{14} \mathcal{Q}_{\iota, k}^{\mathrm{RMST}, h}(\tau, a; \widehat{\eta}_{k}^{\Lambda, h}, \eta_{\infty}^{\Lambda, h})$ into terms which can be squared-bounded in expectation, where
$$
\begin{aligned}
\mathcal{Q}_{1, k}^{\mathrm{RMST}, h}(\tau, a; \widehat{\eta}_{k}^{\Lambda, h}, \eta_{\infty}^{\Lambda, h})=&[P_{n, k}\{\widehat{h}_{k}(X)\}]^{-1}\widehat{h}_{k}(X)\{\mathrm{R}\widehat{\mathrm{MS}}\mathrm{T}_{k}(\tau \mid a, X)-\mathrm{RMST}_{\infty}(\tau \mid a, X)\} \\
\mathcal{Q}_{2, k}^{\mathrm{RMST}, h}(\tau, a; \widehat{\eta}_{k}^{\Lambda, h}, \eta_{\infty}^{\Lambda, h})=&([P_{n, k}\{\widehat{h}_{k}(X)\}]^{-1}\widehat{h}_{k}(X)-[E_{0}\{h_{\infty}(X)\}]^{-1}h_{\infty}(X))\mathrm{RMST}_{\infty}(\tau \mid a, X) \\
\mathcal{Q}_{3, k}^{\mathrm{RMST}, h}(\tau, a; \widehat{\eta}_{k}^{\Lambda, h}, \eta_{\infty}^{\Lambda, h})=&-([P_{n, k}\{\widehat{w}_{k}^{h}(a, X)\}]^{-1}-[E_{0}\{h_{\infty}(X)\}]^{-1}) w_{\infty}^{h}(a, X)\mathrm{RMST}_{\infty}(\tau \mid a, X) \\
&\times \int_{0}^{\tau \wedge \widetilde{T}} \frac{\mathrm{d} M_{\infty}(t \mid a, X)}{S_{\infty}(t \mid a, X)G_{\infty}(t- \mid a, X)} \\
\mathcal{Q}_{4, k}^{\mathrm{RMST}, h}(\tau, a; \widehat{\eta}_{k}^{\Lambda, h}, \eta_{\infty}^{\Lambda, h})=&-[P_{n, k}\{\widehat{w}_{k}^{h}(a, X)\}]^{-1} \{\widehat{w}_{k}^{h}(a, X)-w_{\infty}^{h}(a, X)\}\mathrm{RMST}_{\infty}(\tau \mid a, X) \\
&\times \int_{0}^{\tau \wedge \widetilde{T}} \frac{\mathrm{d} M_{\infty}(t \mid a, X)}{S_{\infty}(t \mid a, X)G_{\infty}(t- \mid a, X)} \\
\mathcal{Q}_{5, k}^{\mathrm{RMST}, h}(\tau, a; \widehat{\eta}_{k}^{\Lambda, h}, \eta_{\infty}^{\Lambda, h})=&-[P_{n, k}\{\widehat{w}_{k}^{h}(a, X)\}]^{-1} \frac{\widehat{w}_{k}^{h}(a, X)I(\widetilde{T} \leq \tau, \Delta=1)}{G_{\infty}(\widetilde{T}- \mid a, X)} \\
&\times \left\{\frac{\mathrm{R}\widehat{\mathrm{MS}}\mathrm{T}_{k}(\tau \mid a, X)}{\widehat{S}_{k}(\widetilde{T} \mid a, X)}-\frac{\mathrm{RMST}_{\infty}(\tau \mid a, X)}{S_{\infty}(\widetilde{T} \mid a, X)}\right\}
\end{aligned}
$$

$$
\begin{aligned}
\mathcal{Q}_{6, k}^{\mathrm{RMST}, h}(\tau, a; \widehat{\eta}_{k}^{\Lambda, h}, \eta_{\infty}^{\Lambda, h})=&-[P_{n, k}\{\widehat{w}_{k}^{h}(a, X)\}]^{-1} \frac{\widehat{w}_{k}^{h}(a, X)I(\widetilde{T} \leq \tau, \Delta=1) \mathrm{R}\widehat{\mathrm{MS}}\mathrm{T}_{k}(\tau \mid a, X)}{\widehat{S}_{k}(\widetilde{T} \mid a, X)} \\
& \times \left\{\frac{1}{\widehat{G}_{k}(\widetilde{T}- \mid a, X)}-\frac{1}{G_{\infty}(\widetilde{T}- \mid a, X)} \right\} \\
\mathcal{Q}_{7, k}^{\mathrm{RMST}, h}(\tau, a; \widehat{\eta}_{k}^{\Lambda, h}, \eta_{\infty}^{\Lambda, h})=&[P_{n, k}\{\widehat{w}_{k}^{h}(a, X)\}]^{-1} \widehat{w}_{k}^{h}(a, X)\mathrm{RMST}_{\infty}(\tau \mid a, X) \int_0^{\tau} \frac{\mathrm{d} \Lambda_{\infty}(t \mid a, X)}{S_{\infty}(t \mid a, X)} \\
&\times \left\{\frac{1}{\widehat{G}_{k}(t- \mid a, X)}-\frac{1}{G_{\infty}(t- \mid a, X)} \right\} \\
\mathcal{Q}_{8, k}^{\mathrm{RMST}, h}(\tau, a; \widehat{\eta}_{k}^{\Lambda, h}, \eta_{\infty}^{\Lambda, h})=&[P_{n, k}\{\widehat{w}_{k}^{h}(a, X)\}]^{-1} \widehat{w}_{k}^{h}(a, X) \left\{\mathrm{R}\widehat{\mathrm{MS}}\mathrm{T}_{k}(\tau \mid a, X) \right. \\
&\left.\times\int_0^{\tau} \frac{\mathrm{d}\widehat{\Lambda}_{k}(t \mid a, X)}{\widehat{S}_{k}(t \mid a, X)\widehat{G}_{k}(t- \mid a, X)}-\mathrm{RMST}_{\infty}(\tau \mid a, X) \int_0^{\tau} \frac{\mathrm{d}\Lambda_{\infty}(t \mid a, X)}{S_{\infty}(t \mid a, X)\widehat{G}_{k}(t- \mid a, X)} \right\} \\
\mathcal{Q}_{9, k}^{\mathrm{RMST}, h}(\tau, a; \widehat{\eta}_{k}^{\Lambda, h}, \eta_{\infty}^{\Lambda, h})=&([P_{n, k}\{\widehat{w}_{k}^{h}(a, X)\}]^{-1}-[E\{h_{\infty}(X)\}]^{-1}) w_{\infty}^{h}(a, X) \\
&\times \int_{0}^{\tau \wedge \widetilde{T}} \frac{\mathrm{RMST}_{\infty}(t \mid a, X)\mathrm{d} M_{\infty}(t \mid a, X)}{S_{\infty}(t \mid a, X)G_{\infty}(t- \mid a, X)} \\
\mathcal{Q}_{10, k}^{\mathrm{RMST}, h}(\tau, a; \widehat{\eta}_{k}^{\Lambda, h}, \eta_{\infty}^{\Lambda, h})=&[P_{n, k}\{\widehat{w}_{k}^{h}(a, X)\}]^{-1} \{\widehat{w}_{k}^{h}(a, X)-w_{\infty}^{h}(a, X)\}\int_{0}^{\tau \wedge \widetilde{T}} \frac{\mathrm{RMST}_{\infty}(t \mid a, X)\mathrm{d} M_{\infty}(t \mid a, X)}{S_{\infty}(t \mid a, X)G_{\infty}(t- \mid a, X)} \\
\mathcal{Q}_{11, k}^{\mathrm{RMST}, h}(\tau, a; \widehat{\eta}_{k}^{\Lambda, h}, \eta_{\infty}^{\Lambda, h})=&[P_{n, k}\{\widehat{w}_{k}^{h}(a, X)\}]^{-1} \frac{\widehat{w}_{k}^{h}(a, X)I(\widetilde{T} \leq \tau, \Delta=1)}{G_{\infty}(\widetilde{T}- \mid a, X)} \\
&\times \left\{\frac{\mathrm{R}\widehat{\mathrm{MS}}\mathrm{T}_{k}(\widetilde{T} \mid a, X)}{\widehat{S}_{k}(\widetilde{T} \mid a, X)}-\frac{\mathrm{RMST}_{\infty}(\widetilde{T} \mid a, X)}{S_{\infty}(\widetilde{T} \mid a, X)}\right\} \\
\mathcal{Q}_{12, k}^{\mathrm{RMST}, h}(\tau, a; \widehat{\eta}_{k}^{\Lambda, h}, \eta_{\infty}^{\Lambda, h})=&[P_{n, k}\{\widehat{w}_{k}^{h}(a, X)\}]^{-1} \frac{\widehat{w}_{k}^{h}(a, X)I(\widetilde{T} \leq \tau, \Delta=1) \mathrm{R}\widehat{\mathrm{MS}}\mathrm{T}_{k}(\widetilde{T} \mid a, X)}{\widehat{S}_{k}(\widetilde{T} \mid a, X)} \\
& \times \left\{\frac{1}{\widehat{G}_{k}(\widetilde{T}- \mid a, X)}-\frac{1}{G_{\infty}(\widetilde{T}- \mid a, X)} \right\} \\
\mathcal{Q}_{13, k}^{\mathrm{RMST}, h}(\tau, a; \widehat{\eta}_{k}^{\Lambda, h}, \eta_{\infty}^{\Lambda, h})=&-[P_{n, k}\{\widehat{w}_{k}^{h}(a, X)\}]^{-1} \widehat{w}_{k}^{h}(a, X) \int_0^{\tau} \frac{\mathrm{RMST}_{\infty}(t \mid a, X)\mathrm{d} \Lambda_{\infty}(t \mid a, X)}{S_{\infty}(t \mid a, X)} \\
&\times \left\{\frac{1}{\widehat{G}_{k}(t- \mid a, X)}-\frac{1}{G_{\infty}(t- \mid a, X)} \right\} \\
\mathcal{Q}_{14, k}^{\mathrm{RMST}, h}(\tau, a; \widehat{\eta}_{k}^{\Lambda, h}, \eta_{\infty}^{\Lambda, h})=&-[P_{n, k}\{\widehat{w}_{k}^{h}(a, X)\}]^{-1} \widehat{w}_{k}^{h}(a, X) \left\{\int_0^{\tau} \frac{\mathrm{R}\widehat{\mathrm{MS}}\mathrm{T}_{k}(t \mid a, X)\mathrm{d}\widehat{\Lambda}_{k}(t \mid a, X)}{\widehat{S}_{k}(t \mid a, X)\widehat{G}_{k}(t- \mid a, X)} \right. \\
&\left.-\int_0^{\tau} \frac{\mathrm{RMST}_{\infty}(t \mid a, X)\mathrm{d}\Lambda_{\infty}(t \mid a, X)}{S_{\infty}(t \mid a, X)\widehat{G}_{k}(t- \mid a, X)} \right\}
\end{aligned}
$$
\end{lemma}

Following the triangle inequality,
$$P_{0}\{\widehat{\phi}^{\mathrm{RMST}, h}(\tau, a; \widehat{\eta}_{k}^{\Lambda, h}, \eta_{\infty}^{\Lambda, h})-\phi_{\infty}^{\mathrm{RMST}, h}(\tau, a; \widehat{\eta}_{k}^{\Lambda, h}, \eta_{\infty}^{\Lambda, h})\}^2 \leq \left(\sum_{\iota=1}^{14}[P_{0}\{\mathcal{Q}_{\iota, k}^{\mathrm{RMST}, h}(\tau, a; \widehat{\eta}_{k}^{\Lambda, h}, \eta_{\infty}^{\Lambda, h})\}^2]^{1 / 2}\right)^2$$

Thus we need to bound $P_{0}\{\mathcal{Q}_{\iota, k}^{\mathrm{RMST}, h}(\tau, a; \widehat{\eta}_{k}^{\Lambda, h}, \eta_{\infty}^{\Lambda, h})\}^2$, $\forall \iota$.
$$
\begin{aligned}
P_{0}\{ \mathcal{Q}_{1, k}^{\mathrm{RMST}, h}(\tau, a; \widehat{\eta}_{k}^{\Lambda, h}, \eta_{\infty}^{\Lambda, h})\}^2=&E_{0} \left| [P_{n, k}\{\widehat{h}_{k}(X)\}]^{-1}\widehat{h}_{k}(X)\{\mathrm{R}\widehat{\mathrm{MS}}\mathrm{T}_{k}(\tau \mid a, X)-\mathrm{RMST}_{\infty}(\tau \mid a, X)\} \right|^2 \\
\leq&\epsilon^2 h_{\mathrm{max}}^2 E_{0} \left| \frac{\mathrm{R}\widehat{\mathrm{MS}}\mathrm{T}_{k}(\tau \mid a, X)}{\widehat{S}_{k}(\tau \mid a, X)}-\frac{\mathrm{RMST}_{\infty}(\tau \mid a, X)}{S_{\infty}(\tau \mid a, X)} \right|^2 \\
\leq&\epsilon^2 h_{\mathrm{max}}^2 E_{0} \left[\sup_{t \in [0, \tau]} \left| \frac{\mathrm{R}\widehat{\mathrm{MS}}\mathrm{T}_{k}(\tau \mid a, X)}{\widehat{S}_{k}(t \mid a, X)}-\frac{\mathrm{RMST}_{\infty}(\tau \mid a, X)}{S_{\infty}(t \mid a, X)} \right| \right]^2 \\
P_{0} \{\mathcal{Q}_{2, k}^{\mathrm{RMST}, h}(\tau, a; \widehat{\eta}_{k}^{\Lambda, h}, \eta_{\infty}^{\Lambda, h})\}^2=&E_{0} \left|([P_{n, k}\{\widehat{h}_{k}(X)\}]^{-1}\widehat{h}_{k}(X)-[E_{0}\{h_{\infty}(X)\}]^{-1}h_{\infty}(X))\mathrm{RMST}_{\infty}(\tau \mid a, X) \right|^2 \\
\leq&\tau^2 E_{0} \left|([P_{n, k}\{\widehat{h}_{k}(X)\}]^{-1}\widehat{h}_{k}(X)-[E_{0}\{h_{\infty}(X)\}]^{-1}h_{\infty}(X)) \right|^2
\end{aligned}
$$

The backwards equation from Theorem 5 of \citet{gill1990survey} suggests that $\int_0^t \frac{S(t \mid a, X)}{S(u \mid, a, X)} \mathrm{d}\Lambda(u \mid a, X)=1-S(t \mid a, X)$ and
$$
\begin{aligned}
&E_{0} \left|\int_{0}^{\tau \wedge \widetilde{T}} \frac{\mathrm{d} M_{\infty}(t \mid a, X)}{S_{\infty}(t \mid a, X)G_{\infty}(t- \mid a, X)} \right|^2 = E_{0} \left|\frac{I(\widetilde{T} \leq \tau, \Delta=1)}{S_{\infty}(\widetilde{T} \mid a, X)G_{\infty}(\widetilde{T}- \mid a, X)}-\int_0^{\tau} \frac{\mathrm{d}\Lambda_{\infty}(t \mid a, X)}{S_{\infty}(t \mid a, X)G_{\infty}(t- \mid a, X)} \right|^2 \\
\leq& E_{0} \left|\epsilon^{2}+\epsilon \int_0^{\tau} \mathrm{d}\left\{\frac{1}{S_{\infty}(t \mid a, X)} \right\} \right|^2 \leq E_{0} \left|\epsilon^{2}+\epsilon(\epsilon-1) \right|^2 \leq 4 \epsilon^{4}
\end{aligned}
$$

$$
\begin{aligned}
P_{0} \{\mathcal{Q}_{3, k}^{\mathrm{RMST}, h}(\tau, a; \widehat{\eta}_{k}^{\Lambda, h}, \eta_{\infty}^{\Lambda, h})\}^2=&E_{0} \left|([P_{n, k}\{\widehat{w}_{k}^{h}(a, X)\}]^{-1}-[E_{0}\{h_{\infty}(X)\}]^{-1}) \right. \\
&\left. \frac{h_{\infty}(X)I(A=a)\mathrm{RMST}_{\infty}(\tau \mid a, X)}{\pi_{\infty}(a \mid X)}\int_{0}^{\tau \wedge \widetilde{T}} \frac{\mathrm{d} M_{\infty}(t \mid a, X)}{S_{\infty}(t \mid a, X)G_{\infty}(t- \mid a, X)} \right|^2 \\
\leq & ([P_{n, k}\{\widehat{w}_{k}^{h}(a, X)\}]^{-1}-[E_{0}\{h_{\infty}(X)\}]^{-1})^2 \epsilon^{2}h_{\mathrm{max}}^2 \tau^2 \times 4 \epsilon^{4}
\end{aligned}
$$

Since
$$
P_{n, k}\{\widehat{w}_{k}^{h}(a, X)\}=P_{n, k}\left\{\frac{\widehat{h}_{k}(X)I(A=a)}{\widehat{\pi}_{k}(a \mid X)}\right\} \geq P_{n, k} \{\widehat{h}_{k}(X)I(A=a)\} \geq P_{n, k} \{\widehat{h}_{k}(X)\} \geq h_{\mathrm{min}}
$$

We have
$$
\begin{aligned}
P_{0} \{\mathcal{Q}_{4, k}^{\mathrm{RMST}, h}(\tau, a; \widehat{\eta}_{k}^{\Lambda, h}, \eta_{\infty}^{\Lambda, h})\}^2=&E_{0} \left|[P_{n, k}\{\widehat{w}_{k}^{h}(a, X)\}]^{-1} \{\widehat{w}_{k}^{h}(a, X)-w_{\infty}^{h}(a, X)\} \mathrm{RMST}_{\infty}(\tau \mid a, X) \right. \\
&\left.\times \int_{0}^{\tau \wedge \widetilde{T}} \frac{\mathrm{d} M_{\infty}(t \mid a, X)}{S_{\infty}(t \mid a, X)G_{\infty}(t- \mid a, X)}\right|^2 \leq h_{\mathrm{min}}^{-2} \tau^2 \times 4 \epsilon^{4} E_{0} \left|\widehat{w}_{k}^{h}(a, X)-w_{\infty}^{h}(a, X) \right|^2 \\
P_{0} \{\mathcal{Q}_{5, k}^{\mathrm{RMST}, h}(\tau, a; \widehat{\eta}_{k}^{\Lambda, h}, \eta_{\infty}^{\Lambda, h})\}^2=&E_{0} \left|[P_{n, k}\{\widehat{w}_{k}^{h}(a, X)\}]^{-1} \frac{\widehat{w}_{k}^{h}(a, X)I(\widetilde{T} \leq \tau, \Delta=1)}{G_{\infty}(\widetilde{T}- \mid a, X)} \right. \\
&\left. \times \left\{\frac{\mathrm{R}\widehat{\mathrm{MS}}\mathrm{T}_{k}(\tau \mid a, X)}{\widehat{S}_{k}(\widetilde{T} \mid a, X)}-\frac{\mathrm{RMST}_{\infty}(\tau \mid a, X)}{S_{\infty}(\widetilde{T} \mid a, X)}\right\} \right|^2 \\
\leq & h_{\mathrm{min}}^{-2} h_{\mathrm{max}}^2 \epsilon^{4} E_{0} \left[\sup_{t \in [0, \tau]} \left| \frac{\mathrm{R}\widehat{\mathrm{MS}}\mathrm{T}_{k}(\tau \mid a, X)}{\widehat{S}_{k}(t \mid a, X)}-\frac{\mathrm{RMST}_{\infty}(\tau \mid a, X)}{S_{\infty}(t \mid a, X)} \right| \right]^2
\end{aligned}
$$

$$
\begin{aligned}
P_{0} \{\mathcal{Q}_{6, k}^{\mathrm{RMST}, h}(\tau, a; \widehat{\eta}_{k}^{\Lambda, h}, \eta_{\infty}^{\Lambda, h})\}^2=&E_{0} \left|[P_{n, k}\{\widehat{w}_{k}^{h}(a, X)\}]^{-1} \frac{\widehat{w}_{k}^{h}(a, X)I(\widetilde{T} \leq \tau, \Delta=1) \mathrm{R}\widehat{\mathrm{MS}}\mathrm{T}_{k}(\tau \mid a, X)}{\widehat{S}_{k}(\widetilde{T} \mid a, X)} \right. \\
&\left. \times \left\{\frac{1}{\widehat{G}_{k}(\widetilde{T}- \mid a, X)}-\frac{1}{G_{\infty}(\widetilde{T}- \mid a, X)} \right\} \right|^2 \\
\leq & h_{\mathrm{min}}^{-2} h_{\mathrm{max}}^2 \tau^2 \epsilon^{4} E_{0} \left[\sup_{t \in [0, \tau]} \left| \frac{1}{\widehat{G}_{k}(t- \mid a, X)}-\frac{1}{G_{\infty}(t- \mid a, X)} \right| \right]^2 \\
P_{0} \{\mathcal{Q}_{7, k}^{\mathrm{RMST}, h}(\tau, a; \widehat{\eta}_{k}^{\Lambda, h}, \eta_{\infty}^{\Lambda, h})\}^2=&E_{0} \left|[P_{n, k}\{\widehat{w}_{k}^{h}(a, X)\}]^{-1} \widehat{w}_{k}^{h}(a, X)\mathrm{RMST}_{\infty}(\tau \mid a, X) \int_0^{\tau} \frac{\mathrm{d} \Lambda_{\infty}(t \mid a, X)}{S_{\infty}(t \mid a, X)} \right. \\
&\left. \times \left\{\frac{1}{\widehat{G}_{k}(t- \mid a, X)}-\frac{1}{G_{\infty}(t- \mid a, X)} \right\} \right|^2 \\
\leq & h_{\mathrm{min}}^{-2} h_{\mathrm{max}}^2 \tau^2 \epsilon^{2} E_{0} \left[\sup_{t \in [0, \tau]} \left| \frac{1}{\widehat{G}_{k}(t- \mid a, X)}-\frac{1}{G_{\infty}(t- \mid a, X)} \right| \int_0^{\tau} \frac{\mathrm{d} \Lambda_{\infty}(t \mid a, X)}{S_{\infty}(t \mid a, X)} \right]^2 \\
\leq & h_{\mathrm{min}}^{-2} h_{\mathrm{max}}^2 \tau^2 \epsilon^{2} (\epsilon-1)^{2} E_{0} \left[\sup_{t \in [0, \tau]} \left| \frac{1}{\widehat{G}_{k}(t- \mid a, X)}-\frac{1}{G_{\infty}(t- \mid a, X)} \right| \right]^2
\end{aligned}
$$

As
$$
\begin{aligned}
&E_{0}\left| \mathrm{R}\widehat{\mathrm{MS}}\mathrm{T}_{k}(\tau \mid a, X) \int_0^{\tau} \frac{\mathrm{d}\widehat{\Lambda}_{k}(t \mid a, X)}{\widehat{S}_{k}(t \mid a, X)\widehat{G}_{k}(t- \mid a, X)} -\mathrm{RMST}_{\infty}(\tau \mid a, X) \int_0^{\tau} \frac{\mathrm{d}\Lambda_{\infty}(t \mid a, X)}{S_{\infty}(t \mid a, X)\widehat{G}_{k}(t- \mid a, X)}\right|^2 \\
=&E_{0}\left|\mathrm{R}\widehat{\mathrm{MS}}\mathrm{T}_{k}(\tau \mid a, X) \left[ \left.\frac{1}{\widehat{S}_{k}(t \mid a, X)\widehat{G}_{k}(t- \mid a, X)}\right|_0^{\tau}-\int_0^{\tau} \frac{1}{\widehat{S}_{k}(t \mid a, X)} \mathrm{d}\left\{\frac{1}{\widehat{G}_{k}(t- \mid a, X)}\right\} \right] \right. \\
&\left.-\mathrm{RMST}_{\infty}(\tau \mid a, X) \left[ \left.\frac{1}{S_{\infty}(t \mid a, X)\widehat{G}_{k}(t- \mid a, X)}\right|_0^{\tau}-\int_0^{\tau} \frac{1}{S_{\infty}(t \mid a, X)} \mathrm{d}\left\{\frac{1}{\widehat{G}_{k}(t- \mid a, X)}\right\} \right] \right|^2 \\
=&E_{0}\left|\left\{\frac{\mathrm{R}\widehat{\mathrm{MS}}\mathrm{T}_{k}(\tau \mid a, X)}{\widehat{S}_{k}(\tau \mid a, X)}-\frac{\mathrm{RMST}_{\infty}(\tau \mid a, X)}{S_{\infty}(\tau \mid a, X)}\right\} \frac{1}{\widehat{G}_{k}(\tau- \mid a, X)} \right. \\
&-\left. \int_0^{\tau} \left\{\frac{\mathrm{R}\widehat{\mathrm{MS}}\mathrm{T}_{k}(\tau \mid a, X)}{\widehat{S}_{k}(t \mid a, X)}-\frac{\mathrm{RMST}_{\infty}(\tau \mid a, X)}{S_{\infty}(t \mid a, X)} \right\} \mathrm{d} \left\{\frac{1}{\widehat{G}_{k}(t- \mid a, X)}\right\} \right|^2 \\
\leq & \epsilon^{2} E_{0} \left\{\sup_{t \in [0, \tau]}\left|\frac{\mathrm{R}\widehat{\mathrm{MS}}\mathrm{T}_{k}(\tau \mid a, X)}{\widehat{S}_{k}(t \mid a, X)}-\frac{\mathrm{RMST}_{\infty}(\tau \mid a, X)}{S_{\infty}(t \mid a, X)}\right|\right\}^2+ E_{0}\left\{ \left.\frac{1}{\widehat{G}_{k}(t- \mid a, X)}\right|_{0}^{\tau} \right. \\
&\left.\times\sup_{t \in [0, \tau]}\left|\frac{\mathrm{R}\widehat{\mathrm{MS}}\mathrm{T}_{k}(\tau \mid a, X)}{\widehat{S}_{k}(t \mid a, X)}-\frac{\mathrm{RMST}_{\infty}(\tau \mid a, X)}{S_{\infty}(t \mid a, X)}\right|\right\}^2 \\
\leq & \{\epsilon^{2}+(\epsilon-1)^{2}\} E_{0} \left\{\sup_{t \in [0, \tau]}\left|\frac{\mathrm{R}\widehat{\mathrm{MS}}\mathrm{T}_{k}(\tau \mid a, X)}{\widehat{S}_{k}(t \mid a, X)}-\frac{\mathrm{RMST}_{\infty}(\tau \mid a, X)}{S_{\infty}(t \mid a, X)}\right|\right\}^2
\end{aligned}
$$

We have
$$
\begin{aligned}
&P_{0} \{\mathcal{Q}_{8, k}^{\mathrm{RMST}, h}(\tau, a; \widehat{\eta}_{k}^{\Lambda, h}, \eta_{\infty}^{\Lambda, h})\}^2 \\
=&E_{0} \left|[P_{n, k}\{\widehat{w}_{k}^{h}(a, X)\}]^{-1} \widehat{w}_{k}^{h}(a, X) \left\{\mathrm{R}\widehat{\mathrm{MS}}\mathrm{T}_{k}(\tau \mid a, X) \right.\right. \\
&\left.\left.\times\int_0^{\tau} \frac{\mathrm{d}\widehat{\Lambda}_{k}(t \mid a, X)}{\widehat{S}_{k}(t \mid a, X)\widehat{G}_{k}(t- \mid a, X)} -\mathrm{RMST}_{\infty}(\tau \mid a, X) \int_0^{\tau} \frac{\mathrm{d}\Lambda_{\infty}(t \mid a, X)}{S_{\infty}(t \mid a, X)\widehat{G}_{k}(t- \mid a, X)} \right\} \right|^2 \\
\leq &h_{\mathrm{min}}^{-2} h_{\mathrm{max}}^2 \epsilon^{2} \{\epsilon^{2}+(\epsilon-1)^{2}\} E_{0} \left\{\sup_{t \in [0, \tau]}\left|\frac{\mathrm{R}\widehat{\mathrm{MS}}\mathrm{T}_{k}(\tau \mid a, X)}{\widehat{S}_{k}(t \mid a, X)}-\frac{\mathrm{RMST}_{\infty}(\tau \mid a, X)}{S_{\infty}(t \mid a, X)}\right|\right\}^2
\end{aligned}
$$

Similar to previous terms, % $P \left\{ \mathcal{Q}_{i, 3}^{\mathrm{RMST}, h}(\tau, a)\}^2$,
$$
\begin{aligned}
P_{0} \{\mathcal{Q}_{9, k}^{\mathrm{RMST}, h}(\tau, a; \widehat{\eta}_{k}^{\Lambda, h}, \eta_{\infty}^{\Lambda, h})\}^2 \leq& ([P_{n, k}\{\widehat{w}_{k}^{h}(a, X)\}]^{-1}-[E_{0}\{h_{\infty}(X)\}]^{-1})^2 \epsilon^{2}h_{\mathrm{max}}^2 \tau^2 \times 4 \epsilon^{4} \\
P_{0} \{\mathcal{Q}_{10, k}^{\mathrm{RMST}, h}(\tau, a; \widehat{\eta}_{k}^{\Lambda, h}, \eta_{\infty}^{\Lambda, h})\}^2 \leq& h_{\mathrm{min}}^{-2} \tau^2 \times 4 \epsilon^{4} E_{0} \left|\widehat{w}_{k}^{h}(a, X)-w_{\infty}^{h}(a, X)\right|^2 \\
P_{0} \{\mathcal{Q}_{11, k}^{\mathrm{RMST}, h}(\tau, a; \widehat{\eta}_{k}^{\Lambda, h}, \eta_{\infty}^{\Lambda, h})\}^2 \leq & h_{\mathrm{min}}^{-2} h_{\mathrm{max}}^2 \epsilon^{4} E_{0} \left[\sup_{t \in [0, \tau]} \left| \frac{\mathrm{R}\widehat{\mathrm{MS}}\mathrm{T}_{k}(t \mid a, X)}{\widehat{S}_{k}(t \mid a, X)}-\frac{\mathrm{RMST}_{\infty}(t \mid a, X)}{S_{\infty}(t \mid a, X)} \right| \right]^2 \\
P_{0} \{\mathcal{Q}_{12, k}^{\mathrm{RMST}, h}(\tau, a; \widehat{\eta}_{k}^{\Lambda, h}, \eta_{\infty}^{\Lambda, h})\}^2\leq& h_{\mathrm{min}}^{-2} h_{\mathrm{max}}^2 \tau^2 \epsilon^{4} E_{0} \left[\sup_{t \in [0, \tau]} \left| \frac{1}{\widehat{G}_{k}(t- \mid a, X)}-\frac{1}{G_{\infty}(t- \mid a, X)} \right| \right]^2 \\
P_{0} \{\mathcal{Q}_{13, k}^{\mathrm{RMST}, h}(\tau, a; \widehat{\eta}_{k}^{\Lambda, h}, \eta_{\infty}^{\Lambda, h})\}^2\leq& h_{\mathrm{min}}^{-2} h_{\mathrm{max}}^2 \tau^2 \epsilon^{2} (\epsilon-1)^{2} E_{0} \left[\sup_{t \in [0, \tau]} \left| \frac{1}{\widehat{G}_{k}(t- \mid a, X)}-\frac{1}{G_{\infty}(t- \mid a, X)} \right| \right]^2
\end{aligned}
$$

As
$$
\begin{aligned}
&E_{0}\left| \int_0^{\tau} \frac{\mathrm{R}\widehat{\mathrm{MS}}\mathrm{T}_{k}(t \mid a, X)\mathrm{d}\widehat{\Lambda}_{k}(t \mid a, X)}{\widehat{S}_{k}(t \mid a, X)\widehat{G}_{k}(t- \mid a, X)}-\int_0^{\tau} \frac{\mathrm{RMST}_{\infty}(t \mid a, X)\mathrm{d}\Lambda_{\infty}(t \mid a, X)}{S_{\infty}(t \mid a, X)\widehat{G}_{k}(t- \mid a, X)}\right|^2 \\
=&E_{0}\left| \left.\frac{\mathrm{R}\widehat{\mathrm{MS}}\mathrm{T}_{k}(t \mid a, X)}{\widehat{S}_{k}(t \mid a, X)\widehat{G}_{k}(t- \mid a, X)} \right|_{0}^{\tau}-\int_0^{\tau} \frac{1}{\widehat{S}_{k}(t \mid a, X)} \mathrm{d}\left\{\frac{\mathrm{R}\widehat{\mathrm{MS}}\mathrm{T}_{k}(t \mid a, X)}{\widehat{G}_{k}(t- \mid a, X)}\right\} \right. \\
&\left.\left.-\frac{\mathrm{RMST}_{\infty}(t \mid a, X)}{S_{\infty}(t \mid a, X)\widehat{G}_{k}(t- \mid a, X)} \right|_{0}^{\tau}+\int_0^{\tau} \frac{1}{S_{\infty}(t \mid a, X)} \mathrm{d}\left\{\frac{\mathrm{RMST}_{\infty}(t \mid a, X)}{\widehat{G}_{k}(t- \mid a, X)}\right\}\right|^2 \\
% =&E\left| \left\{\frac{\mathrm{R}\widehat{\mathrm{MS}}\mathrm{T}_{k}(\tau \mid a, X)}{\widehat{S}_{k}(\tau \mid a, X)}-\frac{\mathrm{RMST}_{\infty}(\tau \mid a, X)}{S_{\infty}(\tau \mid a, X)}\right\}\frac{1}{\widehat{G}_{k}(\tau- \mid a, X)}-\int_0^{\tau} \frac{\mathrm{d}t}{\widehat{G}_{k}(t- \mid a, X)}+\int_0^{\tau} \frac{\mathrm{d}t}{\widehat{G}_{k}(t- \mid a, X)} \right. \\
% &\left.-\int_0^{\tau} \left\{\frac{\mathrm{R}\widehat{\mathrm{MS}}\mathrm{T}_{k}(t \mid a, X)}{\widehat{S}_{k}(t \mid a, X)}-\frac{\mathrm{RMST}_{\infty}(t \mid a, X)}{S_{\infty}(t \mid a, X)}\right\} \mathrm{d} \left\{\frac{1}{\widehat{G}_{k}(t- \mid a, X)}\right\}\right|^2 \\
\leq & \left\{\epsilon^2+(\epsilon-1)^2\right\} E_{0}\left\{\sup_{t \in [0, \tau]} \left| \frac{\mathrm{R}\widehat{\mathrm{MS}}\mathrm{T}_{k}(t \mid a, X)}{\widehat{S}_{k}(t \mid a, X)}-\frac{\mathrm{RMST}_{\infty}(t \mid a, X)}{S_{\infty}(t \mid a, X)}\right|\right\}^2
\end{aligned}
$$

Thus
$$
P_{0} \{\mathcal{Q}_{14, k}^{\mathrm{RMST}, h}(\tau, a; \widehat{\eta}_{k}^{\Lambda, h}, \eta_{\infty}^{\Lambda, h})\}^2\leq h_{\mathrm{min}}^{-2} h_{\mathrm{max}}^2 \epsilon^{2} \{\epsilon^{2}+(\epsilon-1)^{2}\} E_{0} \left\{\sup_{t \in [0, \tau]}\left|\frac{\mathrm{R}\widehat{\mathrm{MS}}\mathrm{T}_{k}(t \mid a, X)}{\widehat{S}_{k}(t \mid a, X)}-\frac{\mathrm{RMST}_{\infty}(t \mid a, X)}{S_{\infty}(t \mid a, X)}\right|\right\}^2
$$

In the competing risks setting, we expand $\widehat{\phi}_{j, k}^{\mathrm{RMTL}, h}(\tau, a; \widehat{\eta}_{j, k}^{\Lambda, h}, \eta_{j, \infty}^{\Lambda, h})-\phi_{j, \infty}^{\mathrm{RMTL}, h}(\tau, a; \widehat{\eta}_{j, k}^{\Lambda, h}, \eta_{j, \infty}^{\Lambda, h})=\sum_{\iota=1}^{36} \mathcal{Q}_{j, \iota, k}^{\mathrm{RMTL}, h}(\tau, a; \widehat{\eta}_{j, k}^{\Lambda, h}, \eta_{j, \infty}^{\Lambda, h})$ which can be squared-bounded in expectation, where
$$
\begin{aligned}
\mathcal{Q}_{j, 1, k}^{\mathrm{RMTL}, h}(\tau, a; \widehat{\eta}_{j, k}^{\Lambda, h}, \eta_{j, \infty}^{\Lambda, h})=&[P_{n, k}\{\widehat{h}_{k}(X)\}]^{-1}\widehat{h}_{k}(X)\{\mathrm{R}\widehat{\mathrm{MT}}\mathrm{L}_{j, k}(\tau \mid a, X)-\mathrm{RMTL}_{j, \infty}(\tau \mid a, X)\} \\
\mathcal{Q}_{j, 2, k}^{\mathrm{RMTL}, h}(\tau, a; \widehat{\eta}_{j, k}^{\Lambda, h}, \eta_{j, \infty}^{\Lambda, h})=&([P_{n, k}\{\widehat{h}_{k}(X)\}]^{-1}\widehat{h}_{k}(X)-[E_{0}\{h_{\infty}(X)\}]^{-1}h_{\infty}(X))\mathrm{RMTL}_{j, \infty}(\tau \mid a, X) \\
\mathcal{Q}_{j, 3, k}^{\mathrm{RMTL}, h}(\tau, a; \widehat{\eta}_{j, k}^{\Lambda, h}, \eta_{j, \infty}^{\Lambda, h})=&([P_{n, k}\{\widehat{w}_{k}^{h}(a, X)\}]^{-1}-[E_{0}\{h_{\infty}(X)\}]^{-1}) w_{\infty}^{h}(a, X)\tau\int_0^{\tau \wedge \widetilde{T}} \frac{\mathrm{d}M_{j, \infty}(t \mid a, X)}{G_{\infty}(t- \mid a, X)} \\
\mathcal{Q}_{j, 4, k}^{\mathrm{RMTL}, h}(\tau, a; \widehat{\eta}_{j, k}^{\Lambda, h}, \eta_{j, \infty}^{\Lambda, h})=&[P_{n, k}\{\widehat{w}_{k}^{h}(a, X)\}]^{-1}\{\widehat{w}_{k}^{h}(a, X)-w_{\infty}^{h}(a, X)\}\tau \int_0^{\tau \wedge \widetilde{T}} \frac{\mathrm{d}M_{j, \infty}(t \mid a, X)}{G_{\infty}(t- \mid a, X)} \\
\mathcal{Q}_{j, 5, k}^{\mathrm{RMTL}, h}(\tau, a; \widehat{\eta}_{j, k}^{\Lambda, h}, \eta_{j, \infty}^{\Lambda, h})=&[P_{n, k}\{\widehat{w}_{k}^{h}(a, X)\}]^{-1} \widehat{w}_{k}^{h}(a, X)I(\widetilde{T} \leq \tau, \widetilde{J}=j) \tau \left\{\frac{1}{\widehat{G}_{k}(\widetilde{T}- \mid a, X)}-\frac{1}{G_{\infty}(\widetilde{T}- \mid a, X)}\right\} \\
\mathcal{Q}_{j, 6, k}^{\mathrm{RMTL}, h}(\tau, a; \widehat{\eta}_{j, k}^{\Lambda, h}, \eta_{j, \infty}^{\Lambda, h})=&-[P_{n, k}\{\widehat{w}_{k}^{h}(a, X)\}]^{-1} \widehat{w}_{k}^{h}(a, X)\tau \int_0^{\tau} \left\{\frac{1}{\widehat{G}_{k}(t- \mid a, X)}-\frac{1}{G_{\infty}(t- \mid a, X)}\right\} \mathrm{d}\Lambda_{j, \infty}(t \mid a, X) \\
\mathcal{Q}_{j, 7, k}^{\mathrm{RMTL}, h}(\tau, a; \widehat{\eta}_{j, k}^{\Lambda, h}, \eta_{j, \infty}^{\Lambda, h})=&-[P_{n, k}\{\widehat{w}_{k}^{h}(a, X)\}]^{-1} \widehat{w}_{k}^{h}(a, X)\tau \int_0^{\tau} \frac{\mathrm{d}\widehat{\Lambda}_{j, k}(t \mid a, X)-\mathrm{d}\Lambda_{j, \infty}(t \mid a, X)}{\widehat{G}_{k}(t- \mid a, X)} \\
\mathcal{Q}_{j, 8, k}^{\mathrm{RMTL}, h}(\tau, a; \widehat{\eta}_{j, k}^{\Lambda, h}, \eta_{j, \infty}^{\Lambda, h})=&-([P_{n, k}\{\widehat{w}_{k}^{h}(a, X)\}]^{-1}-[E_{0}\{h_{\infty}(X)\}]^{-1}) w_{\infty}^{h}(a, X)\int_0^{\tau \wedge \widetilde{T}} \frac{t \mathrm{d}M_{j, \infty}(t \mid a, X)}{G_{\infty}(t- \mid a, X)} \\
\mathcal{Q}_{j, 9, k}^{\mathrm{RMTL}, h}(\tau, a; \widehat{\eta}_{j, k}^{\Lambda, h}, \eta_{j, \infty}^{\Lambda, h})=&-[P_{n, k}\{\widehat{w}_{k}^{h}(a, X)\}]^{-1}\{\widehat{w}_{k}^{h}(a, X)-w_{\infty}^{h}(a, X)\} \int_0^{\tau \wedge \widetilde{T}} \frac{t \mathrm{d}M_{j, \infty}(t \mid a, X)}{G_{\infty}(t- \mid a, X)} \\
\mathcal{Q}_{j, 10, k}^{\mathrm{RMTL}, h}(\tau, a; \widehat{\eta}_{j, k}^{\Lambda, h}, \eta_{j, \infty}^{\Lambda, h})=&-[P_{n, k}\{\widehat{w}_{k}^{h}(a, X)\}]^{-1} \widehat{w}_{k}^{h}(a, X)I(\widetilde{T} \leq \tau, \widetilde{J}=j) \widetilde{T} \left\{\frac{1}{\widehat{G}_{k}(\widetilde{T}- \mid a, X)}-\frac{1}{G_{\infty}(\widetilde{T}- \mid a, X)}\right\} \\
\mathcal{Q}_{j, 11, k}^{\mathrm{RMTL}, h}(\tau, a; \widehat{\eta}_{j, k}^{\Lambda, h}, \eta_{j, \infty}^{\Lambda, h})=&[P_{n, k}\{\widehat{w}_{k}^{h}(a, X)\}]^{-1} \widehat{w}_{k}^{h}(a, X) \int_0^{\tau} \left\{\frac{1}{\widehat{G}_{k}(t- \mid a, X)}-\frac{1}{G_{\infty}(t- \mid a, X)}\right\} t \mathrm{d}\Lambda_{j, \infty}(t \mid a, X) \\
\mathcal{Q}_{j, 12, k}^{\mathrm{RMTL}, h}(\tau, a; \widehat{\eta}_{j, k}^{\Lambda, h}, \eta_{j, \infty}^{\Lambda, h})=&[P_{n, k}\{\widehat{w}_{k}^{h}(a, X)\}]^{-1} \widehat{w}_{k}^{h}(a, X) \int_0^{\tau} \frac{t \{\mathrm{d}\widehat{\Lambda}_{j, k}(t \mid a, X)-\mathrm{d}\Lambda_{j, \infty}(t \mid a, X)\}}{\widehat{G}_{k}(t- \mid a, X)} \\
\mathcal{Q}_{j, 13, k}^{\mathrm{RMTL}, h}(\tau, a; \widehat{\eta}_{j, k}^{\Lambda, h}, \eta_{j, \infty}^{\Lambda, h})=&-([P_{n, k}\{\widehat{w}_{k}^{h}(a, X)\}]^{-1}-[E_{0}\{h_{\infty}(X)\}]^{-1}) w_{\infty}^{h}(a, X)\mathrm{RMTL}_{j, \infty}(\tau \mid a, X) \\
&\times \int_{0}^{\tau \wedge \widetilde{T}} \frac{\mathrm{d} M_{\infty}(t \mid a, X)}{S_{\infty}(t \mid a, X)G_{\infty}(t- \mid a, X)} \\
\mathcal{Q}_{j, 14, k}^{\mathrm{RMTL}, h}(\tau, a; \widehat{\eta}_{j, k}^{\Lambda, h}, \eta_{j, \infty}^{\Lambda, h})=&-[P_{n, k}\{\widehat{w}_{k}^{h}(a, X)\}]^{-1} \{\widehat{w}_{k}^{h}(a, X)-w_{\infty}^{h}(a, X)\} \mathrm{RMTL}_{j, \infty}(\tau \mid a, X) \\
&\times \int_{0}^{\tau \wedge \widetilde{T}} \frac{\mathrm{d} M_{\infty}(t \mid a, X)}{S_{\infty}(t \mid a, X)G_{\infty}(t- \mid a, X)} \\
\mathcal{Q}_{j, 15, k}^{\mathrm{RMTL}, h}(\tau, a; \widehat{\eta}_{j, k}^{\Lambda, h}, \eta_{j, \infty}^{\Lambda, h})=&-[P_{n, k}\{\widehat{w}_{k}^{h}(a, X)\}]^{-1} \frac{\widehat{w}_{k}^{h}(a, X)I(\widetilde{T} \leq \tau, \Delta=1)}{G_{\infty}(\widetilde{T}- \mid a, X)} \\
&\times \left\{\frac{\mathrm{R}\widehat{\mathrm{MT}}\mathrm{L}_{j, k}(\tau \mid a, X)}{\widehat{S}_{k}(\widetilde{T} \mid a, X)}-\frac{\mathrm{RMTL}_{j, \infty}(\tau \mid a, X)}{S_{\infty}(\widetilde{T} \mid a, X)}\right\} \\
\mathcal{Q}_{j, 16, k}^{\mathrm{RMTL}, h}(\tau, a; \widehat{\eta}_{j, k}^{\Lambda, h}, \eta_{j, \infty}^{\Lambda, h})=&-[P_{n, k}\{\widehat{w}_{k}^{h}(a, X)\}]^{-1} \frac{\widehat{w}_{k}^{h}(a, X)I(\widetilde{T} \leq \tau, \Delta=1) \mathrm{R}\widehat{\mathrm{MT}}\mathrm{L}_{j, k}(\tau \mid a, X)}{\widehat{S}_{k}(\widetilde{T} \mid a, X)} \\
& \times \left\{\frac{1}{\widehat{G}_{k}(\widetilde{T}- \mid a, X)}-\frac{1}{G_{\infty}(\widetilde{T}- \mid a, X)} \right\}
\end{aligned}
$$

$$
\begin{aligned}
\mathcal{Q}_{j, 17, k}^{\mathrm{RMTL}, h}(\tau, a; \widehat{\eta}_{j, k}^{\Lambda, h}, \eta_{j, \infty}^{\Lambda, h})=&[P_{n, k}\{\widehat{w}_{k}^{h}(a, X)\}]^{-1} \widehat{w}_{k}^{h}(a, X)\mathrm{RMTL}_{j, \infty}(\tau \mid a, X) \int_0^{\tau} \frac{\mathrm{d} \Lambda_{\infty}(t \mid a, X)}{S_{\infty}(t \mid a, X)} \\
&\times \left\{\frac{1}{\widehat{G}_{k}(t- \mid a, X)}-\frac{1}{G_{\infty}(t- \mid a, X)} \right\} \\
\mathcal{Q}_{j, 18, k}^{\mathrm{RMTL}, h}(\tau, a; \widehat{\eta}_{j, k}^{\Lambda, h}, \eta_{j, \infty}^{\Lambda, h})=&[P_{n, k}\{\widehat{w}_{k}^{h}(a, X)\}]^{-1} \widehat{w}_{k}^{h}(a, X) \left\{\mathrm{R}\widehat{\mathrm{MT}}\mathrm{L}_{j, k}(\tau \mid a, X) \int_0^{\tau} \frac{\mathrm{d}\widehat{\Lambda}_{k}(t \mid a, X)}{\widehat{S}_{k}(t \mid a, X)\widehat{G}_{k}(t- \mid a, X)} \right. \\
&\left.-\mathrm{RMTL}_{j, \infty}(\tau \mid a, X) \int_0^{\tau} \frac{\mathrm{d}\Lambda_{\infty}(t \mid a, X)}{S_{\infty}(t \mid a, X)\widehat{G}_{k}(t- \mid a, X)} \right\} \\
\mathcal{Q}_{j, 19, k}^{\mathrm{RMTL}, h}(\tau, a; \widehat{\eta}_{j, k}^{\Lambda, h}, \eta_{j, \infty}^{\Lambda, h})=&([P_{n, k}\{\widehat{w}_{k}^{h}(a, X)\}]^{-1}-[E_{0}\{h_{\infty}(X)\}]^{-1}) w_{\infty}^{h}(a, X) \int_{0}^{\tau \wedge \widetilde{T}} \frac{\mathrm{RMTL}_{j, \infty}(t \mid a, X)\mathrm{d} M_{\infty}(t \mid a, X)}{S_{\infty}(t \mid a, X)G_{\infty}(t- \mid a, X)} \\
\mathcal{Q}_{j, 20, k}^{\mathrm{RMTL}, h}(\tau, a; \widehat{\eta}_{j, k}^{\Lambda, h}, \eta_{j, \infty}^{\Lambda, h})=&[P_{n, k}\{\widehat{w}_{k}^{h}(a, X)\}]^{-1} \{\widehat{w}_{k}^{h}(a, X)-w_{\infty}^{h}(a, X)\} \int_{0}^{\tau \wedge \widetilde{T}} \frac{\mathrm{RMTL}_{j, \infty}(t \mid a, X)\mathrm{d} M_{\infty}(t \mid a, X)}{S_{\infty}(t \mid a, X)G_{\infty}(t- \mid a, X)} \\
\mathcal{Q}_{j, 21, k}^{\mathrm{RMTL}, h}(\tau, a; \widehat{\eta}_{j, k}^{\Lambda, h}, \eta_{j, \infty}^{\Lambda, h})=&[P_{n, k}\{\widehat{w}_{k}^{h}(a, X)\}]^{-1} \frac{\widehat{w}_{k}^{h}(a, X)I(\widetilde{T} \leq \tau, \Delta=1)}{G_{\infty}(\widetilde{T}- \mid a, X)} \\
&\times \left\{\frac{\mathrm{R}\widehat{\mathrm{MT}}\mathrm{L}_{j, k}(\widetilde{T} \mid a, X)}{\widehat{S}_{k}(\widetilde{T} \mid a, X)}-\frac{\mathrm{RMTL}_{j, \infty}(\widetilde{T} \mid a, X)}{S_{\infty}(\widetilde{T} \mid a, X)}\right\} \\
\mathcal{Q}_{j, 22, k}^{\mathrm{RMTL}, h}(\tau, a; \widehat{\eta}_{j, k}^{\Lambda, h}, \eta_{j, \infty}^{\Lambda, h})=&[P_{n, k}\{\widehat{w}_{k}^{h}(a, X)\}]^{-1} \frac{\widehat{w}_{k}^{h}(a, X)I(\widetilde{T} \leq \tau, \Delta=1) \mathrm{R}\widehat{\mathrm{MT}}\mathrm{L}_{j, k}(\widetilde{T} \mid a, X)}{\widehat{S}_{k}(\widetilde{T} \mid a, X)} \\
& \times \left\{\frac{1}{\widehat{G}_{k}(\widetilde{T}- \mid a, X)}-\frac{1}{G_{\infty}(\widetilde{T}- \mid a, X)} \right\} \\
\mathcal{Q}_{j, 23, k}^{\mathrm{RMTL}, h}(\tau, a; \widehat{\eta}_{j, k}^{\Lambda, h}, \eta_{j, \infty}^{\Lambda, h})=&-[P_{n, k}\{\widehat{w}_{k}^{h}(a, X)\}]^{-1} \widehat{w}_{k}^{h}(a, X) \int_0^{\tau} \frac{\mathrm{RMTL}_{j, \infty}(t \mid a, X)\mathrm{d} \Lambda_{\infty}(t \mid a, X)}{S_{\infty}(t \mid a, X)} \\
&\times \left\{\frac{1}{\widehat{G}_{k}(t- \mid a, X)}-\frac{1}{G_{\infty}(t- \mid a, X)} \right\} \\
\mathcal{Q}_{j, 24, k}^{\mathrm{RMTL}, h}(\tau, a; \widehat{\eta}_{j, k}^{\Lambda, h}, \eta_{j, \infty}^{\Lambda, h})=&-[P_{n, k}\{\widehat{w}_{k}^{h}(a, X)\}]^{-1} \widehat{w}_{k}^{h}(a, X) \left\{\int_0^{\tau} \frac{\mathrm{R}\widehat{\mathrm{MT}}\mathrm{L}_{j, k}(t \mid a, X)\mathrm{d}\widehat{\Lambda}_{k}(t \mid a, X)}{\widehat{S}_{k}(t \mid a, X)\widehat{G}_{k}(t- \mid a, X)} \right. \\
&\left.-\int_0^{\tau} \frac{\mathrm{RMTL}_{j, \infty}(t \mid a, X)\mathrm{d}\Lambda_{\infty}(t \mid a, X)}{S_{\infty}(t \mid a, X)\widehat{G}_{k}(t- \mid a, X)} \right\} \\
\mathcal{Q}_{j, 25, k}^{\mathrm{RMTL}, h}(\tau, a; \widehat{\eta}_{j, k}^{\Lambda, h}, \eta_{j, \infty}^{\Lambda, h})=&([P_{n, k}\{\widehat{w}_{k}^{h}(a, X)\}]^{-1}-[E_{0}\{h_{\infty}(X)\}]^{-1}) w_{\infty}^{h}(a, X)\tau\int_{0}^{\tau \wedge \widetilde{T}} \frac{F_{j, \infty}(t \mid a, X)\mathrm{d} M_{\infty}(t \mid a, X)}{S_{\infty}(t \mid a, X)G_{\infty}(t- \mid a, X)} \\
\mathcal{Q}_{j, 26, k}^{\mathrm{RMTL}, h}(\tau, a; \widehat{\eta}_{j, k}^{\Lambda, h}, \eta_{j, \infty}^{\Lambda, h})=&[P_{n, k}\{\widehat{w}_{k}^{h}(a, X)\}]^{-1} \{\widehat{w}_{k}^{h}(a, X)-w_{\infty}^{h}(a, X)\}\tau \int_{0}^{\tau \wedge \widetilde{T}} \frac{F_{j, \infty}(t \mid a, X)\mathrm{d} M_{\infty}(t \mid a, X)}{S_{\infty}(t \mid a, X)G_{\infty}(t- \mid a, X)} \\
\mathcal{Q}_{j, 27, k}^{\mathrm{RMTL}, h}(\tau, a; \widehat{\eta}_{j, k}^{\Lambda, h}, \eta_{j, \infty}^{\Lambda, h})=&[P_{n, k}\{\widehat{w}_{k}^{h}(a, X)\}]^{-1} \frac{\widehat{w}_{k}^{h}(a, X)I(\widetilde{T} \leq \tau, \Delta=1) \tau}{G_{\infty}(\widetilde{T}- \mid a, X)} \left\{\frac{\widehat{F}_{j, k}(\widetilde{T} \mid a, X)}{\widehat{S}_{k}(\widetilde{T} \mid a, X)}-\frac{F_{j, \infty}(\widetilde{T} \mid a, X)}{S_{\infty}(\widetilde{T} \mid a, X)}\right\} \\
\mathcal{Q}_{j, 28, k}^{\mathrm{RMTL}, h}(\tau, a; \widehat{\eta}_{j, k}^{\Lambda, h}, \eta_{j, \infty}^{\Lambda, h})=&[P_{n, k}\{\widehat{w}_{k}^{h}(a, X)\}]^{-1} \frac{\widehat{w}_{k}^{h}(a, X)I(\widetilde{T} \leq \tau, \Delta=1) \tau F_{j, \infty}(\widetilde{T} \mid a, X)}{\widehat{S}_{k}(\widetilde{T} \mid a, X)} \\
& \times \left\{\frac{1}{\widehat{G}_{k}(\widetilde{T}- \mid a, X)}-\frac{1}{G_{\infty}(\widetilde{T}- \mid a, X)} \right\} \\
\end{aligned}
$$

$$
\begin{aligned}
\mathcal{Q}_{j, 29, k}^{\mathrm{RMTL}, h}(\tau, a; \widehat{\eta}_{j, k}^{\Lambda, h}, \eta_{j, \infty}^{\Lambda, h})=&-[P_{n, k}\{\widehat{w}_{k}^{h}(a, X)\}]^{-1} \widehat{w}_{k}^{h}(a, X)\tau \int_0^{\tau} \frac{F_{j, \infty}(t \mid a, X)\mathrm{d} \Lambda_{\infty}(t \mid a, X)}{S_{\infty}(t \mid a, X)} \\
&\times \left\{\frac{1}{\widehat{G}_{k}(t- \mid a, X)}-\frac{1}{G_{\infty}(t- \mid a, X)} \right\} \\
\mathcal{Q}_{j, 30, k}^{\mathrm{RMTL}, h}(\tau, a; \widehat{\eta}_{j, k}^{\Lambda, h}, \eta_{j, \infty}^{\Lambda, h})=&-[P_{n, k}\{\widehat{w}_{k}^{h}(a, X)\}]^{-1} \widehat{w}_{k}^{h}(a, X)\tau \left\{\int_0^{\tau} \frac{\widehat{F}_{j, k}(t \mid a, X)\mathrm{d}\widehat{\Lambda}_{k}(t \mid a, X)}{\widehat{S}_{k}(t \mid a, X)\widehat{G}_{k}(t- \mid a, X)} \right. \\
&\left.-\int_0^{\tau} \frac{F_{j, \infty}(t \mid a, X)\mathrm{d}\Lambda_{\infty}(t \mid a, X)}{S_{\infty}(t \mid a, X)\widehat{G}_{k}(t- \mid a, X)} \right\} \\
\mathcal{Q}_{j, 31, k}^{\mathrm{RMTL}, h}(\tau, a; \widehat{\eta}_{j, k}^{\Lambda, h}, \eta_{j, \infty}^{\Lambda, h})=&-([P_{n, k}\{\widehat{w}_{k}^{h}(a, X)\}]^{-1}-[E_{0}\{h_{\infty}(X)\}]^{-1}) w_{\infty}^{h}(a, X) \int_{0}^{\tau \wedge \widetilde{T}} \frac{t F_{j, \infty}(t \mid a, X)\mathrm{d} M_{\infty}(t \mid a, X)}{S_{\infty}(t \mid a, X)G_{\infty}(t- \mid a, X)} \\
\mathcal{Q}_{j, 32, k}^{\mathrm{RMTL}, h}(\tau, a; \widehat{\eta}_{j, k}^{\Lambda, h}, \eta_{j, \infty}^{\Lambda, h})=&-[P_{n, k}\{\widehat{w}_{k}^{h}(a, X)\}]^{-1} \{\widehat{w}_{k}^{h}(a, X)-w_{\infty}^{h}(a, X)\} \int_{0}^{\tau \wedge \widetilde{T}} \frac{t F_{j, \infty}(t \mid a, X)\mathrm{d} M_{\infty}(t \mid a, X)}{S_{\infty}(t \mid a, X)G_{\infty}(t- \mid a, X)} \\
\mathcal{Q}_{j, 33, k}^{\mathrm{RMTL}, h}(\tau, a; \widehat{\eta}_{j, k}^{\Lambda, h}, \eta_{j, \infty}^{\Lambda, h})=&-[P_{n, k}\{\widehat{w}_{k}^{h}(a, X)\}]^{-1} \frac{\widehat{w}_{k}^{h}(a, X)I(\widetilde{T} \leq \tau, \Delta=1) \widetilde{T}}{G_{\infty}(\widetilde{T}- \mid a, X)} \left\{\frac{\widehat{F}_{j, k}(\widetilde{T} \mid a, X)}{\widehat{S}_{k}(\widetilde{T} \mid a, X)}-\frac{F_{j, \infty}(\widetilde{T} \mid a, X)}{S_{\infty}(\widetilde{T} \mid a, X)}\right\} \\
\mathcal{Q}_{j, 34, k}^{\mathrm{RMTL}, h}(\tau, a; \widehat{\eta}_{j, k}^{\Lambda, h}, \eta_{j, \infty}^{\Lambda, h})=&-[P_{n, k}\{\widehat{w}_{k}^{h}(a, X)\}]^{-1} \frac{\widehat{w}_{k}^{h}(a, X)I(\widetilde{T} \leq \tau, \Delta=1) \widetilde{T} F_{j, \infty}(\widetilde{T} \mid a, X)}{\widehat{S}_{k}(\widetilde{T} \mid a, X)} \\
& \times \left\{\frac{1}{\widehat{G}_{k}(\widetilde{T}- \mid a, X)}-\frac{1}{G_{\infty}(\widetilde{T}- \mid a, X)} \right\} \\
\mathcal{Q}_{j, 35, k}^{\mathrm{RMTL}, h}(\tau, a; \widehat{\eta}_{j, k}^{\Lambda, h}, \eta_{j, \infty}^{\Lambda, h})=&[P_{n, k}\{\widehat{w}_{k}^{h}(a, X)\}]^{-1} \widehat{w}_{k}^{h}(a, X) \int_0^{\tau} \frac{t F_{j, \infty}(t \mid a, X)\mathrm{d} \Lambda_{\infty}(t \mid a, X)}{S_{\infty}(t \mid a, X)} \\
&\times \left\{\frac{1}{\widehat{G}_{k}(t- \mid a, X)}-\frac{1}{G_{\infty}(t- \mid a, X)} \right\} \\
\mathcal{Q}_{j, 36, k}^{\mathrm{RMTL}, h}(\tau, a; \widehat{\eta}_{j, k}^{\Lambda, h}, \eta_{j, \infty}^{\Lambda, h})=&[P_{n, k}\{\widehat{w}_{k}^{h}(a, X)\}]^{-1} \widehat{w}_{k}^{h}(a, X) \left\{\int_0^{\tau} \frac{t \widehat{F}_{j, k}(t \mid a, X)\mathrm{d}\widehat{\Lambda}_{k}(t \mid a, X)}{\widehat{S}_{k}(t \mid a, X)\widehat{G}_{k}(t- \mid a, X)} \right. \\
&\left.-\int_0^{\tau} \frac{t F_{j, \infty}(t \mid a, X)\mathrm{d}\Lambda_{\infty}(t \mid a, X)}{S_{\infty}(t \mid a, X)\widehat{G}_{k}(t- \mid a, X)} \right\}
\end{aligned}
$$

Similar to $P_{0} \{\mathcal{Q}_{\iota, k}^{\mathrm{RMST}, h}(\tau, a; \widehat{\eta}_{j, k}^{\Lambda, h}, \eta_{j, \infty}^{\Lambda, h})\}^2$
$$
\begin{aligned}
P_{0} \{\mathcal{Q}_{j, 1, k}^{\mathrm{RMTL}, h}(\tau, a; \widehat{\eta}_{j, k}^{\Lambda, h}, \eta_{j, \infty}^{\Lambda, h})\}^2\leq& \epsilon^2 h_{\mathrm{max}}^2 E_{0} \left[\sup_{t \in [0, \tau]} \left| \frac{\mathrm{R}\widehat{\mathrm{MT}}\mathrm{L}_{j, k}(\tau \mid a, X)}{\widehat{S}_{k}(t \mid a, X)}-\frac{\mathrm{RMTL}_{j, \infty}(\tau \mid a, X)}{S_{\infty}(t \mid a, X)} \right| \right]^2 \\
P_{0} \{\mathcal{Q}_{j, 2, k}^{\mathrm{RMTL}, h}(\tau, a; \widehat{\eta}_{j, k}^{\Lambda, h}, \eta_{j, \infty}^{\Lambda, h})\}^2\leq& \tau^2 E_{0} \left|([P_{n, k}\{\widehat{h}_{k}(X)\}]^{-1}\widehat{h}_{k}(X)-[E_{0}\{h_{\infty}(X)\}]^{-1}h_{\infty}(X)) \right|^2
\end{aligned}
$$

Since
$$
\begin{aligned}
&E_{0}\left|\int_0^{\tau \wedge \widetilde{T}} \frac{\mathrm{d}M_{j, \infty}(t \mid a, X)}{G_{\infty}(t- \mid a, X)}\right|^2=E_{0}\left|\frac{I(\widetilde{T} \leq \tau, \widetilde{J}=j)}{G_{\infty}(\widetilde{T}- \mid a, X)}-\int_0^{\tau} \frac{\mathrm{d}F_{j, \infty}(t \mid a, X)}{S_{\infty}(t- \mid a, X)G_{\infty}(t- \mid a, X)} \right|^2 \\
\leq & \epsilon^2+\epsilon^4 E_{0}\left|F_{j, \infty}(\tau \mid a, X) \right|^2 \leq \epsilon^2+\epsilon^4
\end{aligned}
$$

\noindent we have
$$
\begin{aligned}
P_{0} \{\mathcal{Q}_{j, 3, k}^{\mathrm{RMTL}, h}(\tau, a; \widehat{\eta}_{j, k}^{\Lambda, h}, \eta_{j, \infty}^{\Lambda, h})\}^2 \leq& ([P_{n, k}\{\widehat{w}_{k}^{h}(a, X)\}]^{-1}-[E_{0}\{h_{\infty}(X)\}]^{-1})^2 \epsilon^{2}h_{\mathrm{max}}^2 \tau^2 (\epsilon^2+\epsilon^4) \\
P_{0} \{\mathcal{Q}_{j, 4, k}^{\mathrm{RMTL}, h}(\tau, a; \widehat{\eta}_{j, k}^{\Lambda, h}, \eta_{j, \infty}^{\Lambda, h})\}^2\leq& h_{\mathrm{min}}^{-2} \tau^2 (\epsilon^2+\epsilon^4) E_{0} \left|\widehat{w}_{k}^{h}(a, X)-w_{\infty}^{h}(a, X) \right|^2 \\
P_{0} \{\mathcal{Q}_{j, 5, k}^{\mathrm{RMTL}, h}(\tau, a; \widehat{\eta}_{j, k}^{\Lambda, h}, \eta_{j, \infty}^{\Lambda, h})\}^2\leq& h_{\mathrm{min}}^{-2} h_{\mathrm{max}}^2 \tau^2 \epsilon^{2} E_{0} \left[\sup_{t \in [0, \tau]} \left| \frac{1}{\widehat{G}_{k}(t- \mid a, X)}-\frac{1}{G_{\infty}(t- \mid a, X)} \right| \right]^2 \\
P_{0} \{\mathcal{Q}_{j, 6, k}^{\mathrm{RMTL}, h}(\tau, a; \widehat{\eta}_{j, k}^{\Lambda, h}, \eta_{j, \infty}^{\Lambda, h})\}^2\leq& h_{\mathrm{min}}^{-2} h_{\mathrm{max}}^2 \tau^2 \epsilon^{2} E_{0} \left[\sup_{t \in [0, \tau]} \left| \frac{1}{\widehat{G}_{k}(t- \mid a, X)}-\frac{1}{G_{\infty}(t- \mid a, X)} \right| \right]^2 E_{0} \left|\int_0^\tau \frac{\mathrm{d}F_{j, \infty}(t \mid a, X)}{S_{\infty}(t- \mid a, X)} \right|^2 \\
&\leq h_{\mathrm{min}}^{-2} h_{\mathrm{max}}^2 \tau^2 \epsilon^{2} \epsilon^{2} E_{0} \left[\sup_{t \in [0, \tau]} \left| \frac{1}{\widehat{G}_{k}(t- \mid a, X)}-\frac{1}{G_{\infty}(t- \mid a, X)} \right| \right]^2 \\
P_{0} \{\mathcal{Q}_{j, 7, k}^{\mathrm{RMTL}, h}(\tau, a; \widehat{\eta}_{j, k}^{\Lambda, h}, \eta_{j, \infty}^{\Lambda, h})\}^2\leq& h_{\mathrm{min}}^{-2} h_{\mathrm{max}}^2 \tau^2 \epsilon^{2} \epsilon^{2} E_{0} \left[\sup_{t \in [0, \tau]} \left| \widehat{\Lambda}_{j, k}(t \mid a, X)-\Lambda_{j, \infty}(t \mid a, X) \right| \right]^2 \\
P_{0} \{\mathcal{Q}_{j, 8, k}^{\mathrm{RMTL}, h}(\tau, a; \widehat{\eta}_{j, k}^{\Lambda, h}, \eta_{j, \infty}^{\Lambda, h})\}^2 \leq& ([P_{n, k}\{\widehat{w}_{k}^{h}(a, X)\}]^{-1}-[E_{0}\{h_{\infty}(X)\}]^{-1})^2 \epsilon^{2}h_{\mathrm{max}}^2 \tau^2 (\epsilon^2+\epsilon^4) \\
P_{0} \{\mathcal{Q}_{j, 9, k}^{\mathrm{RMTL}, h}(\tau, a; \widehat{\eta}_{j, k}^{\Lambda, h}, \eta_{j, \infty}^{\Lambda, h})\}^2\leq& h_{\mathrm{min}}^{-2} \tau^2 (\epsilon^2+\epsilon^4) E_{0} \left|\widehat{w}_{k}^{h}(a, X)-w_{\infty}^{h}(a, X) \right|^2 \\
P_{0} \{\mathcal{Q}_{j, 10, k}^{\mathrm{RMTL}, h}(\tau, a; \widehat{\eta}_{j, k}^{\Lambda, h}, \eta_{j, \infty}^{\Lambda, h})\}^2\leq& h_{\mathrm{min}}^{-2} h_{\mathrm{max}}^2 \tau^2 \epsilon^{2} E_{0} \left[\sup_{t \in [0, \tau]} \left| \frac{1}{\widehat{G}_{k}(t- \mid a, X)}-\frac{1}{G_{\infty}(t- \mid a, X)} \right| \right]^2 \\
P_{0} \{\mathcal{Q}_{j, 11, k}^{\mathrm{RMTL}, h}(\tau, a; \widehat{\eta}_{j, k}^{\Lambda, h}, \eta_{j, \infty}^{\Lambda, h})\}^2 \leq& h_{\mathrm{min}}^{-2} h_{\mathrm{max}}^2 \tau^2 \epsilon^{2} \epsilon^{2} E_{0} \left[\sup_{t \in [0, \tau]} \left| \frac{1}{\widehat{G}_{k}(t- \mid a, X)}-\frac{1}{G_{\infty}(t- \mid a, X)} \right| \right]^2 \\
P_{0} \{\mathcal{Q}_{j, 12, k}^{\mathrm{RMTL}, h}(\tau, a; \widehat{\eta}_{j, k}^{\Lambda, h}, \eta_{j, \infty}^{\Lambda, h})\}^2\leq& h_{\mathrm{min}}^{-2} h_{\mathrm{max}}^2 \tau^2 \epsilon^{2} \epsilon^{2} E_{0} \left[\sup_{t \in [0, \tau]} \left| \widehat{\Lambda}_{j, k}(t \mid a, X)-\Lambda_{j, \infty}(t \mid a, X) \right| \right]^2 \\
P_{0} \{\mathcal{Q}_{j, 13, k}^{\mathrm{RMTL}, h}(\tau, a; \widehat{\eta}_{j, k}^{\Lambda, h}, \eta_{j, \infty}^{\Lambda, h})\}^2\leq& ([P_{n, k}\{\widehat{w}_{k}^{h}(a, X)\}]^{-1}-[E_{0}\{h_{\infty}(X)\}]^{-1})^2 \epsilon^{2}h_{\mathrm{max}}^2 \tau^2 \times 4 \epsilon^{4} \\
P_{0} \{\mathcal{Q}_{j, 14, k}^{\mathrm{RMTL}, h}(\tau, a; \widehat{\eta}_{j, k}^{\Lambda, h}, \eta_{j, \infty}^{\Lambda, h})\}^2\leq& h_{\mathrm{min}}^{-2} \tau^2 \times 4 \epsilon^{4} E_{0} \left|\widehat{w}_{k}^{h}(a, X)-w_{\infty}^{h}(a, X) \right|^2 \\
P_{0} \{\mathcal{Q}_{j, 15, k}^{\mathrm{RMTL}, h}(\tau, a; \widehat{\eta}_{j, k}^{\Lambda, h}, \eta_{j, \infty}^{\Lambda, h})\}^2\leq& h_{\mathrm{min}}^{-2} h_{\mathrm{max}}^2 \epsilon^{4} E_{0} \left[\sup_{t \in [0, \tau]} \left| \frac{\mathrm{R}\widehat{\mathrm{MT}}\mathrm{L}_{j, k}(\tau \mid a, X)}{\widehat{S}_{k}(t \mid a, X)}-\frac{\mathrm{RMTL}_{j, \infty}(\tau \mid a, X)}{S_{\infty}(t \mid a, X)} \right| \right]^2 \\
P_{0} \{\mathcal{Q}_{j, 16, k}^{\mathrm{RMTL}, h}(\tau, a; \widehat{\eta}_{j, k}^{\Lambda, h}, \eta_{j, \infty}^{\Lambda, h})\}^2 \leq& h_{\mathrm{min}}^{-2} h_{\mathrm{max}}^2 \tau^2 \epsilon^{4} E_{0} \left[\sup_{t \in [0, \tau]} \left| \frac{1}{\widehat{G}_{k}(t- \mid a, X)}-\frac{1}{G_{\infty}(t- \mid a, X)} \right| \right]^2 \\
P_{0} \{\mathcal{Q}_{j, 17, k}^{\mathrm{RMTL}, h}(\tau, a; \widehat{\eta}_{j, k}^{\Lambda, h}, \eta_{j, \infty}^{\Lambda, h})\}^2 \leq& h_{\mathrm{min}}^{-2} h_{\mathrm{max}}^2 \tau^2 \epsilon^{2} (\epsilon-1)^{2} E_{0} \left[\sup_{t \in [0, \tau]} \left| \frac{1}{\widehat{G}_{k}(t- \mid a, X)}-\frac{1}{G_{\infty}(t- \mid a, X)} \right| \right]^2 \\
P_{0} \{\mathcal{Q}_{j, 18, k}^{\mathrm{RMTL}, h}(\tau, a; \widehat{\eta}_{j, k}^{\Lambda, h}, \eta_{j, \infty}^{\Lambda, h})\}^2\leq& h_{\mathrm{min}}^{-2} h_{\mathrm{max}}^2 \epsilon^{2} \{\epsilon^{2}+(\epsilon-1)^{2}\} E_{0} \left\{\sup_{t \in [0, \tau]}\left|\frac{\mathrm{R}\widehat{\mathrm{MT}}\mathrm{L}_{j, k}(\tau \mid a, X)}{\widehat{S}_{k}(t \mid a, X)}-\frac{\mathrm{RMTL}_{j, \infty}(\tau \mid a, X)}{S_{\infty}(t \mid a, X)}\right|\right\}^2 \\
P_{0} \{\mathcal{Q}_{j, 19, k}^{\mathrm{RMTL}, h}(\tau, a; \widehat{\eta}_{j, k}^{\Lambda, h}, \eta_{j, \infty}^{\Lambda, h})\}^2 \leq& ([P_{n, k}\{\widehat{w}_{k}^{h}(a, X)\}]^{-1}-[E_{0}\{h_{\infty}(X)\}]^{-1})^2 \epsilon^{2}h_{\mathrm{max}}^2 \tau^2 \times 4 \epsilon^{4} \\
P_{0} \{\mathcal{Q}_{j, 20, k}^{\mathrm{RMTL}, h}(\tau, a; \widehat{\eta}_{j, k}^{\Lambda, h}, \eta_{j, \infty}^{\Lambda, h})\}^2 \leq& h_{\mathrm{min}}^{-2} \tau^2 \times 4 \epsilon^{4} E_{0} \left|\widehat{w}_{k}^{h}(a, X)-w_{\infty}^{h}(a, X) \right|^2 \\
P_{0} \{\mathcal{Q}_{j, 21, k}^{\mathrm{RMTL}, h}(\tau, a; \widehat{\eta}_{j, k}^{\Lambda, h}, \eta_{j, \infty}^{\Lambda, h})\}^2 \leq & h_{\mathrm{min}}^{-2} h_{\mathrm{max}}^2 \epsilon^{4} E_{0} \left[\sup_{t \in [0, \tau]} \left| \frac{\mathrm{R}\widehat{\mathrm{MT}}\mathrm{L}_{j, k}(t \mid a, X)}{\widehat{S}_{k}(t \mid a, X)}-\frac{\mathrm{RMTL}_{j, \infty}(t \mid a, X)}{S_{\infty}(t \mid a, X)} \right| \right]^2
\end{aligned}
$$

$$
\begin{aligned}
P_{0} \{\mathcal{Q}_{j, 22, k}^{\mathrm{RMTL}, h}(\tau, a; \widehat{\eta}_{j, k}^{\Lambda, h}, \eta_{j, \infty}^{\Lambda, h})\}^2\leq& h_{\mathrm{min}}^{-2} h_{\mathrm{max}}^2 \tau^2 \epsilon^{4} E_{0} \left[\sup_{t \in [0, \tau]} \left| \frac{1}{\widehat{G}_{k}(t- \mid a, X)}-\frac{1}{G_{\infty}(t- \mid a, X)} \right| \right]^2 \\
P_{0} \{\mathcal{Q}_{j, 23, k}^{\mathrm{RMTL}, h}(\tau, a; \widehat{\eta}_{j, k}^{\Lambda, h}, \eta_{j, \infty}^{\Lambda, h})\}^2\leq& h_{\mathrm{min}}^{-2} h_{\mathrm{max}}^2 \tau^2 \epsilon^{2} (\epsilon-1)^{2} E_{0} \left[\sup_{t \in [0, \tau]} \left| \frac{1}{\widehat{G}_{k}(t- \mid a, X)}-\frac{1}{G_{\infty}(t- \mid a, X)} \right| \right]^2
\end{aligned}
$$

As
$$
\begin{aligned}
&E_{0}\left| \int_0^{\tau} \frac{\mathrm{R}\widehat{\mathrm{MT}}\mathrm{L}_{j, k}(t \mid a, X)\mathrm{d}\widehat{\Lambda}_{k}(t \mid a, X)}{\widehat{S}_{k}(t \mid a, X)\widehat{G}_{k}(t- \mid a, X)}-\int_0^{\tau} \frac{\mathrm{RMTL}_{j, \infty}(t \mid a, X)\mathrm{d}\Lambda_{\infty}(t \mid a, X)}{S_{\infty}(t \mid a, X)\widehat{G}_{k}(t- \mid a, X)}\right|^2 \\
=&E_{0}\left| \left.\frac{\mathrm{R}\widehat{\mathrm{MT}}\mathrm{L}_{j, k}(t \mid a, X)}{\widehat{S}_{k}(t \mid a, X)\widehat{G}_{k}(t- \mid a, X)} \right|_{0}^{\tau}-\int_0^{\tau} \frac{1}{\widehat{S}_{k}(t \mid a, X)} \mathrm{d}\left\{\frac{\mathrm{R}\widehat{\mathrm{MT}}\mathrm{L}_{j, k}(t \mid a, X)}{\widehat{G}_{k}(t- \mid a, X)}\right\} \right. \\
&\left.\left.-\frac{\mathrm{RMTL}_{j, \infty}(t \mid a, X)}{S_{\infty}(t \mid a, X)\widehat{G}_{k}(t- \mid a, X)} \right|_{0}^{\tau}+\int_0^{\tau} \frac{1}{S_{\infty}(t \mid a, X)} \mathrm{d}\left\{\frac{\mathrm{RMTL}_{j, \infty}(t \mid a, X)}{\widehat{G}_{k}(t- \mid a, X)}\right\}\right|^2 \\
=&E_{0}\left| \left\{\frac{\mathrm{R}\widehat{\mathrm{MT}}\mathrm{L}_{j, k}(\tau \mid a, X)}{\widehat{S}_{k}(\tau \mid a, X)}-\frac{\mathrm{RMTL}_{j, \infty}(\tau \mid a, X)}{S_{\infty}(\tau \mid a, X)}\right\}\frac{1}{\widehat{G}_{k}(\tau- \mid a, X)}-\int_0^{\tau} \left\{\frac{\widehat{F}_{j, k}(t \mid a, X)}{\widehat{S}_{k}(t \mid a, X)} -\frac{F_{j, \infty}(t \mid a, X)}{S_{\infty}(t \mid a, X)} \right\}\right. \\
& \frac{\mathrm{d}t}{\widehat{G}_{k}(t- \mid a, X)} \left.-\int_0^{\tau} \left\{\frac{\mathrm{R}\widehat{\mathrm{MT}}\mathrm{L}_{j, k}(t \mid a, X)}{\widehat{S}_{k}(t \mid a, X)}-\frac{\mathrm{RMTL}_{j, \infty}(t \mid a, X)}{S_{\infty}(t \mid a, X)}\right\} \mathrm{d} \left\{\frac{1}{\widehat{G}_{k}(t- \mid a, X)}\right\}\right|^2 \\
\leq & \left\{\epsilon^2+(\epsilon-1)^2\right\} E_{0}\left\{\sup_{t \in [0, \tau]} \left| \frac{\mathrm{R}\widehat{\mathrm{MT}}\mathrm{L}_{j, k}(t \mid a, X)}{\widehat{S}_{k}(t \mid a, X)}-\frac{\mathrm{RMTL}_{j, \infty}(t \mid a, X)}{S_{\infty}(t \mid a, X)}\right|\right\}^2 \\
&+ \epsilon^2 \tau^2 E_{0}\left\{\sup_{t \in [0, \tau]} \left|\frac{\widehat{F}_{j, k}(t \mid a, X)}{\widehat{S}_{k}(t \mid a, X)} -\frac{F_{j, \infty}(t \mid a, X)}{S_{\infty}(t \mid a, X)} \right|\right\}^2
\end{aligned}
$$

Thus
$$
\begin{aligned}
P_{0} \{\mathcal{Q}_{j, 24, k}^{\mathrm{RMTL}, h}(\tau, a; \widehat{\eta}_{j, k}^{\Lambda, h}, \eta_{j, \infty}^{\Lambda, h})\}^2\leq& h_{\mathrm{min}}^{-2} h_{\mathrm{max}}^2 \epsilon^{2} \{\epsilon^{2}+(\epsilon-1)^{2}\} E_{0} \left\{\sup_{t \in [0, \tau]}\left|\frac{\mathrm{R}\widehat{\mathrm{MT}}\mathrm{L}_{j, k}(t \mid a, X)}{\widehat{S}_{k}(t \mid a, X)}-\frac{\mathrm{RMTL}_{j, \infty}(t \mid a, X)}{S_{\infty}(t \mid a, X)}\right|\right\}^2 \\
&+h_{\mathrm{min}}^{-2} h_{\mathrm{max}}^2 \epsilon^{2} \epsilon^2 \tau^2 E_{0} \left\{\sup_{t \in [0, \tau]} \left|\frac{\widehat{F}_{j, k}(t \mid a, X)}{\widehat{S}_{k}(t \mid a, X)} -\frac{F_{j, \infty}(t \mid a, X)}{S_{\infty}(t \mid a, X)} \right|\right\}^2 \\
P_{0} \{\mathcal{Q}_{j, 25, k}^{\mathrm{RMTL}, h}(\tau, a; \widehat{\eta}_{j, k}^{\Lambda, h}, \eta_{j, \infty}^{\Lambda, h})\}^2 \leq& ([P_{n, k}\{\widehat{w}_{k}^{h}(a, X)\}]^{-1}-[E_{0}\{h_{\infty}(X)\}]^{-1})^2 \epsilon^{2}h_{\mathrm{max}}^2 \tau^2 \times 4 \epsilon^{4} \\
P_{0} \{\mathcal{Q}_{j, 26, k}^{\mathrm{RMTL}, h}(\tau, a; \widehat{\eta}_{j, k}^{\Lambda, h}, \eta_{j, \infty}^{\Lambda, h})\}^2 \leq& h_{\mathrm{min}}^{-2} \tau^2 \times 4 \epsilon^{4} E_{0} \left|\widehat{w}_{k}^{h}(a, X)-w_{\infty}^{h}(a, X) \right|^2 \\
P_{0} \{\mathcal{Q}_{j, 27, k}^{\mathrm{RMTL}, h}(\tau, a; \widehat{\eta}_{j, k}^{\Lambda, h}, \eta_{j, \infty}^{\Lambda, h})\}^2 \leq& h_{\mathrm{min}}^{-2} h_{\mathrm{max}}^2 \tau^2 \epsilon^{4} E_{0} \left[\sup_{t \in [0, \tau]} \left| \frac{\widehat{F}_{j, k}(t \mid a, X)}{\widehat{S}_{k}(t \mid a, X)}-\frac{F_{j, \infty}(t \mid a, X)}{S_{\infty}(t \mid a, X)} \right| \right]^2 \\
P_{0} \{\mathcal{Q}_{j, 28, k}^{\mathrm{RMTL}, h}(\tau, a; \widehat{\eta}_{j, k}^{\Lambda, h}, \eta_{j, \infty}^{\Lambda, h})\}^2\leq& h_{\mathrm{min}}^{-2} h_{\mathrm{max}}^2 \tau^2 \epsilon^{4} E_{0} \left[\sup_{t \in [0, \tau]} \left| \frac{1}{\widehat{G}_{k}(t- \mid a, X)}-\frac{1}{G_{\infty}(t- \mid a, X)} \right| \right]^2 \\
P_{0} \{\mathcal{Q}_{j, 29, k}^{\mathrm{RMTL}, h}(\tau, a; \widehat{\eta}_{j, k}^{\Lambda, h}, \eta_{j, \infty}^{\Lambda, h})\}^2\leq& h_{\mathrm{min}}^{-2} h_{\mathrm{max}}^2 \tau^2 \epsilon^{2} (\epsilon-1)^{2} E_{0} \left[\sup_{t \in [0, \tau]} \left| \frac{1}{\widehat{G}_{k}(t- \mid a, X)}-\frac{1}{G_{\infty}(t- \mid a, X)} \right| \right]^2
\end{aligned}
$$

Note that
$$
\begin{aligned}
&E_{0} \left|\int_0^{\tau} \left\{\frac{\mathrm{d}\widehat{F}_{j, k}(t \mid a, X)}{\widehat{S}_{k}(t \mid a, X)}-\frac{\mathrm{d}F_{j, \infty}(t \mid a, X)}{S_{\infty}(t \mid a, X)} \right\} \right|^2 \\
=&E_{0} \left|\left.\left\{\frac{\widehat{F}_{j, k}(t \mid a, X)}{\widehat{S}_{k}(t \mid a, X)}\right\} \right|_{0}^{\tau}-\int_0^{\tau} \frac{\widehat{F}_{j, k}(t \mid a, X)}{\widehat{S}_{k}(t \mid a, X)} \mathrm{d}\widehat{\Lambda}_{k}(t \mid a, X)-\left.\frac{F_{j, \infty}(t \mid a, X)}{S_{\infty}(t \mid a, X)} \right|_{0}^{\tau}+\int_0^{\tau} \frac{F_{j, \infty}(t \mid a, X)}{S_{\infty}(t \mid a, X)} \mathrm{d}\Lambda_{\infty}(t \mid a, X)\right|^2 \\
=&E_{0} \left|\frac{\widehat{F}_{j, k}(\tau \mid a, X)}{\widehat{S}_{k}(\tau \mid a, X)}-\frac{F_{j, \infty}(\tau \mid a, X)}{S_{\infty}(\tau \mid a, X)}-\int_0^{\tau}\frac{\widehat{F}_{j, k}(t \mid a, X)}{\widehat{S}_{k}(t \mid a, X)} \{\mathrm{d}\widehat{\Lambda}_{k}(t \mid a, X)-\mathrm{d}\Lambda_{\infty}(t \mid a, X)\} \right. \\
&\left. -\int_0^{\tau}\left\{\frac{\widehat{F}_{j, k}(t \mid a, X)}{\widehat{S}_{k}(t \mid a, X)}-\frac{F_{j, \infty}(t \mid a, X)}{S_{\infty}(t \mid a, X)} \right\} \mathrm{d}\Lambda_{\infty}(t \mid a, X)\right|^2 \\
\leq& E_{0}\left\{\sup_{t \in [0, \tau]} \left|\frac{\widehat{F}_{j, k}(t \mid a, X)}{\widehat{S}_{k}(t \mid a, X)} -\frac{F_{j, \infty}(t \mid a, X)}{S_{\infty}(t \mid a, X)} \right|\right\}^2 + \epsilon^2 E_{0}\left\{\sup_{t \in [0, \tau]} \left|\widehat{\Lambda}_{k}(t \mid a, X)-\Lambda_{\infty}(t \mid a, X)\right|\right\}^2 \\
&+\log^2 \epsilon E_{0}\left\{\sup_{t \in [0, \tau]} \left|\frac{\widehat{F}_{j, k}(t \mid a, X)}{\widehat{S}_{k}(t \mid a, X)} -\frac{F_{j, \infty}(t \mid a, X)}{S_{\infty}(t \mid a, X)} \right|\right\}^2
\end{aligned}
$$

\noindent such that
$$
\begin{aligned}
&E_{0}\left| \int_0^{\tau} \frac{\widehat{F}_{j, k}(t \mid a, X)\mathrm{d}\widehat{\Lambda}_{k}(t \mid a, X)}{\widehat{S}_{k}(t \mid a, X)\widehat{G}_{k}(t- \mid a, X)}-\int_0^{\tau} \frac{F_{j, \infty}(t \mid a, X)\mathrm{d}\Lambda_{\infty}(t \mid a, X)}{S_{\infty}(t \mid a, X)\widehat{G}_{k}(t- \mid a, X)}\right|^2 \\
=&E_{0}\left|\left.\left\{\frac{\widehat{F}_{j, k}(t \mid a, X)}{\widehat{S}_{k}(t \mid a, X)\widehat{G}_{k}(t- \mid a, X)}\right\}\right|_{0}^{\tau}-\int_0^{\tau} \frac{1}{\widehat{S}_{k}(t \mid a, X)} \mathrm{d}\left\{\frac{\widehat{F}_{j, k}(t \mid a, X)}{\widehat{G}_{k}(t- \mid a, X)}\right\}-\left.\frac{F_{j, \infty}(t \mid a, X)}{S_{\infty}(t \mid a, X)\widehat{G}_{k}(t- \mid a, X)} \right|_{0}^{\tau} \right. \\
&\left.+\int_0^{\tau} \frac{1}{S_{\infty}(t \mid a, X)} \mathrm{d}\left\{\frac{F_{j, \infty}(t \mid a, X)}{\widehat{G}_{k}(t- \mid a, X)}\right\}\right|^2 \\
=&E_{0}\left| \left\{\frac{\widehat{F}_{j, k}(\tau \mid a, X)}{\widehat{S}_{k}(\tau \mid a, X)}-\frac{F_{j, \infty}(\tau \mid a, X)}{S_{\infty}(\tau \mid a, X)}\right\}\frac{1}{\widehat{G}_{k}(\tau- \mid a, X)}-\int_0^{\tau} \left\{\frac{\mathrm{d}\widehat{F}_{j, k}(t \mid a, X)}{\widehat{S}_{k}(t \mid a, X)}-\frac{\mathrm{d}F_{j, \infty}(t \mid a, X)}{S_{\infty}(t \mid a, X)} \right\} \right. \\
& \times \frac{1}{\widehat{G}_{k}(t- \mid a, X)} \left.-\int_0^{\tau} \left\{\frac{\widehat{F}_{j, k}(t \mid a, X)}{\widehat{S}_{k}(t \mid a, X)}-\frac{F_{j, \infty}(t \mid a, X)}{S_{\infty}(t \mid a, X)}\right\} \mathrm{d} \left\{\frac{1}{\widehat{G}_{k}(t- \mid a, X)}\right\}\right|^2 \\
\leq & \{\epsilon^2+(\epsilon-1)^2\} E_{0}\left\{\sup_{t \in [0, \tau]} \left| \frac{\widehat{F}_{j, k}(t \mid a, X)}{\widehat{S}_{k}(t \mid a, X)}-\frac{F_{j, \infty}(t \mid a, X)}{S_{\infty}(t \mid a, X)}\right|\right\}^2 \\
&+\epsilon^2 E_{0} \left[\int_0^{\tau} \left\{\frac{\mathrm{d}\widehat{F}_{j, k}(t \mid a, X)}{\widehat{S}_{k}(t \mid a, X)}-\frac{\mathrm{d}F_{j, \infty}(t \mid a, X)}{S_{\infty}(t \mid a, X)} \right\} \right]^2 \\
=&\{2\epsilon^2+(\epsilon-1)^2+\epsilon^2 \log^2 \epsilon\} E_{0}\left\{\sup_{t \in [0, \tau]} \left| \frac{\widehat{F}_{j, k}(t \mid a, X)}{\widehat{S}_{k}(t \mid a, X)}-\frac{F_{j, \infty}(t \mid a, X)}{S_{\infty}(t \mid a, X)}\right|\right\}^2 \\
&+\epsilon^4 E_{0}\left\{\sup_{t \in [0, \tau]} \left|\widehat{\Lambda}_{k}(t \mid a, X)-\Lambda_{\infty}(t \mid a, X)\right|\right\}^2
\end{aligned}
$$

$$
\begin{aligned}
P_{0} \{\mathcal{Q}_{j, 30, k}^{\mathrm{RMTL}, h}(\tau, a; \widehat{\eta}_{j, k}^{\Lambda, h}, \eta_{j, \infty}^{\Lambda, h})\}^2\leq& h_{\mathrm{min}}^{-2} h_{\mathrm{max}}^2 \epsilon^{2} \tau^{2} \left[ \{2\epsilon^2+(\epsilon-1)^2+\epsilon^2 \log^2 \epsilon\} E_{0}\left\{\sup_{t \in [0, \tau]} \left| \frac{\widehat{F}_{j, k}(t \mid a, X)}{\widehat{S}_{k}(t \mid a, X)} \right.\right.\right. \\
&\left.\left.\left. -\frac{F_{j, \infty}(t \mid a, X)}{S_{\infty}(t \mid a, X)}\right|\right\}^2 +\epsilon^4 E_{0}\left\{\sup_{t \in [0, \tau]} \left|\widehat{\Lambda}_{k}(t \mid a, X)-\Lambda_{\infty}(t \mid a, X)\right|\right\}^2 \right] \\
P_{0} \{\mathcal{Q}_{j, 31, k}^{\mathrm{RMTL}, h}(\tau, a; \widehat{\eta}_{j, k}^{\Lambda, h}, \eta_{j, \infty}^{\Lambda, h})\}^2 \leq& ([P_{n, k}\{\widehat{w}_{k}^{h}(a, X)\}]^{-1}-[E_{0}\{h_{\infty}(X)\}]^{-1})^2 \epsilon^{2}h_{\mathrm{max}}^2 \tau^2 \times 4 \epsilon^{4} \\
P_{0} \{\mathcal{Q}_{j, 32, k}^{\mathrm{RMTL}, h}(\tau, a; \widehat{\eta}_{j, k}^{\Lambda, h}, \eta_{j, \infty}^{\Lambda, h})\}^2 \leq& h_{\mathrm{min}}^{-2} \tau^2 \times 4 \epsilon^{4} E_{0} \left|\widehat{w}_{k}^{h}(a, X)-w_{\infty}^{h}(a, X) \right|^2 \\
P_{0} \{\mathcal{Q}_{j, 33, k}^{\mathrm{RMTL}, h}(\tau, a; \widehat{\eta}_{j, k}^{\Lambda, h}, \eta_{j, \infty}^{\Lambda, h})\}^2 \leq& h_{\mathrm{min}}^{-2} h_{\mathrm{max}}^2 \tau^2 \epsilon^{4} E_{0} \left[\sup_{t \in [0, \tau]} \left| \frac{\widehat{F}_{j, k}(t \mid a, X)}{\widehat{S}_{k}(t \mid a, X)}-\frac{F_{j, \infty}(t \mid a, X)}{S_{\infty}(t \mid a, X)} \right| \right]^2 \\
P_{0} \{\mathcal{Q}_{j, 34, k}^{\mathrm{RMTL}, h}(\tau, a; \widehat{\eta}_{j, k}^{\Lambda, h}, \eta_{j, \infty}^{\Lambda, h})\}^2\leq& h_{\mathrm{min}}^{-2} h_{\mathrm{max}}^2 \tau^2 \epsilon^{4} E_{0} \left[\sup_{t \in [0, \tau]} \left| \frac{1}{\widehat{G}_{k}(t- \mid a, X)}-\frac{1}{G_{\infty}(t- \mid a, X)} \right| \right]^2 \\
P_{0} \{\mathcal{Q}_{j, 35, k}^{\mathrm{RMTL}, h}(\tau, a; \widehat{\eta}_{j, k}^{\Lambda, h}, \eta_{j, \infty}^{\Lambda, h})\}^2\leq& h_{\mathrm{min}}^{-2} h_{\mathrm{max}}^2 \tau^2 \epsilon^{2} (\epsilon-1)^{2} E_{0} \left[\sup_{t \in [0, \tau]} \left| \frac{1}{\widehat{G}_{k}(t- \mid a, X)}-\frac{1}{G_{\infty}(t- \mid a, X)} \right| \right]^2
\end{aligned}
$$

Since
$$
\begin{aligned}
&E_{0}\left|\int_0^{\tau} \frac{t \widehat{F}_{j, k}(t \mid a, X)\mathrm{d}\widehat{\Lambda}_{k}(t \mid a, X)}{\widehat{S}_{k}(t \mid a, X)\widehat{G}_{k}(t- \mid a, X)}-\int_0^{\tau} \frac{t F_{j, \infty}(t \mid a, X)\mathrm{d}\Lambda_{\infty}(t \mid a, X)}{S_{\infty}(t \mid a, X)\widehat{G}_{k}(t- \mid a, X)} \right|^2 \\
=&E_{0}\left|\frac{\tau}{\widehat{G}_{k}(\tau- \mid a, X)} \left\{\frac{\widehat{F}_{j, k}(\tau \mid a, X)}{\widehat{S}_{k}(\tau \mid a, X)}-\frac{F_{j, \infty}(\tau \mid a, X)}{S_{\infty}(\tau \mid a, X)}\right\}-\int_0^{\tau} \frac{t}{\widehat{G}_{k}(t- \mid a, X)} \right. \\
&\times\left\{\frac{\mathrm{d}\widehat{F}_{j, k}(t \mid a, X)}{\widehat{S}_{k}(t \mid a, X)}-\frac{\mathrm{d}F_{j, \infty}(t \mid a, X)}{S_{\infty}(t \mid a, X)}\right\} -\int_0^{\tau}\left\{\frac{\widehat{F}_{j, k}(t \mid a, X)}{\widehat{S}_{k}(t \mid a, X)}-\frac{F_{j, \infty}(t \mid a, X)}{S_{\infty}(t \mid a, X)}\right\}\frac{\mathrm{d}t}{\widehat{G}_{k}(t- \mid a, X)} \\
&-\left.\int_0^{\tau}\left\{\frac{\widehat{F}_{j, k}(t \mid a, X)}{\widehat{S}_{k}(t \mid a, X)}-\frac{F_{j, \infty}(t \mid a, X)}{S_{\infty}(t \mid a, X)}\right\}t\mathrm{d}\left\{\frac{1}{\widehat{G}_{k}(t- \mid a, X)}\right\}\right|^2 \\
\leq & \tau^2 \epsilon^2 E_{0}\left\{\sup_{t \in [0, \tau]} \left| \frac{\widehat{F}_{j, k}(t \mid a, X)}{\widehat{S}_{k}(t \mid a, X)}-\frac{F_{j, \infty}(t \mid a, X)}{S_{\infty}(t \mid a, X)}\right|\right\}^2+\tau^2 \epsilon^2 E_{0} \left|\int_0^{\tau} \left\{\frac{\mathrm{d}\widehat{F}_{j, k}(t \mid a, X)}{\widehat{S}_{k}(t \mid a, X)}-\frac{\mathrm{d}F_{j, \infty}(t \mid a, X)}{S_{\infty}(t \mid a, X)} \right\} \right|^2 \\
&+\tau^2 \epsilon^2 E_{0}\left\{\sup_{t \in [0, \tau]} \left| \frac{\widehat{F}_{j, k}(t \mid a, X)}{\widehat{S}_{k}(t \mid a, X)}-\frac{F_{j, \infty}(t \mid a, X)}{S_{\infty}(t \mid a, X)}\right|\right\}^2 \\
&+\tau^2 (\epsilon-1)^2 E_{0}\left\{\sup_{t \in [0, \tau]} \left| \frac{\widehat{F}_{j, k}(t \mid a, X)}{\widehat{S}_{k}(t \mid a, X)}-\frac{F_{j, \infty}(t \mid a, X)}{S_{\infty}(t \mid a, X)}\right|\right\}^2 \\
=&\tau^2(3\epsilon^2+(\epsilon-1)^2+\epsilon^2\log^2\epsilon) E_{0}\left\{\sup_{t \in [0, \tau]} \left| \frac{\widehat{F}_{j, k}(t \mid a, X)}{\widehat{S}_{k}(t \mid a, X)}-\frac{F_{j, \infty}(t \mid a, X)}{S_{\infty}(t \mid a, X)}\right|\right\}^2 \\
&+\tau^2\epsilon^4 E_{0}\left\{\sup_{t \in [0, \tau]} \left|\widehat{\Lambda}_{k}(t \mid a, X)-\Lambda_{\infty}(t \mid a, X)\right|\right\}^2
\end{aligned}
$$

Therefore
$$
\begin{aligned}
&P_{0} \{\mathcal{Q}_{j, 36, k}^{\mathrm{RMTL}, h}(\tau, a; \widehat{\eta}_{j, k}^{\Lambda, h}, \eta_{j, \infty}^{\Lambda, h})\}^2 \\
\leq& h_{\mathrm{min}}^{-2} h_{\mathrm{max}}^2 \epsilon^{2} \tau^{2} \left[ \{3\epsilon^2+(\epsilon-1)^2+\epsilon^2 \log^2 \epsilon\} E_{0}\left\{\sup_{t \in [0, \tau]} \left| \frac{\widehat{F}_{j, k}(t \mid a, X)}{\widehat{S}_{k}(t \mid a, X)}-\frac{F_{j, \infty}(t \mid a, X)}{S_{\infty}(t \mid a, X)}\right|\right\}^2 \right. \\
&\left.+\epsilon^4 E_{0}\left\{\sup_{t \in [0, \tau]} \left|\widehat{\Lambda}_{k}(t \mid a, X)-\Lambda_{\infty}(t \mid a, X)\right|\right\}^2 \right]
\end{aligned}
$$

\subsection{Lemma~\ref{lemma4}}

\begin{lemma} \label{lemma4}
Under Condition~\ref{condition1}-\ref{condition2},
$$
\frac{1}{K} \sum_{k=1}^K \frac{K n_k^{1 / 2}}{n} \mathcal{G}_{n, k} \{\widehat{\phi}_{k}^{\mathrm{RMST}, h}(\tau, a; \widehat{\eta}_{k}^{\Lambda, h})-\phi_{\infty}^{\mathrm{RMST}, h}(\tau, a; \eta_{\infty}^{\Lambda, h})\}=o_{P}(n^{-1 / 2})
$$

If Condition~\ref{condition4} holds additionally,
$$
\frac{1}{K} \sum_{k=1}^K \frac{K n_k^{1 / 2}}{n} \sup_{t \in [0, \tau]} \left| \mathcal{G}_{n, k}\{\widehat{\phi}_{k}^{\mathrm{RMST}, h}(t, a; \widehat{\eta}_{k}^{\Lambda, h})-\phi_{\infty}^{\mathrm{RMST}, h}(t, a; \eta_{\infty}^{\Lambda, h})\}\right|=o_{P}(n^{-1 / 2})
$$

Under Condition~\ref{condition2} and \ref{condition5},
$$
\frac{1}{K} \sum_{k=1}^K \frac{K n_k^{1 / 2}}{n} \mathcal{G}_{n, k} \{\widehat{\phi}_{j, k}^{\mathrm{RMTL}, h}(\tau, a; \widehat{\eta}_{j, k}^{\Lambda, h})-\phi_{j, \infty}^{\mathrm{RMTL}, h}(\tau, a; \eta_{j, \infty}^{\Lambda, h})\}=o_{P}(n^{-1 / 2})
$$

If Condition~\ref{condition7} holds additionally,
$$
\frac{1}{K} \sum_{k=1}^K \frac{K n_k^{1 / 2}}{n} \sup_{t \in [0, \tau]} \left| \mathcal{G}_{n, k}\{\widehat{\phi}_{j, k}^{\mathrm{RMTL}, h}(t, a; \widehat{\eta}_{j, k}^{\Lambda, h})-\phi_{j, \infty}^{\mathrm{RMTL}, h}(t, a; \eta_{j, \infty}^{\Lambda, h})\}\right|=o_{P}(n^{-1 / 2})
$$
\end{lemma}

Proof of Lemma~\ref{lemma4}: Following Lemma 6 of \citet{westling2021inference}, the empirical process term is bounded by
$$
\begin{aligned}
&\left|\frac{1}{K} \sum_{k=1}^K \frac{K n_k^{1 / 2}}{n} \mathcal{G}_{n, k} \{\widehat{\phi}_{k}^{\mathrm{RMST}, h}(\tau, a; \widehat{\eta}_{k}^{\Lambda, h})-\phi_{\infty}^{\mathrm{RMST}, h}(\tau, a; \eta_{\infty}^{\Lambda, h})\}\right| \\
\leq & O_{P}(n^{-1 / 2}) \frac{1}{K} \sum_{k=1}^K\left|\mathcal{G}_{n, k}\{\widehat{\phi}_{k}^{\mathrm{RMST}, h}(\tau, a; \widehat{\eta}_{k}^{\Lambda, h})-\phi_{\infty}^{\mathrm{RMST}, h}(\tau, a; \eta_{\infty}^{\Lambda, h})\}\right|
\end{aligned}
$$

By Lemma~\ref{lemma3},
$$
\frac{1}{K} \sum_{k=1}^K\left|\mathcal{G}_{n, k}\{\widehat{\phi}_{k}^{\mathrm{RMST}, h}(\tau, a; \widehat{\eta}_{k}^{\Lambda, h})-\phi_{\infty}^{\mathrm{RMST}, h}(\tau, a; \eta_{\infty}^{\Lambda, h})\}\right| \leq \max_{k} \sum_{\iota=1}^{6} C_{\iota}^{\mathrm{RMST}, h} \mathcal{B}_{\iota, k}^{\mathrm{RMST}, h}(\widehat{\eta}_{k}^{\Lambda, h}, \eta_{\infty}^{\Lambda, h})
$$

\noindent where $\mathcal{B}_{\iota, k}^{\mathrm{RMST}, h}(\widehat{\eta}_{k}^{\Lambda, h}, \eta_{\infty}^{\Lambda, h})$ is defined in Lemma~\ref{lemma3}.
$$
\begin{aligned}
&[P_0\{\widehat{\phi}_{k}^{\mathrm{RMST}, h}(\tau, a; \widehat{\eta}_{k}^{\Lambda, h})-\phi_{\infty}^{\mathrm{RMST}, h}(\tau, a; \eta_{\infty}^{\Lambda, h})\}^{2}]^{1/2} \leq \sum_{\iota=1}^{6} C_{\iota}^{\mathrm{RMST}, h} \mathcal{B}_{\iota, k}^{\mathrm{RMST}, h}(\widehat{\eta}_{k}^{\Lambda, h}, \eta_{\infty}^{\Lambda, h}) \\
&[P_0\{\sup_{t \in [0, \tau]}|\widehat{\phi}_{k}^{\mathrm{RMST}, h}(t, a; \widehat{\eta}_{k}^{\Lambda, h})-\phi_{\infty}^{\mathrm{RMST}, h}(t, a; \eta_{\infty}^{\Lambda, h})|\}^{2}]^{1/2} \\
&\leq C_{1}^{\mathrm{RMST}, h} \widetilde{\mathcal{B}}_{1, k}^{\mathrm{RMST}, h}(\widehat{\eta}_{k}^{\Lambda, h}, \eta_{\infty}^{\Lambda, h})+\sum_{\iota=2}^{6} C_{\iota}^{\mathrm{RMST}, h} \mathcal{B}_{\iota, k}^{\mathrm{RMST}, h}(\widehat{\eta}_{k}^{\Lambda, h}, \eta_{\infty}^{\Lambda, h}) \\
&[P_0\{\widehat{\phi}_{j, k}^{\mathrm{RMTL}, h}(\tau, a; \widehat{\eta}_{j, k}^{\Lambda, h})-\phi_{j, \infty}^{\mathrm{RMTL}, h}(\tau, a; \eta_{j, \infty}^{\Lambda, h})\}^{2}]^{1/2} \leq \sum_{\iota=1}^{9} C_{j, \iota}^{\mathrm{RMTL}, h} \mathcal{B}_{j, \iota, k}^{\mathrm{RMTL}, h}(\widehat{\eta}_{j, k}^{\Lambda, h}, \eta_{j, \infty}^{\Lambda, h}) \\
&[P_0\{\sup_{t \in [0, \tau]}|\widehat{\phi}_{j, k}^{\mathrm{RMTL}, h}(t, a; \widehat{\eta}_{j, k}^{\Lambda, h})-\phi_{j, \infty}^{\mathrm{RMTL}, h}(t, a; \eta_{j, \infty}^{\Lambda, h})|\}^{2}]^{1/2} \\
&\leq C_{j, 1}^{\mathrm{RMTL}, h} \widetilde{\mathcal{B}}_{j, 1, k}^{\mathrm{RMTL}, h}(\widehat{\eta}_{j, k}^{\Lambda, h}, \eta_{j, \infty}^{\Lambda, h})+\sum_{\iota=2}^{9} C_{j, \iota}^{\mathrm{RMTL}, h} \mathcal{B}_{j, \iota, k}^{\mathrm{RMTL}, h}(\widehat{\eta}_{j, k}^{\Lambda, h}, \eta_{j, \infty}^{\Lambda, h})
\end{aligned}
$$

Based on Condition~\ref{condition1}, $\max_{k} \sum_{\iota=1}^{6} C_{\iota}^{\mathrm{RMST}, h} \mathcal{B}_{\iota, k}^{\mathrm{RMST}, h}(\widehat{\eta}_{k}^{\Lambda, h}, \eta_{\infty}^{\Lambda, h}) \stackrel{P}{\longrightarrow} 0$, such that
$$
\left|\frac{1}{K} \sum_{k=1}^K \frac{K n_k^{1 / 2}}{n} \mathcal{G}_{n, k} \{\widehat{\phi}_{k}^{\mathrm{RMST}, h}(\tau, a; \widehat{\eta}_{k}^{\Lambda, h})-\phi_{\infty}^{\mathrm{RMST}, h}(\tau, a; \eta_{\infty}^{\Lambda, h})\}\right| = o_{P}(n^{-1 / 2})
$$

For simultaneous inference, we bound the empirical stochastic process
$$
\begin{aligned}
&\left|\frac{1}{K} \sum_{k=1}^K \frac{K n_k^{1 / 2}}{n} \sup_{t \in [0, \tau]} \mathcal{G}_{n, k} \{\widehat{\phi}_{k}^{\mathrm{RMST}, h}(t, a; \widehat{\eta}_{k}^{\Lambda, h})-\phi_{\infty}^{\mathrm{RMST}, h}(t, a; \eta_{\infty}^{\Lambda, h})\}\right| \\
\leq & o_{P}(n^{-1 / 2}) \frac{1}{K} \sum_{k=1}^K \left|\sup_{t \in [0, \tau]} \mathcal{G}_{n, k}\{\widehat{\phi}_{k}^{\mathrm{RMST}, h}(t, a; \widehat{\eta}_{k}^{\Lambda, h})-\phi_{\infty}^{\mathrm{RMST}, h}(t, a; \eta_{\infty}^{\Lambda, h})\}\right|
\end{aligned}
$$

Using Theorem 2.14.2 from \citet{vaart1996weak}, we have
$$
\begin{aligned}
&\frac{1}{K} \sum_{k=1}^K \left|\sup_{t \in [0, \tau]} \mathcal{G}_{n, k} \{\widehat{\phi}_{k}^{\mathrm{RMST}, h}(t, a; \widehat{\eta}_{k}^{\Lambda, h})-\phi_{\infty}^{\mathrm{RMST}, h}(t, a; \eta_{\infty}^{\Lambda, h})\}\right| \\
\leq& \max_{k} \left\{C_{1}^{\mathrm{RMST}, h} \widetilde{\mathcal{B}}_{1, k}^{\mathrm{RMST}, h}(\widehat{\eta}_{k}^{\Lambda, h}, \eta_{\infty}^{\Lambda, h})+\sum_{\iota=2}^{6} C_{\iota}^{\mathrm{RMST}, h} \mathcal{B}_{\iota, k}^{\mathrm{RMST}, h}(\widehat{\eta}_{k}^{\Lambda, h}, \eta_{\infty}^{\Lambda, h})\right\}
\end{aligned}
$$

Based on Condition~\ref{condition1} and~\ref{condition4}, $\max_{k} \left\{C_{1}^{\mathrm{RMST}, h} \widetilde{\mathcal{B}}_{1, k}^{\mathrm{RMST}, h}(\widehat{\eta}_{k}^{\Lambda, h}, \eta_{\infty}^{\Lambda, h})+\sum_{\iota=2}^{6} C_{\iota}^{\mathrm{RMST}, h} \mathcal{B}_{\iota, k}^{\mathrm{RMST}, h}(\widehat{\eta}_{k}^{\Lambda, h}, \eta_{\infty}^{\Lambda, h})\right\} \stackrel{P}{\longrightarrow} 0$, such that
$$
\frac{1}{K} \sum_{k=1}^K \left|\sup_{t \in [0, \tau]} \mathcal{G}_{n, k}\{\widehat{\phi}_{k}^{\mathrm{RMST}, h}(t, a; \widehat{\eta}_{k}^{\Lambda, h})-\phi_{\infty}^{\mathrm{RMST}, h}(t, a; \eta_{\infty}^{\Lambda, h})\}\right| = o_{P}(n^{-1 / 2})
$$

Similar arguments also applies to $1/K \sum_{k=1}^K \left|\mathcal{G}_{n, k}\{\widehat{\phi}_{j, k}^{\mathrm{RMTL}, h}(\tau, a; \widehat{\eta}_{j, k}^{\Lambda, h})-\phi_{j, \infty}^{\mathrm{RMTL}, h}(\tau, a; \eta_{j, \infty}^{\Lambda, h})\}\right|$ and $1/K \sum_{k=1}^K \left|\sup_{t \in [0, \tau]} \mathcal{G}_{n, k}\{\widehat{\phi}_{j, k}^{\mathrm{RMTL}, h}(t, a; \widehat{\eta}_{j, k}^{\Lambda, h})-\phi_{j, \infty}^{\mathrm{RMTL}, h}(t, a; \eta_{j, \infty}^{\Lambda, h})\}\right|$ by replacing Condition~\ref{condition1} and Condition~\ref{condition4} with Condition~\ref{condition5} and Condition~\ref{condition7}.

\subsection{Proof of Theorem~\ref{theorem:consistency}}

\begin{condition} \label{condition1}
There exist $\eta_{\infty}^{\Lambda, h}$ such that $\max_{k}\{\mathcal{B}_{\iota, k}^{\mathrm{RMST}, h}(\widehat{\eta}_{k}^{\Lambda, h}, \eta_{\infty}^{\Lambda, h})\}^2\stackrel{P}{\longrightarrow} 0$ for all $a$, $\tau$, and $\iota=1, \ldots, 14$, where $\mathcal{B}_{\iota, k}^{\mathrm{RMST}, h}(\widehat{\eta}_{k}^{\Lambda, h}, \eta_{\infty}^{\Lambda, h})$ is defined in Lemma~\ref{lemma3}. 
\end{condition}

% $\mathrm{RMST}_{\infty}(\tau \mid a, X)$, $S_{\infty}(t \mid a, X)$, $h_{\infty}(X)$, $\pi_{\infty}(a \mid X)$, and $G_{\infty}(t \mid a, X)$, 

% $\text{(1) } \max_{k} E \left[\sup_{t \in [0, \tau]} \left| \frac{\mathrm{R}\widehat{\mathrm{MS}}\mathrm{T}_{k}(\tau \mid a, X)}{\widehat{S}_{k}(t \mid a, X)}-\frac{\mathrm{RMST}_{\infty}(\tau \mid a, X)}{S_{\infty}(t \mid a, X)} \right| \right]^2 \stackrel{P}{\longrightarrow} 0$; \\
% $\text{(2) } \max_{k} E \left|\left([P_{n, k}\{\widehat{h}_{k}(X)\}]^{-1}\widehat{h}_{k}(X)-[E\{h_{\infty}(X)\}]^{-1}h_{\infty}(X)\right) \right|^2 \stackrel{P}{\longrightarrow} 0$; \\
% $\text{(3) } \max_{k} \left(\left[P_{n, k}\left\{\widehat{w}_{k}^{h}(a, X)\right\}\right]^{-1}-[E\{h_{\infty}(X)\}]^{-1} \right)^2 \stackrel{P}{\longrightarrow} 0$; \\
% $\text{(4) } \max_{k} E \left|\widehat{w}_{k}^{h}(a, X)-w_{\infty}^{h}(a \mid X)\right|^2 \stackrel{P}{\longrightarrow} 0$; \\
% $\text{(5) } \max_{k} E \left[\sup_{t \in [0, \tau]} \left| \frac{1}{\widehat{G}_{k}(t- \mid a, X)}-\frac{1}{G_{\infty}(t- \mid a, X)} \right| \right]^2 \stackrel{P}{\longrightarrow} 0$; \\
% $\text{(6) }\max_{k} E \left[\sup_{t \in [0, \tau]} \left| \frac{\mathrm{R}\widehat{\mathrm{MS}}\mathrm{T}_{k}(t \mid a, X)}{\widehat{S}_{k}(t \mid a, X)}-\frac{\mathrm{RMST}_{\infty}(t \mid a, X)}{S_{\infty}(t \mid a, X)} \right| \right]^2 \stackrel{P}{\longrightarrow} 0$.

\begin{condition} \label{condition2}
There exists $\epsilon \in (1, \infty)$ such that $\{\widehat{S}_{k}(t \mid a, X), S_{\infty}(t \mid a, X), \widehat{\pi}_{k}(a \mid X), \pi_{\infty}(a \mid X), \widehat{G}_{k}(t \mid a, X), G_{\infty}(t \mid a, X) \} \geq 1/\epsilon$, $h_{\mathrm{min}}\leq\{\widehat{h}_{k}(X), h_{\infty}(X)\}\leq h_{\mathrm{max}}$, and $P_{n, k}\{\widehat{h}_{k}(X)\}>0$, $E_{0}\{h_{\infty}(X)\}>0$ for all $a$, $X$ and $k$.
\end{condition}

\begin{condition} \label{condition3}
One of the following conditions is satisfied for all $t \in [0, \tau], a \in \{0,1\}$ and almost all $x$: (1) $h_{\infty}(X)=h_{0}(X)$ and $S_{\infty}(t \mid a, X)=S_{0}(t \mid a, X)$; or (2) $h_{\infty}(X)=h_{0}(X)$ and $\pi_{\infty}(a \mid X)=\pi_{0}(a \mid X)$ and $G_{\infty}(t \mid a, X)=G_{0}(t \mid a, X)$.
\end{condition}

\begin{condition} \label{condition4} $\max_{k}\{\widetilde{\mathcal{B}}_{1, k}^{\mathrm{RMST}, h}(\widehat{\eta}_{k}^{\Lambda, h}, \eta_{\infty}^{\Lambda, h})\}^2\stackrel{P}{\longrightarrow} 0$ for every $a$ and $\tau$.
\end{condition}

% $$\max_{k} E \left[\sup_{t \in [0, \tau]} \sup_{u \in [0, t]} \left| \frac{\mathrm{R}\widehat{\mathrm{MS}}\mathrm{T}_{k}(t \mid a, X)}{\widehat{S}_{k}(u \mid a, X)}-\frac{\mathrm{RMST}_{\infty}(t \mid a, X)}{S_{\infty}(u \mid a, X)} \right| \right]^2 \stackrel{P}{\longrightarrow} 0$$

\begin{condition} \label{condition5}
There exist $\eta_{j, \infty}^{\Lambda, h}$ such that $\max_{k}\{\mathcal{B}_{j, \iota, k}^{\mathrm{RMTL}, h}(\widehat{\eta}_{j, k}^{\Lambda, h}, \eta_{j, \infty}^{\Lambda, h})\}^2\stackrel{P}{\longrightarrow} 0$ for cause $j$ of interest, and for all $a$, $\tau$, $\iota=1, \ldots, 36$, where $\mathcal{B}_{j, \iota, k}^{\mathrm{RMTL}, h}(\widehat{\eta}_{j, k}^{\Lambda, h}, \eta_{j, \infty}^{\Lambda, h})$ is defined in Lemma~\ref{lemma3}.
\end{condition}

\begin{condition} \label{condition6}
One of the following conditions is satisfied for all $t \in [0, \tau], a \in \{0,1\}$, almost all $X$, and cause $j$ of interest only: (1) $h_{\infty}(X)=h_{0}(X)$ and $S_{\infty}(t \mid a, X)=S_{0}(t \mid a, X)$ and $F_{j, \infty}(t \mid a, X)=F_{j, 0}(t \mid a, X)$; or (2) $h_{\infty}(X)=h_{0}(X)$ and $\pi_{\infty}(a \mid X)=\pi_{0}(a \mid X)$ and $G_{\infty}(t \mid a, X)=G_{0}(t \mid a, X)$; or (3) $h_{\infty}(X)=h_{0}(X)$ and $F_{j, \infty}(t \mid a, X)=F_{j, 0}(t \mid a, X)$ and $G_{\infty}(t \mid a, X)=G_{0}(t \mid a, X)$. \\
\end{condition}

\begin{condition} \label{condition7}
$\max_{k}\{\widetilde{\mathcal{B}}_{j, 1, k}^{\mathrm{RMTL}, h}(\widehat{\eta}_{j, k}^{\Lambda, h}, \eta_{j, \infty}^{\Lambda, h})\}^2\stackrel{P}{\longrightarrow} 0$ for cause $j$ of interest and every $a$, $\tau$.
\end{condition}

% $$\max_{k} E \left[\sup_{t \in [0, \tau]} \sup_{u \in [0, t]} \left| \frac{\mathrm{R}\widehat{\mathrm{MT}}\mathrm{L}_{j, k}(t \mid a, X)}{\widehat{S}_{k}(u \mid a, X)}-\frac{\mathrm{RMTL}_{j, \infty}(t \mid a, X)}{S_{\infty}(u \mid a, X)} \right| \right]^2 \stackrel{P}{\longrightarrow} 0$$

Under Condition~\ref{condition1}-\ref{condition3},
$$
\widehat{\psi}^{\mathrm{RMST}, h}(\tau, a) \stackrel{P}{\longrightarrow} \psi_{0}^{\mathrm{RMST}, h}(\tau, a), \quad \widehat{\psi}^{\mathrm{RMST}, h}(\tau) \stackrel{P}{\longrightarrow} \psi_{0}^{\mathrm{RMST}, h}(\tau)
$$

If Condition~\ref{condition4} holds additionally, then
$$
\sup_{t \in [0, \tau]}\left|\widehat{\psi}^{\mathrm{RMST}, h}(\tau, a)-\psi_{0}^{\mathrm{RMST}, h}(\tau, a) \right| \stackrel{P}{\longrightarrow} 0, \quad \sup_{t \in [0, \tau]}\left|\widehat{\psi}^{\mathrm{RMST}, h}(\tau)-\psi_{0}^{\mathrm{RMST}, h}(\tau) \right| \stackrel{P}{\longrightarrow} 0
$$

Under Condition~\ref{condition2} and Condition~\ref{condition5}-~\ref{condition6},
$$
\widehat{\psi}_{j}^{\mathrm{RMTL}, h}(\tau, a) \stackrel{P}{\longrightarrow} \psi_{j, 0}^{\mathrm{RMTL}, h}(\tau, a), \quad \widehat{\psi}_{j}^{\mathrm{RMTL}, h}(\tau) \stackrel{P}{\longrightarrow} \psi_{j, 0}^{\mathrm{RMTL}, h}(\tau)
$$

If Condition~\ref{condition7} holds additionally, then
$$
\sup_{t \in [0, \tau]}\left|\widehat{\psi}_{j}^{\mathrm{RMTL}, h}(\tau, a)-\psi_{j, 0}^{\mathrm{RMTL}, h}(\tau, a) \right| \stackrel{P}{\longrightarrow} 0, \quad \sup_{t \in [0, \tau]}\left|\widehat{\psi}_{j}^{\mathrm{RMTL}, h}(\tau)-\psi_{j, 0}^{\mathrm{RMTL}, h}(\tau) \right| \stackrel{P}{\longrightarrow} 0
$$

For Condition~\ref{condition3}, if (1) $h_{\infty}(X)=h_{0}(X)$ and $S_{\infty}(t \mid a, X)=S_{0}(t \mid a, X)$ holds, then $\mathrm{RMST}_{\infty}(t \mid a, X)=\mathrm{RMST}_{0}(t \mid a, X)$, $\Lambda_{\infty}(t \mid a, X)=\Lambda_{0}(t \mid a, X)$, and $\mathrm{d}M_{\infty}(t \mid a, X)=\mathrm{d}M_{0}(t \mid a, X)$. We consider the event martingale representation of $\widehat{\psi}^{\mathrm{RMST}, h}(\tau, a)$ since $E_{0}\{\int_0^{\tau \wedge \widetilde{T}} (\cdot) \mathrm{d} M_{0}(t \mid a, X)\mid A=a, X\}=0$,

$$
\begin{aligned}
&\widehat{\psi}^{\mathrm{RMST}, h}(\tau, a) \stackrel{p}{\longrightarrow} \\
&[E_{0} \{h_{0}(X)\}]^{-1} E_{0} \{h_{0}(X) \mathrm{RMST}_{0}(\tau \mid a, X)\}+\left[E_{0} \left\{\frac{h_{0}(X)I(A=a)}{\pi_{\infty}(a \mid X)}\right\}\right]^{-1} \\
&\times E_{0} \left[ \left\{\frac{h_{0}(X)I(A=a)}{\pi_{\infty}(a \mid X)}\right\} \left\{-\mathrm{RMST}_{0}(\tau \mid a, X)\int_0^{\tau \wedge \widetilde{T}} \frac{\mathrm{d} M_{0}(t \mid a, X)}{S_{0}(t \mid a, X)G_{\infty}(t- \mid a, X)} \right.\right. \\
&+\left.\left.\int_0^{\tau \wedge \widetilde{T}} \frac{\mathrm{RMST}_{0}(t \mid a, X)\mathrm{d} M_{0}(t \mid a, X)}{S_{0}(t \mid a, X)G_{\infty}(t- \mid a, X)} \right\} \right] \\
=&[E_{0} \{h_{0}(X)\}]^{-1} E_{0} \{h_{0}(X) \mathrm{RMST}_{0}(\tau \mid a, X)\}+\left[E_{0} \left\{\frac{h_{0}(X)\pi_{0}(a \mid X)}{\pi_{\infty}(a \mid X)}\right\}\right]^{-1} \\
&\times E_{0} \left( \left\{\frac{h_{0}(X)\pi_{0}(a \mid X)}{\pi_{\infty}(a \mid X)}\right\} \left[-\mathrm{RMST}_{0}(\tau \mid a, X)E\left\{\int_0^{\tau \wedge \widetilde{T}} \frac{\mathrm{d} M_{0}(t \mid a, X)}{S_{0}(t \mid a, X)G_{\infty}(t- \mid a, X)} \mid A=a, X \right.\right.\right\} \\
&+\left.\left.E_{0}\left\{\int_0^{\tau \wedge \widetilde{T}} \frac{\mathrm{RMST}_{0}(t \mid a, X)\mathrm{d} M_{0}(t \mid a, X)}{S_{0}(t \mid a, X)G_{\infty}(t- \mid a, X)}\mid A=a, X\right\} \right] \right) \\
=&[E_{0} \{h_{0}(X)\}]^{-1} E_{0} \{h_{0}(X) \mathrm{RMST}_{0}(\tau \mid a, X)\}=\psi_{0}^{\mathrm{RMST}, h}(\tau, a)
\end{aligned}
$$

If (2) $h_{\infty}(X)=h_{0}(X)$ and $\pi_{\infty}(a \mid X)=\pi_{0}(a \mid X)$ and $G_{\infty}(t \mid a, X)=G_{0}(t \mid a, X)$ holds, then $\Lambda_{\infty}^C(t \mid a, X)=\Lambda_{0}^C(t \mid a, X)$, and $\mathrm{d}M_{\infty}^C(t \mid a, X)=\mathrm{d}M_{0}^C(t \mid a, X)$. We consider the censoring martingale representation of $\widehat{\psi}^{\mathrm{RMST}, h}(\tau, a)$ since $E\{\int_0^{\tau \wedge \widetilde{T}} (\cdot) \mathrm{d} M_{0}^C(t \mid a, X)\mid A=a, X\}=0$,
$$
\begin{aligned}
&\widehat{\psi}^{\mathrm{RMST}, h}(\tau, a) \stackrel{P}{\longrightarrow} \\
=&[E_{0} \{h_{0}(X)\}]^{-1} E_{0}\{h_{0}(X) \mathrm{RMST}_{\infty}(\tau \mid a, X)\}+\left[E_{0} \left\{\frac{h_{0}(X)I(A=a)}{\pi_{0}(a \mid X)}\right\}\right]^{-1} \\
&\times E_{0}\left[\left\{\frac{h_{0}(X)I(A=a)}{\pi_{0}(a \mid X)}\right\} \left\{\frac{\Delta(\tau)\min(\widetilde{T}, \tau)}{G_{0}(\tau \wedge \widetilde{T}- \mid a, X)}-\mathrm{RMST}_{\infty}(\tau \mid a, X) \right.\right. \\
&\left.\left. +\int_{0}^{\tau \wedge \widetilde{T}}\frac{t \mathrm{d} M_{0}^{C}(t \mid a, X)}{G_{0}(t- \mid a, X)}+\int_{0}^{\tau \wedge \widetilde{T}} \frac{\{\mathrm{RMST}_{\infty}(\tau \mid a, X)-\mathrm{RMST}_{\infty}(t \mid a, X)\} \mathrm{d} M_{0}^{C}(t \mid a, X)}{S_{\infty}(t \mid a, X)G_{0}(t- \mid a, X)} \right\}\right] \\
=&[E_{0} \{h_{0}(X)\}]^{-1} E_{0}\{h_{0}(X) \mathrm{RMST}_{\infty}(\tau \mid a, X)\}+\left[E_{0}\{h_{0}(X)\}\right]^{-1} \\
&\times E_{0}\left(h_{0}(X) \left[E\left\{\frac{\Delta(\tau)\min(\widetilde{T}, \tau)}{G_{0}(\tau \wedge \widetilde{T}- \mid a, X)} \mid A=a, X\right\}-\mathrm{RMST}_{\infty}(\tau \mid a, X)\right.\right. \\
&+E_{0}\left\{\int_{0}^{\tau \wedge \widetilde{T}}\frac{t \mathrm{d} M_{0}^{C}(t \mid a, X)}{G_{0}(t- \mid a, X)}\mid A=a, X\right\} \\
&\left.\left. +E_{0}\left\{\int_{0}^{\tau \wedge \widetilde{T}} \frac{\{\mathrm{RMST}_{\infty}(\tau \mid a, X)-\mathrm{RMST}_{\infty}(t \mid a, X)\} \mathrm{d} M_{0}^{C}(t \mid a, X)}{S_{\infty}(t \mid a, X)G_{0}(t- \mid a, X)}\mid A=a, X\right\} \right]\right) \\
=&[E_{0} \{h_{0}(X)\}]^{-1} E_{0} \{h_{0}(X) \mathrm{RMST}_{0}(\tau \mid a, X)\}=\psi_{0}^{\mathrm{RMST}, h}(\tau, a)
\end{aligned}
$$

We used
$$
\begin{aligned}
&E_{0}\left\{\frac{\Delta(\tau)\min(\widetilde{T}, \tau)}{G_{0}(\tau \wedge \widetilde{T}- \mid a, X)} \mid A=a, X\right\} \\
=&E_{0}\left\{\frac{I(C > \tau \wedge \widetilde{T}-)\min(T, \tau)}{G_{0}(\tau \wedge \widetilde{T}- \mid a, X)} \mid A=a, X\right\}=\mathrm{RMST}_{0}(\tau \mid a, X)
\end{aligned}
$$

The proof above also implies $E_0 \{\phi_{0}^{\mathrm{RMST}}(\tau, a; \eta_{0}^{\Lambda}) \mid A=a, X\}=\mathrm{RMST}_{0}(\tau \mid a, X)$.

For Condition~\ref{condition6}, if (1) $h_{\infty}(X)=h_{0}(X)$ and $S_{\infty}(t \mid a, X)=S_{0}(t \mid a, X)$ and $F_{j, \infty}(t \mid a, X)=F_{j, 0}(t \mid a, X)$ holds, then $\mathrm{RMTL}_{j, \infty}(t \mid a, X)=\mathrm{RMTL}_{j, 0}(t \mid a, X)$, $\Lambda_{\infty}(t \mid a, X)=\Lambda_{0}(t \mid a, X)$, $\Lambda_{j, \infty}(t \mid a, X)=\Lambda_{j, 0}(t \mid a, X)$, $\mathrm{d}M_{\infty}(t \mid a, X)=\mathrm{d}M_{0}(t \mid a, X)$, and $\mathrm{d}M_{j, \infty}(t \mid a, X)=\mathrm{d}M_{j, 0}(t \mid a, X)$. We consider the event martingale representation of $\widehat{\psi}_{j}^{\mathrm{RMTL}, h}(\tau, a)$ since $E[\int_0^{\tau \wedge \widetilde{T}} (\cdot) \mathrm{d} M_{0}(t \mid a, X)\mid A=a, X]=0$, and $E[\int_0^{\tau \wedge \widetilde{T}} (\cdot) \mathrm{d} M_{j, 0}(t \mid a, X)\mid A=a, X]=0$,
$$
\begin{aligned}
&\widehat{\psi}_{j}^{\mathrm{RMTL}, h}(\tau, a) \stackrel{P}{\longrightarrow} \\
&[E_{0} \{h_{0}(X)\}]^{-1} E_{0} \{h_{0}(X) \mathrm{RMTL}_{j, 0}(\tau \mid a, X)\}+\left[E_{0} \left\{\frac{h_{0}(X)I(A=a)}{\pi_{\infty}(a \mid X)}\right\}\right]^{-1} \\
&\times E_{0} \left[ \left\{\frac{h_{0}(X)I(A=a)}{\pi_{\infty}(a \mid X)}\right\} \left\{\int_0^{\tau \wedge \widetilde{T}} \frac{(\tau-t)\mathrm{d} M_{j, 0}(t \mid a, X)}{G_{\infty}(t- \mid a, X)}+\int_0^{\tau \wedge \widetilde{T}} \frac{(\tau-t)F_{j, 0}(t \mid a, X)\mathrm{d} M_{0}(t \mid a, X)}{S_{0}(t \mid a, X)G_{\infty}(t- \mid a, X)} \right.\right. \\
&\left.\left.-\int_0^{\tau \wedge \widetilde{T}} \frac{\{\mathrm{RMST}_{0}(\tau \mid a, X)-\mathrm{RMST}_{0}(t \mid a, X)\}\mathrm{d} M_{0}(t \mid a, X)}{S_{0}(t \mid a, X)G_{\infty}(t- \mid a, X)} \right\} \right] \\
=&[E_{0} \{h_{0}(X)\}]^{-1} E_{0} \{h_{0}(X) \mathrm{RMTL}_{j, 0}(\tau \mid a, X)\}+\left[E_{0} \left\{\frac{h_{0}(X)\pi_{0}(a \mid X)}{\pi_{\infty}(a \mid X)}\right\}\right]^{-1} \\
&\times E_{0} \left( \left\{\frac{h_{0}(X)\pi_{0}(a \mid X)}{\pi_{\infty}(a \mid X)}\right\} \left[E_{0}\left\{\int_0^{\tau \wedge \widetilde{T}} \frac{(\tau-t)\mathrm{d} M_{j, 0}(t \mid a, X)}{G_{\infty}(t- \mid a, X)} \mid A=a, X\right\}\right.\right. \\
&+E_{0} \left\{\int_0^{\tau \wedge \widetilde{T}} \frac{(\tau-t)F_{j, 0}(t \mid a, X)\mathrm{d} M_{0}(t \mid a, X)}{S_{0}(t \mid a, X)G_{\infty}(t- \mid a, X)}\mid A=a, X\right\} \\
&\left.\left.-E_{0}\left\{\int_0^{\tau \wedge \widetilde{T}} \frac{\{\mathrm{RMST}_{0}(\tau \mid a, X)-\mathrm{RMST}_{0}(t \mid a, X)\}\mathrm{d} M_{0}(t \mid a, X)}{S_{0}(t \mid a, X)G_{\infty}(t- \mid a, X)} \mid A=a, X \right\} \right] \right) \\
=&[E_{0} \{h_{0}(X)\}]^{-1} E_{0} \{h_{0}(X) \mathrm{RMTL}_{j, 0}(\tau \mid a, X)\}=\psi_{j, 0}^{\mathrm{RMTL}, h}(\tau, a)
\end{aligned}
$$

If (2) $h_{\infty}(X)=h_{0}(X)$ and $\pi_{\infty}(a \mid X)=\pi_{0}(a \mid X)$ and $G_{\infty}(t \mid a, X)=G_{0}(t \mid a, X)$ holds, then $\Lambda_{\infty}^C(t \mid a, X)=\Lambda_{0}^C(t \mid a, X)$, and $\mathrm{d}M_{\infty}^C(t \mid a, X)=\mathrm{d}M_{0}^C(t \mid a, X)$. We consider the censoring martingale representation of $\widehat{\psi}_{j}^{\mathrm{RMTL}, h}(\tau, a)$ since $E\{\int_0^{\tau \wedge \widetilde{T}} (\cdot) \mathrm{d} M_{0}^C(t \mid a, X)\mid A=a, X\}=0$,
$$
\begin{aligned}
&\widehat{\psi}_{j}^{\mathrm{RMTL}, h}(\tau, a) \stackrel{P}{\longrightarrow} \\
&[E_{0} \{h_{0}(X)\}]^{-1} E_{0} \{h_{0}(X) \mathrm{RMTL}_{j, \infty}(\tau \mid a, X)\}+\left[E_{0} \left\{\frac{h_{0}(X)I(A=a)}{\pi_{0}(a \mid X)}\right\}\right]^{-1} E_{0} \left[ \left\{\frac{h_{0}(X)I(A=a)}{\pi_{0}(a \mid X)}\right\} \right. \\
&\times \left\{\frac{(\tau-\min(\widetilde{T}, \tau))I(\widetilde{J}=j)}{G_{0}(\widetilde{T}- \mid a, X)}-\mathrm{RMTL}_{j, \infty}(\tau \mid a, X)-\int_{0}^{\tau \wedge \widetilde{T}} \frac{(\tau-t) F_{j, \infty}(t \mid a, X) \mathrm{d} M^{C}_{0}(t \mid a, X)}{S_{\infty}(t \mid a, X)G_{0}(t- \mid a, X)} \right. \\
&\left.\left.+\int_{0}^{\tau \wedge \widetilde{T}} \frac{\{\mathrm{RMTL}_{j, \infty}(\tau \mid a, X)-\mathrm{RMTL}_{j, \infty}(t \mid a, X)\} \mathrm{d} M_{0}^{C}(t \mid a, X)}{S_{\infty}(t \mid a, X)G_{0}(t- \mid a, X)} \right\}\right]
\end{aligned}
$$

$$
\begin{aligned}
=&[E_{0} \{h_{0}(X)\}]^{-1} E_{0} \{h_{0}(X) \mathrm{RMTL}_{j, \infty}(\tau \mid a, X)\}+[E_{0}\{h_{0}(X)\}]^{-1} \\
&\times E_{0} \left(h_{0}(X) \left[E_{0}\left\{\frac{(\tau-\min(\widetilde{T}, \tau))I(\widetilde{J}=j)}{G_{0}(\widetilde{T}- \mid a, X)} \mid A=a, X \right\}-\mathrm{RMTL}_{j, \infty}(\tau \mid a, X) \right.\right. \\
&-E_{0}\left\{\int_{0}^{\tau \wedge \widetilde{T}} \frac{(\tau-t) F_{j, \infty}(t \mid a, X) \mathrm{d} M^{C}_{0}(t \mid a, X)}{S_{\infty}(t \mid a, X)G_{0}(t- \mid a, X)}\mid A=a, X \right\} \\
&+\left.\left.E_{0}\left\{\int_{0}^{\tau \wedge \widetilde{T}} \frac{\{\mathrm{RMTL}_{j, \infty}(\tau \mid a, X)-\mathrm{RMTL}_{j, \infty}(t \mid a, X)\} \mathrm{d} M_{0}^{C}(t \mid a, X)}{S_{\infty}(t \mid a, X)G_{0}(t- \mid a, X)} \mid A=a, X \right\}\right]\right) \\
=&[E_{0} \{h_{0}(X)\}]^{-1} E_{0} \{h_{0}(X) \mathrm{RMTL}_{j, 0}(\tau \mid a, X)\}=\psi_{j, 0}^{\mathrm{RMTL}, h}(\tau, a)
\end{aligned}
$$

We plugged in
$$
\begin{aligned}
&E_{0}\left\{\frac{(\tau-\min(\widetilde{T}, \tau))I(\widetilde{J}=j)}{G_{0}(\widetilde{T}- \mid a, X)} \mid A=a, X\right\} \\
=&E_{0}\left\{\frac{\Delta(\tau-\min(T, \tau))I(J=j)}{G_{0}(\widetilde{T}- \mid a, X)} \mid A=a, X\right\}=\mathrm{RMTL}_{j, 0}(\tau \mid a, X)
\end{aligned}
$$

If (3) $h_{\infty}(X)=h_{0}(X)$ and $F_{j, \infty}(t \mid a, X)=F_{j, 0}(t \mid a, X)$ and $G_{\infty}(t \mid a, X)=G_{0}(t \mid a, X)$ holds, then $\mathrm{RMTL}_{j, \infty}(t \mid a, X)=\mathrm{RMTL}_{j, 0}(t \mid a, X)$, $\Lambda_{\infty}^C(t \mid a, X)=\Lambda_{0}^C(t \mid a, X)$, and $\mathrm{d}M_{\infty}^C(t \mid a, X)=\mathrm{d}M_{0}^C(t \mid a, X)$. We consider the censoring martingale representation of $\widehat{\psi}_{j}^{\mathrm{RMTL}, h}(\tau, a)$ since $E[\int_0^{\tau \wedge \widetilde{T}} (\cdot) \mathrm{d} M_{0}^C(t \mid a, X)\mid A=a, X]=0$
$$
\begin{aligned}
&\widehat{\psi}_{j}^{\mathrm{RMTL}, h}(\tau, a) \stackrel{P}{\longrightarrow} \\
&[E_{0} \{h_{0}(X)\}]^{-1} E_{0} \{h_{0}(X) \mathrm{RMTL}_{j, 0}(\tau \mid a, X)\}+\left[E_{0} \left\{\frac{h_{0}(X)I(A=a)}{\pi_{\infty}(a \mid X)}\right\}\right]^{-1} E_{0} \left[ \left\{\frac{h_{0}(X)I(A=a)}{\pi_{\infty}(a \mid X)}\right\} \right. \\
&\times \left\{\frac{(\tau-\min(\widetilde{T}, \tau))I(\widetilde{J}=j)}{G_{0}(\widetilde{T}- \mid a, X)}-\mathrm{RMTL}_{j, 0}(\tau \mid a, X)-\int_{0}^{\tau \wedge \widetilde{T}} \frac{(\tau-t) F_{j, 0}(t \mid a, X) \mathrm{d} M^{C}_{0}(t \mid a, X)}{S_{\infty}(t \mid a, X)G_{0}(t- \mid a, X)} \right.\\
&\left.\left.+\int_{0}^{\tau \wedge \widetilde{T}} \frac{\{\mathrm{RMTL}_{j, 0}(\tau \mid a, X)-\mathrm{RMTL}_{j, 0}(t \mid a, X)\} \mathrm{d} M_{0}^{C}(t \mid a, X)}{S_{\infty}(t \mid a, X)G_{0}(t- \mid a, X)} \right\}\right] \\
=&[E_{0} \{h_{0}(X)\}]^{-1} E_{0} \{h_{0}(X) \mathrm{RMTL}_{j, 0}(\tau \mid a, X)\}+\left[E_{0} \left\{\frac{h_{0}(X)\pi_{0}(a \mid X)}{\pi_{\infty}(a \mid X)}\right\}\right]^{-1} \\
&\times E_{0}\left(\left[E_{0}\left\{\frac{h_{0}(X)\pi_{0}(a \mid X)}{\pi_{\infty}(a \mid X)}\right\} \left\{\frac{(\tau-\min(\widetilde{T}, \tau))I(\widetilde{J}=j)}{G_{0}(\widetilde{T}- \mid a, X)} \mid A=a, X \right\}-\mathrm{RMTL}_{j, 0}(\tau \mid a, X) \right.\right. \\
&-E_{0}\left\{\int_{0}^{\tau \wedge \widetilde{T}} \frac{(\tau-t) F_{j, 0}(t \mid a, X) \mathrm{d} M^{C}_{0}(t \mid a, X)}{S_{\infty}(t \mid a, X)G_{0}(t- \mid a, X)} \mid A=a, X \right\} \\
&\left.\left.+E_{0}\left\{\int_{0}^{\tau \wedge \widetilde{T}} \frac{\{\mathrm{RMTL}_{j, 0}(\tau \mid a, X)-\mathrm{RMTL}_{j, 0}(t \mid a, X)\} \mathrm{d} M_{0}^{C}(t \mid a, X)}{S_{\infty}(t \mid a, X)G_{0}(t- \mid a, X)} \mid A=a, X \right\} \right]\right) \\
=&[E_{0} \{h_{0}(X)\}]^{-1} E_{0} \{h_{0}(X) \mathrm{RMTL}_{j, 0}(\tau \mid a, X)\}=\psi_{j, 0}^{\mathrm{RMTL}, h}(\tau, a)
\end{aligned}
$$

The proof above also implies $E_0 \{\phi_{j, 0}^{\mathrm{RMTL}}(\tau, a; \eta_{j, 0}^{\Lambda}) \mid A=a, X\}=\mathrm{RMTL}_{j, 0}(\tau \mid a, X)$.

Using Lemma 2 of \citet{westling2021inference},
$$
\begin{aligned}
&|\widehat{\psi}^{\mathrm{RMST}, h}(\tau, a)-\psi_{0}^{\mathrm{RMST}, h}(\tau, a)| \\
\leq & |P_n\{\varphi_{\infty}^{\mathrm{RMST}, h}(\tau, a; \eta_{\infty}^{\Lambda, h})\}|+\left|\frac{1}{K} \sum_{k=1}^K \frac{K n_k^{1 / 2}}{n} \mathcal{G}_{n, k} \{\widehat{\phi}_{k}^{\mathrm{RMST}, h}(\tau, a; \widehat{\eta}_{k}^{\Lambda, h})-\phi_{\infty}^{\mathrm{RMST}, h}(\tau, a; \eta_{\infty}^{\Lambda, h})\}\right| \\
&+\left|\frac{1}{K} \sum_{k=1}^K \frac{K n_k}{n} P_{0}\{\widehat{\phi}_{k}^{\mathrm{RMST}, h}(\tau, a; \widehat{\eta}_{k}^{\Lambda, h})-\phi_{\infty}^{\mathrm{RMST}, h}(\tau, a; \eta_{\infty}^{\Lambda, h})\}\right|
\end{aligned}
$$

Under Condition~\ref{condition3} and weak law of large numbers, $P_n\{\varphi_{\infty}^{\mathrm{RMST}, h}(t, a; \eta_{\infty}^{\Lambda, h})\}\stackrel{P}{\longrightarrow}0$. With Condition~\ref{condition1}, ~\ref{condition2}, and Lemma~\ref{lemma4}, the second term is $o_{P}(n^{-1 / 2})$. The third term is bounded by
$2 [\max _k P_{0}\{\widehat{\phi}_{k}^{\mathrm{RMST}, h}(\tau, a; \widehat{\eta}_{k}^{\Lambda, h})-\phi_{\infty}^{\mathrm{RMST}, h}(\tau, a; \eta_{\infty}^{\Lambda, h})\}^2]^{1 / 2}$, which is further bounded based on Lemma~\ref{lemma3} and converges to zero in probability. Consequently, $|\widehat{\psi}^{\mathrm{RMST}, h}(\tau, a)-\psi_{0}^{\mathrm{RMST}, h}(\tau, a)|=o_{P}(1)$.

In terms of uniform consistency,
$$
\begin{aligned}
&\sup_{t \in [0, \tau]}|\widehat{\psi}^{\mathrm{RMST}, h}(t, a)-\psi_{0}^{\mathrm{RMST}, h}(t, a)| \\
\leq & \sup_{t \in [0, \tau]}|P_n\{\varphi_{\infty}^{\mathrm{RMST}, h}(t, a; \eta_{\infty}^{\Lambda, h})\}|+\sup_{t \in [0, \tau]}\left|\frac{1}{K} \sum_{k=1}^K \frac{K n_k^{1 / 2}}{n} \mathcal{G}_{n, k} \{\widehat{\phi}_{k}^{\mathrm{RMST}, h}(t, a; \widehat{\eta}_{k}^{\Lambda, h})-\phi_{\infty}^{\mathrm{RMST}, h}(t, a; \eta_{\infty}^{\Lambda, h})\}\right| \\
&+\sup_{t \in [0, \tau]}\left|\frac{1}{K} \sum_{k=1}^K \frac{K n_k}{n} P_{0}\{\widehat{\phi}_{k}^{\mathrm{RMST}, h}(t, a; \widehat{\eta}_{k}^{\Lambda, h})-\phi_{\infty}^{\mathrm{RMST}, h}(t, a; \eta_{\infty}^{\Lambda, h})\}\right|
\end{aligned}
$$

With a Donsker $\phi_{\infty}^{\mathrm{RMST}, h}(t, a; \eta_{\infty}^{\Lambda, h})$, the first term on the RHS is $O_{P}(n^{-1/2})$. The second term is $o_{P}(n^{-1/2})$ when Condition~\ref{condition2} and~\ref{condition4} holds. By Lemma~\ref{lemma3}, the third term converges to zero in probability. Eventually, $\sup_{t \in [0, \tau]}|\widehat{\psi}^{\mathrm{RMST}, h}(t, a)-\psi_{0}^{\mathrm{RMST}, h}(t, a)|=o_{P}(1)$. Similar reasoning applies for $|\widehat{\psi}_{j}^{\mathrm{RMTL}, h}(\tau, a)-\psi_{j, 0}^{\mathrm{RMTL}, h}(\tau, a)|=o_{P}(1)$ and $\sup_{t \in [0, \tau]}|\widehat{\psi}_{j}^{\mathrm{RMTL}, h}(t, a)-\psi_{j, 0}^{\mathrm{RMTL}, h}(t, a)|=o_{P}(1)$.

Condition~\ref{condition1} and~\ref{condition5} require convergence of estimated nuisance functions to their limiting counterparts in probability, which are used to regulate the empirical process terms from the von Mises expansion of $\widehat{\psi}^{\mathrm{RMST}, h}(\tau, a)-\psi_{0}^{\mathrm{RMST}, h}(\tau, a)$ and $\widehat{\psi}_{j}^{\mathrm{RMTL}, h}(\tau, a)-\psi_{j, 0}^{\mathrm{RMTL}, h}(\tau, a)$. Condition~\ref{condition2} bounds conditional distributions, appeared in the denominators after decomposing $P\{\widehat{\phi}^{\mathrm{RMST}, h}(\tau, a; \widehat{\eta}^{\Lambda, h})-\phi_{0}^{\mathrm{RMST}, h}(\tau, a; \eta_{0}^{\Lambda, h})\}^2$ and $P\{\widehat{\phi}_{j}^{\mathrm{RMTL}, h}(\tau, a; \widehat{\eta}_{j}^{\Lambda, h})-\phi_{j, 0}^{\mathrm{RMTL}, h}(\tau, a; \eta_{j, 0}^{\Lambda, h})\}^2$, away from zero in all strata. This condition can be assured empirically by Winsorizing extreme values. Condition~\ref{condition3} and~\ref{condition6} are often referred to as model double robustness where the consistency of the resulting estimators relies only on consistent estimation of a subset of nuisance functions. When $h_{0}(x)$ is a function of $\pi_{0}(a \mid x)$, Condition~\ref{condition3} becomes (1) $\pi_{\infty}(a \mid X)=\pi_{0}(a \mid X)$ and $S_{\infty}(t \mid a, X)=S_{0}(t \mid a, X)$ or (2) $\pi_{\infty}(a \mid X)=\pi_{0}(a \mid X)$ and $G_{\infty}(t \mid a, X)=G_{0}(t \mid a, X)$ while Condition~\ref{condition6} reduces to (1) $\pi_{\infty}(a \mid X)=\pi_{0}(a \mid X)$ and $S_{\infty}(t \mid a, X)=S_{0}(t \mid a, X)$ and $F_{j, \infty}(t \mid a, X)=F_{j, 0}(t \mid a, X)$ or (2) $\pi_{\infty}(a \mid X)=\pi_{0}(a \mid X)$ and $G_{\infty}(t \mid a, X)=G_{0}(t \mid a, X)$. When $h_{0}(x)=1$, Condition~\ref{condition3} reduce to either (1) $S_{\infty}(t \mid a, X)=S_{0}(t \mid a, X)$ or (2) $\pi_{\infty}(a \mid X)=\pi_{0}(a \mid X)$ and $G_{\infty}(t \mid a, X)=G_{0}(t \mid a, X)$. This condition is similar to consistency conditions of doubly-robust estimators for survival functions on ATE \citep{zhang2012double,bai2013doubly} etc. When $h_{0}(x)=1$, Condition~\ref{condition6} weakens to (1) $S_{\infty}(t \mid a, X)=S_{0}(t \mid a, X)$ and $F_{j, \infty}(t \mid a, X)=F_{j, 0}(t \mid a, X)$ or (2) $\pi_{\infty}(a \mid X)=\pi_{0}(a \mid X)$ and $G_{\infty}(t \mid a, X)=G_{0}(t \mid a, X)$ or (3) $F_{j, \infty}(t \mid a, X)=F_{j, 0}(t \mid a, X)$ and $G_{\infty}(t \mid a, X)=G_{0}(t \mid a, X)$, which corresponds to Theorem 1 for cause-specific CIFs in \citet{ozenne2020estimation}. The Donsker conditions that require the nuisance models have bounded complexity, often a bounded entropy integral, are avoided by cross-fitting~\citep{chernozhukov2018double}.

\subsection{Proof of Theorem~\ref{theorem:asymptotic}} \label{proof:asymptotic}

Additional conditions are required for pointwise and uniform asymptotic linearity. % Second order remainder terms $\widehat{r}(\tau, a)$ are provided in the Supplementary Material., Appendix~\ref{proof:asymptotic}.

\begin{condition} \label{condition8}
It holds that $\max_k |[P_{n, k} \{\widehat{w}_{k}^{h}(a, X)\}]^{-1}-[E_{0} \{h_{0}(X)\}]^{-1}|=o_{P}(n^{-1/2})$, $\max_k E_{0}|\widehat{h}_{k}(X)-h_{0}(X)|=o_P(n^{-1/2})$.
\end{condition}

\begin{condition} \label{condition9}
It holds that $\widehat{r}_{1}(\tau, a)=o_{P}(n^{-1/2})$.
\end{condition}

\begin{condition} \label{condition10}
It holds that $\widehat{r}_{2}^{\mathrm{RMST}}(\tau, a)=o_{P}(n^{-1/2})$.
\end{condition}

\begin{condition} \label{condition11}
Furthermore, $\sup_{t \in [0, \tau]} \widehat{r}_{1}(t, a)=o_{P}(n^{-1/2})$ and $\sup_{t \in [0, \tau]} \widehat{r}_{2}^{\mathrm{RMST}}(t, a)=o_{P}(n^{-1/2})$.
\end{condition}

\begin{condition} \label{condition12}
It holds that $\widehat{r}_{2, j}^{\mathrm{RMTL}}(\tau, a)=o_{P}(n^{-1/2})$ and $\widehat{r}_{3, j}^{\mathrm{RMTL}}(\tau, a)=o_{P}(n^{-1/2})$.
\end{condition}

\begin{condition} \label{condition13}
Furthermore, $\sup_{t \in [0, \tau]} \widehat{r}_{1}(t, a)=o_{P}(n^{-1/2})$, $\sup_{t \in [0, \tau]} \widehat{r}_{2, j}^{\mathrm{RMTL}}(t, a)=o_{P}(n^{-1/2})$, and $\sup_{t \in [0, \tau]} \widehat{r}_{3, j}^{\mathrm{RMTL}}(t, a)=o_P(n^{-1/2})$.
\end{condition}

Under Condition~\ref{condition1}, \ref{condition2}, \ref{condition3}, \ref{condition8}, \ref{condition9}, \ref{condition10},
$$
\begin{aligned}
&\widehat{\psi}^{\mathrm{RMST}, h}(\tau, a)=\psi_{0}^{\mathrm{RMST}, h}(\tau, a)+P_n \{\varphi_{0}^{\mathrm{RMST}, h}(\tau, a; \eta_{0}^{\Lambda, h})\}+o_{P}(n^{-1/2}) \\
&n^{1/2}\{\widehat{\psi}^{\mathrm{RMST}, h}(\tau, a)-\psi_{0}^{\mathrm{RMST}, h}(\tau, a)\}\stackrel{D}{\longrightarrow}\mathcal{N}(0, P_{0}[\{\varphi_{0}^{\mathrm{RMST}, h}(\tau, a; \eta_{0}^{\Lambda, h})\}^2])
\end{aligned}
$$

If Condition~\ref{condition4} and~\ref{condition11} hold additionally,
$$
\sup_{t \in [0, \tau]} \left| \widehat{\psi}^{\mathrm{RMST}, h}(t, a)-\psi_{0}^{\mathrm{RMST}, h}(t, a)-P_n \{\varphi_{0}^{\mathrm{RMST}, h}(t, a; \eta_{0}^{\Lambda, h})\} \right| = o_{P}(n^{-1/2})
$$

\noindent such that $n^{1/2}\{\widehat{\psi}^{\mathrm{RMST}, h}(t, a)-\psi_{0}^{\mathrm{RMST}, h}(t, a)\}$ ($t \in [0, \tau]$) converges weakly to a tight mean zero Gaussian process with covariance $\sigma_{0}^{\mathrm{RMST}, h}(u, t, a)=P\{\varphi_{0}^{\mathrm{RMST}, h}(u, a; \eta_{0}^{\Lambda, h})\varphi_{0}^{\mathrm{RMST}, h}(t, a; \eta_{0}^{\Lambda, h})\}$.

Under Condition~\ref{condition2}, \ref{condition5}, \ref{condition6}, \ref{condition8}, \ref{condition9}, \ref{condition12},
$$
\begin{aligned}
&\widehat{\psi}_{j}^{\mathrm{RMTL}, h}(\tau, a)=\psi_{j, 0}^{\mathrm{RMTL}, h}(\tau, a)+P_n \{\varphi_{j, 0}^{\mathrm{RMTL}, h}(\tau, a; \eta_{0}^{\Lambda, h})\}+o_{P}(n^{-1/2}) \\
&n^{1/2}\{\widehat{\psi}_{j}^{\mathrm{RMTL}, h}(\tau, a)-\psi_{j, 0}^{\mathrm{RMTL}, h}(\tau, a)\}\stackrel{D}{\longrightarrow}\mathcal{N}(0, P_{0}[\{\varphi_{j, 0}^{\mathrm{RMTL}, h}(\tau, a; \eta_{0}^{\Lambda, h})\}^2])
\end{aligned}
$$

If Condition~\ref{condition7} and~\ref{condition13} holds additionally,
$$
\sup_{t \in [0, \tau]} \left| \widehat{\psi}_{j}^{\mathrm{RMTL}, h}(t, a)-\psi_{j, 0}^{\mathrm{RMTL}, h}(t, a)-P_n \{\varphi_{j, 0}^{\mathrm{RMTL}, h}(t, a; \eta_{j, 0}^{\Lambda, h})\} \right| = o_{P}(n^{-1/2})
$$

\noindent such that $n^{1/2}\{\widehat{\psi}_{j}^{\mathrm{RMTL}, h}(t, a)-\psi_{j, 0}^{\mathrm{RMTL}, h}(t, a)\}$ ($t \in [0, \tau]$) converges weakly to a tight Gaussian process with mean zero and covariance $\sigma_{j, 0}^{\mathrm{RMTL}, h}(u, t, a)=P\{\varphi_{j, 0}^{\mathrm{RMTL}, h}(u, a; \eta_{j, 0}^{\Lambda, h})\varphi_{j, 0}^{\mathrm{RMTL}, h}(t, a; \eta_{j, 0}^{\Lambda, h})\}$.

Following Proof of Theorem~\ref{theorem:consistency}, and when $\eta_{\infty}^{\Lambda, h}=\eta_{0}^{\Lambda, h}$, we have
% $h_{\infty}(X)=h_{0}(X)$, $S_{\infty}(t \mid a, X)=S_{0}(t \mid a, X)$, $\pi_{\infty}(a \mid X)=\pi_{0}(a \mid X)$, and $G_{\infty}(t \mid a, X)=G_{0}(t \mid a, X)$
$$
\begin{aligned}
&\widehat{\psi}^{\mathrm{RMST}, h}(\tau, a)-\psi_{0}^{\mathrm{RMST}, h}(\tau, a) \\
=& P_n\{\varphi_{0}^{\mathrm{RMST}, h}(\tau, a; \eta_{0}^{\Lambda, h})\}+\frac{1}{K} \sum_{k=1}^K \frac{K n_k^{1 / 2}}{n} \mathcal{G}_{n, k} \{\widehat{\phi}_{k}^{\mathrm{RMST}, h}(\tau, a; \widehat{\eta}_{k}^{\Lambda, h})-\phi_{0}^{\mathrm{RMST}, h}(\tau, a; \eta_{0}^{\Lambda, h})\} \\
&+\frac{1}{K} \sum_{k=1}^K \frac{K n_k}{n} [P_{0}\{\widehat{\phi}_{k}^{\mathrm{RMST}, h}(\tau, a; \widehat{\eta}_{k}^{\Lambda, h})\}-\psi_{0}^{\mathrm{RMST}, h}(\tau, a)]
\end{aligned}
$$

% where the second empirical process term has been studied in Proof of Theorem~\ref{theorem:consistency}.

Under Condition~\ref{condition1}, \ref{condition2}, \ref{condition4}, and Lemma 3,
$$
\frac{1}{K} \sum_{k=1}^K \frac{K n_k^{1 / 2}}{n} \sup_{t \in [0, \tau]} \left|\mathcal{G}_{n, k}\{\widehat{\phi}_{k}^{\mathrm{RMST}, h}(\tau, a; \widehat{\eta}_{k}^{\Lambda, h})-\phi_{0}^{\mathrm{RMST}, h}(\tau, a; \eta_{0}^{\Lambda, h})\}\right|=o_{P}(n^{-1 / 2})
$$

For the third term on the RHS, we substitute the Duhamel equation
$$
\frac{S_{0}(t-\mid a, X)}{\widehat{S}(t \mid a, X)}\{\mathrm{d}\widehat{\Lambda}(t \mid a, X)-\mathrm{d} \Lambda_{0}(t \mid a, X)\}=\mathrm{d}\left\{\frac{S_{0}(t\mid a, X)}{\widehat{S}(t \mid a, X)}-1\right\}
$$

\noindent into Lemma~\ref{lemma2},
$$
\begin{aligned}
&P_{0} \{\widehat{\phi}_{k}^{\mathrm{RMST}, h}(\tau, a; \widehat{\eta}_{k}^{\Lambda, h})\}-\psi_{0}^{\mathrm{RMST}, h}(\tau, a) \\
=&[P_{n, k}\{\widehat{w}_{k}^{h}(a, X)\}]^{-1} E_{0}\left[\frac{\widehat{h}_{k}(X)\pi_{0}(a \mid X)\mathrm{R}\widehat{\mathrm{MS}}\mathrm{T}_{k}(\tau \mid a, X)}{\widehat{\pi}_{k}(a \mid X)} \int_0^{\tau} \left\{\frac{G_{0}(t- \mid a, X)}{\widehat{G}_{k}(t- \mid a, X)}-1\right\}\mathrm{d}\left\{\frac{S_{0}(t\mid a, X)}{\widehat{S}_{k}(t \mid a, X)}-1\right\}\right] \\
&+E_{0}\left[\widehat{h}_{k}(X)\left\{\frac{\pi_{0}(a \mid X)}{\widehat{\pi}_{k}(a \mid X)}-1\right\}\mathrm{R}\widehat{\mathrm{MS}}\mathrm{T}_{k}(\tau \mid a, X) \left\{\frac{S_{0}(\tau \mid a, X)}{\widehat{S}_{k}(\tau \mid a, X)}-1\right\} \right] \\
&-E_{0}\left[\frac{\widehat{h}_{k}(X)\pi_{0}(a \mid X)}{\widehat{\pi}_{k}(a \mid X)} \int_0^{\tau} \mathrm{R}\widehat{\mathrm{MS}}\mathrm{T}_{k}(t \mid a, X) \left\{\frac{G_{0}(t- \mid a, X)}{\widehat{G}_{k}(t- \mid a, X)}-1\right\}\mathrm{d}\left\{\frac{S_{0}(t\mid a, X)}{\widehat{S}_{k}(t \mid a, X)}-1\right\}\right] \\
&-E_{0}\left(\widehat{h}_{k}(X)\left\{\frac{\pi_{0}(a \mid X)}{\widehat{\pi}_{k}(a \mid X)}-1\right\}\left[\mathrm{R}\widehat{\mathrm{MS}}\mathrm{T}_{k}(\tau \mid a, X) \left\{\frac{S_{0}(\tau \mid a, X)}{\widehat{S}_{k}(\tau \mid a, X)}\right\}-\mathrm{RMST}_{0}(\tau \mid a, X) \right] \right) \\
&+([P_{n, k} \{\widehat{h}_{k}(X)\}]^{-1}-[P_{n, k}\{\widehat{w}_{k}^{h}(a, X)\}]^{-1}) E_{0}\{\widehat{h}_{k}(X)\mathrm{R}\widehat{\mathrm{MS}}\mathrm{T}_{k}(\tau \mid a, X)\} \\
&+([P_{n, k}\{\widehat{w}_{k}^{h}(a, X)\}]^{-1}-[E_{0} \{h_{0}(X)\}]^{-1}) E_{0}\{\widehat{h}_{k}(X)\mathrm{RMST}_{0}(\tau \mid a, X)\} \\
&+[E_{0} \{h_{0}(X)\}]^{-1} E_{0}[\{\widehat{h}_{k}(X)-h_{0}(X)\}\mathrm{RMST}_{0}(\tau \mid a, X)]
\end{aligned}
$$

\noindent such that
$$
\begin{aligned}
&\left|P_{0} \{\widehat{\phi}_{k}^{\mathrm{RMST}, h}(\tau, a; \widehat{\eta}_{k}^{\Lambda, h})\}-\psi_{0}^{\mathrm{RMST}, h}(\tau, a)\right| \\
\leq & 2 h_{\mathrm{min}}^{-1} h_{\mathrm{max}}\epsilon \tau E_{0}\left| \int_0^{\tau} \left\{\frac{G_{0}(t- \mid a, X)}{\widehat{G}_{k}(t- \mid a, X)}-1\right\}\mathrm{d}\left\{\frac{S_{0}(t\mid a, X)}{\widehat{S}_{k}(t \mid a, X)}-1\right\}\right| \\
&+ h_{\mathrm{min}}^{-1}h_{\mathrm{max}}\epsilon E_{0} \left|\left\{\widehat{\pi}_{k}(a \mid X)-\pi_{0}(a \mid X)\right\}\{\mathrm{R}\widehat{\mathrm{MS}}\mathrm{T}_{k}(\tau \mid a, X)-\mathrm{RMST}_{0}(\tau \mid a, X)\} \right| \\
&+2 h_{\mathrm{min}}^{-1}h_{\mathrm{max}}\tau h_{\mathrm{max}}\tau \left|[P_{n, k}\{\widehat{w}_{k}^{h}(a, X)\}]^{-1}-[E_{0} \{h_{0}(X)\}]^{-1} \right|+h_{\mathrm{min}}^{-1} \tau E_{0}|\{\widehat{h}_{k}(X)-h_{0}(X)|
\end{aligned}
$$

\noindent where we define
$$
\begin{aligned}
&\widehat{r}_{1}(\tau, a)=\max_k E_{0}\left| \int_0^{\tau} \left\{\frac{G_{0}(t- \mid a, X)}{\widehat{G}_{k}(t- \mid a, X)}-1\right\}\mathrm{d}\left\{\frac{S_{0}(t\mid a, X)}{\widehat{S}_{k}(t \mid a, X)}-1\right\}\right| \\
&\widehat{r}_{2}^{\mathrm{RMST}}(\tau, a)=\max_k E_{0}\left|\{\widehat{\pi}_{k}(a \mid X)-\pi_{0}(a \mid X)\}\{\mathrm{R}\widehat{\mathrm{MS}}\mathrm{T}_{k}(\tau \mid a, X)-\mathrm{RMST}_{0}(\tau \mid a, X)\} \right|
\end{aligned}
$$

Note that $1/K \sum_{k=1}^K K n_k/n \leq 2$, such that
$$
\begin{aligned}
&\left|\frac{1}{K} \sum_{k=1}^K \frac{K n_k}{n} [P_{0}\{\widehat{\phi}_{k}^{\mathrm{RMST}, h}(\tau, a; \widehat{\eta}_{k}^{\Lambda, h})\}-\psi_{0}^{\mathrm{RMST}, h}(\tau, a)]\right| \\
\leq & 4 h_{\mathrm{min}}^{-1} h_{\mathrm{max}}\epsilon \tau \widehat{r}_{1}(\tau, a)+ 2 h_{\mathrm{min}}^{-1}h_{\mathrm{max}}\epsilon \widehat{r}_{2}^{\mathrm{RMST}}(\tau, a) \\
&+4 h_{\mathrm{min}}^{-1}h_{\mathrm{max}}\tau h_{\mathrm{max}}\tau \left|[P_{n, k}\{\widehat{w}_{k}^{h}(a, X)\}]^{-1}-[E_{0} \{h_{0}(X)\}]^{-1} \right|+2 h_{\mathrm{min}}^{-1} \tau E_{0}|\{\widehat{h}_{k}(X)-h_{0}(X)|
\end{aligned}
$$

By Condition~\ref{condition8}, \ref{condition9}, \ref{condition10}, the RHS is $o_{P}(n^{-1/2})$. Hence $\widehat{\psi}^{\mathrm{RMST}, h}(\tau, a)=\psi_{0}^{\mathrm{RMST}, h}(\tau, a)+P_n \{\varphi_{0}^{\mathrm{RMST}, h}(\tau, a; \eta_{0}^{\Lambda, h})\}+o_{P}(n^{-1/2})$. Due to $P_{0}\{\varphi_{0}^{\mathrm{RMST}, h}(\tau, a; \eta_{0}^{\Lambda, h})\}=0$, $P_{0}[\{\varphi_{0}^{\mathrm{RMST}, h}(\tau, a; \eta_{0}^{\Lambda, h})\}]^2<\infty$, and its uniform boundness, $n^{1/2}P_n\{\varphi_{0}^{\mathrm{RMST}, h}(\tau, a; \eta_{0}^{\Lambda, h})\} \stackrel{D}{\longrightarrow}\mathcal{N}(0, P_{0}[\{\varphi_{0}^{\mathrm{RMST}, h}(\tau, a; \eta_{0}^{\Lambda, h})\}^2])$.

When Condition~\ref{condition11} holds
$$
\begin{aligned}
&\sup_{t \in [0, \tau]}\left|\frac{1}{K} \sum_{k=1}^K \frac{K n_k}{n} [P_{0}\{\widehat{\phi}_{k}^{\mathrm{RMST}, h}(t, a; \widehat{\eta}_{k}^{\Lambda, h})\}-\psi_{0}^{\mathrm{RMST}, h}(t, a)]\right| \\
\leq & 4 h_{\mathrm{min}}^{-1} h_{\mathrm{max}}\epsilon \tau \sup_{t \in [0, \tau]}\widehat{r}_{1}(t, a)+ 2 h_{\mathrm{min}}^{-1}h_{\mathrm{max}}\epsilon \sup_{t \in [0, \tau]}\widehat{r}_{2}^{\mathrm{RMST}}(t, a) \\
&+4 h_{\mathrm{min}}^{-1}h_{\mathrm{max}}\tau h_{\mathrm{max}}\tau \left|[P_{n, k}\{\widehat{w}_{k}^{h}(a, X)\}]^{-1}-[E_{0} \{h_{0}(X)\}]^{-1} \right|+2 h_{\mathrm{min}}^{-1} \tau E_{0}|\{\widehat{h}_{k}(X)-h_{0}(X)|=o_{P}(n^{-1/2})
\end{aligned}
$$

As a result,
$$
\sup_{t \in [0, \tau]} \left| \widehat{\psi}^{\mathrm{RMST}, h}(t, a)-\psi_{0}^{\mathrm{RMST}, h}(t, a)-P_n \{\varphi_{0}^{\mathrm{RMST}, h}(t, a; \eta_{0}^{\Lambda, h})\} \right| = o_{P}(n^{-1/2})
$$

As $\varphi_{0}^{\mathrm{RMST}, h}(t, a; \eta_{0}^{\Lambda, h})$ is uniformly bounded for $t \in [0, \tau]$, then $n^{1/2}\{P_n \varphi_{0}^{\mathrm{RMST}, h}(t, a; \eta_{0}^{\Lambda, h})\}$ converges weakly to a tight Gaussian process with mean zero and covariance $\sigma_{0}^{\mathrm{RMST}, h}(u, t, a)=P_{0}\{\varphi_{0}^{\mathrm{RMST}, h}(u, a; \eta_{0}^{\Lambda, h})\varphi_{0}^{\mathrm{RMST}, h}(t, a; \eta_{0}^{\Lambda, h})\}$.

We plug the Duhamel equation into the competing risks setting of Lemma~\ref{lemma2},
$$
\begin{aligned}
&P_{0} \{\widehat{\phi}_{j, k}^{\mathrm{RMTL}, h}(\tau, a; \widehat{\eta}_{j, k}^{\Lambda, h})\}-\psi_{j, 0}^{\mathrm{RMTL}, h}(\tau, a) \\
=&[P_{n, k}\{\widehat{w}_{k}^{h}(a, X)\}]^{-1} \left(-E_{0}\left[\frac{\widehat{h}_{k}(X)\pi_{0}(a \mid X)\tau}{\widehat{\pi}_{k}(a \mid X)} \int_0^{\tau}S_{0}(t \mid a, X)\left\{\frac{G_{0}(t- \mid a, X)}{\widehat{G}_{k}(t- \mid a, X)}-1 \right\} \right.\right. \\
&\left.\times\{\mathrm{d}\widehat{\Lambda}_{j, k}(t \mid a, X)-\mathrm{d}\Lambda_{j, 0}(t \mid a, X)\}\right]-E_{0}\left[\widehat{h}_{k}(X)\left\{\frac{\pi_{0}(a \mid X)}{\widehat{\pi}_{k}(a \mid X)}-1\right\} \right. \\
&\left.\times\tau \int_0^{\tau} S_{0}(t \mid a, X) \{\mathrm{d}\widehat{\Lambda}_{j, k}(t \mid a, X)-\mathrm{d}\Lambda_{j, 0}(t \mid a, X)\} \right] \\
&+E_{0}\left[\frac{\widehat{h}_{k}(X)\pi_{0}(a \mid X)}{\widehat{\pi}_{k}(a \mid X)} \int_0^{\tau}t S_{0}(t \mid a, X)\left\{\frac{G_{0}(t- \mid a, X)}{\widehat{G}_{k}(t- \mid a, X)}-1 \right\} \{\mathrm{d}\widehat{\Lambda}_{j, k}(t \mid a, X)-\mathrm{d}\Lambda_{j, 0}(t \mid a, X)\}\right] \\
&+E_{0}\left[\widehat{h}_{k}(X)\left\{\frac{\pi_{0}(a \mid X)}{\widehat{\pi}_{k}(a \mid X)}-1\right\} \int_0^{\tau} t S_{0}(t \mid a, X) \{\mathrm{d}\widehat{\Lambda}_{j, k}(t \mid a, X)-\mathrm{d}\Lambda_{j, 0}(t \mid a, X)\} \right] \\
&+E_{0}\left[\frac{\widehat{h}_{k}(X)\pi_{0}(a \mid X)\mathrm{R}\widehat{\mathrm{MT}}\mathrm{L}_{j, k}(\tau \mid a, X)}{\widehat{\pi}_{k}(a \mid X)} \int_0^{\tau} \left\{\frac{G_{0}(t- \mid a, X)}{\widehat{G}_{k}(t- \mid a, X)}-1\right\}\mathrm{d}\left\{\frac{S_{0}(t\mid a, X)}{\widehat{S}_{k}(t \mid a, X)}-1\right\}\right] \\
&+E_{0}\left[\widehat{h}_{k}(X)\left\{\frac{\pi_{0}(a \mid X)}{\widehat{\pi}_{k}(a \mid X)}-1\right\}\mathrm{R}\widehat{\mathrm{MT}}\mathrm{L}_{j, k}(\tau \mid a, X) \left\{\frac{S_{0}(\tau \mid a, X)}{\widehat{S}_{k}(\tau \mid a, X)}-1\right\} \right] \\
&-E_{0}\left[\frac{\widehat{h}_{k}(X)\pi_{0}(a \mid X)}{\widehat{\pi}_{k}(a \mid X)} \int_0^{\tau} \mathrm{R}\widehat{\mathrm{MT}}\mathrm{L}_{j, k}(t \mid a, X) \left\{\frac{G_{0}(t- \mid a, X)}{\widehat{G}(t- \mid a, X)}-1\right\}\mathrm{d}\left\{\frac{S_{0}(t\mid a, X)}{\widehat{S}_{k}(t \mid a, X)}-1\right\}\right] \\
&-E_{0}\left(\widehat{h}_{k}(X)\left\{\frac{\pi_{0}(a \mid X)}{\widehat{\pi}_{k}(a \mid X)}-1\right\}\left[\mathrm{R}\widehat{\mathrm{MT}}\mathrm{L}_{j, k}(\tau \mid a, X) \left\{\frac{S_{0}(\tau \mid a, X)}{\widehat{S}_{k}(\tau \mid a, X)}\right\}- \int_0^{\tau} \frac{S_{0}(t \mid a, X)}{\widehat{S}_{k}(t \mid a, X)} \widehat{F}_{j, k}(t \mid a, X) \mathrm{d}t \right] \right) \\
&-E_{0}\left[\frac{\widehat{h}_{k}(X)\pi_{0}(a \mid X)\tau}{\widehat{\pi}_{k}(a \mid X)} \int_0^{\tau} \widehat{F}_{j, k}(t \mid a, X)\left\{\frac{G_{0}(t- \mid a, X)}{\widehat{G}_{k}(t- \mid a, X)}-1\right\}\mathrm{d}\left\{\frac{S_{0}(t\mid a, X)}{\widehat{S}_{k}(t \mid a, X)}-1\right\}\right] \\
&-E_{0}\left[\widehat{h}_{k}(X)\left\{\frac{\pi_{0}(a \mid X)}{\widehat{\pi}_{k}(a \mid X)}-1\right\}\tau \left\{\frac{\widehat{F}_{j, k}(\tau \mid a, X) S_{0}(\tau \mid a, X)}{\widehat{S}_{k}(\tau \mid a, X)}-\int_0^{\tau}S_{0}(\tau \mid a, X)\mathrm{d}\widehat{\Lambda}_{j, k}(t \mid a, X) \right\} \right] \\
&+E_{0}\left[\frac{\widehat{h}_{k}(X)\pi_{0}(a \mid X)}{\widehat{\pi}_{k}(a \mid X)} \int_0^{\tau} t\widehat{F}_{j, k}(t \mid a, X)\left\{\frac{G_{0}(t- \mid a, X)}{\widehat{G}_{k}(t- \mid a, X)}-1\right\}\mathrm{d}\left\{\frac{S_{0}(t\mid a, X)}{\widehat{S}_{k}(t \mid a, X)}-1\right\}\right]
\end{aligned}
$$

$$
\begin{aligned}
&+E_{0}\left[\widehat{h}_{k}(X)\left\{\frac{\pi_{0}(a \mid X)}{\widehat{\pi}_{k}(a \mid X)}-1\right\} \left\{\frac{\widehat{F}_{j, k}(\tau \mid a, X) S_{0}(\tau \mid a, X)\tau}{\widehat{S}_{k}(\tau \mid a, X)}-\int_0^{\tau} \frac{S_{0}(t\mid a, X)}{\widehat{S}_{k}(t \mid a, X)} \widehat{F}_{j, k}(t \mid a, X) \mathrm{d}t \right.\right. \\
&\left.\left.-\int_0^{\tau} t S_{0}(\tau \mid a, X)\mathrm{d}\widehat{\Lambda}_{j, k}(t \mid a, X) \right\} \right] \\
&+([P_{n, k} \{\widehat{h}_{k}(X)\}]^{-1}-[P_{n, k}\{\widehat{w}_{k}^{h}(a, X)\}]^{-1} ) E_{0}\{\widehat{h}_{k}(X)\mathrm{R}\widehat{\mathrm{MT}}\mathrm{L}_{j, k}(\tau \mid a, X)\} \\
&+([P_{n, k}\{\widehat{w}_{k}^{h}(a, X)\}]^{-1}-[E_{0} \{h_{0}(X)\}]^{-1}) E_{0}\{\widehat{h}_{k}(X)\mathrm{RMTL}_{j, 0}(\tau \mid a, X)\} \\
&+[E_{0}\{h_{0}(X)\}]^{-1} E_{0}[\{\widehat{h}_{k}(X)-h_{0}(X)\}\mathrm{RMTL}_{j, 0}(\tau \mid a, X)]
\end{aligned}
$$

After canceling out common terms, we have
$$
\begin{aligned}
&P_{0} \{\widehat{\phi}_{j, k}^{\mathrm{RMTL}, h}(\tau, a; \widehat{\eta}_{j, k}^{\Lambda, h})\}-\psi_{j, 0}^{\mathrm{RMTL}, h}(\tau, a) \\
=&[P_{n, k}\{\widehat{w}_{k}^{h}(a, X)\}]^{-1} \left(-E_{0}\left[\frac{\widehat{h}_{k}(X)\pi_{0}(a \mid X)}{\widehat{\pi}_{k}(a \mid X)} \int_0^{\tau} (\tau-t) S_{0}(t \mid a, X)\left\{\frac{G_{0}(t- \mid a, X)}{\widehat{G}_{k}(t- \mid a, X)}-1 \right\} \right.\right. \\
&\left.\times\{\mathrm{d}\widehat{\Lambda}_{j, k}(t \mid a, X)-\mathrm{d}\Lambda_{j, 0}(t \mid a, X)\}\right]+E_{0}\left[\frac{\widehat{h}_{k}(X)\pi_{0}(a \mid X)}{\widehat{\pi}_{k}(a \mid X)} \int_0^{\tau} \{\mathrm{R}\widehat{\mathrm{MT}}\mathrm{L}_{j, k}(\tau \mid a, X)-\mathrm{R}\widehat{\mathrm{MT}}\mathrm{L}_{j, k}(t \mid a, X)\} \right. \\
&\left.\times\left\{\frac{G_{0}(t- \mid a, X)}{\widehat{G}_{k}(t- \mid a, X)}-1\right\}\mathrm{d}\left\{\frac{S_{0}(t\mid a, X)}{\widehat{S}_{k}(t \mid a, X)}-1\right\}\right]-E_{0}\left[\frac{\widehat{h}_{k}(X)\pi_{0}(a \mid X)}{\widehat{\pi}_{k}(a \mid X)} \int_0^{\tau} (\tau-t) \widehat{F}_{j, k}(t \mid a, X) \right. \\
&\left.\times\left\{\frac{G_{0}(t- \mid a, X)}{\widehat{G}_{k}(t- \mid a, X)}-1\right\}\mathrm{d}\left\{\frac{S_{0}(t\mid a, X)}{\widehat{S}_{k}(t \mid a, X)}-1\right\}\right]+E_{0}\left[\widehat{h}_{k}(X)\left\{\frac{\pi_{0}(a \mid X)}{\widehat{\pi}_{k}(a \mid X)}-1\right\} \right. \\
&\left.\times\{\mathrm{R}\widehat{\mathrm{MT}}\mathrm{L}_{j, k}(\tau \mid a, X)-\mathrm{RMTL}_{j, 0}(\tau \mid a, X)\}\right] \\
&+([P_n \{\widehat{h}_{k}(X)\}]^{-1}-[P_n\{\widehat{w}_{k}^{h}(a, X)\}]^{-1}) E_{0}\{\widehat{h}_{k}(X)\mathrm{R}\widehat{\mathrm{MT}}\mathrm{L}_{j, k}(\tau \mid a, X)\} \\
&+([P_{n, k}\{\widehat{w}_{k}^{h}(a, X)\}]^{-1}-[E_{0} \{h_{0}(X)\}]^{-1}) E_{0}\{\widehat{h}_{k}(X)\mathrm{RMTL}_{j}(\tau \mid a, X)\} \\
&+[E_{0} \{h_{0}(X)\}]^{-1} E_{0}[\{\widehat{h}_{k}(X)-h_{0}(X)\}\mathrm{RMTL}_{j, 0}(\tau \mid a, X)]
\end{aligned}
$$

Finally,
$$
\begin{aligned}
&\left|P_{0} \{\widehat{\phi}_{j, k}^{\mathrm{RMTL}, h}(\tau, a; \widehat{\eta}_{j, k}^{\Lambda, h})\}-\psi_{j, 0}^{\mathrm{RMTL}, h}(\tau, a)\right| \\
=&2 h_{\mathrm{min}}^{-1}h_{\mathrm{max}}\epsilon \tau E_{0}\left|\int_0^{\tau} \left\{\frac{G_{0}(t- \mid a, X)}{\widehat{G}_{k}(t- \mid a, X)}-1 \right\}\{\mathrm{d}\widehat{\Lambda}_{j, k}(t \mid a, X)-\mathrm{d}\Lambda_{j, 0}(t \mid a, X)\}\right| \\
&+4h_{\mathrm{min}}^{-1}h_{\mathrm{max}}\epsilon \tau E_{0} \left|\int_0^{\tau}\left\{\frac{G_{0}(t- \mid a, X)}{\widehat{G}_{k}(t- \mid a, X)}-1\right\}\mathrm{d}\left\{\frac{S_{0}(t\mid a, X)}{\widehat{S}_{k}(t \mid a, X)}-1\right\}\right| \\
&+ h_{\mathrm{min}}^{-1}h_{\mathrm{max}}\epsilon E_{0} \left|\left\{\widehat{\pi}_{k}(a \mid X)-\pi_{0}(a \mid X)\right\}\{\mathrm{R}\widehat{\mathrm{MT}}\mathrm{L}_{j, k}(\tau \mid a, X)-\mathrm{RMTL}_{j, 0}(\tau \mid a, X)\} \right| \\
&+2 h_{\mathrm{min}}^{-1}h_{\mathrm{max}}\tau h_{\mathrm{max}}\tau \left|[P_{n, k}\{\widehat{w}_{k}^{h}(a, X)\}]^{-1}-[E_{0} \{h_{0}(X)\}]^{-1} \right|+h_{\mathrm{min}}^{-1} \tau E_{0}|\{\widehat{h}_{k}(X)-h_{0}(X)|
\end{aligned}
$$

\noindent where we define
$$
\begin{aligned}
\widehat{r}_{2, j}^{\mathrm{RMTL}}(\tau, a)=&\max_k E_{0} \left|\left\{\widehat{\pi}_{k}(a \mid X)-\pi_{0}(a \mid X)\right\}\{\mathrm{R}\widehat{\mathrm{MT}}\mathrm{L}_{j, k}(\tau \mid a, X)-\mathrm{RMTL}_{j, 0}(\tau \mid a, X)\} \right| \\
\widehat{r}_{3, j}^{\mathrm{RMTL}}(\tau, a)=&\max_k E_{0}\left|\int_0^{\tau} \left\{\frac{G_{0}(t- \mid a, X)}{\widehat{G}_{k}(t- \mid a, X)}-1 \right\}\{\mathrm{d}\widehat{\Lambda}_{j, k}(t \mid a, X)-\mathrm{d}\Lambda_{j, 0}(t \mid a, X)\}\right|
\end{aligned}
$$

Remaining proofs are similar to that of $P_{0} \{\widehat{\phi}_{k}^{\mathrm{RMST}, h}(\tau, a; \widehat{\eta}_{k}^{\Lambda, h})\}-\psi_{0}^{\mathrm{RMST}, h}(\tau, a)$.

Regarding efficiency bounds of the proposed estimators, the conditional variances of $\widehat{\psi}^{\mathrm{RMST}, h}(\tau)$ is

% As $n \rightarrow \infty$, the conditional variances of proposed estimators converge
% $$
% \begin{aligned}
% &n \times \operatorname{var}\{\widehat{\psi}^{\mathrm{RMST}, h}(\tau) \mid O_n\} \\
% \rightarrow& \left\{\int h_{0}(x) p_{0}(x) \mathcal{F}(d x)\right\}^{-2} \int\left[\frac{\operatorname{var}\{\phi^{\mathrm{RMST}}_{0}(\tau, 1; o) \mid X\}}{\pi_{0}(1 \mid x)}+\frac{\operatorname{var}\{\phi^{\mathrm{RMST}}_{0}(\tau, 0; o) \mid X\}}{\pi_{0}(0 \mid x)}\right] \\
% &\times h_{0}^2(x) p_{0}(x) \mathcal{F}(d x)
% \end{aligned}
% $$

% and
% $$
% \begin{aligned}
% &n \times \operatorname{var}\{\widehat{\psi}_{j}^{\mathrm{RMTL}, h}(\tau) \mid O_n\} \\
% \rightarrow& \left\{\int h_{0}(x) p_{0}(x) \mathcal{F}(d x)\right\}^{-2} \int\left[\frac{\operatorname{var}\{\phi_{j, 0}^{\mathrm{RMTL}}(\tau, 1; o) \mid X\}}{\pi_{0}(1 \mid x)}+\frac{\operatorname{var}\{\phi_{j, 0}^{\mathrm{RMTL}}(\tau, 0; o) \mid X\}}{\pi_{0}(0 \mid x)}\right] \\
% &\times h_{0}^2(x) p_{0}(x) \mathcal{F}(d x)
% \end{aligned}
% $$

% Under $\operatorname{var}\{\phi_{0}^{\mathrm{RMST}}(\tau, 1; o) \mid X\}=\operatorname{var}\{\phi_{0}^{\mathrm{RMST}}(\tau, 0; o) \mid X\}$, $h_{0}(x)\propto\operatorname{var}_{0}(a \mid x)$ gives the smallest asymptotic variance among the class of balancing IPTW estimators for $\widehat{\psi}^{\mathrm{RMST}, h}(\tau)$. The same applies to $\widehat{\psi}_{j, 0}^{\mathrm{RMTL}, h}(\tau)$ when $\operatorname{var}\{\phi_{j, 0}^{\mathrm{RMTL}}(\tau, 1; o) \mid X\}=\operatorname{var}\{\phi_{j, 0}^{\mathrm{RMTL}}(\tau, 0; o) \mid X\}$.

$$
\begin{aligned}
&\operatorname{var}\{\widehat{\psi}^{\mathrm{RMST}, h}(\tau) \mid O_n\} \\
=&\frac{1}{n}[P_n\{\widehat{h}(X)\}]^{-2} P_n[\widehat{h}^2(X)\operatorname{var}\{\mathrm{R}\widehat{\mathrm{MS}}\mathrm{T}(\tau \mid 1, X)\mid O_n\}]+\frac{1}{n}[P_n \{\widehat{w}^{h}(1, X)\}]^{-2} \\
&\times P_n\left[\frac{\widehat{h}^2(X) I(A=1)}{\widehat{\pi}^2(1 \mid X)}\operatorname{var}\left\{-\int_0^{\tau \wedge \widetilde{T}} \frac{\{\mathrm{R}\widehat{\mathrm{MS}}\mathrm{T}(\tau \mid 1, X)-\mathrm{R}\widehat{\mathrm{MS}}\mathrm{T}(t \mid 1, X)\} \mathrm{d} \widehat{M}(t \mid 1, X)}{\widehat{S}(t \mid 1, X)\widehat{G}(t- \mid 1, X)} \mid O_n\right\} \right] \\
&+\frac{1}{n}[P_n\{\widehat{h}(X)\}]^{-2} P_n[\widehat{h}^2(X)\operatorname{var}\{\mathrm{R}\widehat{\mathrm{MS}}\mathrm{T}(\tau \mid 0, X)\mid O_n\}]+\frac{1}{n}[P_n \{\widehat{w}^{h}(0, X)\}]^{-2} \\
&\times P_n\left[\frac{\widehat{h}^2(X) I(A=0)}{\widehat{\pi}^2(0 \mid X)}\operatorname{var}\left\{-\int_0^{\tau \wedge \widetilde{T}} \frac{\{\mathrm{R}\widehat{\mathrm{MS}}\mathrm{T}(\tau \mid 0, X)-\mathrm{R}\widehat{\mathrm{MS}}\mathrm{T}(t \mid 0, X)\} \mathrm{d} \widehat{M}(t \mid 0, X)}{\widehat{S}(t \mid 0, X)\widehat{G}(t- \mid 0, X)} \mid O_n\right\} \right]
\end{aligned}
$$

By the Slutsky's theorem,
$$
\begin{aligned}
&n \times \operatorname{var}\{\widehat{\psi}^{\mathrm{RMST}, h}(\tau) \mid O_n\} \\
\rightarrow& \left\{\int h_{0}(x) p_{0}(x) \mathcal{F}(d x)\right\}^{-2} \int\left[\frac{\operatorname{var}\{\phi_{0}^{\mathrm{RMST}}(\tau, 1; \eta_{0}^{\Lambda}) \mid x\}}{\pi_{0}(1 \mid x)}+\frac{\operatorname{var}\{\phi_{0}^{\mathrm{RMST}}(\tau, 0; \eta_{0}^{\Lambda}) \mid x\}}{\pi_{0}(0 \mid x)}\right] h_{0}^2(x) p_{0}(x) \mathcal{F}(d x)
\end{aligned}
$$

where $p(x)$ is the probability density function of $X$ with respect to a measure $\mathcal{F}(\cdot)$. Assuming untilted uncentered true EIFs are conditionally homoskedastic, i.e., $\operatorname{var}\{\phi_{0}^{\mathrm{RMST}}(\tau, 1; \eta_{0}^{\Lambda}) \mid x\}=\operatorname{var}\{\phi_{0}^{\mathrm{RMST}}(\tau, 0; \eta_{0}^{\Lambda}) \mid x\}=C_{\mathrm{var}}^{\mathrm{RMST}}$, we have
$$
n \times \operatorname{var}\{\widehat{\psi}^{\mathrm{RMST}, h}(\tau) \mid O_n\}
\rightarrow C_{\mathrm{var}}^{\mathrm{RMST}} \left\{\int h_{0}(x) p_{0}(x) \mathcal{F}(d x)\right\}^{-2} \int\left[\frac{h_{0}^2(x) p_{0}(x) \mathcal{F}(d x)}{\pi_{0}(1 \mid x)\pi_{0}(0 \mid x)}\right]
$$

Following the Cauchy-Schwarz inequality,
$$
\begin{aligned}
&\left\{\int h_{0}(x) p_{0}(x) \mathcal{F}(d x)\right\}^{2}=\left\{\int \frac{h_{0}(x)}{\sqrt{\pi_{0}(1 \mid x)\pi_{0}(0 \mid x)}} \sqrt{\pi_{0}(1 \mid x)\pi_{0}(0 \mid x)} p_{0}(x) \mathcal{F}(d x)\right\}^{2} \\
\leq&\int \frac{h_{0}^2(x)}{\pi_{0}(1 \mid x)\pi_{0}(0 \mid x)} p_{0}(x) \mathcal{F}(d x) \times \int \pi_{0}(1 \mid x)\pi_{0}(0 \mid x) p_{0}(x) \mathcal{F}(d x)
\end{aligned}
$$

The equal sign holds if and only if $h_{0}(x)/\{\sqrt{\pi_{0}(1 \mid x)\pi_{0}(0 \mid x)}\}\propto\sqrt{\pi_{0}(1 \mid x)\pi_{0}(0 \mid x)}$, identically $h_{0}(x)\propto\pi_{0}(1 \mid x)\pi_{0}(0 \mid x)\propto\operatorname{var}(a \mid x)$. This is when the proposed estimators reach the semiparametric efficiency bounds for the WATE under conditional homoscedasticity of untilted uncentered EIFs. The proof for $\operatorname{var}\{\widehat{\psi}_{j}^{\mathrm{RMTL}, h}(\tau) \mid O_n\}$ is similar.

Note that when $h_{0}(x)=\pi(1 \mid x)\pi(0 \mid x)$, we have
$$\psi_{0}^{\mathrm{RMST}, \text{ATO}}(\tau)=\frac{E_{0}[\operatorname{cov}\{A, \min(T, \tau) \mid X\}]}{E_{0}\{\operatorname{var}(A \mid X)\}}$$

and
$$\psi_{j, 0}^{\mathrm{RMTL}, \text{ATO}}(\tau)=\frac{E_{0}(\operatorname{cov}[A, \{\tau-\min(T, \tau)\}I(J=j) \mid X])}{E_{0}\{\operatorname{var}(A \mid X)\}}$$

\citet{vansteelandt2020assumption} discussed adaptive estimation of a partially linear model for continuous and binary outcomes. However, as the survival time is subject to censoring, the structural nested model requires additional semiparametric assumptions on the restricted survival time \citep{hagiwara2020g} and hence is not pursued in this paper.